\pdfoutput=1
\documentclass[aps,prd,preprint, onecolumn,superscriptaddress,preprintnumbers,floatfix,nofootinbib,longbibliography,eqsecnum]{revtex4-2}
\usepackage[colorlinks,allcolors=blue]{hyperref}
\usepackage[german,english]{babel}
\usepackage{amsmath,amssymb}
\usepackage{bm}
\usepackage{xcolor}
\usepackage{bbm}
\usepackage{bbold}
\usepackage{ulem}
\usepackage{braket}
\usepackage{tablefootnote}
\usepackage{slashed}
\usepackage{multirow}
\usepackage{diagbox}
\usepackage[mathlines]{lineno}
\usepackage[utf8]{inputenc}
\usepackage{comment}
\usepackage{graphicx}
\usepackage{scalerel}

\newcommand{\subsubsubsection}[1]{\paragraph{#1}\mbox{}\\}
     \def\dag{\dagger}

\def\L{\Lambda_{\mathrm {QCD}}}
\def\als{\alpha_s}

\newcommand{\RN}[1]{%
  \textup{\uppercase\expandafter{\romannumeral#1}}%
}

\newcommand{\beq}{\begin{equation}}
\newcommand{\eeq}{\end{equation}}

\newcommand{\be}{\begin{equation}}
\newcommand{\ee}{\end{equation}}
\newcommand{\bea}{\begin{eqnarray}}
\newcommand{\eea}{\end{eqnarray}}
\newcommand{\nn}{\nonumber}

\makeatletter
\let\LN@align\align
\let\LN@endalign\endalign
\renewcommand{\align}{\linenomath\LN@align}
\renewcommand{\endalign}{\LN@endalign\endlinenomath}
\let\LN@gather\gather
\let\LN@endgather\endgather
\renewcommand{\gather}{\linenomath\LN@gather}
\renewcommand{\endgather}{\LN@endgather\endlinenomath}
\makeatother
\allowdisplaybreaks
\begin{document}

\preprint{TUM-EFT 185/23}

\title{One Born--Oppenheimer Effective Theory to rule them all: hybrids, tetraquarks, pentaquarks, doubly heavy baryons and quarkonium}

\author{Matthias Berwein}
\affiliation{Technical University of Munich,\\
TUM School of Natural Sciences, Physics Department,\\   
James-Franck-Str.~1, 85748 Garching, Germany.}

\author{Nora Brambilla}
\affiliation{Technical University of Munich,\\
TUM School of Natural Sciences, Physics Department,\\   
James-Franck-Str.~1, 85748 Garching, Germany.}
\affiliation{Technical University of Munich, Institute for Advanced Study, \\ 
Lichtenbergstrasse 2 a, 85748 Garching, Germany.}
\affiliation{Technical University of Munich, Munich Data Science Institute, \\ 
Walther-von-Dyck-Strasse 10, 85748 Garching, Germany.}

\author{Abhishek Mohapatra}
\email{abhishek.mohapatra@tum.de}
\affiliation{Technical University of Munich,\\
TUM School of Natural Sciences, Physics Department,\\   
James-Franck-Str.~1, 85748 Garching, Germany.}

\author{Antonio Vairo}
\affiliation{Technical University of Munich,\\
TUM School of Natural Sciences, Physics Department,\\   
James-Franck-Str.~1, 85748 Garching, Germany.}

\begin{abstract}
The discovery of XYZ exotic states in the hadronic sector with two heavy quarks, represents a significant challenge in particle theory. Understanding and predicting their nature remains an open problem. In this work, we demonstrate how the Born--Oppenheimer (BO) effective field theory (BOEFT), derived from Quantum Chromodynamics (QCD) on the basis of  scale separation and symmetries, can address XYZ exotics of any composition. We derive the Schrödinger coupled equations that describe hybrids, tetraquarks, pentaquarks, doubly heavy baryons, and quarkonia at leading order, incorporating nonadiabatic terms, and present the predicted multiplets.
We define the static potentials in terms of the QCD static energies for all relevant cases. We provide the precise form of the nonperturbative low-energy gauge-invariant correlators required for the BOEFT: static energies, generalized Wilson loops, gluelumps, and adjoint mesons. These are to be calculated on the lattice and we calculate here their short-distance behavior.
Furthermore, we outline how spin-dependent corrections and mixing terms can be incorporated using matching computations. Lastly, we discuss how static energies with the 
same BO quantum numbers  mix at large distances
leading  to the phenomenon of avoided level crossing. This effect is crucial to understand the emergence of exotics with molecular characteristics, such as the $\chi_{c1}(3872)$. With BOEFT  both the tetraquark and the molecular picture 
appear as part of the same description.
 \end{abstract}
 

\maketitle
\clearpage

\tableofcontents 

\section{Introduction}
The XYZs are exotic states found in the last twenty years in the sector of the hadron spectrum with two heavy quarks
at or above the strong decay threshold to heavy-light meson pairs.
They are manifestly  exotic due to electric charge and isospin quantum numbers or they display
indirectly other exotic characteristics in the masses, decay widths and production patterns.
They  may involve tetraquark,  pentaquark, and  hybrid compositions that go beyond 
what has been observed in the standard quark model  of
mesons  (quark-antiquark pairs)  and  baryons  (combinations of  three
quarks)~\cite{GellMann:1964nj,Zweig:1964CERN,Jaffe:1975fd}.

In the  last two decades, since  the discovery of the 
$\chi_{c1}(3872)$ (also known as $X(3872)$) by Belle \cite{Belle:2003nnu}, dozens of new XYZ states have
been observed  and confirmed by different  experimental groups:  B-factories (BaBar,
Belle, Belle2 and  CLEO), $\tau$-charm  facilities (CLEO-c, BES, BESIII),  and also
proton-(anti)proton  colliders  (CDF,  D0,   LHCb,  ATLAS,  CMS)  (see
Refs.~\cite{Guo:2017jvc,          Ali:2017jda,          Olsen:2017bmm,
	Brambilla:2019esw,  Liu:2019zoy,  Chen:2022asf}  for  reviews).  The
charged states  observed within  the charm and  bottom sector  such as
$Z_c$,  $Z_b$,  and  $T_{cc}\left(3875\right)^+$  are  obvious candidates  for
tetraquarks \cite{BESIII:2013ris, Belle:2013yex, Belle:2011aa,Belle:2013urd,BESIII:2015cld, Belle:2007hrb, Belle:2014nuw, BESIII:2020qkh, LHCb:2021uow, LHCb:2023hxg, LHCb:2021vvq}, while  isospin
$1/2$ baryons such as  $P_{c\bar{c}}(4380)^+$, $P_{c\bar{c}}(4312)^+$, and $P_{c\bar{c}}(4440)^+$,
discovered by LHCb in the  charm sector,  are  candidates for  pentaquarks containing a charm-anticharm  pair  and   three  light  quarks  \cite{LHCb:2015yax, LHCb:2019kea}.

The existence of these exotic hadrons presents a unique opportunity to study the strong force in nature in ways that were previously unexplored.
The implications of such studies extend beyond fundamental particle theory and have the potential to impact various other fields dealing with strongly correlated systems.

However, due to  the nonperturbative  regime of  QCD, it  is difficult  to make
first-principle predictions of spectrum,  widths and cross sections of
such   multiquark   states.   Various  models   have   been   proposed
\cite{Brambilla:2022ura,Brambilla:2019esw,Brambilla:2014jmp,Brambilla:2010cs,Guo:2017jvc,Karliner:2017qhf,Ali:2017jda,Bodwin:2013nua,Maiani:2020pur,Ali:2019roi,Godfrey:2008nc,Lebed:2016hpi,Mezzadri:2022loq,Gross:2022hyw,Olsen:2017bmm}.
A priori,  the simplest system consisting  of only two quarks  and two
antiquarks (tetraquarks) is  already a very complicated  object and it
is unclear whether  any kind of clustering occurs in  it.  To simplify
the problem, models focus on certain substructures, investigating their
implications.       This       includes       hadronic       molecules
\cite{Tornqvist:2004qy,Close:2003sg}, composed of color-singlet mesons
bound together  by residual nuclear forces, tetraquarks, bound states
between        a        diquark       and        an        antidiquark
\cite{Jaffe:1976ig,Maiani:2004vq,Jaffe:2003sg},  and hadro-quarkonium,
a cloud of light  quarks and gluons bound to a  heavy $Q \bar{Q}$ core
singlet  state  via  van  der  Waals  forces  \cite{Dubynskiy:2008mq}.
Additionally, threshold effects such as threshold
cusps  \cite{Bugg:2011jr} and rescattering processes  \cite{Pakhlov:2014qva}  have been suggested as possible explanations for the
observed enhancements in the XYZ particles.
While models played a pioneering role, they are based on a somewhat ad
hoc  choice of  dominant configurations  and interaction  Hamiltonian.
Although inspired  by QCD, they  are not derived  from QCD nor  can they be
systematically improved.  An exception is the molecular description of
states very  close to  the strong decay  threshold like  the $X(3872)$
where,  due to the small  binding  energy  or large  scattering  length,
universal characteristics  emerge and an effective  field theory (EFT)
description can be introduced  to systematically calculate corrections
\cite{Guo:2017jvc,Braaten:2007dw,Braaten:2004rn,Fleming:2011xa,Fleming:2007rp,Canham:2009zq,Baru:2021ldu},
thanks to universal properties. 
Also for lattice QCD  an ab
initio   calculation   of   the  XYZ   spectra   remains   challenging
\cite{Prelovsek:2022eie,Brambilla:2021mpo},  as  it requires  studying
the   coupled   channel  scattering   of   hadrons   on  the   lattice
\cite{Luscher:1990ux,Briceno:2017max}.  Pioneering calculations are found in
\cite{Prelovsek:2013cra,Prelovsek:2020eiw,Prelovsek:2014swa,Cheung:2017tnt,CLQCD:2019npr,Ryan:2020iog,Padmanath:2015era,Colquhoun:2022dte,Leskovec:2019ioa,Junnarkar:2018twb,Alexandrou:2023cqg,Meinel:2022lzo,Padmanath:2022cvl}.

In this paper, we show how a QCD derived effective field theory called
Born--Oppenheimer EFT (BOEFT) can address all these states in a unified framework,
without making any assumption on their configurations. The BOEFT does need lattice input
on some static energies and some generalized Wilson loops but, thanks to factorization,
these are only few universal (i.e.\ non flavor-dependent) nonperturbative correlators,
which hugely simplifies  the problem. The BOEFT is derived from QCD on the basis of symmetries and
scale separation and all the relevant potentials and correlators are obtained in
the (nonperturbative) matching procedure. At the large scale of the  heavy quark mass, perturbative calculations
and resummation are possible, while at the low energy scale, nonperturbative calculations of the matching
coefficients are necessary. Independently of this, the BOEFT is supplying the
full structure of the  Schr\"odinger coupled equations, the corrections in
$1/m_Q$, where $m_Q$ is the heavy quark mass, and the mixing terms. 
The structure of the coupled equations is largely determined by symmetry, while the dynamics are set by the perturbative or nonperturbative matching coefficients.

The idea to use a {\it Born--Oppenheimer picture} \cite{Born-Oppenheimer, Landau:1991wop} to describe states with two heavy quarks has been put forward some time ago, especially in relation to hybrids
\cite{Griffiths:1983ah,  Juge:1999ie,   Juge:2002br, Braaten:2013boa, Braaten:2013boa,  Braaten:2014qka, Braaten:2014ita, Braaten:2015pqm,  Meyer:2015eta}. It also has been underlying  the construction of strongly coupled potential nonrelativistic QCD (pNRQCD)    \cite{Brambilla:2000gk,Brambilla:2004jw,Brambilla:2004jw,Pineda:2000sz}. It has been cast for the first time in the form of a proper effective field theory  for the
description of hybrids in   \cite{Berwein:2015vca} and it has been used to calculate the hybrid multiplets ~\cite{Braaten:2014qka,Oncala:2017hop,Berwein:2015vca}.
Hybrid spin separations \cite{Oncala:2017hop,Brambilla:2018pyn,Brambilla:2019jfi,Soto:2023lbh}
and some  decays of hybrids to quarkonium  \cite{Oncala:2017hop,Brambilla:2022hhi,TarrusCastella:2021pld}  have been calculated.
General applications of the BOEFT have been put forward in \cite{Brambilla:2017uyf,Soto:2020xpm}, the doubly heavy baryon case has been
studied in   \cite{Soto:2020pfa,Soto:2021cgk,Brambilla:2005yk}
and the effect of the interaction with the open-flavor threshold has been addressed in 
\cite{TarrusCastella:2022rxb,Bruschini:2023zkb,Bruschini:2023tmm, TarrusCastella:2024zps, Braaten:2024stn}.

Lattice calculations of some Born--Oppenheimer (BO) static energies have been addressed 
for hybrids in a comprehensive way  \cite{Juge:1999ie,Juge:2002br,Bali:2000vr,Bali:2003jq,Capitani:2018rox,Schlosser:2021wnr, Bicudo:2021tsc, Sharifian:2023idc, Hollwieser:2023bud}, while for tetraquarks they are still
sparse \cite{Brown:2012tm, Bicudo:2015kna, Prelovsek:2019ywc,Sadl:2021bme,Mueller:2023wzd}. As we will discuss,  the BOEFT identifies the best operators to be used to calculate the BO static energies.

In this paper, based on  the treatment  in~\cite{Berwein:2015vca},
our primary objective  is to expand the BOEFT framework
to analyze  tetraquark and  pentaquark states  consisting of  either a
heavy quark-antiquark pair or two  heavy quarks and light degrees of freedom (LDF) with arbitrary
quantum numbers,  including isospin.  
A  fundamental aspect  of  understanding  the spectrum  of tetraquarks  and  pentaquarks  is  the application  of  the  adiabatic expansion principle, which distinguishes  the dynamics of heavy quarks
from  that  of  the  LDF.  Within this  context,  the  tetraquark  and
pentaquark  are  bound states  of  heavy  quarks  in the  spectrum  of
BO-potentials (referred to as static energies or adiabatic surfaces at
leading-order  in the $1/m_Q$  expansion) associated to the LDF.
We present here  for the first time the explicit form of the Schr\"odinger coupled equations
in a unified form for hybrids, tetraquarks, pentaquarks, quarkonia and doubly heavy baryons.
We construct the corresponding multiplets, explain general selection rules,
give the explicit operators for the lattice calculation of all the potentials,
calculate their perturbative expression and the 
short distance multipole expansion.
Besides gluelump masses for hybrids, we show that such multipole expansions feature
adjoint mesons for tetraquarks. 

This paper  is organized  as follows. 
In  Sec.~\ref{sec:effective}, we describe the physical picture underlying  quarkonium and the XYZ exotic states, 
and the regime in which the soft scale is perturbative. 
In  Sec.~\ref{sec:BO}, we introduce the NRQCD static energies and  most general Born--Oppenheimer effective field theory description (BOEFT), and in Sec.~\ref{sec_mixing}, we obtain  the coupled  Schr\"odinger equations 
and  the  general expression of the mixing matrices for each type of exotic state. 
We derive the corresponding multiplets.
In Sec.~\ref{sec:characterization},  we list  the interpolating  operators suitable
to obtain  the exotic static energies  and discuss their short and  long   distance  behavior. 
We discuss the properties of these operators with respect to what is currently used in lattice calculations.
In Sec.~\ref{sec:LD}, we discuss the mixing at large distances, the overlap of the  tetraquark 
interpolating operators with static heavy-light states and the possible emergence of molecular states.
In  Sec.~\ref{sec:spin}, we comment on the size of the heavy-quark spin dependent effects
and their phenomenological impact and we summarize how decays and transitions are calculated in the BOEFT.
We also comment on multichannel BO equations.

Section~\ref{sec:phen} contains some phenomenological identifications of the exotics while
Sec.~\ref{sec:conclusion} gives conclusions and an outlook. 
Seven appendices complement this work, presenting the details of some calculations, 
in particular about gauge-invariance checks, explicit forms of the projection operators, mixing matrices and details about the construction of pentaquarks. 
Appendices~\ref{app:Overlap} and~\ref{app:Schreodinger} present the overlap of our interpolating operators with heavy-light and with quarkonium plus pion states and the explicit coupled Schr\"odinger equations to be used in  phenomenological applications.

\section{Scales, physical picture and  short distance regime} \label{sec:effective}

Quarkonium ($Q\bar{Q}$), hybrids ($Q\bar{Q}g$), tetraquarks ($Q\bar{Q} q \bar{q}$, $QQ \bar{q} \bar{q} $), pentaquarks  ($Q \bar{Q} qqq$, $QQqq\bar{q}$) and doubly heavy baryons
($Q\bar{Q}q$), all involve a heavy quark $Q$ and a heavy antiquark $\bar{Q}$ or two heavy quarks  bound together with  some gluonic $g$ or light (anti)quark $q$ ($\bar{q}$) degrees of freedom,\footnote{An exception is a system like 
	the $X(6900)$ composed by four heavy quarks. We will comment about this in the conclusions.} which from now on will be abbreviated with LDF (light degrees of freedom).  
The presence of a large scale, which is the mass $m_Q$ of the heavy quark, allows some simplification in the problem.  
The large scale,  on one hand,  is perturbative ($m_Q \gg \L$, $\L$ being the typical hadronic scale), on the other, it ensures that the heavy quark is moving nonrelativistically.

However, as a matter of fact, even the treatment of quarkonium  is challenging, due to the presence of several energy scales.
Quarkonium  is  a nonrelativistic bound system  with a   small  relative velocity $v$ between
the quark and the antiquark. In the rest frame of the meson: 
$v^2 \sim 0.1$ for $b\bar{b}$, $v^2 \sim 0.3$ for $c\bar{c}$ systems. Besides the heavy-quark mass (hard scale),
a hierarchy of dynamically generated energy scales is induced: the typical relative momentum $p \sim m_Q v$, corresponding to the inverse Bohr radius $r \sim 1/(m_Q v)$ (soft scale), and the  typical binding  energy $E \sim m_Q v^2$ (ultrasoft scale).  
This is similar to what happens for positronium in QED, but  for QCD there is also the complication of the nonperturbative low energy behaviour.
Apart from the hard scale of the mass of the quark,  the treatment of the other scales depends on their relation to $\L$, as getting closer to it implies that nonperturbative methods have to be used.
The problem posed by the existence of the different entangled  energy scales characterizing the nonrelativistic bound state in QCD  has been  addressed by substituting QCD with simpler but equivalent nonrelativistic effective field theories (NREFTs) \cite{Brambilla:2004jw}.
A hierarchy of  NREFTs may be constructed by systematically integrating out modes associated with high-energy scales. 
Such integration is made in a well defined  matching procedure that 
enforces the equivalence between QCD and the NREFT at any given order of accuracy. 
The NREFT Lagrangian is factorized in matching coefficients,
encoding the high energy degrees of freedom,  and low energy operators that remain dynamical.
It displays a power counting in  the small ratios of high energy over low energy scales,  allowing 
to assign a definite size to the contribution of each operator to physical observables. 
At leading order in the NREFT, some underlying symmetries are exposed, allowing 
model independent predictions.
By integrating out modes associated with the scale of the heavy quark  mass, 
nonrelativistic QCD (NRQCD)~\cite{Bodwin:1994jh,Thacker:1990bm, Lepage:1992tx} is obtained. 
It displays still  entangled soft and ultrasoft degrees of freedom. 
The matching to NRQCD can be done in perturbation theory, i.e. within an expansion in $\als$. 
The NRQCD matching coefficients encode the nonanalytic dependence on the scale $m_Q$.

In order to obtain the Schr\"odinger equation as a zeroth order problem and the potentials from QCD, 
it is necessary to integrate out modes associated to the relative momentum $p$. 
In the case in which $p \sim 1/r \gg \L$, this integration can be done within perturbation theory arriving 
at a lower energy EFT called  weakly coupled  potential nonrelativistic QCD (pNRQCD)~\cite{Pineda:1997bj,Brambilla:1999xf,Brambilla:2004jw}. 
The potentials  are matching coefficients calculated with a standard  matching in perturbation theory, comparing and expanding in $\als$  diagrams in NRQCD and in pNRQCD.
The potentials acquire a proper, field theoretical definition: they are  matching  coefficients   encoding  the soft scale contributions, undergo  renormalization, develop scale dependence and satisfy renormalization group equations, which allow a resummation of potentially large logarithms. 

Weakly coupled pNRQCD has facilitated the calculation of  the physical properties, like spectra and decays, of several quarkonium systems with small radius and in particular the calculation of the quarkonium static potential and the  static energy at next-to-next-to-next-to leading logarithmic (NNNLL) order \cite{Brambilla:2006wp, Brambilla:2009bi},
which enabled precise extractions of $\als$ from the QCD static energy \cite{Bazavov:2019qoo}.
What is important here is that weakly coupled pNRQCD can be used also to describe the properties of any XYZ system at a distance scale smaller than $\L^{-1}$ and this will be exploited in this paper, especially in Sec.~\ref{subsec:potential}.

However, for soft scales comparable to  $\L$ a full nonperturbative approach should be used to arrive from NRQCD to the lower energy EFT featuring the Schr\"odinger equation as the zeroth order problem and to obtain the interaction potentials.

\section{The Born--Oppenheimer effective field theory description}\label{sec:BO}

\subsection{The BO picture}
When  the typical  quarkonium radius  is larger, $p\sim 1/r \ge \Lambda_{\mathrm {QCD}}\gg E$, the soft scale of the binding  and the  matching have to be treated nonperturbatively.
The same should be done for exotic systems for which the LDF are part of the binding at the soft scale.
Still, a low energy  EFT description can be constructed, based on the presence of two heavy quarks with a large mass scale  $m_Q \gg \L$, an underlying symmetry and  a residual scale  separation $\L\gg E$ between the energy and momentum of the LDF and the energy of the heavy quarks. 
On one hand, this is similar to the case of heavy-light mesons and baryons where the Heavy Quark Effective Field Theory (HQET) can be constructed based on the factorization of the heavy $\sim m_Q$ and the light $\sim \L$  degrees of freedom. 
On the other hand, the situation is more intricated here due to the fact that there is another scale hierarchy, $\L\gg E$.

The  scale separation $\L\gg E$ between the energy and momentum of the LDF and the energy of the heavy quarks is suggestive of the Born--Oppenheimer description of diatomic molecules, whose electrons (fast degrees of freedom) adjust adiabatically to the motion of the heavier nuclei (slow degrees of freedom) \cite{Born-Oppenheimer, Landau:1991wop}.
The Born--Oppenheimer description exploits the fact that the masses of the nuclei are much larger than the electron masses and, consequently, the time scales for the dynamics of the two types of particles are very different. 
It entails no restriction on the strength of the coupling between the slow and the fast degrees of freedom.
Concretely, the BO approximation  gives a method to obtain the molecular energy levels by solving the Schr\"odinger equation for the nuclei with a potential  given 
by the electronic static energies at fixed nucleus positions, where the static energies are labeled by the molecular quantum numbers corresponding to the symmetries of the diatomic molecules. 

\begin{figure}[!tbp]
	\begin{minipage}[t]{0.48\textwidth}
		\includegraphics[width=1\textwidth]{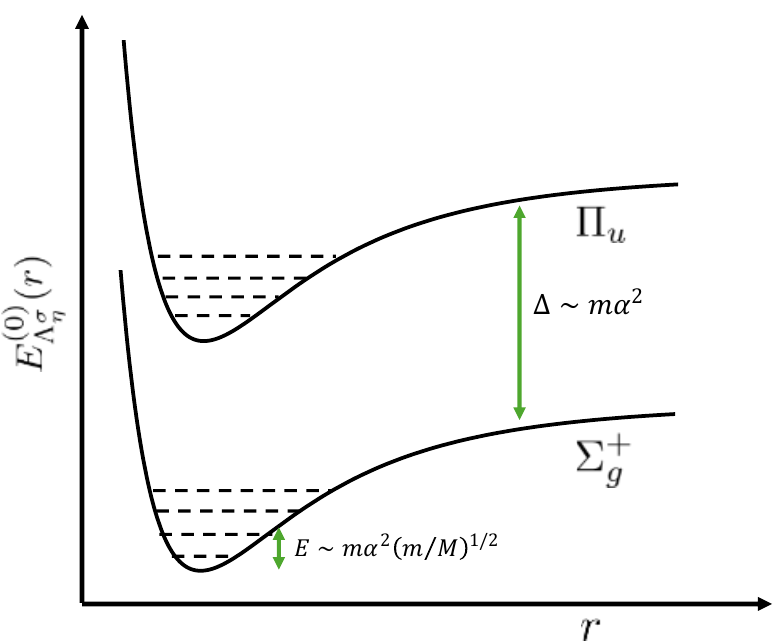}
		\caption{Born--Oppenheimer picture in QED: Sketch of energy levels (dashed lines) in the electronic static energies (BO-potentials) for molecular systems. 
			The energy levels solution of the Schr\"odinger equation are at the energy scale $m\alpha^2(m/M)^{1/2}$.} \label{fig:qed}
	\end{minipage}
	\hfill
	\begin{minipage}[t]{0.48\textwidth}
		\includegraphics[width=1\textwidth]{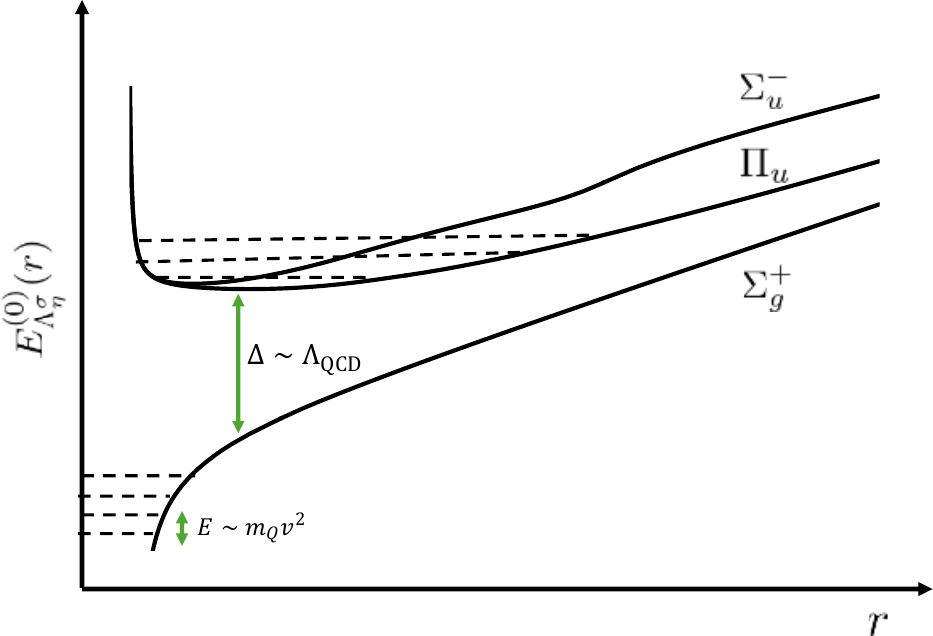}
		\caption{Born--Oppenheimer picture in QCD: Sketch of energy levels (dashed lines) in the LDF static energies (BO-potentials). 
			The energy levels solution of the Schr\"odinger equation are at the energy scale $m_Qv^2$.}\label{fig:qcd}
	\end{minipage}
\end{figure}

In the QED case, different BO-potentials (static energies) are separated by a gap of order $m\alpha^2$ while the energy levels in these BO-potentials are separated by a much smaller gap of order $m\alpha^2(m/M)^{1/2}$ \cite{Brambilla:2017uyf}, where $m$ is the electron mass, $M$ is the nucleus mass in atomic or molecular systems and $\alpha$ is the fine structure constant, see Fig.~\ref{fig:qed}.
In the same way, nonperturbative static energies can be defined in NRQCD and calculated on the lattice. The BOEFT then allows to obtain the actual form of  coupled Schr\"odinger equations.  
Solving such equations gives the quarkonium and the XYZ  energy levels and multiplets, see Fig.~\ref{fig:qcd}.
In Figs.~\ref{fig:qed} and \ref{fig:qcd}, the dashed lines refer to the energy levels obtained by solving the Schr\"odinger equation with the potential corresponding to the given static energy.  
In the QCD case, the typical separation of such energy levels is of order $E \sim m_Qv^2$.

\subsection{The (NR)QCD static energies}
\label{sec:NRQCDstatic}
Since $m_Q \gg \L$, an appropriate starting point is NRQCD.  
The matching to the lower EFT can be performed order by order in the $1/m_Q$ expansion.
NRQCD is obtained from QCD by integrating out the hard modes associated to the heavy quark mass.  
This amounts to expanding in inverse powers of the mass and including the nonanalytic dependence on the quark mass in the NRQCD matching coefficients.
All degrees of freedom at the hard scale, including light quarks and gluons, are systematically accounted for in the matching coefficients.
The NRQCD Hamiltonian  in the heavy-quark--heavy-antiquark sector of the Fock space reads
\begin{align}
	H &=  H^{(0)}+\frac{1}{m_Q}H^{(1,0)}+\frac{1}{m_{\bar{Q}}}H^{(0,1)} + \dots\,, \label{HH}\\
	H^{(0)} &= \int d^3x\, \frac{1}{2}\left( \bm{E}^a\cdot\bm{E}^a +\bm{B}^a\cdot\bm{B}^a \right)-\sum_{j=1}^{n_f} \int d^3x\, \bar{q}_j \, (i \bm{D}\cdot \bm{\gamma} -m_j)\, q_j \,,\label{H0}\\
	H^{(1,0)} &= - \frac{1}{2} \int d^3x\, \psi^\dagger \left( \bm{D}^2 + g c_F\, \bm{\sigma} \cdot \bm{B}\right) \psi\,, \label{H10}\\
	H^{(0,1)} &=\frac{1}{2}\int d^3x\, \chi^\dagger \left(\bm{D}^2+ g c_F\, \bm{\sigma} \cdot \bm{B} \right) \chi\,,\label{H01}\\
	\dots \quad & \quad \dots \nonumber 
\end{align}
where $m_Q$ and $m_{\bar{Q}}$ are the  heavy quark and heavy antiquark pole masses respectively and we have shown only terms up to first order in the heavy quark mass expansion. 
Terms of first order in the mass contain the kinetic energy.
The NRQCD Hamiltonian  in the heavy-quark--heavy-quark sector of the Fock space is given by Eqs.~\eqref{H0} and~\eqref{H10} without the heavy antiquark terms~\eqref{H01}.
For simplification, we refer from now on to the equal heavy quark mass case only.  
Equations and results may be extended straightforwardly to the case of two different heavy quark masses, i.e. states containing a bottom and a charm quark.
The field $\psi$ is the Pauli spinor  that annihilates the heavy quark, the field
$\chi$ is the Pauli spinor  that creates the heavy antiquark; 
they satisfy  canonical equal time anticommutation relations.
The fields $q_j$ are $n_f$ Dirac spinor fields that annihilate a light quark of flavor~$j$ and mass $m_j$, $iD^0=i\partial_0-gA^0$, $i\bm{D}=i\bm{\nabla}+g\bm{A}$, and  the chromoelectric and  chromomagnetic fields are defined as the components ${\bm E}^i=G^{i0}$ and ${\bm B}^i=-\epsilon_{ijk}G^{jk}/2$ ($\epsilon_{123}=1$) of the field-strength tensor $G^{\mu\nu}$. The matching coefficient $c_F$ is known at 3 loops and it  
is equal to one at tree level. 
The physical states are constrained by the Gauss law:
\begin{equation}
	(\bm{D}\cdot \bm{E})^a \vert {\rm phys} \rangle =
	g \left(\psi^\dagger T^a \psi + \chi^\dagger T^a \chi + \sum_{j=1}^{n_f} 
	\bar{q}_j \gamma^0 T^a q_j\right) \vert {\rm phys} \rangle.
	\label{gausslaw}
\end{equation}

We restrict ourselves to the one-quark--one-antiquark (one-quark) sector of the NRQCD Fock space, where quarkonium and XYZ states live. 
In this sector, we denote an energy eigenstate of the NRQCD Hamiltonian by $| \underline{\rm n}; \bm{x}_1, \bm{x}_2 \rangle$,
where $n$ represents a generic set of conserved quantum numbers, and $\bm{x}_1$ and $\bm{x}_2$ are the positions of the quark and antiquark, respectively.
We normalize the states as
$\langle \underline{\rm n}; \bm{x}_1, \bm{x}_2 | \underline{\rm m}; \bm{x}_1',\bm{x}_2' \rangle = \delta_{nm} \delta^{(3)} (\bm{x}_1- \bm{x}_1') \delta^{(3)} (\bm{x}_2- \bm{x}_2')$. 
The heavy quark and antiquark positions are conserved quantum numbers in the static limit
$m_Q,\,m_{\bar{Q}} \to \infty$, where we have also 
\begin{equation}
	H = H^{(0)}\,,
	\label{eq:H-static}
\end{equation}
which still contains the kinetic terms associated with gluons and light quarks.
The eigenvalue equation is 
\begin{equation}
	H^{(0)} | \underline{\rm n}; \bm{x}_1, \bm{x}_2 \rangle^{(0)} = E^{(0)}_n (\bm{x}_1, \bm{x}_2) 
	| \underline{\rm n}; \bm{x}_1, \bm{x}_2 \rangle^{(0)} \,.
\end{equation}
In the static limit, the $Q\bar{Q}$ or $QQ$  sector of the Fock space is spanned  by 
\begin{align}
	\vert \underline{n}; \bm{x}_1 ,\bm{x}_2 \rangle^{(0)} & = \psi^{\dagger}(\bm{x}_1) \chi (\bm{x}_2)
	|n;\bm{x}_1 ,\bm{x}_2\rangle^{(0)},&& \forall \bm{x}_1,\bm{x}_2\,,\label{basis0}\\
	\vert \underline{n}; \bm{x}_1 ,\bm{x}_2 \rangle^{(0)} & = \psi^{\dagger}(\bm{x}_1) \psi^{\dagger}(\bm{x}_2)
	|n;\bm{x}_1 ,\bm{x}_2\rangle^{(0)},&& \forall \bm{x}_1,\bm{x}_2\,,\label{basis1}
\end{align}
where $|\underline{n}; \bm{x}_1 ,\bm{x}_2\rangle^{(0)} $ is a gauge-invariant eigenstate of $H^{(0)}$ (defined up to a phase and satisfying the Gauss law) with eigenenergy $E_{n}^{(0)}(\bm{x}_1 ,\bm{x}_2)$ that transforms like $3_{\bm{x}_1}\otimes 3_{\bm{x}_2}^{\ast}$ or $3_{\bm{x}_1}\otimes 3_{\bm{x}_2}$  under color $SU(3)$. 
The  state $|n;\bm{x}_1 ,\bm{x}_2\rangle^{(0)}$  encodes the purely LDF content of the state, and it is annihilated by the heavy quark fields for any $\bm{x}$;
its normalization is $^{(0)}\langle {n}; \bm{x}_1 ,\bm{x}_2|{m}; \bm{x}_1 ,\bm{x}_2\rangle^{(0)} =\delta_{nm}$.
Since the static Hamiltonian $ H^{(0)}$ does not contain any heavy fermion field, the state $|n;\bm{x}_1,\bm{x}_2\rangle^{(0)}$ is also an eigenstate of $H^{(0)}$ with energy $E_{n}^{(0)}(\bm{x}_1 ,\bm{x}_2)$. 
Translational invariance requires that $E_{n}^{(0)}(\bm{x}_1 ,\bm{x}_2)=E_{n}^{(0)}(r)$ with $\bm{r}=\bm{x}_1-\bm{x}_2$.
In the following, we often use the center of mass $\bm{R}=\left(\bm{x}_1+\bm{x}_2\right)/2$ and the relative coordinate $\bm{r}=\bm{x}_1-\bm{x}_2$ in place of the individual coordinates 
of the two heavy quarks, and we mostly work in the center of mass frame $\bm{R}=0$ unless explicitly specified otherwise. 

Since the static states $\left|\underline{n};\,\bm{x}_1,\,\bm{x}_2\right\rangle^{(0)}$ form a complete basis, any state  $|X_n \rangle$
can be written as an expansion:  
\begin{equation}
	|X_n\rangle=
	c_n\,\left|\underline{n};\,\bm{x}_1,\,\bm{x}_2\right\rangle^{(0)}
	+\dots\,,
	\label{x}
\end{equation} 
where $\left|\underline{n};\,\bm{x}_1,\,\bm{x}_2\right\rangle^{(0)}$ 
is the state with the lowest energy eigenvalue that overlaps with $|X_n\rangle$
and the 
dots denote excited states also overlapping with $|X_n \rangle$.
Using the time evolution of the states,
\begin{equation}
	^{(0)}\left\langle \underline{n};\,\bm{x}_1,\,\bm{x}_2,\,T/2\right|\left.\underline{n};\,\bm{x}_1,\,\bm{x}_2,\,-T/2\right\rangle^{(0)}=\mathcal{N}\,\exp\left[-iE_n^{(0)}(r)\,T\right]\,,
	\label{timecorr}
\end{equation}
where the normalization constant ${\cal N}$ does not depend on time, it follows that
\begin{equation}
	E_n^{(0)}(r)=\lim_{T\to\infty}\frac{i}{T}\log\langle X_n,\,T/2|X_n,\,-T/2\rangle\,.
	\label{eelwl}
\end{equation}
So 
$|X_n\rangle$ 
is not required to be an eigenstate of the static Hamiltonian,
it just needs to have non-vanishing overlap with the 
eigenstate of energy $E_n^{(0)}$,
i.e. $c_n\neq0$, and this should be the eigenstate with lowest energy overlapping with $|X_n\rangle$.
This is the technique used to extract the static energies $E_n^{(0)}(r)$ in lattice QCD and it reduces to the calculation of generalized static Wilson loops. In Sec.~\ref{sec:characterization}, we present a convenient choice for the interpolating gauge invariant operators  to be used on the lattice to obtain the static energies for  hybrids, tetraquarks, pentaquarks, and doubly heavy baryons.
The ground state, $n=0$, is the Fock state made of a heavy quark-antiquark pair
without LDF,\footnote{Notice that such a state does not exist for $QQ$, since $QQ$ is not a color singlet.}
other values of $n$ identify heavy quark-antiquark or  heavy quark-quark pairs in the presence of such excitations.

\begin{figure}[t!]
	\begin{minipage}[b]{0.45\textwidth}
		\includegraphics[width=1.1\textwidth]{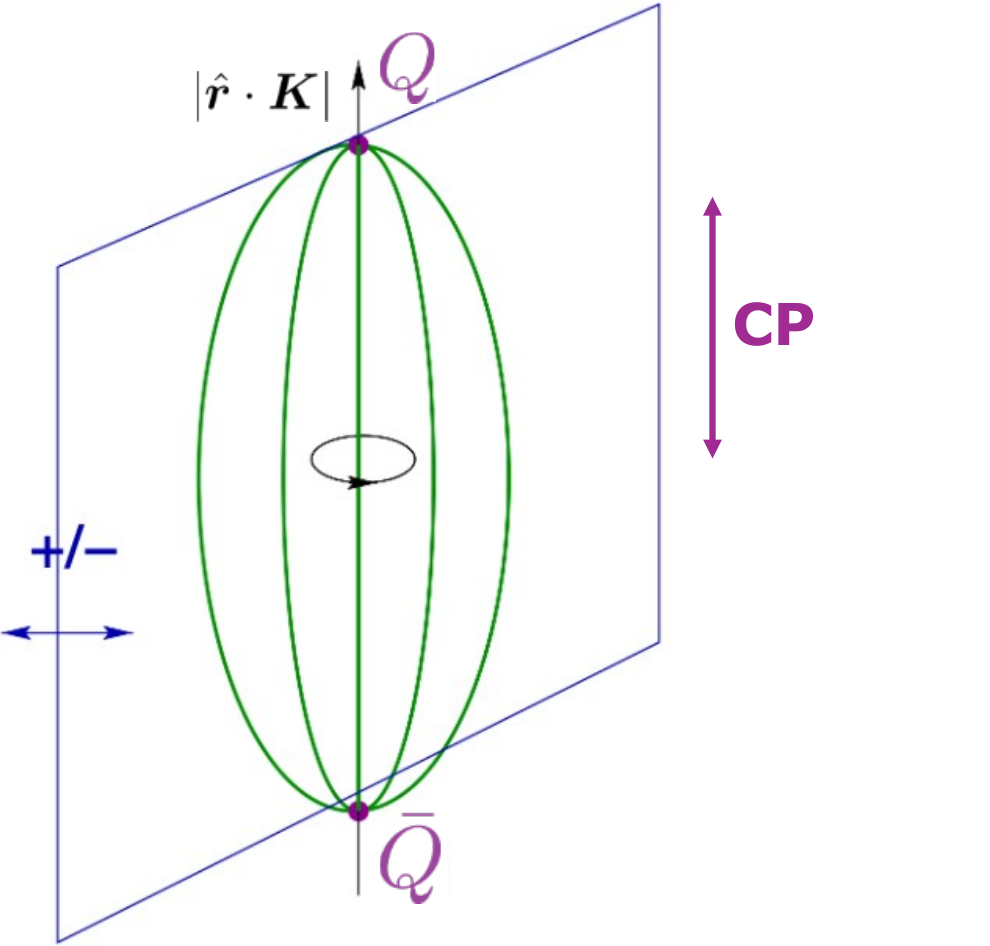}
		\caption{Symmetries of a system made by a heavy quark, a heavy antiquark and nonperturbative constitutents at the scale $\L$.} \label{sym1}
	\end{minipage}
	\hfill
	\begin{minipage}[b]{0.45\textwidth}
		\hspace*{0mm}
		\vspace*{0.02in} 
		\includegraphics[width=1.1\textwidth]{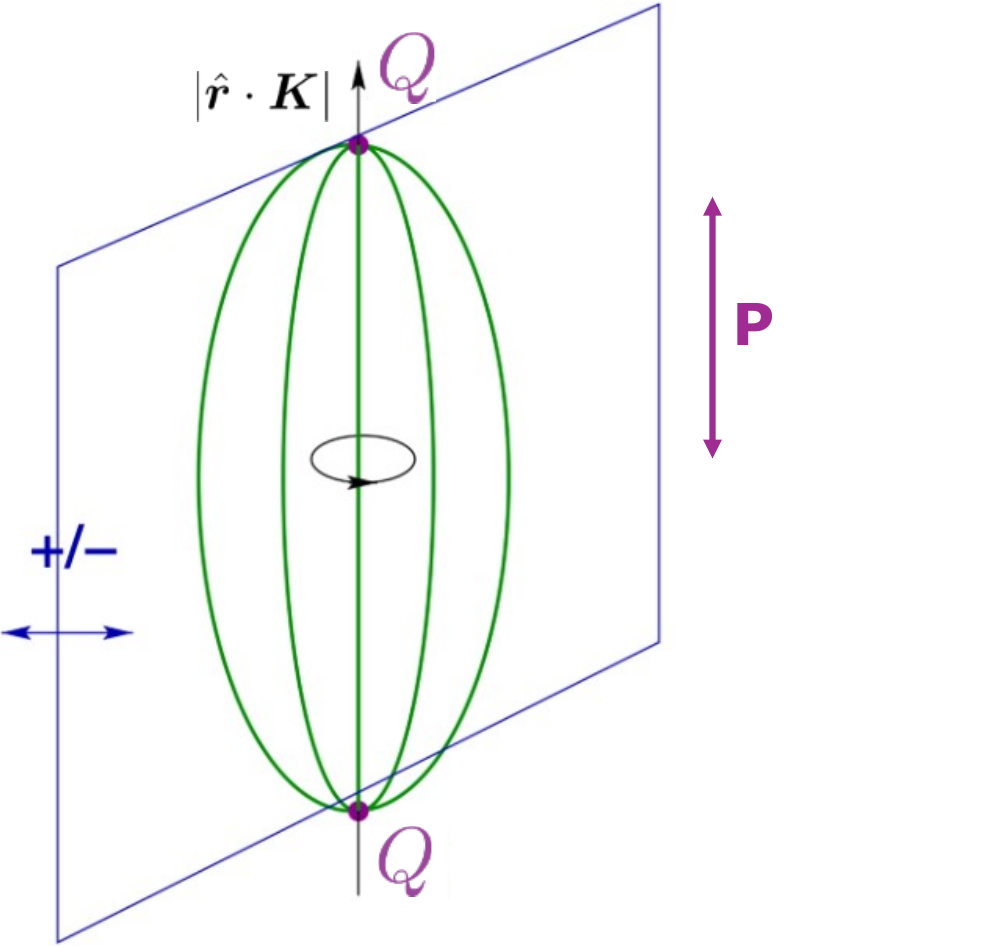}
		\caption{Symmetries of a system made by two heavy quarks and nonperturbative constituents at the scale $\L$.}\label{sym2}
	\end{minipage}
	\vspace*{-0.1in} 
\end{figure}

The set of conserved numbers $n$ is related to the symmetries of the $Q\bar{Q}$-LDF system, see Fig.~\ref{sym1}, and of the $QQ$-LDF system, see Fig.~\ref{sym2}. 
The heavy quarks are static and supply a color source, while the different configurations of the LDF identify different static energies, on the one hand similarly to what happens in HQET, and on the other hand similarly to what happens in a diatomic molecule.
The states are characterized by the flavor composition of the LDF, given in terms of isospin quantum numbers $I$, $m_I$ and baryon number $b$, and the projections of the total angular momentum ${\bm K}$ of the LDF along the $\hat{r}$-axis joining the two heavy quarks.\footnote{In the context of light quarks, we also refer to the angular momentum ${\bm K}$ as the spin.}  
These are labeled by the representations of the cylindrical symmetry group $D_{\infty h}$ with 
{\it BO quantum numbers} $\Lambda_\eta^\sigma$.
The projection of the angular momentum $\bm{K}\cdot\hat{\bm{r}}$ has eigenvalues $\lambda$, and $\Lambda \equiv |\lambda|$.
We denote integer values of $\Lambda$ by capital Greek letters: $\Sigma$, $\Pi$, $\Delta,...$ for $\Lambda=0,1,2,...$. The index $\eta$ is the $CP$ eigenvalue in case of $Q\bar{Q}$, and just parity, $P$, in case of $QQ$ and is denoted by $g = + 1$ and $u = - 1$. 
Finally, the index $\sigma=\pm 1$ is the eigenvalue of the reflection operator  with respect to a plane passing through the $\hat{\bm{r}}$ axis. 
The index $\sigma$ is explicitly written only for $\Sigma$ states, which are not degenerate with respect to the reflection symmetry.
The ground state with given BO quantum numbers is labeled $\Lambda_\eta^\sigma$;
excited states with the same quantum numbers are labeled 
$\Lambda_\eta^{\sigma\prime}$, $\Lambda_\eta^{\sigma\prime\prime}$, \ldots.

\begin{figure}[t]
	\includegraphics[width=0.45\textwidth]{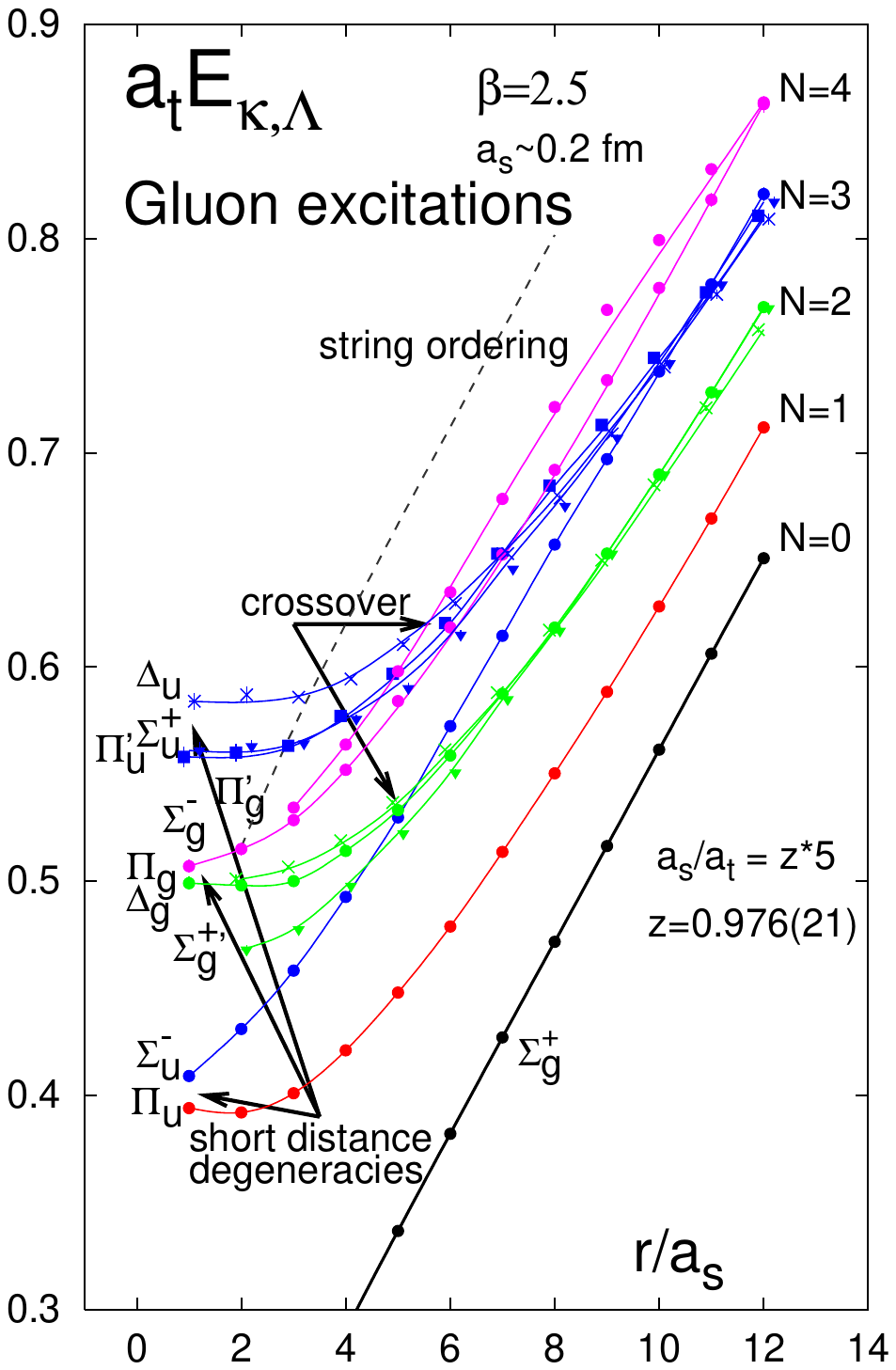}
	\caption{Tower of hybrid static energies in the $I=0$ sector with only gluonic excitations as LDF (quenched approximation) from \cite{Juge:2002br}. Static energies with the same value of $\kappa$ are degenerate at short distance $r$ between the static quark and antiquark.} \label{morning}
\end{figure}

Assuming that the angular momentum operator $\bm{K}^2$ has 
eigenvalues $k(k+1)$, which restricts the projection to $\Lambda\leq k$, and introducing the shorthand notation
\begin{equation}
	\kappa\equiv\{k^{P[C]}, f\}\,,
	\label{label-n}
\end{equation}
where $P[C]$ is defined to be $PC$ in case of $Q\bar{Q}$ and just $P$ in case of $QQ$, and $f$ denotes the flavor indices of the LDF,\footnote{For notational ease, we will explicitly mention whenever we suppress the flavor index $f$.} 
we can indicate  the static energies  $E^{\Lambda_\eta^\sigma}$ as 
$E^{\left(0\right)}_{\kappa, |\lambda|}\left({\bm x}_1, {\bm x}_2\right) \equiv  E^{\left(0\right)}_{\kappa, |\lambda|}\left(r\right)$.  Based on the quantum number $\kappa$, we have the  static energy $E^{\left(0\right)}_{\kappa, |\lambda|}\left(r\right)$ for any state with at least two heavy quarks: quarkonium $\left(Q\bar{Q}\right)$, hybrids $\left(Q\bar{Q}g\right)$, doubly heavy baryons $\left(QQq\right)$, tetraquarks $\left(Q\bar{Q}q\bar{q},\,QQ\bar{q}\bar{q}\right)$, and pentaquarks $\left(Q\bar{Q}qqq,\,QQqq\bar{q}\right)$. It should be kept in mind, however, that $k$ is a good quantum number at short distances, where the symmetry of the system is $O\left(3\right)\otimes C$, 
but not at large distances, where the symmetry of the system is $D_{\infty h}$.
This labeling applies to any possible XYZ state.
Note that the sign of $\lambda=\pm\Lambda$ does not affect the static energies, since they are degenerate in it, 
but for several expressions appearing in the following sections it makes sense to keep it as a label of the states.

In the static limit, the BO quantum numbers are exactly conserved. 
The static energies are nonperturbative quantities at the scale $\L$ 
and must be calculated on the lattice or in QCD vacuum models  \cite{Isgur:1984bm,Brambilla:1999ja,Brambilla:1996aq,Baker:1996mk,Soto:2021cgk,Brambilla:2014eaa,Perez-Nadal:2008wtr}.
In the $I=0$ sector, for the quenched case (pure $SU(3)$ gauge theory)
and with only gluonic excitations in the LDF, several calculations of the tower of energy levels exist \cite{Juge:1999ie,Juge:2002br,Bali:2000vr,Bali:2003jq,Capitani:2018rox,Schlosser:2021wnr, Bicudo:2021tsc, Sharifian:2023idc, Hollwieser:2023bud}  starting from the one reported in Fig.~\ref{morning}.
In the short distance limit,  the static energies $E_{\kappa, |\lambda|}^{(0)}(r)$ with the same $\kappa$ become degenerate and the cylindrical symmetry group $D_{\infty h}$ is expanded to the spherical symmetry group $O\left(3\right)\otimes C$.
This fact is understood and predicted in weakly coupled pNRQCD \cite{Brambilla:1999xf},
which can be applied to short distances, see Sec.~\ref{sec:characterization} for a detailed discussion. 

\subsection{Born Oppenheimer EFT (BOEFT)}
The static Hamiltonian does not contain the kinetic energy. 
In order to obtain from NRQCD the BOEFT, i.e. an effective field theory that has the Schr\"odinger equation as zeroth order equation of motion, we need to go beyond the static limit and consider the correction in $1/m_Q$ to the NRQCD Hamiltonian. 
For our purposes, it is sufficient to consider the relative kinetic energy and to work in the center of mass frame.\footnote{We do not consider the spin-dependent term in the NRQCD Hamiltonian $H^{(1,0)}$ or $H^{(0,1)}$, nor we do consider $r$-dependent potentials generated at order $1/m_Q$ on the BOEFT side, 
	see Sec.~\ref{sec:spin} for a discussion of these effects.}

To match BOEFT to NRQCD, we take advantage of the nonperturbative quantum matching developed in Ref.\ \cite{Brambilla:2000gk,Pineda:2000sz,Brambilla:2002nu,Brambilla:2020xod,Berwein:2015vca,Soto:2020xpm,Soto:2020pfa}, which is performed perturbing order by order in $1/m_Q$  around the static limit.
The $1/m_Q$ expansion provides a convenient way to organize the matching calculation. 
The relative size of the different contributions in the BOEFT Lagrangian is, however, not always trivially dictated by the $1/m_Q$ expansion, but by the power counting of the EFT.
The BOEFT  has only ultrasoft dynamical degrees of freedom at the scale $E$ and therefore all the physical degrees of freedom  with energies larger than $E$ should be integrated out.

An ideal situation in 
the BO picture is when the difference in energy among states 
associated with BO potentials labeled with different quantum numbers $n$ is much larger than the difference in energy among states obtained from solving the Schr\"odinger equation of a given BO potential.
The gap between different static energies  
is 
a consequence of the nonperturbative dynamics, 
therefore it can be evaluated only after having obtained the static energies from a lattice calculation.
The ideal situation is, however, realized only by quarkonium states.
Higher excitations in the quarkonium spectrum show various degeneration patterns.
At short distance,  
the BO energies sharing the same $\kappa$   
become degenerate,
hence, eigenstates of these energies mix and satisfy coupled Schr\"odinger equations.
At large distance, 
static energies with the same BO quantum numbers
may become close at some distance $r$;
if they also develop non zero transition amplitudes then the states mix.
Diagonalizing the transition matrix leads to the 
avoided level crossing phenomenon illustrated in Appendix~\ref{app_crossing}. 
This phenomenon has been observed on the lattice in the calculation of the static energy of quarkonium and $I=0$ tetraquarks~\cite{Bali:2005fu, Bulava:2019iut, Bulava:2024jpj}.\footnote{The operators used on the lattice up to now are heavy-light states and not tetraquarks, we will comment on this in Secs.~\ref{sec:characterization}, \ref{sec:LD} and in Appendix~\ref{app:Overlap}} 
A similar effect has been seen in the hybrid static energies~\cite{Alasiri:2024nue,Juge:2002br}.
Hence, also in the long distance regime, coupled Schr\"odinger equations 
are responsible for the low energy dynamics of the states, 
as we discuss in Sec.~\ref{sec:LD}.

To see the how degeneracies affect the effective field theories, 
we consider the NRQCD Hamiltonian $H$ at some order of the $1/m_Q$ expansion (for the purposes of this paper, first order is sufficient).
Then $H$ is not diagonal in the basis $|\underbar{n}\rangle^{(0)}$ of $H^{(0)}$.  
We consider instead a basis of states,  
\be
|\underbar{n}; \bm{x}_1 ,\bm{x}_2\rangle ,
\label{basis1-full}
\ee 
normalized as 
\be
\langle \underbar{m}; \bm{x}_1 ,\bm{x}_2| \underbar{n}; {\bf y}_1,{\bf y}_2\rangle =
\delta_{mn}\delta^{(3)} (\bm{x}_1 -{\bf y}_1)\delta^{(3)} (\bm{x}_2 -{\bf y}_2) \,,
\label{cond2}
\ee
such that the Hamiltonian $H$ is diagonal with respect to these states 
\be
\langle \underbar{m}; \bm{x}_1 ,\bm{x}_2|H| \underbar{n}; {\bf y}_1
,{\bf y}_2\rangle =\delta_{mn}E_{n}(\bm{x}_1 ,\bm{x}_2,\bm{p}_1,\bm{p}_2)
\delta^{(3)} (\bm{x}_1 -{\bf y}_1)\delta^{(3)} (\bm{x}_2 -{\bf y}_2) \,,
\label{cond1}
\ee
where $E_{n}(\bm{x}_1 ,\bm{x}_2,\bm{p}_1,\bm{p}_2)$ is an analytic function in $\bm{p}_1$, $\bm{p}_2$.
Conditions (\ref{cond1}) and (\ref{cond2}) give 
\be
H| \underbar{n}; {\bf y}_1 ,{\bf y}_2\rangle
= \int d^3x_1d^3x_2| \underbar{n}; \bm{x}_1 ,\bm{x}_2\rangle 
E_{n}(\bm{x}_1 ,\bm{x}_2,\bm{p}_1,\bm{p}_2)
\delta^{(3)} (\bm{x}_1 -{\bf y}_1)\delta^{(3)} (\bm{x}_2 -{\bf y}_2) \,.
\label{cond1p}
\ee
The positions $\bm{x}_1$ and $\bm{x}_2$  are no longer good quantum numbers since the position operator does not commute with $H$ beyond the static limit. 
This is reflected in the dependence of $E_n$ on the momenta $\bm{p}_1$ and $\bm{p}_2$, which are represented by derivatives, making $E_n$ a differential operator. 
The diagonalization of $H$ in Eq.~\eqref{cond1} is meant in this operator sense.
A set of states $|\underbar{n}\rangle$ and an operator $E_{n}$ that satisfy 
Eqs.~\eqref{cond2} and~\eqref{cond1p}
can be obtained from the static solutions $|n\rangle^{(0)}$ and $E_{n}^{(0)}$ to any desired
order of accuracy by working out formulas analogous to the ones used in standard quantum mechanical  
perturbation theory. 
We can write $|\underbar{n}\rangle$ and $E_{n}$ as an expansion in $1/m_Q$ 
(again we assume for simplicity that quark and antiquark have the same mass $m_Q$):
\be
|\underbar{n} \rangle = |\underbar{n} \rangle^{(0)} 
+ \frac{1}{m_Q}|\underbar{n}\rangle^{(1)} 
+ \dots\,, 
\label{fullstate}
\ee
\be
E_n =E_n^{(0)} + \frac{1}{m_Q}E_n^{(1)}
+ \dots\,.
\label{fullenergy}
\ee

States and matrix elements in NRQCD are matched onto states and matrix elements in the BOEFT, 
which is the effective field theory that follows from NRQCD by integrating out modes of order $\L$ and 
describes excitations associated to a given (set of) static energies.
The matching is nonperturbative (no expansion in $\als$ involved).
The state $|\underbar{n}; \bm{x}_1 ,\bm{x}_2\rangle$ in NRQCD is matched to the state
$\Psi_n^{\dagger}(\bm{x}_1 ,\bm{x}_2) |\mathrm{vac}\rangle$ in the BOEFT,  
\be 
|\underbar{n}; \bm{x}_1 ,\bm{x}_2\rangle \to \Psi_n^{\dagger}(\bm{x}_1 ,\bm{x}_2) |\mathrm{vac}\rangle,
\ee
where $\Psi_n(\bm{x}_1 ,\bm{x}_2)$ is a color singlet composed field that annihilates 
all excitations associated with the static energy $E_n$ and $|\mathrm{vac}\rangle$ is the vacuum state of the effective field theory.
The field satisfies 
the canonical equal time commutation relation: $[\Psi_n(\bm{x}_1',\bm{x}_2'),\Psi_m^\dagger(\bm{x}_1,\bm{x}_2)] = \delta_{nm} \delta(\bm{x}_1'-\bm{x}_1) \delta(\bm{x}_2'-\bm{x}_2)$.
The static energy $E_n$ of NRQCD is matched to the BOEFT Hamiltonian, $h_n$, 
\be 
E_{n}(\bm{x}_1 ,\bm{x}_2,\bm{p}_1,\bm{p}_2) = h_n(\bm{x}_1 ,\bm{x}_2,\bm{p}_1,\bm{p}_2) \,.
\ee

In the absence of mixing with excitations from different $n$, 
the BOEFT Lagrangian describing ultrasoft excitations of $h_n$ is then 
\begin{equation}
	L_\text{EFT}^{(n)}
	=\int d^3x_1\int d^3x_2\, \Psi_n^\dagger(\bm{x}_1,\bm{x}_2,t)[i\partial_t-h_n(\bm{x}_1,\bm{x_2},\bm{p}_1,\bm{p}_2)]\Psi_n(\bm{x}_1,\bm{x}_2,t)\,.
\end{equation}
The equation of motion is in this case 
an uncoupled Schr\"odinger equation.

In the case in which the static energy $E_n$ becomes close or degenerate
with some other static energy, we have, for instance, that 
\begin{equation}
	| \underline{\rm m}; \bm{x}_1, \bm{x}_2 \rangle^{(1)} = - \sum_{n \neq m} \int d^3x_1' d^3x_2'\, | \underline{\rm n} ; \bm{x}_1', \bm{x}_2' \rangle^{(0)} 
	\frac{{}^{(0)} \langle \underline{\rm n}; \bm{x}_1', \bm{x}_2' | H^{(1)} | \underline{\rm m}; \bm{x}_1, \bm{x}_2 \rangle^{(0)}}{E_n^{(0)} (\bm{x}_1', \bm{x}_2') - E_m^{(0)} (\bm{x}_1, \bm{x}_2)} 
	\label{FirstOrderUndegenerate}
\end{equation}
becomes large and possibly divergent.
This signals the break down of non degenerate perturbation theory and the necessity to
resolve to degenerate perturbation theory, 
in which case 
$h_n$ is replaced by a non-diagonal matrix computed on
all degenerate states.
The solution of the ensuing coupled Schr\"odinger equations provides eigenstates, eigenvalues 
and eventually diagonallizes the Hamiltonian.
The full eigenstates at leading order,
$\ket{N}^{(0)}$,  
are no longer a single unperturbed state but a superposition of the (approximately) degenerate states and the coefficients of this superposition 
are eigenfunctions of the BOEFT Hamiltonian 
in the degenerate sector:
\begin{equation}
	\ket{N}^{(0)}
	=\sum_{n\in\{N\}}\int d^3x'_1\int d^3x'_2\ket{\underline{n};\bm{x}'_1,\bm{x}'_2}^{(0)}\phi_n^{(N)}(\bm{x}'_1,\bm{x}'_2)\,,
	\label{FirstOrderDegenerate}
\end{equation}
where the sum runs over 
$\{N\}$, which identifies the set of states belonging to the degenerate multiplet associated with the energy labeled $N$.
These states are labeled by new quantum numbers $N$, which no longer contain labels related to degenerate states, whereas those unrelated to the degeneracy are still good quantum numbers and are included in $N$ without change; see Sec.~\ref{sec_mixing} for details.
Diagonalizing the first order Hamiltonian then corresponds to finding the sets of eigenfunctions $\phi_n^{(N)}$ satisfying
\begin{equation}
	\sum_{n\in\{N\}}\left(\delta_{mn}E_n^{(0)}+\frac{1}{m_Q}E_{mn}^{(1)}\right)\phi_n^{(N)}=\mathcal{E}_N^{(1)}\,\phi_{m}^{(N)}\,,
	\label{degpert}
\end{equation}
where the matrix elements are given through 
($H^{(1)} \equiv H^{(1,0)} + H^{(0,1)}$ in the equal mass case)
\begin{equation}
	{}^{(0)}\bra{\underline{m};\bm{x}_1,\bm{x}_2}H^{(1)}\ket{\underline{n};\bm{y}_1,\bm{y}_2}^{(0)}=E_{mn}^{(1)}(\bm{x}_1,\bm{x}_2,\bm{p}_1,\bm{p}_2)\delta^{3}(\bm{x}_1-\bm{y}_1)\delta^{(3)}(\bm{x}_2-\bm{y}_2)\,.
\end{equation}
The eigenvalue $\mathcal{E}_N^{(1)}$ finally gives the mass of the XYZ state as $m=2m_Q+\mathcal{E}_N^{(1)}+\mathcal{O}\left(m_Q^{-2}\right)$.

\subsubsection{BOEFT for quarkonium: strongly coupled pNRQCD}
\begin{figure}[t]
	\centerline{\includegraphics[width=.5\textwidth]{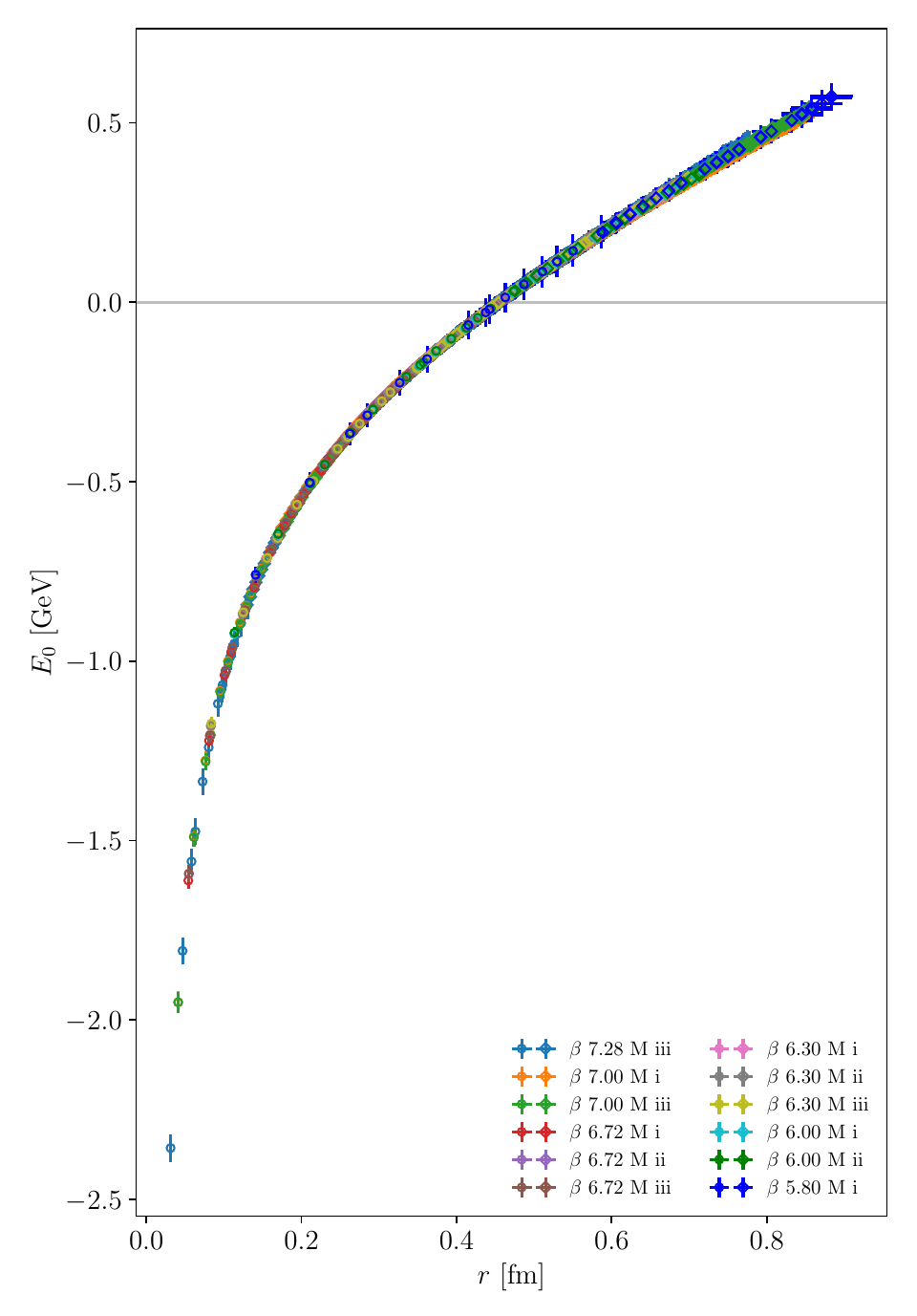}}
	\caption{Results for the lattice QCD calculation of the quarkonium static energy performed with a QCD action with  $2+1+1$  dynamical light flavors \cite{Brambilla:2022het}.
		The lattice data is obtained from twelve ensembles of varying lattice spacing (denoted by $\beta$) and three choices of light quark masses. 
		The overall normalization has been fixed by setting $E_{0}(r_{0})=0$, 
		where $r_0 \approx 0.5$~fm is the so-called Sommer scale.  
	}
	\label{fig:Sigma_gpl}
\end{figure}

The most well-known QCD static energy is the one with $n=0$ corresponding to quarkonium: 
it identifies the ground state static energy between a heavy quark and a heavy antiquark in a color singlet configuration, which is well approximated by a Cornell-like potential.  
This static energy 
can be extracted from the large time behaviour of 
a static Wilson loop~\cite{Wilson:1974sk} 
and it has been measured with more and more accuracy 
over the last 50 years.
It represents 
an 
evident proof of confinement 
in the heavy quark sector 
since it shows a linear rising behavior with the distance $r$. 
We 
show 
in Fig.~\ref{fig:Sigma_gpl} 
its 
most recent and state-of-the-art calculation  performed with a QCD action with $2+1+1$  dynamical light flavors  \cite{Brambilla:2022het}. 
This static energy has been identified with the $Q\bar{Q} $ potential $V(r)$; 
a
proper theoretical justification 
in the framework of nonrelativistic effective field theories of QCD
is
provided  
by 
strongly coupled pNRQCD~\cite{Brambilla:2000gk,Pineda:2000sz}.\footnote{In the short distance,  the static energy $E^{(0)}_0(r)$ and the static potential $V(r)$ differ by ultrasoft contributions, the energy being a physical quantity and the potential being cutoff at the ultrasoft scale, 
	which is compensated in the static energy by the ultrasoft degrees of freedom 
	that remain dynamical in the short distance regime~\cite{Brambilla:1999qa}.} 

\begin{figure}[!tbp]
	\includegraphics[width=0.5\textwidth]{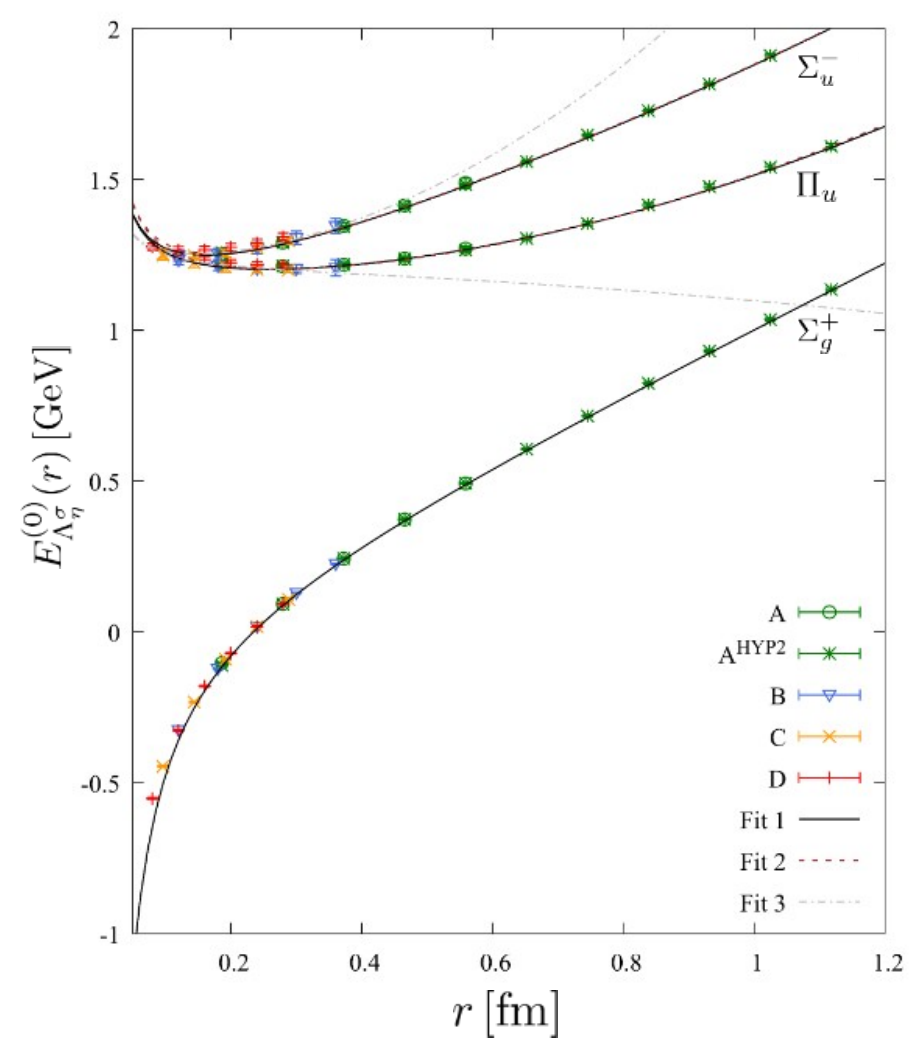}
	\caption{The lowest hybrid static energies $\Pi_u$ and $\Sigma_u^-$ and the quarkonium static energy $\Sigma_g^+$, shown as a function of the quark-antiquark separation $r$ in fm from short to large distances~\cite{Schlosser:2021wnr}. 
		The gap between quarkonium and hybrid static energies is of order $\L$.} \label{fig:hyb-Wagner}
\end{figure}

Strongly coupled pNRQCD is  
the BOEFT restricted to  
quarkonium. 
In this case, there is a gap of order $\L$ between the static energy with $n=0$ (the $\Sigma_g^+$ BO-state) and the static energies with $n \neq 0$, see e.g. Fig.~\ref{morning} and, for more details on the lowest lying states, Fig.~\ref{fig:hyb-Wagner}.
Because of the assumed energy gap of order $\L$ between the ground state and the higher excitations of the heavy quark-antiquark pair, these 
are
integrated out when matching 
NRQCD 
to strongly coupled pNRQCD. 
Moreover, since 
the $\Sigma_g^+$ static energy is not mixing with any other static energy at short distance,
the NRQCD ground state $|\underline{\rm 0}; \bm{x}_1, \bm{x}_2 \rangle$ simply matches 
the pNRQCD state made of one color singlet heavy quark-antiquark pair, 
created by the field $S^\dagger(\bm{x}_1, \bm{x}_2)$ acting on the pNRQCD vacuum $|{\rm vac} \rangle$,
\begin{equation}
	| \underline{0}; \bm{x}_1, \bm{x}_2 \rangle \to S^\dagger(\bm{x}_1, \bm{x}_2)|{\rm vac} \rangle.
\end{equation}
The 
Hamiltonian
for strongly coupled pNRQCD 
is 
\be
E_0(\bm{x}_1,\bm{x}_2, 
\bm{p}_1, \bm{p}_2) = h_s(\bm{x}_1,\bm{x}_2, \bm{p}_1, \bm{p}_2) = E_0^{(0)}(\bm{x}_1, \bm{x}_2) + {\bm{p}_1^2 \over 2 m_Q} +      { \bm{p}_2^2\over 2 m_Q}\,,
\label{eq:E0-QQbar}
\ee
which 
leads to 
the strongly coupled pNRQCD Lagrangian 
for the singlet composite field $S$
\begin{equation}
	L_{\mathrm {pNRQCD}}=\int d^3{R}\int d^3{r} \, \mathrm{Tr}\Bigg\{{S}^{\dagger}(\bm{r},\,\bm{R},\,t) \left(i\partial_t\,+ \frac{\bm{\nabla}^2}{m_Q}
	-V^{(0)}(r)\right) {S}(\bm{r},\,\bm{R},\,t)\Bigg\}, 
	\label{eq:LQQuar}
\end{equation}
where 
we have dropped the center of mass kinetic energy as we are working in the center of mass frame of the quarkonium, $\bm{r} = \bm{x}_1-\bm{x}_2$,  $\bm{R} = (\bm{x}_1+\bm{x}_2)/2$, and 
$V^{(0)}(r)=  E_0^{(0)}(r)$
is the quarkonium singlet static potential,\footnote{For quarkonium $k^{PC}= 0^{++}$.} 
which can be 
measured by lattice QCD and 
is 
well approximated by a Cornell type potential. 
The color singlet field $S$ is proportional to the identity in color space;
it satisfies the canonical equal time commutation relation, $[S(\bm{r}',\bm{R}'),S^\dagger(\bm{r},\bm{R})] = \delta^{(3)}(\bm{r}'-\bm{r}) \delta^{(3)}(\bm{R}'-\bm{R})$.
The 
trace is over color and spin.
The contributions from the LDF of energy and momentum of order $\L$ 
are encoded in the  potential between the two heavy quarks, which is a matching coefficient of pNRQCD.
Higher order relativistic corrections to the $Q\bar{Q}$ singlet potential have been calculated using the same framework in~\cite{Brambilla:2000gk,Pineda:2000sz} and are given in terms of generalized static Wilson loops with chromoelectric and chromomagnetic insertions~\cite{Brambilla:2004jw}. 
They have been calculated on the lattice mainly in the quenched approximation and in QCD vacuum models.
The quarkonium spectrum is the spectrum of the Hamiltonian $h_s$.

Strongly coupled pNRQCD 
provides a field theoretical justification 
of why potential models have been successful in describing quarkonium, and it relates the potentials directly to 
QCD. 
As all the dependence on the heavy flavor has been factorized,
this  enables 
predictions in different heavy flavor sectors based on the same universal potentials.
The imaginary part of the potential contains information on the inclusive decays and has been calculated in the same framework  \cite{Brambilla:2002nu}.

Light 
quarks
are taken into account in three ways.
(1) Light quarks at the hard scale $m_Q$ are integrated out in 
NRQCD, 
they contribute to the 
matching coefficients 
of NRQCD 
and appear 
as such 
in the relativistic corrections to the potential. 
The 
renormalization scale dependence of the NRQCD matching coefficients cancels against the soft part of the pNRQCD potential. 
(2) Light quarks at the soft scale enter the lattice calculation of the potential with the unquenched QCD action, see Fig.~\ref{fig:Sigma_gpl}. 
(3) Ultrasoft light fermions 
become pions.
They enter as ultrasoft degrees of freedom and could be explicitly added to the Lagrangian in Eq.~\eqref{eq:LQQuar}, 
see e.g. \cite{Brambilla:2015rqa}, 
but we do not consider them here.
For quarkonium, there are no light 
quark and gluon excitations contributing to the quantum numbers of the static energy, i.e. what we have called LDF.

The calculation of the quarkonium potential from the static Wilson loop in full QCD does not show conclusive evidence of {\it string breaking} \cite{Bali:2000vr}.
In the BOEFT, string breaking comes from the fact that the quarkonium static energy  and the first tetraquark ($I=0$) static energy,\footnote{The $I=0$ tetraquark static energy evolves at large distance in a heavy-light static meson-antimeson pair, see Sec.~\ref{sec:LD}.}
which have the same BO quantum numbers, become  close at a distance of about 1.2~fm called the string breaking length. 
Because the transition amplitudes are different from zero at the string breaking length, 
this leads to a non diagonal matrix of static energies, whose eigenvalues show the avoided level crossing behavior (see Appendix~\ref{app_crossing}).
The avoided level crossing phenomenon is realized in QCD, 
see Fig.~\ref{fig:string-breaking}.\footnote{Fig.~\ref{fig:string-breaking} contains three static energies because there also the strange quark is considered.}
Moreover, avoided level crossing 
modifies the 
Schr\"odinger equation for quarkonium from one channel [as one would obtain from Eq.~\eqref{eq:LQQuar}] 
to  the coupled equations given in Eq.~\eqref{eq:VQQbar-threshold}.

\subsubsection{BOEFT for hybrids, tetraquarks, pentaquarks and doubly heavy baryons}
Here we obtain the general BOEFT Lagrangian for the case of hybrids, tetraquarks. pentaquarks and heavy baryons.  
when 
the difference among energies labeled with different $\kappa$ is much larger than the difference in energy among states labeled within the same $\kappa$. 
The LDF responsible for the bound state are characterized by the energy scale $\L$, 
while the energy separations of the exotic states are of order $E\ll  \L$.
Since, at least parametrically, there is no mixing between states separated by a gap of order $\L$,  
we can restrict to a BOEFT of static energies with the same $\kappa$.
%
We call this the single-channel approximation.
Differently from the quarkonium case, the BOEFT for states with the same $\kappa$ leads to  coupled  Schr\"odinger equations 
and the coupling is induced by the kinetic energy term contained in $H_1$, see Eq.~\eqref{degpert}, and it is related 
to the degeneration appearing in some of the BO static energies with the same $\kappa$ and different $\lambda$ 
at short distance.
Hence, we work considering BO nonadiabatic couplings,  i.e.\ non diagonal terms in the kinetic energy. 
At the end of this section, as well as in Sec.~\ref{sec:spin}, 
we comment about when a multichannel BO description, involving static energies with different $\kappa$ becomes important. 

The matching condition for 
the states reads 
\begin{equation}
	| \underline{\kappa, \lambda}; \bm{x}_1, \bm{x}_2 \rangle \to {\Psi}^{\dagger}_{\kappa\lambda}(\bm{x}_1, \bm{x}_2)|{\rm vac} \rangle,
\end{equation}
where 
$|{\rm vac} \rangle$ is the BOEFT vacuum.
As the static energies above the quarkonium ground state are approximately degenerate in the short-distance limit with respect to the quantum number $\lambda$, 
we need to calculate the matrix elements of $H^{(1)}$ between those states, 
in particular the 
relative kinetic energy,
which is dominant in the center of mass 
frame.
To this aim we need to express 
\begin{equation}
	{\Psi}_{\kappa\lambda}
	= \sum_{\alpha} P^{\alpha\dagger}_{\kappa\lambda}\,\Psi^\alpha_\kappa\,,
	\quad \quad \Psi^\alpha_\kappa = \sum_{\lambda} P^{\alpha}_{\kappa\lambda}\,{\Psi}_{\kappa\lambda}\,,
\end{equation}
where the projection vectors $P^{\alpha}_{\kappa\lambda}$ project the eigenstates of $\bm{K}\cdot\bm{\hat{r}}$ with eigenvalue $\lambda$ generated by $\Psi_{\kappa\lambda}^\dagger$ onto the eigenstates of $K_z$ with eigenvalue $\alpha$ generated by $\Psi_\kappa^{\alpha\,\dagger}$.
We list these projectors for all the cases in Appendix~\ref{app:projectors}. 
In the $\Psi_\kappa^\alpha$ basis, the LDF operators do not depend on the $\bm{\hat{r}}$ axis. Therefore, the $\bm{\nabla}_r^2$ part of $H^{(1)}$ is diagonal in this basis, whereas in the $\Psi_{\kappa\lambda}$ basis one needs to take into account the fact that the derivatives also act on the projection vectors.

Keeping at order $1/m_Q$ only the relative kinetic energy, 
the BOEFT Lagrangian describing the heavy exotic states (hybrids, tetraquarks, pentaquarks  and doubly heavy baryons) 
with quantum number $\kappa$
can be written as
\begin{align}
	L_{\mathrm {BOEFT}}=&\int d^3{\bm R}\int d^3{\bm r} \, \sum_{\lambda\lambda^{\prime}}\mathrm{Tr}\Bigg\{{\Psi}^{\dagger}_{\kappa\lambda}(\bm{r},\,\bm{R},\,t) \bigg[i\partial_t\,\delta_{\lambda\lambda'}-
	V_{\kappa, \lambda\lambda^\prime }(r)   \nonumber\\
	&\hspace{4 cm}+ \sum_\alpha P^{\alpha\dag}_{\kappa\lambda}\left(\theta, \varphi\right)\frac{\bm{\nabla}^2_r}{m_Q}P^{\alpha}_{\kappa\lambda^{\prime}}\left(\theta, \varphi\right)\bigg]{\Psi}_{\kappa\lambda^{\prime}}(\bm{r},\,\bm{R},\,t)\Bigg\}, 
	\label{eq:LQQ}
\end{align}
where the trace is over color, spin and isospin indices.
Here the potential $V_{\kappa\lambda\lambda^{\prime}}$ in Eq.~\eqref{eq:LQQ} is 
\begin{align}
	&V_{\kappa, \lambda\lambda^{\prime}}(r) 
	=V^{(0)}_{\kappa, \vert\lambda \vert}(r)\delta_{\lambda\lambda^{\prime}}+{\cal O}\left(1/m_Q\right)\,,
	\label{eq:VQQ}
\end{align}
where $V^{(0)}_{\kappa, \vert\lambda \vert}(r)$
is equal to  the static energy $E^{(0)}_{\kappa, \vert\lambda \vert}(r)$  and can be calculated on the lattice or in QCD vacuum models. See Sec.~\ref{sec:characterization} for a concrete definition in terms of Wilson loops with appropriate interpolating operators.
As we discuss in Sec.~\ref{sec:spin},  the potential $V_{\kappa, \lambda\lambda^{\prime}}$ in Eq.~\eqref{eq:LQQ} can be organized as an expansion in $1/m_Q$ and subleading corrections can be calculated.
The $1/m_Q$ term in NRQCD  generates a spin-dependent and a spin-independent correction in $1/m_Q$ in the potential.
The fact that, for the exotics, the spin potential already starts at order $1/m_Q$, differently from quarkonium, is due to the interaction of the spin of the heavy quark with the spin $\bm{K} \neq 0$ of the LDF, which is not suppressed by a mass term~\cite{Oncala:2017hop,Brambilla:2018pyn,Brambilla:2019jfi}.
This feature has a prominent impact on the spectrum.

As we will see in Sec.~\ref{subsec:potential}, the potential $V^{\left(0\right)}_{\kappa, \vert\lambda\vert}$ can be  calculated at short distances in weakly coupled pNRQCD.
The  kinetic energy is not commuting with the projection vectors $P^\alpha_{\kappa\lambda}$, which generates coupled Schr\"odinger equations and the 
$\Lambda$ doubling effect seen in molecular physics, breaking the degeneration between opposite parity multiplets~\cite{ Berwein:2015vca}.
This effect is  considerably enhanced in QCD with respect to QED \cite{Brambilla:2017uyf} and has an important impact on the structure of the multiplets.

The light quark degrees of freedom are taken into account in four ways. 
(1)~Light quarks  in the LDF: they are part of the binding  at the scale $\L$ for states like the tetraquarks, the pentaquarks, and the doubly heavy baryons.
(2)~Light quarks at the hard scale $m_Q$: they  are integrated out in NRQCD, encoded in its matching coefficients,  
and appear as such in the relativistic corrections of the potential. 
The renormalization scale dependence of the NRQCD matching coefficients cancels against the soft part of the potential. 
(3)~Sea light quarks at the soft scale: they enter the lattice calculation of the potential with the unquenched QCD action.
Differently from the quarkonium case, see Fig.~\ref{fig:Sigma_gpl}, few full lattice QCD calculations are available for $QQ$ and $Q\bar{Q}$ static potentials with non trivial LDF.
(4)~Ultrasoft light fermions: they are the ones that will become pions.
They enter as ultrasoft degrees of freedom; they could be explicitly added to the Lagrangian in~\eqref{eq:LQQ}, but we do not consider them here. 
Ultrasoft degrees of freedom associated with heavy-light thresholds
may be also added to the Lagrangian depending on the processes that one would like to study.

Note that the coupled BOEFT Lagrangian given in Eq.~\eqref{eq:LQQ} contains only the short distance mixing: in Sec.~\ref{sec_mixing} we work out explicitly the form of the coupled Schr\"odinger equations following from such Lagrangian  for each type of exotics using a short distance description.
The avoided level crossing or large $r$ mixing is introduced in Sec.~\ref{sec:LD}.

\subsection{Mixing}
Let us summarize the types of mixing  that can occur among the static energies and in the BOEFT equation of motions.
We have three types of mixing.
\begin{itemize} 
	\item{\it Mixing induced by the $1/m_Q$ kinetic term.}
	It originates from the kinetic term in Eq.~\eqref{eq:LQQ}, whose 
	non diagonal part is called nonadiabatic coupling.
	It is a leading order effect induced by BO static energies with the same $\kappa$ 
	that become degenerate at short distance.
	In Eq.~\eqref{eq:LQQ} we show the general form of the mixing in the 
	{\it adiabatic picture}, which is the picture where the static potential is diagonal.
	After a unitary transformation, we can make the kinetic energy diagonal 
	and generate non diagonal terms in the potential; this alternative 
	but equivalent picture is called {\it diabatic picture}.
	In Sec.~\ref{sec_mixing}, we calculate the precise form of the mixing matrix 
	appearing in the Schr\"odinger equation for each type of exotics using a 
	short-range calculation, which is a good approximation in this case.
	In particular, we obtain exotic multiplets that should be 
	eventually identified with XYZ states.
	\begin{figure}[ht]
		\includegraphics[width=0.7\textwidth]{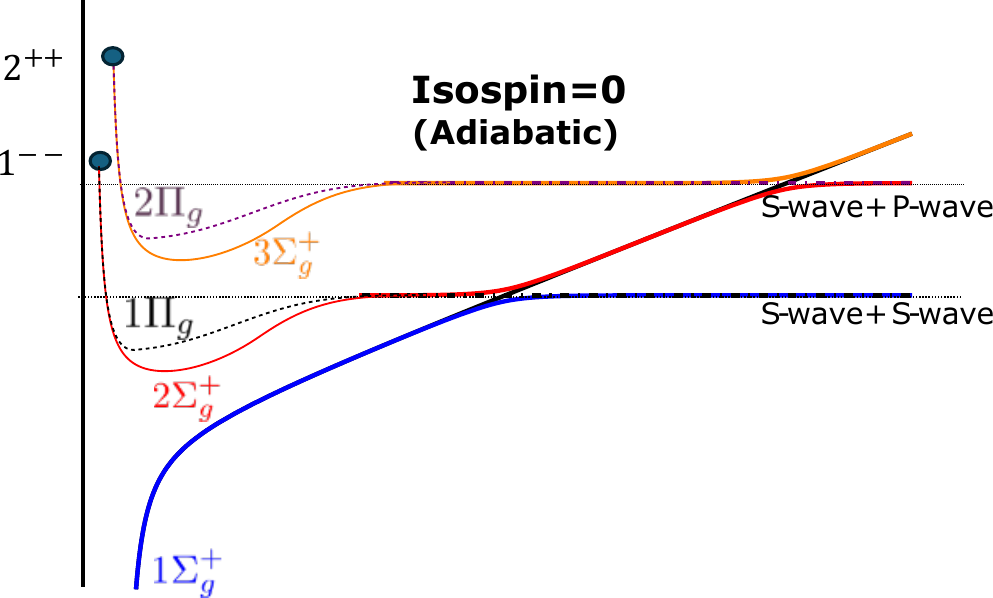}
		\caption{ Avoided levels crossing between the quarkonium potential with BO quantum number  $\Sigma_g^+$ and 
			the  first two tetraquark potentials with BO quantum numbers  $\Sigma_g^{+\prime}$, $\Sigma_g^{+\prime \prime}$
			in the isospin-singlet $I=0$ case. All the tetraquarks evolve in the static heavy-light 
			meson-antimeson threshold at large distance, see Sec.~\ref{sec:LD}. The  adiabatic static energies $\left(1\Sigma_g^+, 2\Sigma_g^+, 3\Sigma_g^+, 1\Pi_g, 2\Pi_g\right)$ are shown. At short distances $\left(r\rightarrow0\right)$, the adiabatic energy $1\Sigma_g^+$ has attractive behavior corresponding to quarkonium $\Sigma_g^+$ BO potential. At short distances $\left(r\rightarrow0\right)$, the adiabatic energies $\left(2\Sigma_g^+, 1\Pi_g\right)$ and $\left(3\Sigma_g^+, 2\Pi_g\right)$ have repulsive behavior corresponding to tetraquark BO potentials $\left(\Sigma_g^{+\prime}, \Pi_g\right)$ and $\left(\Sigma_g^{+\prime\prime}, \Pi_g^{\prime}\right)$ respectively and form degenerate multiplets  corresponding to $1^{--}$ and $2^{++}$ adjoint mesons respectively. At distances of about 1.2~fm, where there is the avoided level crossing, the $1\Sigma_g^+$ adiabatic   static energy bends down and goes asymptotically to the S-wave plus S-wave static heavy-light
			meson-antimeson
			threshold, while the $2\Sigma_g^+$ static energy assumes the confining behavior of quarkonium up to the next avoided level crossing. Note that the avoided crossing is not affecting the BO potentials $\Pi_g$ and the $\Pi_g^\prime$ and therefore $\Pi_g = 1\Pi_g$ and $\Pi_g^\prime = 2\Pi_g$. The expressions of the adiabatic energies $\left(1\Sigma_g^+, 2\Sigma_g^+, 1\Pi_g\right)$ are given by Eqs.~\eqref{eq:V1S}, \eqref{eq:V2S}, and \eqref{eq:V1P}. The details of the mixing due to avoided crossing are discussed in Sec.~\ref{sec:LD} and appendix~\ref{app_crossing}.
		}
		\label{tetra01}
		\vspace*{-0.1in} 
	\end{figure}
	\item{\it Mixing at long distance.}
	It occurs  when two (or more) BO static energies with the same quantum numbers, one the excitation of the other,
	become close one to the other, in some range of $r$ at large distance. 
	In such a case, we have to consider mixing with states with different $\kappa$ quantum numbers.
	It is convenient to start with the diabatic picture, i.e. static energies 
	with off-diagonal transition amplitudes and with diagonal entries that 
	intersect one the other at some large distance $r$.
	One recovers the adiabatic picture upon diagonalization of the 
	static energy matrix.
	The static energy eigenvalues show the typical avoided level crossing 
	behaviour discussed in appendix~\ref{app_crossing}:
	at large distance, the ground state energy bends down following the pattern
	of the smaller diagonal entry in the diabatic picture, while
	the excited state energy grows following the pattern of 
	the larger diagonal entry in the diabatic picture.
	This happens e.g. between quarkonium or hybrid and isospin zero 
	tetraquark static energies,  see Fig.~\ref{tetra01}.
	It is a nonperturbative phenomenon arising only at large distances,
	which may be responsible for exotic states of a prominent molecular nature
	like the $\chi_{c1}(3872)$.
	We discuss this effect in detail in Sec.~\ref{sec:LD} using the 
	diabatic description to make contact with existing lattice calculations.
	\item{\it Mixing between ultrasoft states that exist in the same energy range.}
	It arises in the case in which  some static energies in different $\kappa$ 
	representations do not have a large gap between them,
	which is the situation we have considered so far, 
	or in the case in which one considers very excited states from the lower 
	static energy that may overlap with the states that live in the upper  
	static energy~\cite{Oncala:2017hop}.
	It is therefore involving more than one BO channel.
	For spin-flipping processes, it is mediated by operators suppressed by at 
	least one power in the inverse of the mass.
	If mediated by a chromoelectric field, it is not suppressed by inverse 
	powers of the mass. 
	It is not important when considering only  the lowest resonances in each 
	static energy.
	We discuss this mixing and list the relevant operators inducing it in Sec.~\ref{sec:spin}.
\end{itemize}

\section{Coupled equations and multiplets}
\label{sec_mixing}

\subsection{Coupled Schr\"odinger equations, adiabatic and diabatic representation}
\label{subsec:Sch}
In this section, we  obtain the BOEFT coupled Schr\"odinger equations that describe in a unified way hybrids, tetraquarks, pentaquarks and doubly heavy baryons, i.e.\ any 
states with LDF in presence of two heavy quarks (antiquark).  This derivation is new.
The results on the coupled Schr\"odinger equations are new 
for tetraquarks  and pentaquarks, while they reproduce
the results of \cite{Berwein:2015vca,Oncala:2017hop}
for hybrids and the results of \cite{Soto:2020pfa}
for doubly heavy baryons.

For the derivation, it is convenient to express a general  eigenstate of the NRQCD Hamiltonian $H^{(0)}$ plus the kinetic term in $H^{(1)}$ as  
as a linear combination of the static states convoluted with a wave function, following Ref.~\cite{Berwein:2015vca} and Eq. \eqref{FirstOrderDegenerate}:
\begin{equation}
	\ket{N}=\sum_\lambda\int d^3r\ket{\bm{r}}\ket{
		\kappa,
		\lambda}\phi^{(N)}_{\kappa\lambda}(\bm{r})\,,
	\label{eq:Exotic-wf}
\end{equation}
where the wavefunction $\phi^{(N)}_{\kappa\lambda}$ is given by
\begin{equation}
	\phi^{(N)}_{\kappa\lambda}(\bm{r}) =\langle \underline{\kappa, \lambda}; {\bf r}/2, -{\bf r}/2\,|\,N\rangle\,.
	\label{eq:Psiwf}
\end{equation}
Here $N$ is a shorthand notation for all quantum numbers of the system described by the full Hamiltonian and it contains a fixed $\kappa$.
The state $\ket{\bm{r}}$ denotes the creation operator of a static heavy quark pair at the center of mass with relative distance $\bm{r}$ between the heavy quarks acting on the vacuum, while $\ket{\kappa,\lambda}$ gives the corresponding configuration of the LDF.
In the short-distance limit $r\to0$, the static states are approximate eigenstates of $\bm{K}^2$ as well, whose eigenvalues are labeled by $k$. The different $\Lambda=|\lambda|$ 
static energies 
are 
degenerate for the same $k$ at 
order $r^0$ 
in the 
multipole expansion. 
The differences in the static energies appear only at higher orders in $r$, where also the symmetry under $\bm{K}^2$ is broken. 

The equation of motion of the BOEFT Lagrangian in Eq.~\eqref{eq:LQQ} results in the following Schr\"odinger equation for the wavefunction $\phi^{(N)}_{\kappa\lambda}\left(\bm {r}\right)$: 
\begin{align}
	\sum_{\lambda}  \left[-P^{\alpha\dag}_{\kappa\lambda'}\left(\theta, \varphi\right)\frac{\bm{\nabla}^2_r}{m_Q}\,P^{\alpha}_{\kappa\lambda}\left(\theta, \varphi\right)\,+ V_{\kappa, \lambda'\lambda}\right]\phi_{\kappa\lambda}^{(N)}\left({\bm r}\right)= {\cal E}_N\, \phi_{\kappa\lambda^\prime}^{(N)}\left({\bm r}\right),
	\label{eq:Sch1}
\end{align}
where ${\cal E}_N$ is the energy of the bound state. The kinetic term in the Schr\"odinger equation  can be split into a radial and an angular (orbital) term.  The angular term, proportional to $1/r^2$ is the most important term in the short-distance limit $r\ll\Lambda_\mathrm{QCD}^{-1}$. 
Equation \eqref{eq:Sch1} couples the LDF to the heavy quark dynamics through the angular part of the heavy quark kinetic operator, whose derivatives act on both the wave function and projectors $P^{\alpha}_{\kappa\lambda}\left(\theta, \varphi\right)$. 

The potential $V_{\kappa, \lambda^\prime\lambda}$  in Eq.~\eqref{eq:Sch1} is a potential matrix (in the $\lambda$ and $\lambda^\prime$ index) with static energies in the diagonal entries [see Eq.~\eqref{eq:VQQ}]. 
Since the exotic eigenstates are superpositions of different static states as in Eq.~\eqref{eq:Exotic-wf}, which are degenerate in the short-distance limit, this leads to the appearance of off-diagonal matrix elements of the heavy quark kinetic energy operator, more specifically in the angular part\footnote{
	The radial piece of the derivative acting on the projection operators $P^i_{\kappa\lambda}\left(\theta, \varphi\right)$ vanishes: $\partial_r P^i_{\kappa\lambda}=0$.} 
between different LDF static states, as in Eq.~\eqref{eq:Mlambdalambdap}, 
which are referred to as the {\it non-adiabatic coupling terms (NACTs)} \cite{Brambilla:2017uyf}. 
Ideally, the NACTs could be computed on the lattice, but currently, they remain unknown. 
Nonetheless, assuming the states to be eigenstates of ${\bm K}^2$, which holds exactly in the short distance limit, allows to estimate the NACT matrix elements (see Appendix~\ref{app_Wigner_matrix}) responsible for mixing various contributions of static energies in exotic states. 
As outlined below, the shape of these mixing matrices is determined by other quantum numbers (parity in particular), which leads to small differences between energy eigenvalues $\mathcal{E}_N$, which would otherwise be degenerate. 
This is recognized as the $\Lambda$-doubling effect in molecular physics. 

Separating the radial and the angular parts in spherical coordinates (no summation over $\lambda$ implied here),
\be
\phi_{\kappa\lambda}^{(N)}(\bm{r})=v_{\lambda}^{(N)}(\theta,\varphi)\psi_{\kappa\lambda}^{(N)}(r)\,,
\ee
the radial Schr\"odinger equation, whose orbital term takes on a matrix form mixing static states,\footnote{Ultimately due to the static states having cylindrical symmetry, 
	while the kinetic energy term of the heavy quark pairs in the Schr\"odinger equation has spherical symmetry.} can be written as \cite{Berwein:2015vca}
\begin{align}
	\sum_{\lambda}  \left[-\frac{1}{m_Qr^2}\partial_r\,r^2\partial_r+\frac{1}{m_Qr^2}\,M_{\lambda'\lambda}+ V_{\kappa, \lambda'\lambda}\right]\psi_{\kappa\lambda}^{(N)}\left(r\right)= {\cal E}_N\, \psi_{\kappa\lambda'}^{(N)}\left(r\right),  \label{eq:Sch2}  
\end{align}
where $M_{\lambda'\lambda}$ are the mixing matrices and $\psi_{\kappa\lambda'}^{(N)}\left(r\right)$ are the radial wavefunctions. 
We derive the mixing matrix $M_{\lambda'\lambda}$  in Appendix~\ref{app_Wigner_matrix}. 
The above set of coupled equations [Eqs.~\eqref{eq:Sch1} and \eqref{eq:Sch2}] are the Schr\"odinger equations in the \textit{adiabatic Born--Oppenheimer} framework \cite{Born-Oppenheimer, Landau:1991wop}, wherein the kinetic part has off-diagonal terms (corresponding to NACTs) and the potential matrix is diagonal.

The adiabatic Schrödinger equation~\eqref{eq:Sch2} can be recast into another coupled Schrödinger equation through a unitary transformation. In this transformation, the kinetic matrix [specifically the mixing matrix $M_{\lambda'\lambda}$ in Eq.~\eqref{eq:Sch2}] becomes diagonal, while the potential matrix $V_{\kappa, \lambda\lambda'}$ acquires off-diagonal terms to accommodate the mixing between different states. This is the \textit{diabatic Born$-$Oppenheimer} framework, with the resulting potential matrix termed as the diabatic potential matrix. The  diabatic Schr\"odinger equation is given by
\begin{align}
	\sum_{\lambda}  \left[-\frac{1}{m_Qr^2}\delta_{\lambda^{\prime}\lambda}\partial_r\,r^2\partial_r+\frac{1}{m_Qr^2}\,{\cal M}_{\lambda'\lambda}+ {\cal V}_{\kappa, \lambda'\lambda}\right]\psi'^{(N)}_{\kappa\lambda}\left(r\right)= {\cal E}_N\, \psi'^{(N)}_{\kappa\lambda'}\left(r\right),  \label{eq:Sch2-diabatic}  
\end{align}
where the matrix ${\cal M}_{\lambda'\lambda}$ is diagonal with the eigenvalues of the mixing matrix $M_{\lambda'\lambda}$ in the diagonal entries, while ${\cal V}_{\kappa, \lambda'\lambda}$ has off-diagonal terms. The wavefunctions 
$\psi'^{(N)}_{\kappa\lambda}$ and $\psi^{\left(N\right)}_{\kappa\lambda}$, [$\lambda$ is the index for the column matrix $\psi_{\kappa}^{(N)}$ and  $\lambda=\pm \Lambda$, where $\Lambda \in \left(0,1,\cdots,k\right)$] are related by
\begin{equation}
	\psi'^{(N)}_{\kappa}=U \psi^{(N)}_{\kappa}, 
	\label{eq:U-adia-dia}
\end{equation}
where $U$ is a unitary matrix \cite{doi:https://doi.org/10.1002/0471780081.ch2}. The relation in Eq.~\eqref{eq:U-adia-dia} implies
\begin{equation}
	{\cal M }= U M U^{-1},\qquad\qquad {\cal V}_{\kappa}= U V_{\kappa} U^{-1}.
\end{equation}
The diabatic Born-Oppenheimer framework has been used to study the hybrid spectrum and mixing with quarkonium \cite{Oncala:2017hop, Andrea}, quarkonium mixing with the meson-antimeson pair threshold \cite{Bruschini:2020voj, Bruschini:2023zkb, Bruschini:2021cty}, and exotic hadron mixing with the meson-antimeson pair threshold in the dynamical diquark model \cite{Lebed:2022vks, Lebed:2023kbm}.

\subsection{Mixing matrix and discrete symmetries}\label{subsec:Mlambdalambda'}
Let us look only at the angular part of the exotic wavefunction $\ket{N}$, which we denote by $\ket{l,m;k,\lambda}$:
\begin{equation}
	\ket{l,m;k,\lambda}=\int d\Omega\ket{\theta,\varphi}\ket{k,\lambda}v_{lm}^\lambda(\theta,\varphi)\,,
	\label{eq:angular}
\end{equation}
where $v_{lm}^\lambda(\theta,\varphi)$ is in general not given by the usual spherical harmonics, a consequence of the static states $\ket{k,\lambda}$ depending implicitly on the quark-antiquark axis $\bm{\hat{r}}$ \cite{Berwein:2015vca}. 
Instead, with this angular wavefunction one obtains an eigenstate of the combined angular momentum $\bm{L}=\bm{K}+\bm{L}_Q$, where $\bm{L}_Q$ is the angular momentum for the heavy quark pair. The quantum numbers are defined as follows: 
$k\left(k+1\right)$ is the eigenvalue of ${\bm K}^2$, 
$\lambda$ is the eigenvalue of $\bm{K}\cdot\bm{\hat{r}}$, 
and  $l(l+1)$ and $m$ are the  eigenvalues of $\bm{L}^2$ and $L_z$ respectively. See appendix~\ref{app_Wigner_matrix} for more details.

The mixing matrix in the Schr\"odinger equation results from the matrix elements of $\bm{L}_Q^2$ between the combined angular momentum eigenstates:
\begin{align}
	M_{\lambda'\lambda}& \approx\bra{l,m;k,\lambda'}\bm{L}_Q^2\ket{l,m;k,\lambda}\nonumber\\
	&=\bigl(l(l+1)-2\lambda^2+k(k+1)\bigr)\delta_{\lambda'\lambda}-\sqrt{k(k+1)-\lambda(\lambda+1)}\sqrt{l(l+1)-\lambda(\lambda+1)}\delta_{\lambda'\lambda+1}\notag\\
	&\quad-\sqrt{k(k+1)-\lambda(\lambda-1)}\sqrt{l(l+1)-\lambda(\lambda-1)}\delta_{\lambda'\lambda-1}\,,
	\label{eq:Mlambdalambdap}
\end{align}
which is valid under the assumption that $k$ is a good quantum number. This assumption is smoothly broken at large distances.
This $\min\left(2k+1, 2l+1\right)$  dimensional mixing matrix can be put into block-diagonal form, where each of the two blocks corresponds to either positive or negative parity states.

The transformation of the angular momentum eigenstates under parity follows from that of their different components: 
the static quark state $\ket{\theta,\varphi}$, the light quark state $\ket{k,\lambda}$, and the orbital wave function $v_{lm}^\lambda\left(\theta, \varphi\right)$. 
The parity 
operator 
acts on the 
kets
\begin{equation}
	\ket{l,m;k,\lambda}\xrightarrow{P}\int d\Omega\,\,\sigma_Q\ket{-\theta+\pi, \varphi-\mathrm{sgn}^-(\varphi)\pi}\sigma_T(-1)^k\ket{k,\lambda}v_{lm}^\lambda(\theta,\varphi)\,,
	\label{eq:P}
\end{equation}
where $\sigma_Q=\pm1$ is the sign of the intrinsic parity of two heavy quarks (heavy quark-quark pair or heavy quark-antiquark pair), the sign function\footnote{The $\pm$ in the superscript of  the $\mathrm{sgn}$ function in Eq.~\eqref{eq:P} gives the value at zero: $\mathrm{sgn}^\pm(0)=\pm1$.} 
in the shift of the $\varphi$ coordinate has been introduced in order to ensure that the new coordinate still lies within the range $(-\pi,\pi]$, and $\sigma_T$ is associated with the intrinsic parity of the light state given by $\sigma_T(-1)^k$. 
For light states with integer quantum numbers $k$, we have $\sigma_T=+1$ for a tensor representation and $\sigma_T=-1$ for a pseudotensor representation~\cite{ARFKEN2013205}. 
For light states with half-integer quantum numbers such as  a light quark, the intrinsic parity  is positive in the ground state with a minus for each orbital excitation. This implies that $\sigma_T=(-1)^{-1/2}$ if the corresponding $(k-1/2)$ representation is a tensor and $\sigma_T=(-1)^{+1/2}$ if the corresponding $(k-1/2)$ representation is a pseudotensor. 
Note that here we rely again on the short-distance limit, where the light states are eigenstates of parity and charge conjugation separately. 
At higher orders in the 
multipole expansion, contributions from states with opposite parity will appear, as long as the product with charge conjugation remains the same.

The transformation 
of the orbital wave function 
comes into play when we shift the integration variables in $\displaystyle \int d\Omega$ in order to express the transformed state again as an integral over $\ket{\theta,\varphi}$.
We shift the coordinates in the following way:
\begin{equation}
	\theta\to-\theta+\pi\,,\qquad\varphi\to\varphi-\mathrm{sgn}^-(\varphi)\pi\,.
	\label{eq:coordinate-change}
\end{equation}
Under this change of coordinates, the orbital wave functions are given by 
\begin{align}
	v_{lm}^\lambda(-\theta+\pi,\varphi-\mathrm{sgn}^-(\varphi)\pi)
	=(-1)^{\mathrm{sgn}^-(\varphi)\,l}v_{lm}^{-\lambda}(\theta,\varphi)\,.
\end{align}
The $\text{sgn}^-(\varphi)$ factor matters only if $l$ is a half-integer.
Also, the light states $\ket{k,\lambda}$
are affected by the coordinate changes in Eq.~\eqref{eq:coordinate-change}, and with Eqs.~\eqref{Lstate-extended} and \eqref{eq:lambda-m-expansion}, it is straightforward to see that they transform in the same way as the orbital wave functions but complex conjugated. 
Ultimately, we obtain a rather straightforward transformation 
for the eigenstates of angular momentum under space inversion:
\begin{align}
	\ket{l,m;k,\lambda}\xrightarrow{P}{}&\int\,d\Omega\,\sigma_Q\,\ket{\theta,\varphi}\,\sigma_T(-1)^k\,\ket{k,-\lambda}(-1)^{\text{sgn}^-(\varphi)(l-k)}\,v_{lm}^{-\lambda}(\theta,\varphi)\notag\\
	&=\sigma_Q\sigma_T(-1)^l\ket{l,m;k,-\lambda}\equiv\sigma_P\ket{l,m;k,-\lambda}\,.
	\label{eq:Parity-state}
\end{align}
Since $(l-k)$ is always an integer, the $\text{sgn}^-(\varphi)$ can be neglected in the first line.

The eigenstates of combined angular momentum transform under parity with the usual sign $(-1)^l$ 
multiplied by 
$\sigma_T$ to distinguish between tensor and pseudotensor and $\sigma_Q$ 
to account for the intrinsic parity of the heavy quarks. 
But since the sign of $\lambda$ is reversed, these are not yet eigenstates of parity.

Since the static energies appearing in the radial Schr\"odinger equations depend solely on 
the absolute value of $\lambda$, i.e, $\Lambda \equiv |\lambda|$ and not its sign, we may combine both states with the same $\Lambda>0$ to parity eigenstates ($\Lambda=0$ already is an eigenstate):
\begin{equation}
	\ket{l,m;k,\Lambda;\epsilon}\equiv\frac{1}{\sqrt{2}}\bigl(\ket{l,m;k,\Lambda}+\epsilon\sigma_P\ket{l,m;k,-\Lambda}\bigr)\,,
	\label{eq:CPstates}
\end{equation}
where the $\epsilon=\pm1$ sign is now the eigenvalue of parity. The action of $\bm{L}_Q^2$ on these parity eigenstates for $\Lambda>1$ with $k,l\ge\Lambda$ is given by:
\begin{align}
	\bm{L}_Q^2\ket{l,m;k,\Lambda;\epsilon}={}&\left(l(l+1)-2\Lambda^2+k(k+1)\right)\ket{l,m;k,\Lambda;\epsilon}\notag\\
	&-\sqrt{k(k+1)-\Lambda(\Lambda-1)}\sqrt{l(l+1)-\Lambda(\Lambda-1)}\,\ket{l,m;k,\Lambda-1;\epsilon}\notag\\
	&-\sqrt{k(k+1)-\Lambda(\Lambda+1)}\sqrt{l(l+1)-\Lambda(\Lambda+1)}\,\ket{l,m;k,\Lambda+1;\epsilon}.
	\label{eq:Mlambdalambdap-1}
\end{align}
For $\Lambda=k$ or $\Lambda=l$, a $\Lambda+1$ state does not exist, but since its coefficient vanishes in Eq.~\eqref{eq:Mlambdalambdap-1}, 
there is no need to exclude explicitly that state from the equation.

For $\Lambda=0$, there are no two different states with $\lambda=\pm\Lambda$ to be used in Eq.~\eqref{eq:CPstates}; instead, the original state for $\lambda=0$ is already a parity eigenstate with eigenvalue $\sigma_P$ and does not need to be redefined. The operator $\bm{L}_Q^2$ does not change parity, so when it acts on a $\Lambda=1$ state, it can only produce a $\Lambda=0$ contribution if $\epsilon=\sigma_P$, or vice versa. We write
\begin{align}
	\bm{L}_Q^2\ket{l,m;k,1;\epsilon}={}&\left(l(l+1)-2+k(k+1)\right)\,\ket{l,m;k,1;\epsilon}\notag\\
	&-\delta_{\epsilon,\sigma_P}\sqrt{2}\sqrt{k(k+1)}\sqrt{l(l+1)}\,\ket{l,m;k,0}\notag\\
	&-\sqrt{k(k+1)-2}\sqrt{l(l+1)-2}\,\ket{l,m;k,2;\epsilon}\,,& k,l&\ge1\,,\\
	\bm{L}_Q^2\ket{l,m;k,0}={}&\left(l(l+1)+k(k+1)\right)\ket{l,m;k,0}\notag\\
	&-\sqrt{2}\sqrt{k(k+1)}\sqrt{l(l+1)}\,\ket{l,m;k,1;\sigma_P}\,,& k,l&\ge0\,.
\end{align}
In the case $\Lambda=1/2$, the $\Lambda-1$ contribution is not a separate state; instead, if we insert $\Lambda-1=-1/2$ on the right-hand-side of Eq.~\eqref{eq:CPstates}, we obtain again the $\Lambda=1/2$ state, but with a sign $\epsilon\sigma_P$. So we have
\begin{align}
	\bm{L}_Q^2\ket{l,m;k,1/2;\epsilon}={}&\left(l(l+1)-\frac{1}{2}+k(k+1)-\epsilon\sigma_P\left(l+\frac{1}{2}\right)\left(k+\frac{1}{2}\right)\right)\ket{l,m;k,1/2;\epsilon}\notag\\
	&-\sqrt{k(k+1)-\frac{3}{4}}\sqrt{l(l+1)-\frac{3}{4}}\,\ket{l,m;k,3/2;\epsilon}\,,\quad k,l\ge1/2\,.
\end{align}
In this basis, the mixing matrices are evidently block-diagonal, with one block for each parity eigenvalue. 
Labeling them as $M_{lk\epsilon}^{\Lambda'\Lambda}$, the indices take values $0,1\leq\Lambda',\Lambda\leq\text{min}(l,k)$ in the integer case (where the lower boundary is 0 for $\epsilon=\sigma_P$ and 1 for $\epsilon=-\sigma_P$), or $1/2\leq\Lambda',\Lambda\leq\text{min}(l,k)$ in the half-integer case.

Here, we explicitly list the first few mixing matrices. For integers with $l\geq k$ we have
\begin{align}
	M_{l1\sigma_P}&=\begin{pmatrix} l(l+1)+2 & -2\sqrt{l(l+1)} \\ -2\sqrt{l(l+1)} & l(l+1) \end{pmatrix}\,,\label{eq:M1}\\
	M_{l2\sigma_P}&=\begin{pmatrix} l(l+1)+6 & -2\sqrt{3l(l+1)} & 0 \\ -2\sqrt{3l(l+1)} & l(l+1)+4 & -2\sqrt{(l-1)(l+2)} \\ 0 & -2\sqrt{(l-1)(l+2)} & (l-1)(l+2) \end{pmatrix}\,\label{eq:M2}.
\end{align}
The row and column indices for these matrices correspond to $\Lambda'+1$ and $\Lambda+1$, respectively. For the opposite parity $\epsilon=-\sigma_P$, the first row and column have to be deleted, and $\Lambda',\Lambda$ correspond to the row and column indices directly.
In the half-integer case, also for $l\geq k$, we have
\begin{align}
	M_{l1/2\epsilon}&=\left(l-\frac{\epsilon\sigma_P}{2}\right)\left(l-\frac{\epsilon\sigma_P}{2}+1\right)\,,\label{eq:M1/2}\\
	M_{l3/2\epsilon}&=\begin{pmatrix} (l-\epsilon\sigma_P)(l-\epsilon\sigma_P+1)+\dfrac{9}{4} & -\sqrt{3\left(l-\dfrac{1}{2}\right)\left(l+\dfrac{3}{2}\right)} \\ -\sqrt{3\left(l-\dfrac{1}{2}\right)\left(l+\dfrac{3}{2}\right)} & \left(l-\dfrac{1}{2}\right)\left(l+\dfrac{3}{2}\right) \end{pmatrix}\,,\label{eq:M3/2}
\end{align}
Here, the row and column indices correspond to $\Lambda'+1/2$ and $\Lambda+1/2$. 
If $l<k$, then the mixing matrices can be obtained from the identity $M_{lk\epsilon}=M_{kl\epsilon'}$ (where $\epsilon'=(-1)^{l-k}\epsilon$, since $\sigma_P$ also depends on $l$), which is identical to deleting all rows and columns with $\Lambda',\Lambda>l$. 
We note that a similar result for the mixing matrices has been obtained for the doubly heavy baryons in Ref~\cite{Soto:2020pfa}.
In Appendix~\ref{app:Schreodinger}, we write down explicitly the radial coupled Schr\"odinger equations for the lowest hybrid, tetraquark, pentaquark, and doubly heavy baryons states.

For electrical neutral states in the $Q\bar{Q}$ sector, the parity eigenstates are also eigenstates under charge conjugation because charge conjugation  $C$ interchanges the positions of the quark and antiquark and thus reverses the direction of $\hat{\bm{r}}$. 
Its effect on the coordinates is the same as for space inversion, including the change of the sign of $\lambda$. 
The combination $CP$ accordingly leaves the coordinates invariant, and the light state has the eigenvalue $\eta=\pm1$, while the static quark-antiquark pair state transforms with $(-1)^{S_Q+1}$ from the heavy spin configuration $S_Q=0,1$ and the intrinsic parity $\sigma_Q=-1$. 
We can, therefore, obtain the transformation of the angular momentum eigenstates under charge conjugation simply by performing a space inversion first and then a $CP$ transformation:
\begin{equation}
	\ket{l,m;k,\lambda}\xrightarrow{C}\sigma_P(-1)^{S_Q+1}\eta\ket{l,m;k,-\lambda}=\sigma_T(-1)^{S_Q+l}\eta\ket{l,m;k,-\lambda}\,.
\end{equation}
Accordingly, the parity eigenstates transform under charge conjugation as
\begin{equation}
	|l,m;k,\Lambda;\epsilon\rangle\xrightarrow{C}(-1)^{S_Q+1}\eta\,\epsilon\,|l,m;k,\Lambda;\epsilon\rangle\,.
	\label{ConLstates}
\end{equation}
This relation stays true beyond the short-distance approximation.

Finally, let us also discuss briefly the last of the quantum numbers for the static states at finite distances, which is the reflection symmetry with respect to an arbitrary plane passing through the $\bm{\hat{r}}$ axis, which we call $M$. 
We have introduced this symmetry in Sec.~\ref{sec:NRQCDstatic}.  
The reflection $M$ 
is supposed to act only on the light degrees of freedom. 
However, a transformation of the light coordinates is equivalent to an inverse transformation of the rotated coordinate system, see appendix~\ref{app_Wigner_matrix}.
So performing a reflection on the light degrees of freedom is the same as performing a parity transformation  and shifting the coordinates  $\left(\theta,\varphi\right)$ in the same way as for the parity transformation. 
Using Eq.~\eqref{eq:lambda-m-expansion},
\begin{align}
	\ket{k,0}\xrightarrow{M}{}&\sqrt{\frac{4\pi}{2k+1}}\sum_{m_k=-k}^kv_{km_k}^{0*}\left(-\theta+\pi,\varphi-\mathrm{sgn}^-(\varphi)\pi\right)\,\sigma_T\,(-1)^k\,\ket{k,m_k}\notag\\
	&=\sigma_T\ket{k,0}\,,
	\label{eq:reflection}
\end{align}
where $\ket{k, m_k}$ are eigenstates of ${\bm K}^2$ and $K_z$. 
Hence, the sign $\sigma$ under reflections $M$ is the same as the sign $\sigma_T$ introduced to distinguish between tensor and pseudotensor\footnote{We refer the reader to Appendix~\ref{app_reflection} for further details.}.
Hence, $+$ stands for tensor states and $-$ for pseudotensor states.

With these relations, we have established how the 
quantum numbers $\Lambda_\eta^\sigma$ correspond to those of the degenerate $k^{PC}$ multiplet in the short-distance limit: the largest value of $\Lambda$ appearing in the same degenerate multiplet is equal to $k$; the value of $\sigma$ determines whether it is a tensor or a pseudotensor, so $P=\sigma(-1)^k$; and $\eta$ determines the eigenvalue of charge conjugation relative to the parity eigenvalue, so $C=\eta P=\eta\sigma(-1)^k$. 
Based on this, in Sec.~\ref{sec:characterization} we give in all exotic cases tables of the short distance degeneration of the static energies, the corresponding $k^{PC}$ numbers and the related LDF operators.

\subsection{Flavored states and multiplets}
\label{sec_flav}
When we consider the inclusion of light quarks, additional quantum numbers appear. 
If we restrict ourselves to up and down quarks, then these new flavor quantum numbers can be parametrized by isospin and baryon number: $|I,m_I;b\rangle$. 
We list here the flavor quantum numbers associated with up and down quarks and antiquarks:
\begin{align}
	|u\rangle&=|1/2,+1/2;+1/3\rangle\,, &
	|\bar{u}\rangle&=|1/2,-1/2;-1/3\rangle\,,\notag\\
	|d\rangle&=|1/2,-1/2;+1/3\rangle\,, &
	|\bar{d}\rangle&=|1/2,+1/2;-1/3\rangle\,.
\end{align}

The introduction of light quarks allows for the appearance of electrically charged states. In terms of the flavor quantum numbers, the electric charge $Ze$ of the light degrees of freedom is given by
\begin{equation}
	Z=m_I+\frac{b}{2}\,.
\end{equation}
However, in the previous section we have constructed the exotic states as eigenstates of charge conjugation $C$. Electrically charged states can obviously not be eigenstates of charge conjugation, since $C$ reverses the signs of $m_I$ and $b$. 
Still, the classification of the eigenstates can be based on quantum numbers associated with either charge conjugation or the electric charge operator (as both commute with the Hamiltonian).

The relation between both charge conjugation and electrically charged eigenstates is straightforward. For each electrically charged state, there exists another state with an opposite charge but otherwise identical properties (a result of the Hamiltonian commuting with charge conjugation). 
Charge conjugation eigenstates can then be expressed as symmetric or antisymmetric linear combinations of these two electrically charged states. 
Similarly, electric charge eigenstates can be formed as symmetric or antisymmetric combinations of states with opposite sign under charge conjugation but otherwise identical properties.

In a basis where isospin and baryon number are good quantum numbers, we can define the sign under which $C$ transforms a state into its opposite-charge state:
\begin{equation}
	C|I,m_I;b\rangle=\sigma_C|I,-m_I,-b\rangle\,.
\end{equation}
With this extended definition in mind, we can continue to use Eq.~\eqref{ConLstates}, even though the charged states are not eigenstates of $C$. 
For electrically neutral  states (which includes $b=0$) this sign $\sigma_C$ is a good quantum number, and by extension it also contains nontrivial information for charged states within the same isospin multiplet. 
However, if a multiplet does not contain an uncharged state, then $\sigma_C$ is arbitrary, as we may freely change the phases of opposite-charge states.

In the following, we discuss the  nontrivial cases of light flavor quantum numbers, including baryonic ($QQq$), tetraquark ($Q\bar{Q}q\bar{q}$ or $QQ\bar{q}\bar{q}$) and pentaquark ($Q\bar{Q}qqq$ or $QQqq\bar{q}$) configurations. 
We refer the reader to Ref.~\cite{Brambilla:2004jw} for discussions on quarkonium and  Refs.~\cite{Berwein:2015vca, Oncala:2017hop} for discussions on quarkonium hybrids. 
In general, an even number of light quarks will have integer isospin, while an odd number corresponds to a half-integer isospin.

\subsubsection{Integer isospin case}
The pairing of an up and a down (anti)quark can give rise to both an isospin singlet and an isospin triplet. 
The triplet is even under interchange of flavors, the singlet is odd. 
The $q\bar{q}$ combination has $0$ baryon number, while $\bar{q}\bar{q}$ has $-2/3$ (compensated in the tetraquark $QQ\bar{q}\bar{q}$ by the $2/3$ of the heavy $QQ$ pair). 
Considering color, a $q\bar{q}$ combination can form a singlet or an octet, while in the $\bar{q}\bar{q}$ case they can form a triplet or an antisextet. 
These color configurations can be combined with the respective color configurations  of the $Q\bar{Q}$ or $QQ$ to form a color-neutral meson state.\\

%

\subsubsubsection{$Q\bar{Q}q\bar{q}$ tetraquarks}
\label{subsubsec:QQbar}
We discuss the $q\bar{q}$ case first. Quark and antiquark  are distinguishable, so any possible combination of spin, isospin, and color quantum numbers may appear. 
The combination of  the color quantum numbers of the  quark and antiquark results in both color singlet and color octet configurations:
\begin{align}
	\langle q_i, \bar{q}_j| 1\rangle=\frac{1}{\sqrt{3}}\delta_{ij}, \qquad \langle q_i, \bar{q}_j| 8\rangle^a=\sqrt{2}\,T^a_{ij},
\end{align}
where $T^a\,\left(a=1,\cdots,8\right)$
are the generators of $SU(3)$ in the fundamental representation and the color states are given by 
\begin{equation}
	\ket{1}=\frac{1}{\sqrt{3}}\,\delta_{ji}|q_i, \bar{q}_j\rangle\qquad\mathrm{and}\qquad\ket{8}^a=\sqrt{2}\,T_{ji}^a |q_i, \bar{q}_j\rangle\,.
	\label{eq:qqbarcolor}
\end{equation}

The other quantum numbers are not affected by the color state. Parity is also not affected by spin or isospin. Without orbital excitations, it holds that $\sigma_T=\left(-1\right)^{k-1}$, where $k$ is the total angular momentum of the LDF and $\sigma_T$ has been defined in Eq.~\eqref{eq:P}. 
It follows that the parity of the $q\bar{q}$ pair is given by $\sigma_T \left(-1\right)^k = -1$, 
i.e. the product of the intrinsic parities of the quark and antiquark.
Higher orbital configurations, which may be excited by adding covariant derivatives to the generating operators, will add more minus signs to $\sigma_T$ in the usual fashion. 
One may also think of combinations with gauge fields with which any set of light quantum numbers may be excited.

Focusing on the ground state, both the isospin singlet and the triplet contain a neutral state, which means that $\sigma_C$ is a good quantum number. Charge conjugation acts on a single light (anti)quark by reversing the signs of $m_I$ and $b$, but does not produce additional signs (like the intrinsic parity) from the particle transformations themselves. However, charge conjugation also exchanges the positions of particle and antiparticle operators, so returning them to their original positions gives a minus sign from the commutation of the Grassmann fields and a plus or minus sign depending on the symmetry of the spin configuration. So the composite states transform as
\begin{equation}
	\ket{I,m_I;b}\ket{k,m_k}\xrightarrow{C}(-1)^{k}\ket{I,-m_I;-b}\ket{k,m_k}\,.
\end{equation}
For the static quarks $Q\bar{Q}$ the same arguments apply, except for the absence of isospin, so the total sign under charge conjugation of a $Q\bar{Q}q\bar{q}$ state is given by  $\sigma_C=(-1)^{l_Q+S_Q+k}$, 
where $l_Q$ is 
the 
eigenvalue of the 
orbital angular momentum of the heavy  quark-antiquark pair ${\bm L}_Q$, $S_Q$ is the total spin of the heavy quark-antiquark pair, and $k$ is the total angular momentum of LDF.

For light quarks in a $q\bar{q}$ configuration without orbital excitation, we have $k^{PC}=0^{-+}$ and $1^{--}$ which correspond to the BO-potentials (static energies)  $\Sigma_u^-$ and $\{\Sigma_g^{+\prime},\Pi_g\}$, respectively. 
We assume that $0^{-+}$ and $1^{--}$ are the lowest $q\bar{q}$ states, as in the quarkonium case.  
For $k^{PC}= 0^{-+}$ corresponding to the $\Sigma_u^-$ BO-potential, $\sigma = \sigma_T =-1$. 
Since $\lambda =0$,  we already have the parity eigenstate [see Eq.~\eqref{eq:Parity-state}] which implies  all the states have parity $\epsilon$ given by  $\epsilon = \sigma_P = (-1)^l$. 
The charge conjugation is given by $\left(-1\right)^{l+S_Q} = \left(-1\right)^{l_Q+S_Q}$, as $k=0$. For $k^{PC} = 1^{--}$, corresponding to the $\{\Sigma_g^{+\prime},\Pi_g\}$ BO-potentials, $\sigma = \sigma_T =1$ and $\lambda = 0, \pm 1$.  
This implies that for $\lambda=0$, the parity  $\epsilon$ is given by $\epsilon = \sigma_P = (-1)^{l+1} $ while for $\lambda = \pm 1$  using Eq.~\eqref{eq:CPstates}, we can construct parity even and odd eigenstates (with respect to $\lambda =0$) given by $\epsilon = \sigma_P = (-1)^{l+1} $ and $\epsilon =- \sigma_P = (-1)^{l} $ respectively. 
The corresponding charge conjugations are  $\left(-1\right)^{l+S_Q}$ (for $\epsilon = \sigma_P$) and $\left(-1\right)^{l+S_Q+1}$ (for $\epsilon = -\sigma_P$), both of which agree with $\sigma_C = \left(-1\right)^{l_Q+S_Q+k}$.  
As a result, states with parity $\epsilon =\sigma_p = (-1)^{l+1}$ involve mixing of $\Sigma_g^{+\prime}$ and $\Pi_g$ static energies in the Schr\"odinger equation [see Eqs.~\eqref{eq:M1} and \eqref{eq:diffeq3}], while states with  $\epsilon =-\sigma_p = (-1)^{l}$ involve only $\Pi_g$ static energy in the Schr\"odinger equation [see Eq.~\eqref{eq:diffeq4}].

\begin{table}[t]
	\begin{tabular}{||c|c|c||c|c|c||}
		\hline\hline
		\multirow{2}{*}{\hspace{2pt}$\begin{array}{c} Q\bar{Q}\\\text{color state}\end{array}$\hspace{2pt}} & \multirow{2}{*}{\hspace{2pt}$\begin{array}{c}\text{Light spin}\\k^{PC}\end{array}$\hspace{2pt}} & \multirow{2}{*}{\hspace{2pt} $\begin{array}{c} \text{BO quantum \#}\\\Lambda^\sigma_\eta\end{array}$\hspace{2pt}}&\multirow{2}{*}{\hspace{2pt} $l$\hspace{2pt}}& \multirow{2}{*}{\hspace{2pt} $\begin{array}{c}J^{PC}\\\{S_Q=0, S_Q=1\}\end{array}$\hspace{2pt}}& \multirow{2}{*}{\hspace{2pt}Multiplets\hspace{2pt}}\\
		& & & & &  \\
		\hline\hline
		\multirow{6}{*} {\hspace{2pt}$\begin{array}{c} \text{Octet}\\\mathbf{8}\end{array}$\hspace{2pt}}
		&\multirow{3}{*}{$0^{-+}$} & \multirow{3}{*}{$\Sigma_u^-$}  & \hspace{2pt}$0$\hspace{2pt} & \hspace{2pt}$\{0^{++}, 1^{+-}\}$\hspace{2pt}
		&\hspace{2pt}$T_1^0$\hspace{2pt}\\
		\cline{4-6}
		& & & \hspace{2pt}$1$\hspace{2pt} & \hspace{2pt}$\{1^{--}, \left(0, 1, 2\right)^{-+}\}$\hspace{2pt}&\hspace{2pt}$T_2^0$\hspace{2pt}\\
		\cline{4-6}
		& & & \hspace{2pt}$2$\hspace{2pt} & \hspace{2pt}$\{2^{++}, \left(1, 2, 3\right)^{+-}\}$\hspace{2pt}&\hspace{2pt}$T_3^0$\hspace{2pt}
		\\
		\cline{2-5}\cline{2-6}
		&\multirow{4}{*}{$1^{--}$} & ${\Sigma_g^{+\prime},\Pi_g}$  & \hspace{2pt}$1$\hspace{2pt} & \hspace{2pt}$\{1^{+-}, (0,1,2)^{++}\}$\hspace{2pt}&\hspace{2pt}$T_1^1$\hspace{2pt} \\
		\cline{3-6}
		& & ${\Sigma_g^{+\prime}}$  & \hspace{2pt}$0$\hspace{2pt} & \hspace{2pt}$\{0^{-+},  1^{--}\}$\hspace{2pt}&\hspace{2pt}$T_2^1$\hspace{2pt} \\
		\cline{3-6}
		& & ${\Pi_g}$  & \hspace{2pt}$1$\hspace{2pt} & \hspace{2pt}$\{1^{-+},  (0,1,2)^{--}\}$\hspace{2pt}&\hspace{2pt}$T_3^1$\hspace{2pt} \\
		\cline{3-6}
		& & ${\Sigma_g^{+\prime},\Pi_g}$  & \hspace{2pt}$2$\hspace{2pt} & \hspace{2pt}$\{2^{-+}, (1,2, 3)^{--}\}$\hspace{2pt} &\hspace{2pt}$T_4^1$\hspace{2pt}\\
		\hline\hline
	\end{tabular}
	\caption{$J^{PC}$ multiplets for the lowest $Q\bar{Q}q\bar{q}$ tetraquarks. The third
		column shows the  BO-potentials that appear in the Schr\"odinger equation of the respective
		multiplet.  We label the multiplet as $T_i^k$ in the last column in the order of increasing energies, where $k$ is the eigenvalue corresponding to the LDF total angular momentum ${\bm K}$. Since the quarkonium BO quantum number is $\Sigma_g^+$, we denote the tetraquark BO quantum number as $\Sigma_g^{+\prime}$.}
	\label{tab:QQbarqqbar}
\end{table}

The results for the $J^{PC}$ multiplets for tetraquark states in the $\Sigma_u^-$ or $\{\Sigma_g^{+\prime}, \Pi_g\}$ BO-potentials are shown in Table~\ref{tab:QQbarqqbar}. 
The lowest tetraquark spin-symmetry multiplet in the $\{\Sigma_u^-\}$ BO-potential has quantum numbers $\{0^{++}, 1^{+-}\}$ corresponding to the $l=0$ ground state. 
The lowest tetraquark spin-symmetry multiplet in the $\{\Sigma_g^{+\prime}, \Pi_g\}$ BO-potentials has quantum numbers $\{1^{+-},(0,1,2)^{++}\}$ corresponding to the first $l=1$ mixed state. 
We have assumed that the ordering of the states  reflects the one of the hybrids. 
In analogy with the multiplet $\{\Sigma_u^-, \Pi_u\}$ being lower than $\Pi_u$, we assume the multiplet $\{\Sigma_g^{+\prime}, \Pi_g\}$ lower than $\Pi_g$. 
Furthermore, because in Fig.~\ref{morning} and in Refs.~\cite{Juge:1999ie, Juge:2002br, Bali:2003jq, Capitani:2018rox}, the $\Sigma_g^{+\prime}$ potential is lower than $\Pi_g$, we place $\Sigma_g^{+\prime}$ in between $\{\Sigma_g^{+\prime}, \Pi_g\}$ and $\Pi_g$. For the rest, we order the states according to the quantum number $l$. 
The quantum numbers are independent of the isospin configuration, while the BO-potentials $\Sigma_u^-, \Sigma_g^{+\prime}, \Pi_g$ themselves depend on the isospin $I$.

We would like to point out that conceptually, there is no difference between a $Q\bar{Q}q\bar{q}$ tetraquark with isospin $I=0$ and a hybrid with the same quantum numbers. 
Once light quarks are introduced, there can be transitions between the gluonic configuration and the light quark configuration, so a proper eigenstate of the static Hamiltonian will have overlap with both of them. 
Transitions may however be suppressed, which would still allow an identification of states as primarily hybrid or tetraquark.\\

\subsubsubsection{$QQ\bar{q}\bar{q}$ tetraquarks}
\label{subsubsec:QQ}
Let us turn now to the case of $QQ\bar{q}\bar{q}$ tetraquarks. The two heavy quarks and two light antiquarks are indistinguishable, and therefore, the color, spin, and isospin quantum numbers are constrained by the Pauli exclusion principle.
The combination of  the color quantum numbers of the two quarks (antiquarks)  results in both color antitriplet (triplet) and color sextet (antisextet) configurations:
\begin{align}
	\langle q_i, q_j| \bar{3}\rangle^{\ell}=\underline{T}^{\ell}_{ij}=\frac{1}{\sqrt{2}}\epsilon_{\ell ij}, \qquad \langle q_i, q_j| 6\rangle^{\sigma}=\underline{\Sigma}^{\sigma}_{ij},
\end{align}
where we use the notation from Ref.~\cite{Brambilla:2005yk} for the antitriplet (triplet) color and sextet (antisextet) tensor invariants $\underline{T}^{\ell}_{ij}\,\left(i,j,\ell= 1, 2, 3\right)$ and $\underline{\Sigma}^{\sigma}_{ij},\,\left(i,j=1,2,3;\,\sigma=1,\cdots, 6\right)$ respectively [see appendix~\ref{app:group}]. The color  states are given by 
\begin{equation}
	\ket{\bar{3}}^{\ell}=\underline{T}^{\ell}_{ij}\,|q_i, q_j\rangle\qquad\mathrm{and}\qquad\ket{6}^{\sigma}=\underline{\Sigma}^{\sigma}_{ij}\,|q_i, q_j\rangle\,.
	\label{eq:qqcolor}
\end{equation}
Both the tensor invariants $\underline{T}^{\ell}_{ij}$ and $\underline{\Sigma}^{\sigma}_{ij}$ are real; $\underline{T}^{\ell}_{ij}$ is totally antisymmetric and $\underline{\Sigma}^{\sigma}_{ij}$  symmetric in the $i$ and $j$ indices.

The two light antiquarks (or quarks) can also form spin and isospin singlets and triplets just like in the quark-antiquark case (only the baryon number is different). 
However, the Pauli principle, expressed in the Grassmann nature of the light quark fields, forbids both fields to have exactly the same quantum numbers when evaluated at the same point. 
Thus, out of the $12^2=144$ angular momentum-isospin-color combinations that appear in the $q\bar{q}$ case, only $\displaystyle\binom{12}{2}=66$ remain for $\bar{q}\bar{q}$. Since the light quark fields anticommute, only antisymmetric combinations of the angular momentum-isospin-color indices survive. 
The color triplet and the (iso)spin singlet are antisymmetric, while the color antisextet and (iso)spin triplet are symmetric. This means that we can have $(I=0, k=0)$ and $(I=1, k=1)$ combinations in the triplet sector, or $(I=0, k=1)$ and $(I=1, k=0)$ combinations in the antisextet sector. 
Combining the multiplicities for color, isospin, and angular momentum, we get $3\times1\times1+3\times3\times3+6\times1\times3+6\times3\times1=66$, which reproduces exactly the predicted number of combinations.

Adding orbital excitations lifts these restrictions somewhat; for instance, with a single covariant derivative, we may construct a symmetric and an antisymmetric combination: $q_i(\bm{D}q)_j\pm(\bm{D}q)_iq_j$. 
There may also be higher excited static states where the Pauli principle limits the allowed combinations, but in general, there are sufficient operator combinations available to generate any desired quantum numbers. 
Again, we discuss only the lowest case without orbital excitations.

The intrinsic parity of two antiquarks (or two quarks) is positive, which means that depending on the spin configuration, we either have a scalar or a pseudovector representation. 
The sign under reflections is thus given by $\sigma=(-1)^{k}$. $CP$ is not a symmetry of the static $QQ$ system, which instead is invariant under parity plus interchange of the static particle indices (such that $\bm{r}=\bm{x}_1-\bm{x}_2\xrightarrow[1\leftrightarrow2]{P}-\bm{x}_2+\bm{x}_1=\bm{r}$ stays invariant), and the $\eta$ quantum number corresponds to the sign under parity acting only on the light antiquarks. 
So the static energies appear in $\Sigma_g^+$ and $\{\Sigma_g^-,\Pi_g\}$ multiplets.

The static quarks are distinguished by their positions, hence a priori, any spin or color combination is allowed. 
However, when we construct the angular momentum eigenstates $\ket{l,m;k,\lambda}$, we integrate over all possible orientations of the quark axis. 
Since for any orientation, there is another one where the quark positions are exchanged, the Pauli principle  restricts also for the heavy quarks the final states that we can construct from the static states by limiting the polar angle integration to a half sphere.

Let us introduce the sign $\sigma_{SC}$ for the exchange in spin and color of the two static quark fields. 
We have $\sigma_{SC}=1$ for singlet-antitriplet and triplet-sextet spin-color configurations, and $\sigma_{SC}=-1$ for singlet-sextet and triplet-antitriplet spin-color configurations. 
With this, it follows that exchanging the two static quarks has the same effect on the angles as applying a parity transformation:
\begin{equation}
	\ket{\theta,\varphi}\rightarrow\left(-1\right)^{l_Q}\ket{\pi-\theta,\varphi-\mathrm{sgn}(\varphi)\pi}=-\sigma_{SC}\ket{\pi-\theta,\varphi-\mathrm{sgn}^{-1}(\varphi)\pi}\,,
\end{equation}
where $l_Q$ is the eigenvalue of the angular momentum operator of the heavy quark pair and we have used that $QQ$ states are antisymmetric, while even $l_Q$ describes symmetric wavefunctions and odd $l_Q$ describes antisymmetric wavefunctions. 
Accordingly, because angular momentum eigenstates remain the same after
exchanging the two static quarks, we have that
\begin{align}
	\ket{l,m;k,\lambda}&=\int d\Omega\ket{\theta,\varphi}\ket{k,\lambda}v_{lm}^\lambda(\theta,\varphi)\notag\\
	&=-\sigma_{SC}\int d\Omega\ket{\pi-\theta, \varphi-\mathrm{sgn}^{-1}(\varphi)\pi}\,\ket{k,\lambda}\,v_{lm}^\lambda\left(\theta,\varphi\right)\notag\\
	&=-\sigma_{SC}\int d\Omega\ket{\theta, \varphi}\,\ket{k,-\lambda}(-1)^{-k}\,v_{lm}^\lambda\left(\pi-\theta, \varphi-\mathrm{sgn}^{-1}(\varphi)\pi\right)\notag\\
	&=-\sigma_{SC}\,(-1)^{l-k}\ket{l,m;k,-\lambda}\,,
	\label{QQexchange}
\end{align}
where in the third line, we have redefined the integration variables in $d\Omega$ to express the transformed state again as $|\theta, \varphi\rangle$; 
note that the LDF state $\ket{k, \lambda}$ is affected by the coordinate changes, 
since $\ket{k, \lambda}$ parametrically depends on heavy quark coordinate $\theta$ and $\varphi$.
This, in particular, implies that the sign of $\lambda$ is not a good quantum number in the $QQ$ case. The reason for this is immediately apparent: without the distinction of particle versus antiparticle, there is no way to unambiguously define the direction of the quark axis.

When constructing the parity eigenstates, we used the symmetric or the antisymmetric combination of $\Lambda$ states, where $\Lambda \equiv |\lambda|$ as in Eq.~\eqref{eq:CPstates}. 
However, in view of Eq.~\eqref{QQexchange}, only one of such combinations survives, as the other
\begin{equation}
	\ket{l,m;k,\Lambda}+\sigma_{SC}(-1)^{l-k}\ket{l,m;k,-\Lambda},
\end{equation}
cancels trivially when we replace the $\Lambda$ state by $-\Lambda$. This means that the parity $\epsilon$ of a state and its mixing properties are already fully determined by the other quantum numbers.

Here we have $\sigma=\sigma_T=(-1)^{k}$, so $\sigma_P=(-1)^{l-k}$, and the remaining linear combination of opposite parity states is
\begin{equation}
	\ket{l,m;k, \Lambda;-\sigma_{SC}}=\frac{1}{2}\left(\ket{l,m;k,\Lambda}-\sigma_{SC}\,\sigma_P\ket{l,m;k,-\Lambda}\right)\,.
\end{equation}
The resulting tetraquark states all have  parity $\epsilon=-\sigma_{SC}$, so for the $QQ$ color antitriplet $\epsilon=(-1)^{S_Q+1}$, while for the color sextet $\epsilon=(-1)^{S_Q}$. Since mixing only occurs when $\epsilon=\sigma_P$ [see Eq.~\eqref{eq:Mlambdalambdap-1}], we no longer have states with opposite parity where one is mixed and the other is not. Instead, it is already determined by the other quantum numbers whether a state mixes $\Sigma$ and $\Pi$ potentials or not. In the case of the $\Sigma_g^+$ potential or $l=0$ states, only $\lambda=0$ is allowed, so several combinations of quantum numbers where $\epsilon$ would not be equal to $\sigma_P$ are excluded. For example, in the $QQ$ color antitriplet sector, the combination $(I=0, k=0, S_Q=0)$ allows only odd values for $l$, while $(I=0, k=0, S_Q=1)$ allows only even values.

\begin{table}[t]
	\begin{tabular}{||c|c|c|c|c||c|c||}
		\hline\hline
		\multirow{2}{*}{\hspace{2pt}$\begin{array}{c} QQ\\\text{ color state}\end{array}$\hspace{2pt}} & \multirow{2}{*}{\hspace{2pt}$\begin{array}{c} \text{Light spin}\\ k^{P}\end{array}$\hspace{2pt}} & \multirow{2}{*}{\hspace{2pt} $\begin{array}{c} \text{BO quantum \#}\\\Lambda^\sigma_\eta\end{array}$\hspace{2pt}} & \multirow{2}{*}{\hspace{2pt}$\begin{array}{c} \text{Isospin}\\I\end{array}$ \hspace{2pt}}& \multirow{2}{*}{\hspace{2pt} $l$\hspace{2pt}} & \multicolumn{2}{|c||}{$J^{P}$}\\
		\cline{6-7}
		& & & & & \hspace{2pt}$S_Q=0$\hspace{2pt} & \hspace{2pt}$S_Q=1$\hspace{2pt} \\
		\hline\hline
		\multirow{4}{*}{\hspace{2pt}$\begin{array}{c} \text{Antitriplet}\\\bar{\mathbf{3}}\end{array}$\hspace{2pt}} &\multirow{2}{*}{$0^+$} & \multirow{2}{*}{${\Sigma_g^+}$} & \multirow{2}{*}{$0$}& $0$ & \hspace{2pt} $\cdots$ & \hspace{2pt}$1^+$\hspace{2pt} \\
		\cline{5-7}
		& & & & 1 &\hspace{2pt}$1^-$\hspace{2pt}&$\cdots$\\
		\cline{2-7}
		& \multirow{2}{*}{$1^+$} & \multirow{2}{*}{${\Sigma_g^-, \Pi_g}$} & \multirow{2}{*}{$1$}& $0$ &  \hspace{2pt}$0^-$\hspace{2pt}& $\cdots$ \\
		\cline{5-7}
		& & & & 1 & $1^{-}$&\hspace{2pt}$\left(0, 1, 2\right)^+$\hspace{2pt}\\
		\cline{5-7}
		\hline\hline
		\multirow{4}{*}{\hspace{2pt}$\begin{array}{c} \text{Sextet}\\\mathbf{6}\end{array}$\hspace{2pt}} &\multirow{2}{*}{$0^+$} & \multirow{2}{*}{${\Sigma_g^+}$} & \multirow{2}{*}{$1$}& $0$  & \hspace{2pt}$0^+$\hspace{2pt}& \hspace{2pt} $\cdots$ \\
		\cline{5-7}
		& & & & 1 &$\cdots$&\hspace{2pt}$\left(0, 1, 2\right)^-$\hspace{2pt}\\
		\cline{2-7}
		& \multirow{2}{*}{$1^+$} & \multirow{2}{*}{${\Sigma_g^-, \Pi_g}$} & \multirow{2}{*}{$0$}& $0$ & $\cdots$ &  \hspace{2pt}$1^-$\hspace{2pt} \\
		\cline{5-7}
		& & & & 1 &\hspace{2pt}$ 1^+$\hspace{2pt}&$\left(0, 1, 2\right)^-$\\
		\cline{5-7}
		\hline\hline
	\end{tabular}
	\caption{Spin-symmetry multiplets for the lowest $QQ\bar{q}\bar{q}$ tetraquarks. 
		The angular momentum quantum number $l$ is the sum of the orbital angular momentum of the heavy quark pair and the light quark spin. 
		In the last column, the dotted entry means that that particular state is not allowed due to the Pauli exclusion principle.  }
	\label{QQqqnumbers}
\end{table}

In Table~\ref{QQqqnumbers}, we show the lowest expected $QQ\bar{q}\bar{q}$ tetraquark states for each possible set of quantum numbers. If also here we assume that  states with a coupled Schr\"odinger equation have lower energies, then there is only one candidate: the $(0,1,2)^+$ multiplet corresponding to the quantum numbers $(I=1, k=1, S_Q=1, l=1)$ in the color antitriplet sector. In the color sextet sector, it would be the $1^+$ state with $(I=0, k=1, S_Q=0, l=1)$, but since the sextet potential is repulsive at short distances, these are not expected to be low-lying tetraquarks. The ground states in the two $\Sigma$ potentials ($\Sigma_g^+$ and $\Sigma_g^-$) of the color antitriplet sector without mixing appear with the quantum numbers $1^+$ for $(I=0, k=0, S_Q=1, l=0)$ and $0^-$ for $(I=1, k=1, S_Q=0, l=0)$. The $1^+$ could actually be the lowest state in case the $\Sigma_g^+$ potential is lying lower than both $\{\Sigma_g^-,\Pi_g\}$.

\subsubsection{Half-integer isospin case}
\subsubsubsection{ $QQq$ Baryons} 
Half-integer isospin is encountered exclusively in configurations with a non-zero baryon number. The simplest of such configurations are $QQq$ baryons. 
In this case, the only viable color combination involves pairing the light color triplet with the  color antitriplet from the heavy quarks. 
We are not aware of a generally used notation for half-integer representations of the cylindrical symmetry group $D_{\infty h}$. 
Therefore, we resort to defining these representations by specifying the relevant quantum numbers here. Notably, the concept of reflection and the associated quantum number $\sigma$ is inapplicable for half-integer values of $k$ since there is no $\lambda=0$ state. 
Moreover, the light quark spin can be aligned parallel or antiparallel to the angular momentum. 
In the parallel case, we get $k^P$ values $\left(1/2\right)^+, \left(3/2\right)^-, \cdots$ so $\sigma_T = (-1)^{-1/2}$, $\sigma_P = (-1)^{l-1/2}$, and $\eta = (-1)^{k-1/2}$, 
whereas in the antiparallel case, we have $\left(1/2\right)^-, \left(3/2\right)^+, \cdots$, and $\sigma_T = (-1)^{+1/2}$, $\sigma_P = (-1)^{l+1/2}$ and $\eta = (-1)^{k+1/2}$. 
The ground state has $\left(1/2\right)^+$, the first orbital excitation $\left(3/2\right)^-$ and $\left(1/2\right)^-$, and so on.

Regarding the spin and color configuration of the heavy quarks, the same restrictions based on the Pauli principle that follow from Eq.~\eqref{QQexchange} apply. 
Focusing on the ground state, which has $k^P =\left(1/2\right)^+$, the parity of the final states is given by $\epsilon=-\sigma_{SC}(-1)^{k-1/2}$. 
So for the ground state $k=1/2$, the heavy spin singlet has negative parity and the triplet has positive parity. 
The radial Schr\"odinger equation is not coupled, but the orbital term depends on the angular momentum and other quantum numbers as $\left(l+\dfrac{\sigma_{SC}}{2}(-1)^{l-1/2}\right)\left(l+\dfrac{\sigma_{SC}}{2}(-1)^{l-1/2}+1\right)$. 
The lowest value corresponds to $\sigma_{SC}=-1$ (i.e.\ the heavy spin triplet) and $l=1/2$, for which the orbital term vanishes, giving a $\{(1/2,3/2)^+\}$ spin-symmetry multiplet as the lowest $QQq$ baryon. 
For the heavy spin singlet $(\sigma_{SC}=1)$, the $l=1/2$ and $l=3/2$ configurations are actually degenerate (before heavy spin interactions are considered), leading to a $\{(1/2,3/2)^-\}$ multiplet from a radial Schr\"odinger equation with a coefficient $2$ for the orbital term. 
Isospin does not affect any of these quantum numbers, but it does determine the electric charge of the resulting states (based on the flavor of the heavy quarks). 
For ground state doubly heavy baryons $QQq$ multiplets, we refer to Table.~II in Ref.~\cite{Soto:2020pfa}.

\subsubsubsection{$Q\bar{Q} qqq$, $QQqq\bar{q}$ pentaquarks}
The next possible half-integer isospin configurations involve pentaquarks, which can manifest in $Q\bar{Q}qqq$ or $QQqq\bar{q}$ combinations, including their respective antiparticles. We will first discuss the case of $Q\bar{Q}qqq$, again ignoring the orbital excitations of the light quarks. Just like in the case of light baryons, the light quarks may form spin and isospin doublet or quartet combinations (i.e., $I=1/2$ or $I=3/2$). Furthermore, the three light quarks can exist in either color singlet or color octet configurations to combine with the corresponding color states of the $Q\bar{Q}$ pair to form a color-neutral pentaquark state.  The heavy quark-antiquark pair is distinguishable whereas in the light sector, the Pauli principle applies, which means that the whole color-spin-isospin combination needs to be fully antisymmetric regarding particle exchange. As a consequence, spin and isospin have to be both doublets or both quartets in the color-singlet sector, while in the color-octet sector any combination of spin and isospin is allowed except for both being quartets (cf.\ Appendix~\ref{SICcombinations}).

Similar to the baryonic case, the cylindrical symmetry notation $\Lambda_\eta^\sigma$ for static states with integer rotational quantum numbers is not applicable 
as there are no established labels for $\Lambda$, the reflection quantum number $\sigma$ does not appear without a $\lambda=0$ state, and 
$\eta$ is useless for $Q\bar{Q}qqq$ states. 
In case of $QQq$ baryons, $\eta$ corresponds to the parity $P$ of the light state since that is the symmetry of the static $QQ$. In case of $Q\bar{Q}qqq$ pentaquarks,  $\eta$ corresponds to the $CP$ eigenvalue of the light state, since that is the symmetry of the static $Q\bar{Q}$. 
Nevertheless, $CP$ transforms a light $qqq$ state into a $\bar{q}\bar{q}\bar{q}$ state, 
so the $CP$ eigenstates are even and odd linear combinations of 
the two.
Since the electric charge or baryon number operator transforms one of these combinations into the other while commuting with the static Hamiltonian, the static energies of both combinations have to be degenerate. 
Therefore, $\eta$ can no longer be used to distinguish between static energies. For notation, we choose to label the static states by $k^P$, where $k$ is the LDF angular momentum.

For a heavy quark and antiquark, there is no Pauli exclusion principle, and their spins may either form a singlet or a triplet, irrespective of their color configurations. 
The light quarks without orbital excitations have positive parity. 
Therefore, for $k^P = \left(1/2\right)^+$, 
$\sigma_T=(-1)^{-1/2}$ and for $k^P = \left(3/2\right)^+$, $\sigma_T=(-1)^{1/2}$. 
A simple expression for $\sigma_T$ that encompasses both cases is $\sigma_T = \left(-1\right)^{k-1}$. With this we have $\sigma_P=(-1)^{l+k}$, where $k=1/2$ for the isospin doublet and quartet and $k=3/2$ for the isospin doublet (cf.\ Appendix~\ref{SICcombinations}). 
We expect the lowest $Q\bar{Q}qqq$ pentaquarks to form spin-symmetry multiplets $\{(1/2)^-,(1/2,3/2)^-\}$ for $k^P=\left(1/2\right)^+$ and $\{(3/2)^-,(1/2,3/2,5/2)^-\}$ for $k^P=\left(3/2\right)^+$, where again the first entries correspond to a heavy spin-singlet and the second to a triplet. The ground state $Q\bar{Q}qqq$ multiplets are shown in Table.~\ref{tab:QQbarpenta}.

\begin{table}[t]
	\begin{tabular}{||c|c|c||c|c||}
		\hline\hline
		\multirow{2}{*}{\hspace{2pt}$\begin{array}{c} Q\bar{Q}\\\text{color state}\end{array}$\hspace{2pt}} & \multirow{2}{*}{\hspace{2pt}$\begin{array}{c}\text{Light spin}\\k^{P}\end{array}$\hspace{2pt}} & \multirow{2}{*}{\hspace{2pt} $\begin{array}{c}\text{BO quantum \#}\\D_{\infty h}\end{array}$\hspace{2pt}} &\multirow{2}{*}{\hspace{2pt} $l$\hspace{2pt}}& \multirow{2}{*}{\hspace{2pt} $\begin{array}{c}J^{P}\\\{S_Q=0, S_Q=1\}\end{array}$\hspace{2pt}}\\
		& &  &  &  \\
		\hline\hline
		\multirow{2}{*} {\hspace{2pt}$\begin{array}{c} \text{Octet}\\\mathbf{8}\end{array}$\hspace{2pt}} &\multirow{1}{*}{$(1/2)^{+}$}& \multirow{1}{*}{$\left(1/2\right)_g$}\hspace{2pt}  & \hspace{2pt}$1/2$\hspace{2pt} & \hspace{2pt}$\{1/2^{-}, \left(1/2, 3/2\right)^-\}$\hspace{2pt}
		\\
		\cline{2-5}
		&\multirow{1}{*}{$(3/2)^{+}$}  & \multirow{1}{*}\{$\left(1/2\right)_g^{\prime}$, $\left(3/2\right)_g$\}\hspace{2pt} & \hspace{2pt}$3/2$\hspace{2pt} & \hspace{2pt}$\{3/2^{-}, (1/2, 3/2, 5/2)^{-}\}$\hspace{2pt}\\
		\hline\hline
	\end{tabular}
	\caption{$J^{PC}$ multiplets for the lowest $Q\bar{Q}qqq$ pentaquark states. In the third column, we represent the BO quantum number ($D_{\infty h}$ representation) with $\left(k\right)_\eta$ instead of $\Lambda^{\sigma}_{\eta}$ and  $\eta = g$ denotes positive parity.}
	\label{tab:QQbarpenta}
\end{table}

The other case for a doubly-heavy pentaquark consists in $QQqq\bar{q}$ (or its charge conjugate). 
In the light sector, the Pauli principle requires an antisymmetric configuration for the two quarks, while the antiquark configuration is independent. 
The color antitriplet from the heavy quarks can couple to the triplet of the light quarks or the heavy sextet to the light antisextet. 
As explained in Appendix~\ref{SICcombinations}, the Pauli principle is less restrictive in this case. In the color-triplet sector, there are four doublet-doublet spin-isospin combinations, two doublet-quartet and two quartet-doublet combinations, as well as a quartet-quartet combination. 
In the color-antisextet sector, there are two doublet-doublet combinations, as well as one doublet-quartet, quartet-doublet, or quartet-quartet combination.

We assume again that the lowest pentaquarks in a $QQqq\bar{q}$ configuration come from the color-(anti)triplet sector because of its attractive short-range potential. 
Irrespective of their spin or isospin configuration, the parity of the light quarks is always negative (because of the intrinsic parity of the antiquark and no orbital excitations). 
Therefore, for $k^P = \left(1/2\right)^-$, 
$\sigma_T=(-1)^{1/2}$ and for $k^P = \left(3/2\right)^-$, $\sigma_T=(-1)^{-1/2}$. 
A simple expression for $\sigma_T$ that encompasses both cases is $\sigma_T = \left(-1\right)^{1-k}$. 
With this we have $\sigma_P=(-1)^{l-k+1}$ and $\epsilon=\sigma_{SC}$. 
Depending on the values of $\epsilon$ and $\sigma_P$, the orbital terms may take larger or lower values, and depending on $l$ and $k$ there may or may not be mixing. 
For $k=1/2$, the lowest energy eigenvalues come from the smallest orbital term, for $k=3/2$ they come from the mixed equation with the smallest orbital terms. 
We have listed the different sets of quantum numbers in table~\ref{QQpenta}.

\begin{table}[t]
	\centering
	\resizebox{0.95\columnwidth}{!}{%
		\begin{tabular}{||c|c|c||c|c||}
			\hline\hline
			\multirow{2}{*}{\hspace{2pt}$\begin{array}{c} QQ\\\text{color state}\end{array}$\hspace{2pt}} & \multirow{2}{*}{\hspace{2pt}$\begin{array}{c}\text{Light spin}\\k^{P}\end{array}$\hspace{2pt}} & \multirow{2}{*}{\hspace{2pt} $\begin{array}{c} \text{BO quantum \#}\\D_{\infty h}\end{array}$\hspace{2pt}}  & \multicolumn{2}{|c||}{$J^P$}\\
			\cline{4-5}
			& & & \hspace{2pt}$S_Q=0$\hspace{2pt} & \hspace{2pt}$S_Q=1$\hspace{2pt} \\
			\hline\hline
			\multirow{2}{*}{\hspace{2pt}$\begin{array}{c} \text{Antitriplet}\\\bar{\mathbf{3}}\end{array}$\hspace{2pt}} & $(1/2)^-$ & $\left(1/2\right)_u$ & \hspace{2pt}$(1/2)^+,\,(3/2)^+$\hspace{2pt} & \hspace{2pt}$(1/2,3/2)^-$\hspace{2pt} \\
			\cline{2-5}
			& $(3/2)^-$ & \{$\left(1/2\right)^{\prime}_u, \left(3/2\right)_u$\} & \hspace{2pt}$(1/2)^+, (3/2)^+, (5/2)^+$\hspace{2pt} & \hspace{2pt}$(1/2,3/2,5/2)^-$\hspace{2pt} \\
			\hline\hline
			\multirow{3}{*}{\hspace{2pt}$\begin{array}{c} \text{Sextet}\\\mathbf{6}\end{array}$\hspace{2pt}} & $(1/2)^-$ & $\left(1/2\right)_u$ & \hspace{2pt}$(1/2)^-$\hspace{2pt} & \hspace{2pt}$(1/2,3/2)^+,(1/2,3/2,5/2)^+$\hspace{2pt} \\
			\cline{2-5}
			& $(3/2)^-$ & \hspace{2pt}\{$\left(1/2\right)^{\prime}_u, \left(3/2\right)_u$\}\hspace{2pt}  & $\left(3/2\right)^-$ & \hspace{2pt}$\begin{array}{c}(1/2,3/2)^+, (1/2,3/2,5/2)^+,\\(3/2,5/2,7/2)^+\end{array}$ \hspace{2pt} \\
			\hline\hline
	\end{tabular}}
	\caption{Spin-symmetry multiplets for the lowest $QQqq\bar{q}$ pentaquarks. Where two multiplets are given in the same set (identified by entries within a parenthesis), they correspond to degenerate values of $l$. 
		Where two or more sets are given, they are not degenerate, but it is not clear which lies lower without knowing the actual shape of the potentials.}
	\label{QQpenta}
\end{table}

\section{Exotic static energies: operator definition and behavior at short distances}
\label{sec:characterization}

\subsection{Matching operators for static energies}
\label{subsec:operators}
As discussed in Sec.~\ref{sec:BO}, the static energies for all exotics are given by Eq.~\eqref{eelwl}. 
Here, we characterize the state $|X_n,\,-T/2\rangle$ using gauge-invariant interpolators that select the quantum numbers of the exotic BO static energy in consideration.
While, in general, the interpolator can be anything that has a good overlap with the state whose static energy we want to measure, we point out that the operators that we are suggesting here are well suited for QCD lattice calculations for two reasons. 
On the one hand, they are appropriate to calculate the $r$ dependence of the static energy also in the relatively short distance region. 
Moreover, since the short distance behavior of the correlators is calculable using weakly coupled pNRQCD, as we show in Sec.~\ref{subsec:potential}, this is a considerable guidance and test of the lattice calculation.
On the other hand, our interpolating operators do not overlap with states containing quarkonium and pions, a problem that has plagued some of the lattice calculations in the $Q\bar{Q}$ sector \cite{Prelovsek:2019ywc, Bali:2000vr, Bulava:2019iut, Bulava:2024jpj}. 
In fact, in the $Q\bar{Q}$ sector, our interpolating operators contain the $Q\bar{Q}$ pair in a color octet configuration, as it should be from BOEFT previously presented.
We show in Appendix~\ref{app:Overlap} that the projection of such operators on quarkonium plus pion states is zero.

\subsubsection{Form of the interpolating operators}
We consider a bound system made of two heavy quarks: $Q\bar{Q}$ or $QQ$ pair along with the light degrees of freedom (LDF).  
As we have seen in Sec.~\ref{sec:BO}, the static energies for quantum numbers $\kappa$ and $\lambda$, $E_{\kappa,|\lambda|}^{\left(0\right)}\left(r\right)$, can be determined  from the large time logarithms of the relevant Wilson loop obtained from the gauge-invariant correlators in NRQCD:
\begin{align}
	E_{\kappa, |\lambda|}^{\left(0\right)}(r)=\lim_{T\to\infty}\frac{i}{T}\,\log\left[ \langle \mathrm{vac}| \mathcal{O}^h_{\kappa, \lambda}(T/2,\,\bm{r},\,\bm{R})\,\mathcal{O}^{h\dagger}_{\kappa, \lambda}(-T/2,\,\bm{r},\,\bm{R})|\mathrm{vac}\rangle\right],\,\, h \in \left\{Q\bar{Q},\,QQ\right\},\label{eq:En}
\end{align}
where $|\mathrm{vac}\rangle$ denotes the NRQCD vacuum and $\mathcal{O}_{\kappa, \lambda}$ is an interpolating gauge-invariant operator. 
The isospin quantum numbers of the LDF are not explicitly written.

In terms of NRQCD fields, we define the gauge-invariant interpolating operator for exotic hadrons with a heavy  (octet) quark-antiquark pair or  (antitriplet or sextet) heavy quark-quark pair as 
\begin{align}
	\mathcal{O}^{(Q\bar{Q})_8}_{\kappa, \lambda}\left(t, \bm{r}, \bm{R}\right)&=\nonumber\\
	&\hspace{-1.7 cm}\chi^{\dagger}\left(t, \bm{R}+\bm{r}/2\right)\phi\left(t; \bm{R}+\bm{r}/2,\bm{R}\right)P^{\alpha \dag}_{\kappa, \lambda}H_{8,\,\kappa}^{\alpha,\,a}(t, \bm{R})\,T^a\,\phi\left(t; \bm{R},\bm{R}-\bm{r}/2\right)\psi\left(t, \bm{R}-\bm{r}/2\right),
	\label{eq:intop-1}\\
	\mathcal{O}^{(QQ)_{\bar{3}}}_{\kappa, \lambda}\left(t, \bm{r}, \bm{R}\right)&=\nonumber\\
	&\hspace{-1.7 cm}\psi^T\left(t, \bm{R}+\bm{r}/2\right)\phi^T(t; \bm{R}+\bm{r}/2,\bm{R})P^{\alpha \dag}_{\kappa, \lambda}H_{3,\,\kappa}^{\alpha,\,\ell}\left(t, \bm{R}\right)\underline{T}^\ell\,\phi\left(t; \bm{R}, \bm{R}-\bm{r}/2\right)\psi\left(t,  \bm{R}-\bm{r}/2\right),
	\label{eq:intop-2}\\
	\mathcal{O}^{(QQ)_6}_{\kappa, \lambda}\left(t, \bm{r}, \bm{R}\right)&=\nonumber\\
	&\hspace{-1.7 cm}\psi^T\left(t, \bm{R}+\bm{r}/2\right)\phi^T(t; \bm{R}+\bm{r}/2,\bm{R})P^{\alpha \dag}_{\kappa, \lambda}H_{\bar{6},\,\kappa}^{\alpha,\,\sigma}\left(t, \bm{R}\right)\underline{\Sigma}^\sigma\,\phi\left(t; \bm{R}, \bm{R}-\bm{r}/2\right)\psi\left(t,  \bm{R}-\bm{r}/2\right),
	\label{eq:intop-3}
\end{align}
where  $P^{\alpha}_{\kappa, \lambda}$ are the projectors that are given in Appendix~\ref{app:projectors} (with $\alpha$ as the vector or spin index); 
the $SU(3)$ generators $T^a\,\left(a=1,\cdots,8\right)$ project on a color octet $\left(Q\bar{Q}\right)_8$ state,  
while $\underline{T}^\ell\,\left(\ell = 1, 2, 3\right)$ on a color triplet $\left(\bar{Q}\bar{Q}\right)_{3}$ 
or antitriplet $\left(QQ\right)_{\bar{3}}$ state, 
and $\underline{\Sigma}^\sigma\,\left(\sigma=1, \cdots, 6\right)$ on a color sextet $\left(QQ\right)_6$ or antisextet $\left(\bar{Q}\bar{Q}\right)_{\bar{6}}$ state.
The  tensors $\underline{T}^\ell$  and $\underline{\Sigma}^\sigma$, which are respectively antisymmetric and symmetric, are defined in Appendix~\ref{app:group}. 
The Wilson line $\phi\left(\bm{x}, \bm{y}\right)$ is
\begin{align}
	\phi\left(t; \bm{x}, \bm{y}\right)=\mathcal{P}\exp\left[-ig\int_{\bm{y}}^{\bm{x}}\,d\bm{z}\cdot \bm{A}\left(t, \bm{z}\right)\right]
\end{align}
where $\mathcal{P}$ denotes the path ordering. 
The operators $H_{8,\,\kappa}^{\alpha,\,a}(\bm{R})$  are related to gluelump operators 
when the LDF are gluons and to adjoint mesons when the LDF are light quarks~\cite{Foster:1998wu,Michael:1985ne, Campbell:1985kp}.
The operators $H_{3,\,\kappa}^{\alpha,\,\ell}(\bm{R})$ and $H_{\bar{6},\,\kappa}^{\alpha,\,a}(\bm{R})$ are related to baryons and tetraquarks. 
These relations will be spelled out in Sec.\ref{subsec:potential}. 
The operators transform under color gauge transformations $\mathcal{U}\left(t, \bm {R}\right)$ as follows
\begin{align}
	& H_{8,\,\kappa}^{\alpha,\,a}\left(t, \bm{R}\right)\, T^a\longrightarrow \mathcal{U}\, H_{8,\,\kappa}^{\alpha,\,a}\left(t,\,\bm{R}\right)\,T^a\,\mathcal{U}^\dagger,\label{eq:HnQQbar}\\
	& H_{3,\,\kappa}^{\alpha,\,\ell}\left(t, \bm{R}\right)\underline{T}^\ell\longrightarrow 
	\mathcal{U}^*\, H_{3,\,\kappa}^{\alpha,\,\ell}(t, \bm{R})\,\underline{T}^\ell\,\mathcal{U}^\dagger, 
	\label{eq:HnQQT}\\
	& H_{\bar{6},\,\kappa}^{\alpha,\,\sigma}\left(t, \bm{R}\right)\underline{\Sigma}^\sigma\longrightarrow 
	\mathcal{U}^*\, H_{\bar{6},\,\kappa}^{\alpha,\,\sigma}(t, \bm{R})\,\underline{\Sigma}^\sigma\,\mathcal{U}^\dagger, 
	\label{eq:HnQQS}
\end{align}
where we have used the relation: $\mathcal{U}\mathcal{U}^{\dagger}=\mathcal{U}^*\mathcal{U}^T=\mathbb{1}$.
The quarkonium static energy is obtained by substituting $P^\alpha_{\kappa, \lambda}\,H_{8,\,\kappa}^{\alpha,\,a}(t, \bm{R})\,T^a$ with the identity matrix, 
which leads to an ordinary static Wilson loop in the right-hand side of Eq.~\eqref{eq:En}.

\begin{table}[h!]
	\begin{tabular}{||c|c|c|c|c||}
		\hline\hline
		\multirow{2}{*}{\hspace{2pt}$\begin{array}{c} \Lambda^{\sigma}_{\eta}\end{array}$\hspace{2pt}} & \multirow{2}{*}{\hspace{2pt}$\begin{array}{c} k^{PC}\end{array}$\hspace{2pt}} & \multirow{2}{*}{\hspace{2pt}$\begin{array}{c} \text{Representation}\end{array}$\hspace{2pt}} &\multirow{2}{*}{\hspace{2pt}$\begin{array}{c} \text{ Operator Examples}\\H^{\alpha, a}_{8, \kappa} T^a\end{array}$\hspace{2pt}}&\multirow{2}{*}{\hspace{2pt}$\begin{array}{c} \text{ Projectors}\\P^{\alpha}_{\kappa\lambda} \end{array}$\hspace{2pt}}\\
		&  &  & & \\
		\hline\hline
		$\Sigma_g^+$ & $0^{++}$ & scalar & $\mathbbm{1}$\footnote{Note that $P^\alpha_{\kappa, \lambda}\,H_{8,\,\kappa}^{\alpha,\,a}(t, \bm{R})\,T^a = \mathbbm{1}$ identifies the color singlet $\left(Q\bar{Q}\right)_1$ state corresponding to quarkonium.} & $1$ \\
		$\Sigma_u^+$ & $0^{+-}$ & scalar & $\bm{D}\cdot\bm{E}$ &$1$\\
		$\Sigma_g^-$ & $0^{--} $ & pseudoscalar & $[\bm{E}\cdot,\bm{B}]$ & $1$\\
		$\Sigma_u^-$ & $0^{-+} $& pseudoscalar &  $\{\bm{E}\cdot,\bm{B}\}$ & $1$\\
		$\displaystyle\left\{\Sigma_g^+,\Pi_g\right\}$ & $1^{--}$ & vector & $E^i$ & $\{\hat{r}^i, \hat{r}^i_{\pm}\}$\\
		$\displaystyle\left\{\Sigma_u^+,\Pi_u\right\}$ & $1^{-+}$ & vector & $\left([\bm{E}\times,\bm{B}]\right)^i$ & $\{\hat{r}^i, \hat{r}^i_{\pm}\}$\\
		$\displaystyle\left\{\Sigma_g^-,\Pi_g\right\}$ & $1^{++}$ & pseudovector & $\left(\bm{D}\times[\bm{E}\times,\bm{B}]\right)^i$ & $\{\hat{r}^i, \hat{r}^i_{\pm}\}$ \\
		$\displaystyle\left\{\Sigma_u^-,\Pi_u\right\}$ & $1^{+-}$& pseudovector  & $B^i$ & $\{\hat{r}^i, \hat{r}^i_{\pm}\}$\\
		$\displaystyle\left\{\Sigma_g^+,\Pi_g,\Delta_g\right\}$ & $2^{++}$ & tensor & $E_{\{i}E_{j\}}$ & $\{\hat{r}^i \hat{r}^j, \frac{1}{2}\left(\hat{r}^i\hat{r}_{\pm}^j+\hat{r}_{\pm}^i\hat{r}^j\right), \hat{r}_{\pm}^i \hat{r}_{\pm}^j\}$ \\
		$\displaystyle\left\{\Sigma_u^+,\Pi_u,\Delta_u\right\}$ & $2^{+-}$ & tensor & $D_{\{i}E_{j\}}$ & $\{\hat{r}^i \hat{r}^j, \frac{1}{2}\left(\hat{r}^i\hat{r}_{\pm}^j+\hat{r}_{\pm}^i\hat{r}^j\right), \hat{r}_{\pm}^i \hat{r}_{\pm}^j\}$\\
		$\displaystyle\left\{\Sigma_g^-,\Pi_g,\Delta_g\right\}$ & $2^{--}$ & pseudotensor & $D_{\{i}B_{j\}}$ & $\{\hat{r}^i \hat{r}^j, \frac{1}{2}\left(\hat{r}^i\hat{r}_{\pm}^j+\hat{r}_{\pm}^i\hat{r}^j\right), \hat{r}_{\pm}^i \hat{r}_{\pm}^j\}$ \\
		$\displaystyle\left\{\Sigma_u^-,\Pi_u,\Delta_u\right\}$ & $2^{-+}$ & pseudotensor  & $\{E_{\{i},B_{j\}}\}$ & $\{\hat{r}^i \hat{r}^j, \frac{1}{2}\left(\hat{r}^i\hat{r}_{\pm}^j+\hat{r}_{\pm}^i\hat{r}^j\right), \hat{r}_{\pm}^i \hat{r}_{\pm}^j\}$\\
		\hline
	\end{tabular}
	\caption{In the first column, we list the BO quantum numbers $\Lambda_\eta^\sigma$ for the hybrid static states. The second column shows the $k^{PC}$ quantum numbers that correspond to these BO quantum numbers in the short distance limit. For convenience, we have grouped together within curly brackets those static states that belong to the same short-distance multiplet. The representation in the third column refers to the value of the reflection quantum number $\sigma$ [or $\sigma_T$ as defined in Eq.~\eqref{eq:reflection}], which is applicable only for $\Sigma$. In the fourth column, we list the LDF operators, and in the fifth column, we show the projection vectors that, after combining with the corresponding LDF operators in the fourth column, give the interpolating operators in Eq.~\eqref{eq:intop-1}. Substituting the interpolating operator into Eq.~\eqref{eq:En} gives the static energy corresponding to the BO quantum number $\Lambda^\sigma_\eta$. When we list two or more projectors within curly brackets, their order corresponds to the order of the $\Lambda^\sigma_\eta$ states in the first column. In the fourth column, the curly brackets around the indices stand for the symmetric traceless tensor: $U_{\{i}V_{j\}}\equiv(U_iV_j+U_jV_i)/2-\delta_{ij}\bm{U}\cdot\bm{V}$. Up to mass dimension $3$, the table agrees with the one in~\cite{Brambilla:1999xf}. Note that we are using a Cartesian basis to write the LDF operators and projectors; the indices are $i,\,j = 1, 2, 3$ and the projection vectors $\hat{{\bm r}}_{\pm}$ are given in Eq.~\eqref{eq:P_1}. 
	}
	\label{tab:Reps}
\end{table}

\subsubsection{Hybrids and $Q\bar{Q}$ tetraquarks and pentaquarks}
In the case of heavy quark-antiquark $Q\bar{Q}$ pairs in a color octet configuration, 
the color singlet states that we consider are quarkonium hybrids $\left(Q\bar{Q}g\right)$, quarkonium tetraquarks $\left(Q\bar{Q}q\bar{q}\right)$, and quarkonium pentaquarks $\left(Q\bar{Q}qqq\right)$. 
For hybrids, the LDF operator  $H_{8,\,\kappa}^{\alpha,\,a}$  are given in Table \ref{tab:Reps}.

\begin{table}[t!]
	\begin{tabular}{||c|c|c|c|c||}
		\hline\hline
		\multirow{2}{*}{\hspace{2pt}$\begin{array}{c} \Lambda^{\sigma}_{\eta}\end{array}$\hspace{2pt}} & \multirow{2}{*}{\hspace{2pt}$\begin{array}{c} k^{PC}\end{array}$\hspace{2pt}} & \multirow{2}{*}{\hspace{2pt}$\begin{array}{c} \text{Representation}\end{array}$\hspace{2pt}} &\multirow{2}{*}{\hspace{2pt}$\begin{array}{c} \text{ Operator Examples}\\H^{\alpha, a}_{8,\,\kappa} \left(I=0\right) \end{array}$\hspace{2pt}}&\multirow{2}{*}{\hspace{2pt}$\begin{array}{c} \text{ Projectors}\\P^{\alpha}_{\kappa\lambda} \end{array}$\hspace{2pt}}\\
		&  &  & & \\
		\hline\hline
		$\Sigma_g^+$ & $0^{++}$ & scalar & $\bar{q} T^a q$ & $1$\\
		$\Sigma_u^-$ & $0^{-+}$ & pseudoscalar & $\bar{q}\gamma^5 T^a q$ & $1$\\
		$\displaystyle\left\{\Sigma_g^+,\Pi_g\right\}$ & $1^{--}$ & vector & $\bar{q}\,{\bm \gamma}^i\,T^a q$ & $\{\hat{r}^i, \hat{r}^i_{\pm}\}$\\
		$\displaystyle\left\{\Sigma_g^-,\Pi_g\right\}$ & $1^{++}$ & pseudovector & $\bar{q}\,{\bm \gamma}^i\,\gamma^5\,T^a q$ & $\{\hat{r}^i, \hat{r}^i_{\pm}\}$ \\
		$\displaystyle\left\{\Sigma_u^-,\Pi_u\right\}$ & $1^{+-}$& pseudovector  & $\bar{q}\,\left({\bm \gamma}\times{\bm \gamma}\right)^i\,\gamma^5\,T^a q$ & $\{\hat{r}^i, \hat{r}^i_{\pm}\}$\\
		\hline
	\end{tabular}
	\caption{ In the first column, we list the BO quantum numbers $\Lambda_\eta^\sigma$ for the quarkonium tetraquark $Q\bar{Q}q\bar{q}$ static states. The second column shows the $k^{PC}$ quantum numbers that correspond to these BO quantum numbers in the short distance limit. For convenience, we have grouped together within curly brackets those static states that belong to the same short-distance multiplet. The representation in the third column refers to the value of the reflection quantum number $\sigma$ [or $\sigma_T$ as defined in Eq.~\eqref{eq:reflection}], which is only applicable for $\Sigma$. In the fourth column, we list the isopsin singlet ($I=0$) LDF operators. The isospin $I=1$ operators can be obtained by adding $\bm{e}_{I_3}\cdot\bm{\tau}$ in between the light quark fields. In the fifth column, we show the projection vectors that, after combining with the corresponding LDF operators in the fourth column, give the interpolating operators in  Eq.~\eqref{eq:intop-1}. Substituting the interpolating operator into Eq.~\eqref{eq:En} gives the static energy corresponding to the BO quantum number $\Lambda^\sigma_\eta$. When we list two or more projectors within curly brackets, their order corresponds to the order of the $\Lambda^\sigma_\eta$ states in the first column. Also here we are using a Cartesian basis to write the LDF operators and projectors; the indices are $i,\,j = 1, 2, 3$ and the projection vectors $\hat{{\bm r}}_{\pm}$ are given in Eq.~\eqref{eq:P_1}. 
	}
	\label{tab:Reps-QQbar}
\end{table}

For quarkonium tetraquark states (Table~\ref{tab:QQbarqqbar}), the LDF operator $H_{8,\,\kappa}^{\alpha,\,a}$  consists of light-quark fields 
with
quantum number $\kappa$, which includes the spin $ k$ and flavor (isospin)  quantum numbers. 
Examples of light-quark operators interpolating isospin $I=0$  tetraquark $Q\bar{Q}q\bar{q}$ states are given in Table~\ref{tab:Reps-QQbar}.
The isospin $I=1$ tetraquark states can be interpolated by similar operators by just adding $\bm{e}_{I_3}\cdot\bm{\tau}$ in between the light quark fields, where $\tau^i$ are the isospin Pauli matrices and $\bm{e}_{I_3}$ are the spherical basis coordinate vectors:
\begin{equation}
	{\bm e}_0=\left(0, 0, 1\right),\qquad {\bm e}_{-1}=-\left(1, i, 0\right)/\sqrt{2},\qquad {\bm e}_{+1}=\left(1, -i, 0\right)/\sqrt{2}.
	\label{eq:eI3}
\end{equation}
The same set of operators as in Table~\ref{tab:Reps-QQbar} has been obtained in Ref.~\cite{Soto:2020xpm}. 

For quarkonium pentaquark states  with $I=1/2$, $I_3=\pm 1/2$, and $k=1/2$, the light-quark interpolating operator is  given by 
\begin{align}
	&H^{\alpha,\,a}_{8,I_3=\pm 1/2, (1/2)^{+}}(t,\bm{x})=\nonumber\\
	&\hspace{1.5 cm}\Bigg[\left(\delta_{\alpha\beta_1}\sigma^2_{\beta_2\beta_3}+\delta_{\alpha\beta_2}\sigma^2_{\beta_1\beta_3}+\delta_{\alpha\beta_3}\sigma^2_{\beta_1\beta_2}\right)\left(\delta_{I_3f_1}\tau^2_{f_2f_3}+\delta_{I_3f_2}\tau^2_{f_1f_3}+\delta_{I_3f_3}\tau^2_{f_1f_2}\right)\left(T_2\right)^a_{l_1,l_2,l_3}\nonumber\\
	&\hspace{1.5 cm}+\left(\delta_{\alpha\beta_1}\sigma^2_{\beta_2\beta_3}+\delta_{\alpha\beta_2}\sigma^2_{\beta_3\beta_1}+\delta_{\alpha\beta_3}\sigma^2_{\beta_2\beta_1}\right)\left(\delta_{I_3f_1}\tau^2_{f_2f_3}+\delta_{I_3f_2}\tau^2_{f_3f_1}+\delta_{I_3f_3}\tau^2_{f_2f_1}\right)\left(T_3\right)^a_{l_1,l_2,l_3}\nonumber\\
	&\hspace{1.5 cm}+\left(\delta_{\alpha\beta_1}\sigma^2_{\beta_3\beta_2}+\delta_{\alpha\beta_2}\sigma^2_{\beta_3\beta_1}+\delta_{\alpha\beta_3}\sigma^2_{\beta_1\beta_2}\right)\left(\delta_{I_3f_1}\tau^2_{f_3f_2}+\delta_{I_3f_2}\tau^2_{f_3f_1}+\delta_{I_3f_3}\tau^2_{f_1f_2}\right)\left(T_1\right)^a_{l_1,l_2,l_3}\Bigg]\nonumber\\
	&\hspace{2.5 cm}\left(P_+q_{l_1f_1}(t,\bm{x})\right)^{\beta_1} \left(P_+q_{l_2f_2}(t,\bm{x})\right)^{\beta_2} \left(P_+q_{l_3f_3}(t,\bm{x})\right)^{\beta_3}\,,
	\label{eq:OpQQbarqqq}
\end{align}
where repeated indices have been summed over, $\alpha, \beta_i\,\left(i=1, 2, 3\right)$ are the spin or vector indices, $f_i\,\left(i=1, 2, 3\right)$ are the isospin or flavor index, $l_i\,\left(i=1, 2, 3\right)$ are the color index, and the color matrices $\left(T_i\right)^a\,\left(i=1, 2, 3\right)$ are given by Eq.~\eqref{eq:T1T2T3}. 
The projector $P_{+}=\left(1+\gamma^0\right)/2$ is required due to positive parity \cite{Soto:2020xpm, Sadl:2021bme}, $\sigma^2$ is the antisymmetric spin  Pauli matrix, and $\tau^2$ is the antisymmetric isospin  Pauli matrix .

\subsubsection{Doubly-heavy  baryons and $Q{Q}$ tetraquarks and pentaquarks}
In the case of heavy quark-quark $QQ$ pairs, the color singlet states that we consider are the doubly heavy baryons $\left(QQq\right)$, doubly heavy tetraquarks $\left(QQ\bar{q}\bar{q}\right)$ and doubly heavy pentaquarks $\left(QQqq\bar{q}\right)$. 
For doubly heavy baryons with color antitriplet $\left(QQ\right)_{\bar{3}}$ pair, the interpolating operator is given in Table~\ref{tab:Reps-QQq}. 
The same set of operators  has been obtained in Ref.~\cite{Soto:2020xpm}.

\begin{table}[h!]
	\begin{tabular}{||c|c|c|c|c||}
		\hline\hline
		\multirow{2}{*}{\hspace{2pt}$\begin{array}{c} \text {BO quantum \#}\\D_{\infty h}\end{array}$\hspace{2pt}} & \multirow{2}{*}{\hspace{2pt}$\begin{array}{c} k^{P}\end{array}$\hspace{2pt}} & \multirow{2}{*}{\hspace{2pt}$\begin{array}{c} \left(k-1/2\right)\\ \text{Representation}\end{array}$\hspace{2pt}} &\multirow{2}{*}{\hspace{2pt}$\begin{array}{c} \text{ Operator Examples}\\H^{\alpha, \ell}_{3,\kappa} \end{array}$\hspace{2pt}}&\multirow{2}{*}{\hspace{2pt}$\begin{array}{c} \text{ Projectors}\\P^{\alpha}_{\kappa\lambda} \end{array}$\hspace{2pt}}\\
		&  &  & & \\
		\hline\hline
		$\left(1/2\right)_g$ & $\left(1/2\right)^{+}$ & scalar & $\left[P_+q^{a}\right]^\alpha$ & $P^{\alpha}_{1/2,\,\pm1/2}$\\
		$\left(1/2\right)_u^{\prime}$ & $\left(1/2\right)^{-}$ & pseudoscalar & $\left[P_+\gamma^5q^a\right]^\alpha$ & $P^{\alpha}_{1/2,\,\pm1/2}$\\
		$\displaystyle\left\{\left(1/2\right)_u, \left(3/2\right)_u\right\}$ & $\left(3/2\right)^{-}$ & vector & ${\cal C}^{3/2\,\alpha}_{1\,m\,1/2\,\beta}\left[\left(\bm{e}_{m}\cdot\bm{D}\right) \left(P_+q^a\right)^\beta\right]$ & $\{P^{\alpha}_{3/2,\,\pm1/2},\,P^{\alpha}_{3/2,\,\pm3/2}\}$\\
		\hline
	\end{tabular}
	\caption{ In the first column, we list the BO quantum numbers ($D_{\infty h}$ representation) denoted as $\left(k\right)_{\eta}$, where $\eta = \left(g, u\right)$ indicates parity, for the doubly heavy baryon $QQq$ static states. The second column shows the $k^{P}$ quantum numbers that correspond to the BO quantum numbers in the short distance limit. For convenience, we have grouped together within curly brackets those static states that belong to the same short-distance multiplet. The representation in the third column refers to whether $\left(k-1/2\right)^P$ is a tensor or pseudotensor. In the fourth column, we list the LDF operators. In the fifth column, we show the projectors, which, after combining with the corresponding LDF operator in the fourth column, give the interpolating operators in Eq.~\eqref{eq:intop-2}. Substituting the interpolating operator into Eq.~\eqref{eq:En} gives the static energy. When we list two or more projectors within curly brackets, their order corresponds to the order of the $\Lambda^\sigma_\eta$ states in the first column. Note that we are using a spherical basis $\left(\alpha = k, \cdots, 0, \cdots -k\right)$ to write the LDF operators and projectors. The projectors $P^{\alpha}_{1/2, \pm 1/2}$ are given in Eqs.~\eqref{eq:P1/2-1} and \eqref{eq:P1/2-2} and $P^{\alpha}_{3/2, \pm 1/2}$ and $P^{\alpha}_{3/2, \pm 3/2}$ can be obtained from Eq.~\eqref{eq:Pdef2}. ${\cal C}^{3/2\,\alpha}_{1\,m\,1/2\,\beta}$ is the Clebsch--Gordan coefficient. We list only the operators for the lowest $k^{P}$.}
	\label{tab:Reps-QQq}
\end{table}

\begin{table}[h!]
	\begin{tabular}{||c|c|c|cc||c||}
		\hline\hline
		\multirow{2}{*}{\hspace{2pt}$\begin{array}{c} \Lambda^{\sigma}_{\eta}\end{array}$\hspace{2pt}} & \multirow{2}{*}{\hspace{2pt}$\begin{array}{c} k^{P}\end{array}$\hspace{2pt}} & \multirow{2}{*}{\hspace{2pt}$\begin{array}{c} \text{Representation}\end{array}$\hspace{2pt}} &\multicolumn{2}{c||}{\multirow{2}{*} {\hspace{2pt}$\begin{array}{c} \text{ Operator Examples}\\ \!\!\! H^{\alpha, \ell}_{3,\kappa} \left(I=0\right) ~~~~H^{\alpha, \sigma}_{\bar{6},\kappa} \left(I=0\right) \end{array}$\hspace{2pt}}}&\multirow{2}{*}{\hspace{2pt}$\begin{array}{c} \text{ Projectors}\\P^{\alpha}_{\kappa\lambda} \end{array}$\hspace{2pt}} \\
		&  &  & & & \\
		\hline\hline
		$\Sigma_g^+$ & $0^{+}$ & scalar & $\bar{q} \gamma^5 \gamma^2 \tau^2\,\underline{T}^a\, q^*$  &\,$\cdots$ & 1\\
		$\Sigma_u^-$ & $0^{-}$ & pseudoscalar & $\cdots$ & $\bar{q} \gamma^2 \tau^2\,\underline{\Sigma}^a\, q^*$ & 1\\
		$\displaystyle\left\{\Sigma_u^+,\Pi_u\right\}$ & $1^{-}$ & vector & $\bar{q}\bm{\gamma}^{i}\,\gamma^5\gamma^2 \tau^2\,\underline{T}^a\, q^*$&\,  $\bar{q}\bm{\gamma}^{i}\,\gamma^5\gamma^2 \tau^2\,\underline{\Sigma}^a\, q^*$ & $\{\hat{r}^i, \hat{r}^i_{\pm}\}$\\
		$\displaystyle\left\{\Sigma_g^-,\Pi_g\right\}$ & $1^{+}$ & pseudovector & $\cdots$ &\,$\bar{q}\bm{\gamma}^{i}\,\gamma^2 \tau^2 \,\underline{\Sigma}^a\, q^*$ & $\{\hat{r}^i, \hat{r}^i_{\pm}\}$\\
		\hline
	\end{tabular}
	\caption{In the first column, we list the BO quantum numbers $\Lambda_\eta^\sigma$ for the isospin singlet $(I=0)$ doubly heavy tetraquark $QQ\bar{q}\bar{q}$ static states. The second column shows the $k^{P}$ quantum numbers that correspond to the BO quantum numbers in the short distance limit. For convenience, we have grouped together within curly brackets those static states that belong to the same short-distance multiplet. The representation in the third column refers to the value of the reflection quantum number $\sigma$ [or $\sigma_T$ as defined in Eq.~\eqref{eq:reflection}], which is only applicable for $\Sigma$. In the fourth column, we list the $I=0$ LDF operators on the left for the color triplet and on the right for the color antisextet. A dotted entry means that that particular entry is not allowed due to the Pauli exclusion principle.\footnote{For quantum numbers $k^P = 0^-$ and $k^P = 1^-$, there can be one unit of orbital angular momentum between the light quark pair or between the heavy quark pair and the light quark pair. We consider the latter case as the states are expected to be lower in energy \cite{Manohar:2000dt, Narodetskii:2008pn}.} We list only the operators for the lowest $k^{P}$. In the fifth column, we show the projectors  which, after combining with the corresponding LDF  operators in the fourth column, give the interpolating operators in Eqs.~\eqref{eq:intop-2} and \eqref{eq:intop-3}. Substituting the interpolating operator into Eq.~\eqref{eq:En} gives the static energy corresponding to the BO quantum number $\Lambda^\sigma_\eta$. When we list two or more projectors within curly brackets, their order corresponds to the order of the $\Lambda^\sigma_\eta$ states in the first column. We are using a Cartesian basis to write the LDF operators and projectors; the indices are $i,\,j = 1, 2, 3$ and the projection vectors $\hat{{\bm r}}_{\pm}$ are given in Eq.~\eqref{eq:P_1}. }
	\label{tab:Reps-QQ-I_0}
\end{table}

\begin{table}[h!]
	\centering
	\resizebox{0.95\columnwidth}{!}{%
		\begin{tabular}{||c|c|c|cc||c||}
			\hline\hline
			\multirow{2}{*}{\hspace{2pt}$\begin{array}{c} \Lambda^{\sigma}_{\eta}\end{array}$\hspace{2pt}} & \multirow{2}{*}{\hspace{2pt}$\begin{array}{c} k^{P}\end{array}$\hspace{2pt}} & \multirow{2}{*}{\hspace{2pt}$\begin{array}{c} \text{Representation}\end{array}$\hspace{2pt}} &\multicolumn{2}{c||}{\multirow{2}{*} {\hspace{2pt}$\begin{array}{c} \text{ Operator Examples}\\  \!\!\! H^{\alpha, \ell}_{3,\kappa} \left(I=1\right) ~~~~~~~~~~~~~~~~H^{\alpha, \sigma}_{\bar{6},\kappa} \left(I=1\right) \end{array}$\hspace{2pt}}}&\multirow{2}{*}{\hspace{2pt}$\begin{array}{c} \text{ Projectors}\\P^{\alpha}_{\kappa\lambda} \end{array}$\hspace{2pt}} \\
			&  &  & &  &\\
			\hline\hline
			$\Sigma_g^+$ & $0^{+}$ & scalar & $\cdots$  &\,$\bar{q} \gamma^5 \gamma^2 \bm{e}_{I_3}\cdot\left(\tau^2\bm{\tau}\right)\,\underline{\Sigma}^a\, q^*$ & 1\\
			$\Sigma_u^-$ & $0^{-}$ & pseudoscalar & $\bar{q} \gamma^2 \bm{e}_{I_3}\cdot\left(\tau^2\bm{\tau}\right)\,\underline{T}^a\, q^*$ &\,$\cdots$ & 1\\
			$\displaystyle\left\{\Sigma_u^+,\Pi_u\right\}$ & $1^{-}$ & vector & $\bar{q}\bm{\gamma}^{i}\,\gamma^5\gamma^2 \bm{e}_{I_3}\cdot\left(\tau^2\bm{\tau}\right)\,\underline{T}^a\, q^*$ &\,$\bar{q}\bm{\gamma}^{i}\,\gamma^5\gamma^2 \bm{e}_{I_3}\cdot\left(\tau^2\bm{\tau}\right)\,\underline{\Sigma}^a\, q^*$ & $\{\hat{r}^i, \hat{r}^i_{\pm}\}$\\
			$\displaystyle\left\{\Sigma_g^-,\Pi_g\right\}$ & $1^{+}$ & pseudovector & $\bar{q}\bm{\gamma}^{i}\,\gamma^2 \bm{e}_{I_3}\cdot\left(\tau^2\bm{\tau}\right) \,\underline{T}^a\, q^*$ &\,$\cdots$ & $\{\hat{r}^i, \hat{r}^i_{\pm}\}$\\
			\hline
	\end{tabular}}
	\caption{In the first column, we list the BO quantum numbers $\Lambda_\eta^\sigma$ for the isospin triplet $(I=1)$ doubly heavy tetraquark $QQ\bar{q}\bar{q}$ static states. The second column shows the $k^{P}$ quantum numbers that correspond to these BO quantum numbers in the short distance limit. For convenience, we have grouped together within curly brackets those static states that belong to the same short-distance multiplet. The representation in the third column refers to the value of the reflection quantum number $\sigma$ [or $\sigma_T$ as defined in Eq.~\eqref{eq:reflection}], which is only applicable for $\Sigma$. In the fourth column, we list the $I=1$ LDF operators on the left for the color triplet and on the right for the color antisextet. A dotted entry means that that particular entry is not allowed due to the Pauli exclusion principle.\footnote{For quantum numbers $k^P = 0^-$ and $k^P = 1^-$, there can be one unit of orbital angular momentum between the light quark pair or between the heavy quark pair and the light quark pair. We consider the latter case as the states are expected to be lower in energy \cite{Manohar:2000dt, Narodetskii:2008pn}.} We list only the operators for the lowest $k^{P}$. In the fifth column, we show the projectors that, after combining with the corresponding LDF operators in the fourth column, give the interpolating operators in Eqs.~\eqref{eq:intop-2} and \eqref{eq:intop-3}. Substituting the interpolating operator into Eq.~\eqref{eq:En} gives the static energy corresponding to the BO quantum number $\Lambda^\sigma_\eta$. When we list two or more projectors within curly brackets, their order corresponds to the order of the $\Lambda^\sigma_\eta$ states in the first column. Also here we are using a Cartesian basis to write the LDF operators and projectors; the indices are $i,\,j = 1, 2, 3$ and the projection vectors $\hat{{\bm r}}_{\pm}$ are  given in Eq.~\eqref{eq:P_1}. }
	\label{tab:Reps-QQ-I_1}
\end{table}

For doubly heavy tetraquark states  in Table~\ref{QQqqnumbers} with color antitriplet $\left(QQ\right)_{\bar{3}}$ and sextet $\left(QQ\right)_{6}$ pairs, the light-quark interpolating isospin $I=0$ and $I=1$ operators that transform like Eqs.~\eqref{eq:HnQQT} 
and~\eqref{eq:HnQQS} under gauge transformations are given in Tables~\ref{tab:Reps-QQ-I_0} and \ref{tab:Reps-QQ-I_1}, respectively.
The Dirac matrix $\gamma^2$ is required by Lorentz invariance \cite{Peskin:1995ev}.  
The same set of operators as in Tables~\ref{tab:Reps-QQ-I_0} and~\ref{tab:Reps-QQ-I_1}, 
after correcting for a missing isospin factor in the color antitriplet $\left(QQ\right)_{\bar{3}}$ case, has been obtained in Ref.~\cite{Soto:2020xpm}.

For doubly heavy pentaquark states with color antitriplet $\left(QQ\right)_{\bar{3}}$ pair and quantum numbers: $I=1/2$, $I_3=\pm 1/2$, and $k=1/2$, the light-quark interpolating operator is 
\begin{align}
	&H^{\alpha,\,\ell}_{3,I_3=\pm1/2, (1/2)^{+}}(t,\bm{x})=\nonumber\\
	&\hspace{1.5 cm}\Bigg[\left(\delta_{\alpha\beta_1}\sigma^2_{\beta_2\beta_3}+\delta_{\alpha\beta_2}\sigma^2_{\beta_3\beta_1}+\delta_{\alpha\beta_3}\sigma^2_{\beta_2\beta_1}\right)\left(\delta_{I_3f_1}\tau^2_{f_2f_3}+\delta_{I_3f_2}\tau^2_{f_3f_1}+\delta_{I_3f_3}\tau^2_{f_2f_1}\right)\underline{T}^i_{l_1,l_2}\underline{T}^\ell_{i, l_3}\Bigg]\nonumber\\
	&\hspace{2.5 cm}\left(P_+q_{l_1f_1}(t,\bm{x})\right)^{\beta_1} \left(P_+q_{l_2f_2}(t,\bm{x})\right)^{\beta_2} \left(\bar{q}_{l_3f_3}(t,\bm{x})P_-\right)^{\beta_3},
	\label{eq:OpQQqqqbar}
\end{align}
where $P_-=\left(1-\gamma^0\right)/2$ and repeated indices have been summed over.

\subsubsection{Matching}
The matching procedure from NRQCD to BOEFT relates the interpolating operator for exotic hadrons in NRQCD, Eqs.~\eqref{eq:intop-1}-\eqref{eq:intop-3}, to the field $\Psi_{\kappa\lambda}$ in BOEFT:
\begin{align}
	\mathcal{O}^h_{\kappa, \lambda}(t,\,\bm{r},\,\bm{R})\longrightarrow \sqrt{Z_{\kappa,\,\lambda}^{h}}\,\Psi^h_{\kappa, \lambda}(t,\,\bm{r},\,\bm{R})\,,\quad h\in \left\{Q\bar{Q},\,QQ\right\},
	\label{eq:match-1}
\end{align}
where $Z_{\kappa,\,\lambda}^{h}$ is a field normalization factor, in general a function of ${\bm r}$ and ${\bm R}$.\footnote{
	In general, the field normalization factor depends also on the relative momentum ${\bm p}$ 
	and the center of mass momentum ${\bm P}$; 
	however, we do not mention this here as we are considering the static limit.}
Based on Eq.~\eqref{eq:match-1}, the matching condition between NRQCD and BOEFT correlators reads
\begin{align}
	&\langle \mathrm{vac}|\mathcal{O}^h_{\kappa, \lambda}(T/2,\,\bm{r},\,\bm{R})\mathcal{O}_{\kappa, \lambda}^{h\dag}(-T/2,\,\bm{r},\,\bm{R})|\mathrm{vac}\rangle=\nonumber\\
	&\hspace{3.0 cm}\sqrt{Z_{\kappa,\,\lambda}^{h}}\langle \mathrm{vac}|\Psi^h_{\kappa, \lambda}(T/2,\, \bm{r},\,\bm{R})\Psi_{\kappa, \lambda}^{h\dag}(-T/2,\,\bm{r},\,\bm{R})|\mathrm{vac}\rangle \sqrt{Z_{\kappa,\,\lambda}^{h\dagger}}\,.\label{eq:corr}
\end{align}
From \eqref{eq:En}, it follows that the BOEFT static potentials $V^{\left(0\right)}_{\kappa|\lambda|}$, defined in Eq.~\eqref{eq:VQQ}, can be determined from the static energies $E_{\kappa, |\lambda|}^{\left(0\right)}$:
\begin{equation}
	E_{\kappa, |\lambda|}^{\left(0\right)}\left(r\right) = V_{\kappa, |\lambda|}^{\left(0\right)}\left(r\right).
	\label{eq:pot-match}
\end{equation}

\subsection{Static energies: behavior at short distances and form of the potential}
\label{subsec:potential}
In this subsection, we show how we can use weakly coupled pNRQCD (cf.\ Sec.~\ref{sec:effective}) to calculate the short range behavior of the static energies, which, in turn, corresponds to the  short range behavior of the static potential $V^{\left(0\right)}_{\kappa, |\lambda|}$ in Eq.~\eqref{eq:VQQ},
to be used in the BOEFT coupled Schr\"odinger equations.
In the limit of short distance $r$ between the heavy quarks, i.e.\ for $r \Lambda_{\rm QCD} \ll 1$, the soft scale related to the relative momentum transfer between the heavy quarks is perturbative. 
In weakly coupled pNRQCD, the degrees of freedom are color singlet $(S)$ and  color octet $(O^a)$ fields for heavy quark-antiquark pairs \cite{Brambilla:1999xf,Brambilla:2004jw}
or color antitriplet $(T^\ell)$ and color sextet $(\Sigma^\sigma)$ fields for heavy quark-quark pairs \cite{Brambilla:2005yk}.
Additionally, the effective theory incorporates low energy (ultrasoft) LDF (gluons and light quarks), which are multipole expanded with respect to the relative coordinate ${\bm r}$, and  depend only on the center of mass coordinate ${\bm R}$ and time~$t$.
Integrating out the soft scale generates a perturbative static potential, while integrating out also $\L \gg E$ generates nonperturbative short range corrections to the static potential.
The nonperturbative short range corrections to the static potential that emerge from the multipole expansion can be extracted from the large time limit of appropriate correlators.
Some of these correlators have already been computed in lattice QCD.

\subsubsection{$Q\bar{Q}$ systems}
\label{subsubsec:QQbar-match}
We consider here bound states of a $Q\bar{Q}$ pair and LDF.
If the heavy quarks are static, i.e. at zeroth order in a $1/m_Q$ expansion, the static energy of the system depends only on the distance $r$ between the heavy quark and antiquark.
At short distances $r$, the $Q\bar{Q}$ pair may be in a color singlet configuration, the corresponding field being $S$, and bind with LDF that are also in a color singlet configuration to form color singlet hadrons,
or it may be in a color octet configuration, the corresponding fields being $O^a$ with $a=1$, $2$, $\dots$, $8$, and bind with LDF that are in the adjoint representation of $SU(3)$ to form color singlet hadrons.
The first case describes the short-distance behavior of quarkonia $\left[\left(Q\bar{Q}\right)_1\right]$, quarkonia with glueballs $\left[\left(Q\bar{Q}\right)_1+g_1\right]$
and quarkonia with hadrons $\left[\left(Q\bar{Q}\right)_1+\left(q\bar{q}\right)_1\right]$ or $\left[\left(Q\bar{Q}\right)_1+\left(qqq\right)_1\right]$.
The static energy at short distances is the sum of the color singlet potential $V_s$, glueball or hadron masses, and higher-order corrections suppressed by powers of $\Lambda_{\rm QCD}r$.
The second case describes the short-distance behavior of quarkonium hybrids $\left[\left(Q\bar{Q}\right)_8+g_8\right]$, tetraquarks $\left[\left(Q\bar{Q}\right)_8+\left(q\bar{q}\right)_8\right]$
and pentaquarks $\left[\left(Q\bar{Q}\right)_8+\left(qqq\right)_8\right]$.
In this case, the static energy at short distances is the sum of the color octet potential $V_o$, gluelump or adjoint meson or adjoint baryon masses, respectively, 
and higher-order corrections suppressed by powers of $\Lambda_{\rm QCD}r$.
At leading order, the singlet and octet static potentials are of Coulombic form and read
\begin{equation}
	V_s(r) = -\frac{4}{3}\frac{\alpha_s}{ r}, \qquad\qquad  
	V_o(r) = \frac{\alpha_s}{ 6r},
	\label{eq:VsVo}
\end{equation}
where $\alpha_s$ is the strong coupling constant computed at a typical scale of order $1/r$.
The singlet and octet potentials up to next-to-next-to-next-to leading order in $\alpha_s$ can be found in Ref.~\cite{Anzai:2009tm,Smirnov:2009fh,Anzai:2013tja}.
From an effective field theory perspective, the potentials encode the contributions coming from the soft gluons.

In the following, we concentrate on the short-distance behavior of quarkonium hybrids, adjoint tetraquarks and pentaquarks.
The matching relation between the interpolating operators and BOEFT, Eq. \eqref{eq:match-1}, becomes at short distance 
\begin{align}
	\mathcal{O}_{\kappa, \lambda}^{(Q\bar{Q})_8}(t,\,\bm{r},\,\bm{R})\longrightarrow \sqrt{Z_{\kappa,\,\lambda}}\, O^a(t,\,\bm{r},\,\bm{R})\,P^{\alpha \dag}_{\kappa, \lambda}\,H^{\alpha, \,a}_{8,\,\kappa}(t,\,\bm{R})+{\cal O}\left(r\right),
	\label{eq:matchQQbar}
\end{align} 
where  $Z_{\kappa,\,\lambda}$ is the field normalization factor and $H^{\alpha, \,a}_{8,\,\kappa}$ is the interpolating operator made of light fields (light quarks and gluons).
For quarkonium hybrids, $H^{\alpha, \,a}_{8,\,\kappa}$ is given by the gluonic operators shown in Table~\ref{tab:Reps},
for quarkonium tetraquarks, it is given in Table~\ref{tab:Reps-QQbar}, 
and for quarkonium pentaquarks, it is given in Eq.~\eqref{eq:OpQQbarqqq}.
The operators on the right-hand side describe quarkonium hybrids, tetraquarks, and pentaquarks in BOEFT in the short-distance limit,
where $\Psi^{(Q\bar{Q})_8}_{\kappa, \lambda} \to O^a\,P^{\alpha \dag}_{\kappa, \lambda}\,H^{\alpha, \,a}_{8,\,\kappa}$.
Indeed BOEFT in the short distance limit coincides with weakly coupled pNRQCD,
and  gluelumps, adjoint mesons and adjoint baryons are the short distance limit of 
the LDF in static hybrids, tetraquarks and pentaquarks, respectively.\footnote{See Eq.~(20) in Ref.~\cite{Berwein:2015vca}}

Using the matching condition in Eq.~\eqref{eq:matchQQbar}, the two-point correlator in Eq.~\eqref{eq:En} is then given, up to corrections of order $\Lambda_{\rm QCD}^3r^2$ in the exponent, by
\begin{align}
	&\langle \mathrm{vac}|\mathcal{O}^{(Q\bar{Q})_8}_{\kappa, \lambda}(T/2,\,\bm{r},\,\bm{R})\mathcal{O}_{\kappa, \lambda'}^{(Q\bar{Q})_8\,\dag}(-T/2,\,\bm{r},\,\bm{R})|\mathrm{vac}\rangle\nonumber\\
	&\hspace{2cm} = \sqrt{Z_{\kappa,\,\lambda} Z^{\dag}_{\kappa,\,\lambda^{\prime}} } \,e^{-i V_o(r)T}\, P^{\alpha \dag}_{\kappa\lambda}\,P^{\alpha^\prime}_{\kappa\lambda^{\prime}} \nonumber\\
	&\hspace{3.5cm} \times \langle \mathrm{vac}|\,H^{\alpha, \,a}_{8,\,\kappa}(T/2,\,\bm{R})\,\phi^{ab}_{\mathrm{adj}}\left(T/2,-T/2\right)\,\,H^{\alpha^{\prime}, \,b \dag}_{8,\,\kappa}(-T/2,\,\bm{R})|\mathrm{vac}\rangle \nonumber\\
	&\hspace{2cm}  = \sqrt{Z_{\kappa,\,\lambda} Z^{\dag}_{\kappa,\,\lambda^{\prime}} } \,\delta_{\lambda\lambda'}\,e^{-i 
		\left(V_o(r)\,+\,\Lambda_{\kappa}\right)T},
	\label{eq:matchQQbar-2}
\end{align}
where to obtain the last equality we have used that the matrix element is diagonal in the Lorentz indices, the following Eq.~\eqref{eq:LDF-mass}, 
and the projection vector identity $P^{\dagger \alpha}_{\kappa,\,\lambda} P^{\alpha}_{\kappa,\,\lambda^{\prime}}  = \delta_{\lambda\lambda^{\prime}}$.
The adjoint static Wilson line $\phi^{ab}_{\mathrm{adj}}\left(t, t'\right)$ is given by 
\be
\phi^{ab}_{\mathrm{adj}}(t_1,t_2) \equiv {\mathcal{P}}\exp\left[-ig\int^{t_1}_{t_2}dt\, A_0^{\rm adj}(t, \bm{R})\right]_{ab}\,.
\label{eq:TWilson}
\ee
The temporal Wilson line in the LDF correlator in the third line in Eq.~\eqref{eq:matchQQbar-2} ensures the gauge invariance of the expression.
The LDF correlator can only be evaluated nonperturbatively since it depends on 
$\Lambda_\mathrm{QCD}$. 
The large time limit of the LDF correlator yields 
\begin{equation}
	\langle \mathrm{vac}|H^{\alpha, \,a }_{8,\,\kappa}(T/2,\,\bm{R})\,\phi^{ab}_{\mathrm{adj}}\left(T/2,-T/2\right)H^{\alpha^{\prime}, \,b \dag}_{8,\,\kappa}(-T/2,\,\bm{R})|\mathrm{vac}\rangle=\delta_{\alpha\alpha^{\prime}}\,e^{-i\Lambda_{\kappa}T},
	\label{eq:LDF-mass}
\end{equation}
where $\delta_{\alpha\alpha^{\prime}}$ comes from the normalization condition and depends on the LDF in consideration, $\Lambda_{\kappa}$ is the gluelump,
the adjoint meson or the adjoint baryon mass. As an example, the LDF correlators that 
yield the $1^{+-}$ gluelump mass $\Lambda^g_{1^{+-}}$, and the $0^{-+}$ and $1^{--}$ adjoint meson masses $\Lambda^a_{0^{-+}}$ and $\Lambda^a_{1^{--}}$
in the large time limit are\footnote{For clarity and to differentiate between gluelump ($g$) and adjoint meson ($a$), we denote their masses by $\Lambda^{i=\left(g,a\right)}_{\kappa}$ wherever necessary. No sum over the index $i$ is implied here.} 
\begin{align}
	\langle \mathrm{vac}|B^{i,\,a}(T/2,\,\bm{R})\,\phi^{ab}_{\mathrm{adj}}\left(T/2,-T/2\right)\,B^{i,\,b}(-T/2,\,\bm{R})|\mathrm{vac}\rangle&=e^{-i\Lambda^g_{1^{+-}}T},\nonumber\\
	\langle \mathrm{vac}|\left[\bar{q}\,\gamma^5\,T^a\, q\right](T/2,\,\bm{R})\,\phi^{ab}_{\mathrm{adj}}\left(T/2,-T/2\right)\,\left[\bar{q}\,\gamma^5\,T^a\, q\right](-T/2,\,\bm{R})|\mathrm{vac}\rangle&=e^{-i\Lambda^a_{0^{-+}}T},\nonumber\\
	\langle \mathrm{vac}|\left[\bar{q}\,\gamma^i\,T^a\, q\right](T/2,\,\bm{R})\,\phi^{ab}_{\mathrm{adj}}\left(T/2,-T/2\right)\,\left[\bar{q}\,\gamma^i\,T^a\, q\right](-T/2,\,\bm{R})|\mathrm{vac}\rangle&=e^{-i\Lambda^a_{1^{--}}T}.
	\label{eq:glue-adj-mass}
\end{align}

Based on Eq.~\eqref{eq:En}, taking the logarithm of both sides of  Eq.~\eqref{eq:matchQQbar-2}, and from the matching conditions in Eqs.~\eqref{eq:corr} and \eqref{eq:pot-match},
we obtain the following relation between the static energy computed in NRQCD (left), the short-distance potential in pNRQCD (middle),  and the BOEFT potential (right)
\begin{align}
	E^{(0)}_{\kappa, |\lambda|}(r) = V_o(r) + \Lambda_{\kappa} + {\cal O}(r^2) = V^{(0)}_{\kappa, |\lambda|}(r).
	\label{eq:QQbarpot-short}
\end{align}
Note that the static energy and the BOEFT potential are approximated by the middle equality only for small $r$.  
The ${\cal O}\left(r^2\right)$ terms arise from multipole expanding the pNRQCD interaction vertices between the heavy quark pair and the LDF.
Equation~\eqref{eq:QQbarpot-short} implies that at leading order in the multipole expansion, several static energies are degenerate since they only depend on the gluelump or adjoint meson or adjoint baryon quantum number $\kappa\equiv k^{PC}$ and not on $\lambda$.
This is seen very clearly in the lattice calculations of the hybrid static energies shown in Figs.~\ref{morning} and \ref{fig:hyb-Wagner}. 
The higher order corrections in the multipole expansion break this degeneracy starting from order $r^2$.
The repulsive octet behavior at short distance is also seen in lattice calculations.
The short-distance behavior of the BO-potential given in Eq.~\eqref{eq:QQbarpot-short} differs from the analytic form used to fit the tetraquark static potential from lattice QCD in Ref.~\cite{Prelovsek:2019ywc}. 

The gluelump masses are the energy levels associated with the binding of gluon fields (isospin singlet) to a static color-octet source in the short distance limit.
Initially computed through quenched lattice QCD by Campbell, Jorysz, and Michael \cite{Campbell:1985kp}, the lowest gluelump had quantum numbers $k^{PC}=1^{\pm-}$.
Foster and Michael later provided more precise results for the gluelump masses and ordering of the low-lying gluelumps based on their masses: $k^{PC}=1^{+-}, 1^{--}, 2^{--},\cdots$.
They found that the excited gluelumps $1^{--}$ and $2^{--}$ have masses $368\left(7\right)$~MeV and $584\left(10\right)$~MeV, respectively, above the 
lowest 
gluelump $1^{+-}$ \cite{Foster:1998wu}.
The first computation of the gluelump spectrum using lattice QCD with $2+1$ flavors of dynamical light quarks was performed by Marsh and Lewis \cite{Marsh:2013xsa}.
They set the strange quark mass to its physical value, while the up and down quark masses were heavy, the pion mass being approximately 3.5 times its actual value.
Their results indicate that the inclusion of dynamical light quarks appears to decrease the  energy for the $1^{--}$ gluelump,
while increasing that of the $2^{--}$ gluelump with respect to the lowest $1^{+-}$ gluelump: the $1^{--}$ and $2^{--}$ gluelump masses are $300$~MeV and $700$~Mev higher than the $1^{+-}$ gluelump mass.
Several of the gluelump masses in the renormalon-subtracted scheme (RS-scheme), which is useful for perturbative calculations, have been estimated in Ref.~\cite{Bali:2003jq}.
A more recent and precise computation of the gluelump spectrum through quenched lattice QCD was performed by Herr, Schlosser, and Wagner;
their result for the lowest gluelump mass in the RS-scheme at the subtraction scale $\nu_f\approx 1~\mathrm{GeV}$ is $\Lambda_{1^{+-}}^{g}\left(\nu_f\approx 1~\mathrm{GeV}\right)=0.857(3)(143)$~GeV
\cite{Herr:2023xwg}, which is consistent with the estimate in ~\cite{Bali:2003jq}.

The adjoint meson masses are the energy levels associated with the binding of light quarks, which could also have nonsinglet flavor (isospin) quantum numbers, to a static color-octet source.
Our understanding of their spectrum primarily relies on the work of Foster and Michael, who computed the spectrum by employing  quenched lattice QCD with a light valence quark and antiquark.
The energies of the adjoint mesons were computed for two distinct mass values of the light valence quarks: one closer to the strange quark mass and another larger.
They also had two different lattice spacings: $\beta=5.7$ and $\beta=6.0$.
Subsequently, these results were extrapolated to the chiral limit.
The adjoint mesons with the lowest energies were a vector meson with quantum numbers $k^{PC}=1^{--}$ and a pseudoscalar meson with $k^{PC}=0^{-+}$, with the vector meson found to have the lower energy: 
$\Lambda^a_{0^{-+}}-\Lambda^a_{1^{--}}=50 \pm 70$~MeV for $s\bar{s}$ adjoint mesons.
At $\beta=5.7$, after taking the chiral extrapolation, they found that the $1^{--}$ and $0^{-+}$ adjoint mesons have masses shifted by $-10\left(103\right)$~MeV and $34\left(161\right)$~MeV, respectively, with respect to the $1^{+-}$ gluelump mass.
Combining results for both lattice spacings $\beta=5.7$ and $\beta=6.0$,
they found that both adjoint mesons were heavier than the lowest  $1^{+-}$ gluelump state: $\Lambda^a_{1^{--}}-\Lambda^g_{1^{+-}}=120\pm 70$~MeV for $s\bar{s}$ adjoint meson
and $\Lambda^a_{1^{--}}-\Lambda^g_{1^{+-}}=47\pm 90$~MeV for $q\bar{q}$ adjoint mesons, where $q$ denotes the light quark $u$ or $d$.
Nonetheless, the presence of sizable error bars in relation to the central values implies that the exact ordering of gluelump, pseudoscalar and vector adjoint mesons in lattice QCD for systems involving $u$ and $d$ quarks remains unclear.
There are no lattice QCD calculations for adjoint baryon states.

In general, the static energy for adjoint mesons in Eq.~\eqref{eq:QQbarpot-short} will also depend on the isospin quantum number $I$.
In QCD with two flavors of light quarks, $u$ and $d$, the lightest adjoint mesons may be in an isospin triplet, $I=1$, or an isospin singlet, $I=0$, configuration.
The $k^{PC}$ quantum numbers for the isospin singlet and the neutral member of the isospin triplet are $1^{--}$ for the vector and $0^{-+}$ for the pseudoscalar.
The appropriate quantum numbers for the charged adjoint mesons are expressed as $I^G(k^{P})$, where $G = (-1)^I C$ and $C$ represents the charge conjugation quantum number of the neutral member of the multiplet.
The vector adjoint mesons have quantum numbers $0^-(1^-)$ and $1^+(1^-)$, while the pseudoscalar adjoint mesons have quantum numbers $0^+(0^-)$ and $1^-(0^-)$.
To establish the ordering of the ground state gluelump and the four lowest energy adjoint mesons with quantum numbers $0^-(1^-), 1^+(1^-), 0^+(0^-)$, and $1^-(0^-)$ according to their masses,
computations utilizing lattice QCD with dynamical light quarks are indispensable.

\subsubsection{$QQ$ systems}
\label{subsubsec:QQ-match}
We consider now bound states of a $QQ$ pair and LDF.
At short distances $r$ between the two heavy quarks, the  $QQ$ pair may be in a color antitriplet representation of $SU(3)$, the corresponding field being  $T^\ell$ with $\ell =1$, 2, 3, and bind with LDF that are in a color triplet representation to form color singlet hadrons,
or it may be in a color sextet representation, the corresponding field being $\Sigma^\sigma$ with $\sigma=$ 1, 2, ..., 6, and bind with LDF that are in a color antisextet representation to form color singlet hadrons.
The first case describes the short distance behavior of doubly heavy baryons $\left[\left(QQ\right)_{\bar{3}}+q_3\right]$,
tetraquark states of the type $\left[\left(QQ\right)_{\bar{3}}+\left(\bar{q}\bar{q}\right)_3\right]$,
and pentaquark states of the type $\left[\left(QQ\right)_{\bar{3}}+\left(qq\bar{q}\right)_3\right]$, 
The static energy at short distances is the sum of the color triplet potential $V_T$, a constant that accounts for the mass shift induced by the LDF, 
which may be $q_3$, $(\bar{q}\bar{q})_3$ and $(qq\bar{q})_3$ or the corresponding charge conjugated states, respectively, and higher-order corrections suppressed by powers of $\Lambda_{\rm QCD}r$.
The second case describes the short distance behavior of sextet tetraquarks $\left[\left(QQ\right)_6+\left(\bar{q}\bar{q}\right)_{\bar{6}}\right]$,
and sextet pentaquarks  $\left[\left(QQ\right)_6+\left(qq\bar{q}\right)_{\bar{6}}\right]$.
In this case, the static energy at short distances is the sum of the color sextet potential $V_\Sigma$, a constant that accounts for the mass shift induced by the LDF,
which may be $(\bar{q}\bar{q})_{\bar{6}}$ or $(qq\bar{q})_{\bar{6}}$ or the corrsponding charge conjugated states, respectively, and higher-order corrections suppressed by powers of $\Lambda_{\rm QCD}r$.
At leading order, the triplet and sextet potentials read
\begin{equation}
	V_T(r)= -\frac{2}{3}\frac{\alpha_s}{ r}, \qquad\qquad 
	V_{\Sigma}(r) = \frac{\alpha_s}{3 r} ,
	\label{eq:VtVsex}
\end{equation}
where $\alpha_s$ is the strong coupling constant computed at a typical scale of order $1/r$.
The triplet and sextet potentials up to next-to-next-to leading order in $\alpha_s$ can be found in Ref.~\cite{Assi:2023cfo}.
Since the antitriplet potential is attractive and the sextet is repulsive, we shall assume that a bound state with triplet potential is lower in energy, i.e.\ more bound, 
than the corresponding one with sextet potential.
Potential NRQCD in the  weak-coupling regime has been used earlier to study doubly heavy baryons and tetraquark states \cite{Brambilla:2005yk, Mehen:2019cxn, Braaten:2020nwp}.

The matching relations between the interpolating operators and BOEFT, Eq.~\eqref{eq:match-1}, become at short distance 
\begin{align}
	\mathcal{O}_{\kappa, \lambda}^{(QQ)_{\bar{3}}}(t,\,\bm{r},\,\bm{R}) &\longrightarrow \sqrt{Z^{\prime}_{\kappa,\,\lambda}}\,T^\ell(t,\,\bm{r},\,\bm{R})\,P^{\alpha \dag}_{\kappa, \lambda}\,H_{3,\,\kappa}^{\alpha,\,\ell}(t,\,\bm{R})+{\cal O}\left(r\right),
	\label{eq:matchQQT}\\
	\mathcal{O}_{\kappa, \lambda}^{(QQ)_6}(t,\,\bm{r},\,\bm{R}) &\longrightarrow \sqrt{Z^{\prime\prime}_{\kappa,\,\lambda}}\,\Sigma^\sigma(t,\,\bm{r},\,\bm{R})\,P^{\alpha \dag}_{\kappa, \lambda}\,H_{\bar{6},\,\kappa}^{\alpha,\,\sigma}(t,\,\bm{R})+{\cal O}\left(r\right),
	\label{eq:matchQQS}
\end{align} 
where  $Z^\prime_{\kappa,\,\lambda}$ and $Z^{\prime\prime}_{\kappa,\,\lambda}$ are field normalization factors, 
$\ell=1, 2, 3$ and $\sigma = 1, \dots, 6$ are color indices,  
and $H_{3,\,\kappa}^{\alpha,\,\ell}$ and $H_{\bar{6},\,\kappa}^{\alpha,\,\sigma}$ are operators representing the light fields 
in the different color representations.
For doubly heavy baryons, $H^{\alpha, \,\ell}_{3,\,\kappa}$ is given in Table~\ref{tab:Reps-QQq}, for doubly heavy tetraquarks, 
$H_{3,\,\kappa}^{\alpha,\,\ell}$ and $H_{\bar{6},\,\kappa}^{\alpha,\,\sigma}$ are given in Tables~\ref{tab:Reps-QQ-I_0} and \ref{tab:Reps-QQ-I_1},
and for pentaquarks,  $H^{\alpha,\,\ell}_{3,I_3=\pm1/2, (1/2)^{+}}$ is given in Eq.~\eqref{eq:OpQQqqqbar}.
The operators in the right-hand side describe baryons, tetraquarks and pentaquarks in BOEFT in the short distance limit,
when the BOEFT fields $\Psi^{QQ}_{\kappa, \lambda}$ reduce either to $T^\ell\,P^{\alpha \dag}_{\kappa, \lambda}\,H_{3,\,\kappa}^{\alpha,\,\ell}$ or $\Sigma^\sigma\,P^{\alpha \dag}_{\kappa, \lambda}\,H_{\bar{6},\,\kappa}^{\alpha,\,\sigma}$;
the fields $T^\ell$ and $\Sigma^\sigma$ may be also understood as fields in weakly coupled pNRQCD.

Following a similar matching procedure as the one in Sec.~\ref{subsubsec:QQbar-match} [see Eq.~\eqref{eq:matchQQbar-2}], 
we obtain the following relation between the static energy computed in NRQCD (left), the short-distance potential in pNRQCD (middle),  and the BOEFT potential (right)
\begin{align}
	E^{(0), F}_{\kappa, |\lambda|}(r) = V_F(r) + \Lambda^F_{\kappa} + {\cal O}\left(r^2\right) = V^{(0), F}_{\kappa, |\lambda|}(r) \qquad\qquad  (F = T, \Sigma),
	\label{eq:QQpot-short}
\end{align}
where $V_F(r)$ $\left(F= T, \Sigma\right)$ is the color triplet or sextet potential given at leading order in Eq.~\eqref{eq:VtVsex},
$\Lambda^F_{\kappa}\,\left(F= T, \Sigma\right)$ is a constant of mass dimension one that accounts for the mass shift induced by the LDF.
Note that for the ground state doubly heavy tetraquarks, the Pauli principle requires that 
the antitriplet and sextet color configurations in Eq.~\eqref{eq:QQpot-short} do not contribute to the same quantum number $\kappa$
that includes isospin [see Eq.~\eqref{label-n}], see Tables~\ref{tab:Reps-QQ-I_0} and \ref{tab:Reps-QQ-I_1}.
The large time limits of the LDF correlators that define $\Lambda^T_{\kappa}$ and $\Lambda^\Sigma_{\kappa}$ are
\begin{align}
	& \langle \mathrm{vac}|H^{\alpha, \,\ell }_{3,\,\kappa}(T/2,\,\bm{R})\,\phi^{\ell \ell'}_{\bar{3}}\left(T/2,-T/2\right)H^{\alpha^{\prime}, \,\ell' \dag}_{3,\,\kappa}(-T/2,\,\bm{R})\}|\mathrm{vac}\rangle=\delta_{\alpha\alpha^{\prime}}\,e^{-i\Lambda^T_{\kappa}T},
	\label{eq:LDF-mass-QQT}\\
	& \langle \mathrm{vac}|H^{\alpha, \,\sigma }_{\bar{6},\,\kappa}(T/2,\,\bm{R})\,\phi^{\sigma \sigma'}_{6}\left(T/2,-T/2\right)H^{\alpha^{\prime}, \,\sigma' \dag}_{\bar{6},\,\kappa}(-T/2,\,\bm{R})\}|\mathrm{vac}\rangle=\delta_{\alpha\alpha^{\prime}}\,e^{-i\Lambda^\Sigma_{\kappa}T},
	\label{eq:LDF-mass-QQS}
\end{align}
where $\delta_{\alpha\alpha^{\prime}}$ comes from the normalization condition.
The Wilson line $\phi^{\ell \ell'}_{\bar{3}}\left(T/2,-T/2\right)$ with $\ell,\, \ell'= 1, 2, 3$ is in the antitriplet representation of $SU(3)$,
\begin{equation}
	\phi^{\ell\ell'}_{\bar{3}}\left(T/2,-T/2\right) \equiv {\mathcal{P}}\exp\left[-ig\int^{t_1}_{t_2}dt\, A_0^{a}(t, \bm{R})\,T_{\bar{3}}^a\right]_{\ell\ell'}\,,
	\label{eq:TWilson-bar3}
\end{equation}
while the Wilson line $\phi^{\sigma \sigma'}_{6}\left(T/2,-T/2\right)$ with $\sigma,\, \sigma'= 1, 2, \dots, 6$ is in the sextet $6$ representation of $SU(3)$,
\begin{equation}
	\phi^{\sigma\sigma'}_{6}\left(T/2,-T/2\right) \equiv {\mathcal{P}}\exp\left[-ig\int^{t_1}_{t_2}dt\, A_0^{a}(t, \bm{R})\,T^a_{6}\right]_{\sigma\sigma'}\,,
	\label{eq:TWilson-bar6}
\end{equation}
where
\begin{equation}
	T_{\bar{3}}^a = -\frac{\lambda_G^{*a}}{2},\qquad\qquad \left(T_6^a\right)_{\sigma\sigma'} = \sum_{ijk}^3\,\underline{\Sigma}_{ij}^\sigma\,\left(\lambda^a_G\right)_{jk}\,\underline{\Sigma}_{ki}^{\sigma'},
	\label{eq:Reps-T-S}
\end{equation}
with $\lambda_G^a$ the Gell-Mann matrices and the symmetric tensor $\underline{\Sigma}^\sigma$ defined in Appendix~\ref{app:group}.

The static energies and the BOEFT potentials are approximated by the middle equality in \eqref{eq:QQpot-short} only for small $r$.  
The ${\cal O}\left(r^2\right)$ terms arise from multipole expanding the pNRQCD interaction vertices between the heavy quark pair and the LDF.
Equation~\eqref{eq:QQpot-short} implies that at leading order in the multipole expansion, several static energies are degenerate since they only depend on the quantum number $\kappa\equiv \{k^{PC}, f\}$ and not on $\lambda$.
We are not aware of any explicit lattice QCD
extraction of $\Lambda^T_{\kappa}$ and $\Lambda^\Sigma_{\kappa}$ for tetraquarks and pentaquarks.

In Refs.~\cite{Bicudo:2015kna,Mueller:2023wzd}, lattice computations were performed for several static energies with BO quantum number \(\Lambda^\sigma_{\eta^\prime}\) for both the color antitriplet \(\left(QQ\right)_{\bar{3}}\) and the color sextet \(\left(QQ\right)_{6}\) in isospin singlet \(\left(I=0\right)\) and isospin triplet \(\left(I=1\right)\) configurations. Note that to compare with the static energies with BO quantum number \(\Lambda^\sigma_\eta\) in Tables~\ref{QQqqnumbers}, \ref{tab:Reps-QQ-I_0}, and \ref{tab:Reps-QQ-I_1}, one must use \(\eta^\prime = -\eta\) for \(\left(QQ\right)_{\bar{3}}\) and \(\eta^\prime = \eta\) for \(\left(QQ\right)_{6}\) due to the different parity convention in Refs.~\cite{Bicudo:2015kna,Mueller:2023wzd}. For the color antitriplet \(\left(QQ\right)_{\bar{3}}\), the lowest static energies with BO quantum number \(\Sigma_g^+\) (\(I=0\)) and \(\{\Sigma_g^-, \Pi_g\}\) (\(I=1\)) computed on the lattice exhibit attractive behavior at short distances,
which aligns with our short distance attractive prediction for the \(\Sigma_g^+\) (\(I=0\)) and \(\{\Sigma_g^-, \Pi_g\}\) (\(I=1\)) BO potentials in
Tables \ref{tab:Reps-QQ-I_0} and \ref{tab:Reps-QQ-I_1}.
Other excited static energies in \(I=0\) with BO quantum numbers \(\{\Sigma_u^+, \Pi_u\}\) and in \(I=1\) with BO quantum numbers \(\Sigma_u^-\) and \(\{\Sigma_u^+, \Pi_u\}\) also show attractive behavior at short distances in Refs.~\cite{Bicudo:2015kna,Mueller:2023wzd}, which is consistent with our short-distance predictions for these BO potentials in 
Tables \ref{tab:Reps-QQ-I_0} and \ref{tab:Reps-QQ-I_1}.
For the color sextet \(\left(QQ\right)_{6}\), the lowest static energies with BO quantum numbers \(\{\Sigma_g^-, \Pi_g\}\) (\(I=0\)) and \(\Sigma_g^+\) (\(I=1\))  exhibit repulsive behavior at short distances in Refs.~\cite{Bicudo:2015kna,Mueller:2023wzd}, which  agrees with our short-distance repulsive prediction for the \(\{\Sigma_g^-, \Pi_g\}\) (\(I=0\)) and \(\Sigma_g^+\) (\(I=1\)) BO potentials in 
Tables \ref{tab:Reps-QQ-I_0} and \ref{tab:Reps-QQ-I_1}. 
Similarly, our predictions for the excited static energies of the color sextet 
(\(\Sigma_u^-\) and  \(\{\Sigma_u^+, \Pi_u\}\)  (\(I=0\)),  and \(\{\Sigma_u^+, \Pi_u\}\) (\(I=1\)))
also qualitatively agree with the results presented in Refs.~\cite{Bicudo:2015kna,Mueller:2023wzd}.

\section{Mixing at large distance and heavy-light thresholds}
\label{sec:LD}
\begin{figure}[!tbp]
    \includegraphics[width=0.8\textwidth]{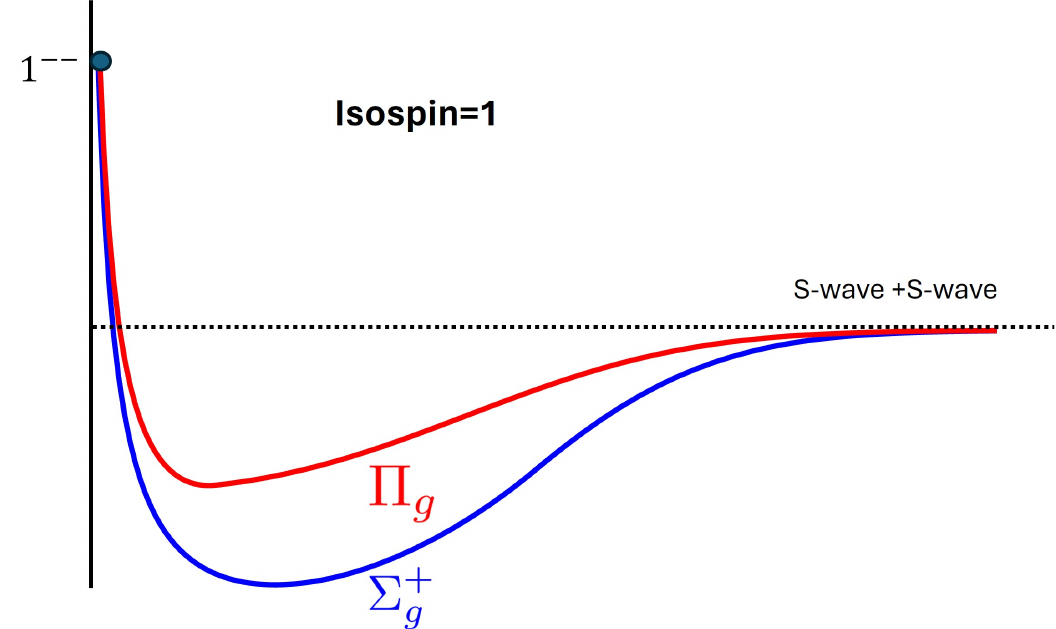}
    \caption{Example of a possible evolution of a 
    $I=1$ tetraquark static energy into a S-wave plus S-wave heavy-light static energy with the same BO quantum number. 
    Both the $\Pi_g$ and the $\Sigma_g^+$ evolve to the same heavy-light pair that has both BO quantum numbers. 
    The short distance and long distance behavior are fixed by symmetry in the BOEFT.} \label{tetra1}
  \end{figure}

\begin{figure}[!tbp]
    \includegraphics[width=0.8\textwidth]{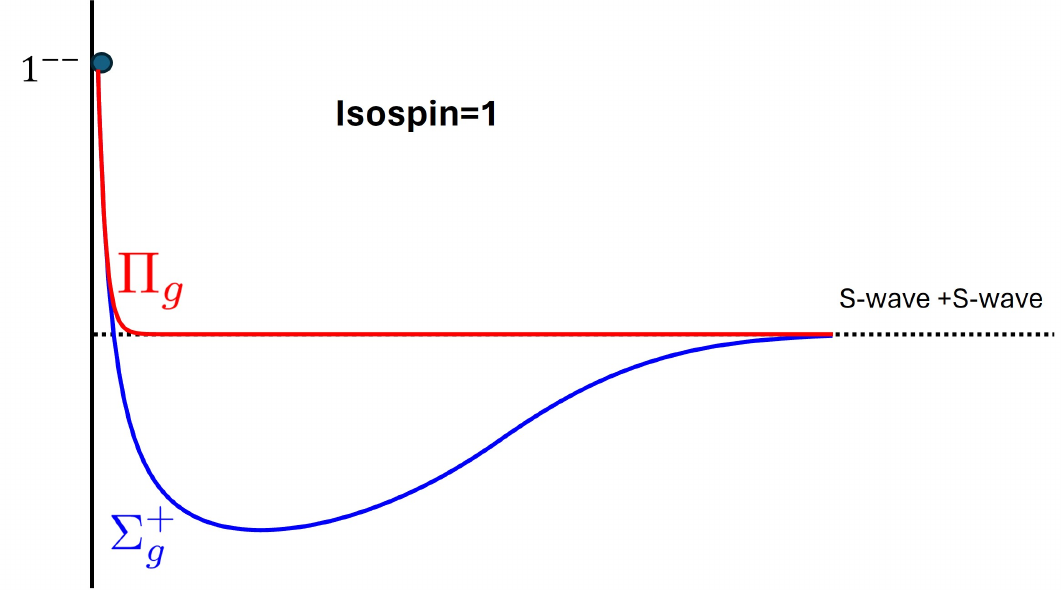}
    \caption{Same as Fig.~\ref{tetra1} with a different possible intermediate distance behavior for the $\Pi_g$ static energy. 
    The short distance and long distance behavior are fixed by symmetry in the BOEFT.} \label{tetra2}
  \end{figure}

\begin{figure}[!tbp]
    \includegraphics[width=0.8\textwidth]{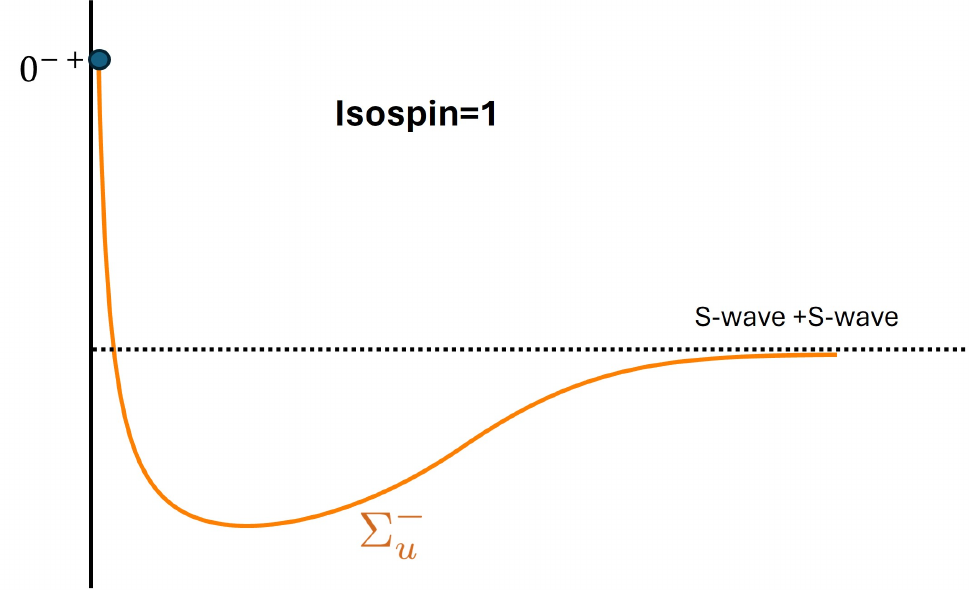}
    \caption{Example of an $I=1$ tetraquark static energy $\Sigma_u^-$ 
    evolving into a S-wave plus S-wave  heavy-light static energy with the same BO quantum numbers.} \label{tetra0}
  \end{figure}

The behavior of the static energies at large distances $r$ shows two different effects that are intertwined. 
On one hand, at large distance in the presence of light quarks, we should consider static heavy-light states (meson-(anti)meson, baryon-(anti)meson). Such static states can also be catalogued in terms of the BO quantum numbers as we show below.
Now, such static heavy-light mesons already exist in the list of BO static energies that 
we have discussed in Sec.~\ref{sec:BO}: 
indeed the tetraquark static energies with $I=1$ at large distance evolve directly into the heavy-light pairs with the same BO quantum numbers, see Figs.~\ref{tetra1}, \ref{tetra2} and \ref{tetra0}, hinting at the complementarity of the tetraquark  model and the molecular model used in the literature.
In Appendix~\ref{app:Overlap}, we show that the tetraquark interpolating operators that we presented in Sec.~\ref{sec:characterization} do indeed have 
an overlap with the heavy-light pair state, but that they have no overlap with quarkonium plus pion(s),
which makes such operators suitable to extract the whole behavior of the static energies at short and long distance. 
This is what we need as input to the BO equations presented in Sec.~\ref{sec:BO}.

\begin{figure}[!tbp]
    \includegraphics[width=0.9\textwidth]{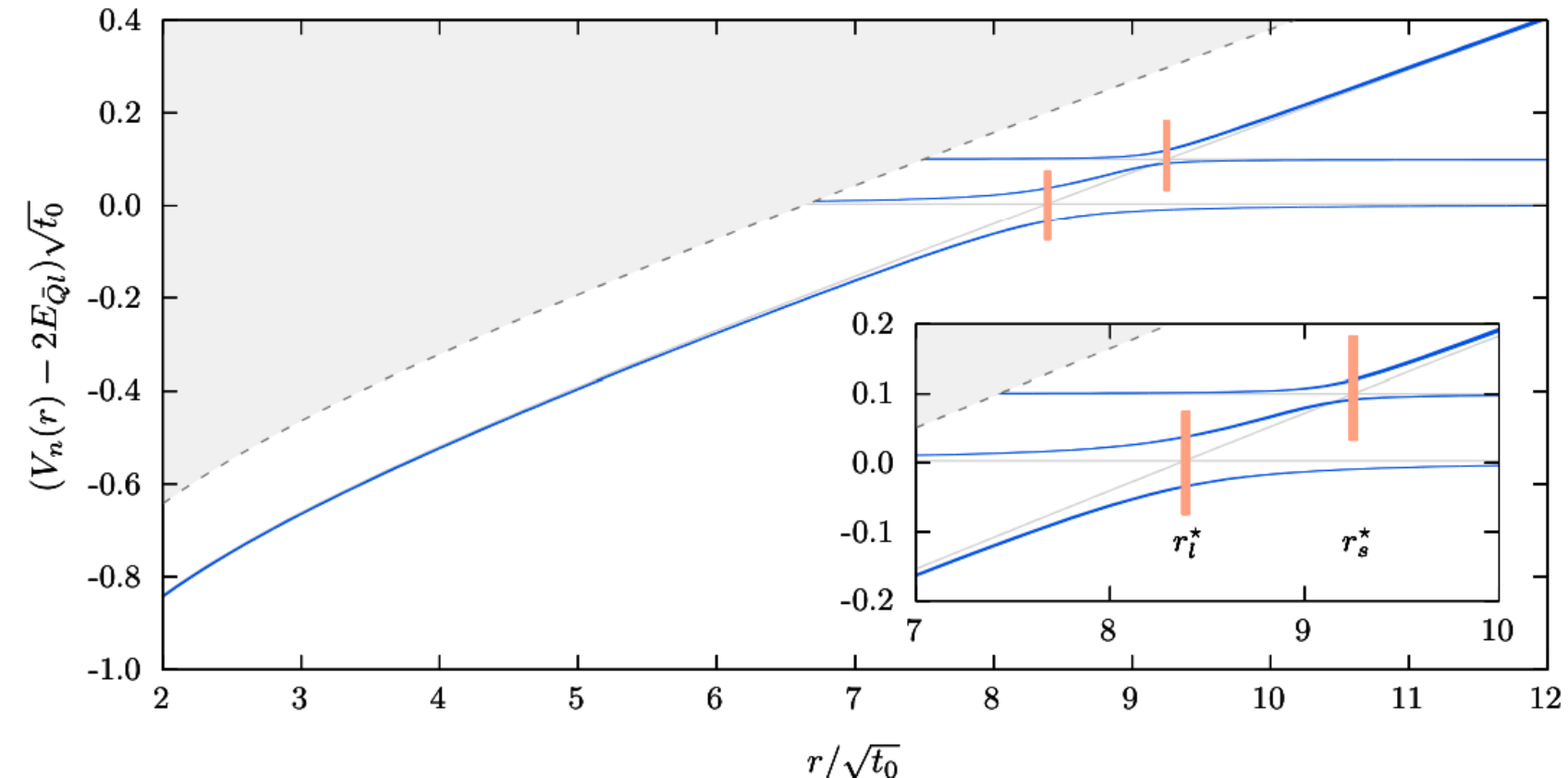}
    \caption{Plot showing the avoided level crossing of the quarkonium $1\Sigma_g^+$ static energy with the static heavy-light and static heavy-strange meson pair energies. 
    The figure shows the three lowest-lying static energies in the adiabatic basis  (blue lines). 
    Energy levels are normalized with respect to twice the energy of the static heavy-light pair (denoted as $E_{\bar{Q}l}$ in the figure). 
    The grey region denotes the energy regime where two-pion states are of relevance. 
    The inset shows the zoomed-in region of the string breaking or the avoided crossing region. 
    The orange bars denote the string breaking radii: $r^*_l = 1.211(13)$~fm and $r^*_s = 1.336(14)$~fm, since $\sqrt{t_0}  = 0.1443(15)$~fm. 
    The picture is taken from Ref.~\cite{Bulava:2024jpj}.} 
    \label{fig:string-breaking}
 \end{figure}

On the other hand, in the $I=0$ sector, things are even more interesting.
In this case, it has been explicitly observed on the lattice that the quarkonium static 
energy and the heavy-light static energy show an avoided level crossing, see Fig.~\ref{fig:string-breaking}.\footnote{The reason for which in Fig.~\ref{fig:string-breaking}  there are three static energies is that also the $s$ quark is considered.}
In our description, this means that two static energies with the same BO quantum numbers, namely
the diabatic static energy of quarkonium $\Sigma_g^+$
and the first tetraquark diabatic static energy $\Sigma_g^{+\prime}$, become close in a 
range of $r$ of about the distance of string breaking and in that region their transition amplitude 
is different from zero. 
Hence, we need to diagonalize the diabatic energy matrix obtaining the avoided level crossing  effect between the adiabatic static energies shown in Fig.~\ref{fig:string-breaking} (see also 
Appendix~\ref{app_crossing}).

To explain in a deeper way  the effect, let us consider Fig.~\ref{tetra01}.
There, we see that the adiabatic quarkonium static energy with BO
quantum number $1\Sigma_g^+$ shows an avoided level crossing with the tetraquark static energy 
$2\Sigma_g^{+}$. 
This produces the effect that the quarkonium static energy will continue at large $r$ as a heavy-light static meson pair (in S-wave plus S-wave), 
while the tetraquark static energy will show a long range confining behavior of quarkonium until the 
avoided level crossing with another excited tetraquark static energy with the same quantum numbers.  
In the same figure, we show also the tetraquark  adiabatic static energy $1\Pi_g$, which is degenerate  with the tetraquark adiabatic static energy $2\Sigma_g^+$ at short distance. 
We see that, due to the different BO quantum numbers, the $1\Pi_g$ static energy does not show avoided level crossing with the quarkonium $1\Sigma_g^+$ static energy and directly evolves into the heavy light pair in S-wave plus S-wave at large distance.

This changes our perception of the confining behavior of the strong force.
Although it was known that the confining string of quarkonium breaks into two heavy-light pairs in the large distance, 
the details of this mechanism in relation to the tower of QCD static energies had not been elucidated. 
The introduction of the tetraquark static energies brings light on this.

This same effect appears for hybrid and $I=0$ tetraquark static energies.
For $I=0$ states,  the avoided level crossing between quarkonium and tetraquark static energies 
or hybrid and tetraquark static energies is related to the decay of quarkonium and hybrids into heavy-light pairs.
For a quantitative description, a diabatic picture may be more useful, as we discuss below.

\subsection{BO quantum numbers of static heavy-light pairs}
Here, we consider heavy-light meson-(anti)meson pairs and baryon-(anti)meson pairs, 
we derive their BO quantum numbers and determine which static energies evolve to a given heavy-light pair.
As a start, however, we need to identify the quantum number $\kappa \equiv \{ k^{P[C]}, f \} $ given in Eq.~\eqref{label-n} for the heavy-light pair.
The charge conjugation $C$ of the LDF is a good quantum number only for the  static heavy-light meson-antimeson threshold. 
For the static heavy-light meson-meson or meson-baryon threshold, the static energies are characterized by $k$ and parity $P$.
We can identify the LDF quantum numbers from the heavy meson or baryon multiplets. For the heavy meson-antimeson or heavy meson-meson pair thresholds, we consider the mesons in the ground and first excited states. 
For the heavy baryon-meson or heavy baryon-antimeson pair thresholds, we only consider the ground state for simplicity.  
The ground state heavy meson $Q\bar{q}$  (antimeson $\bar{Q}q$) corresponds to light antiquark $k^P=(1/2)^-$  (quark $k^P=(1/2)^+$), followed by two states of similar mass corresponding to light antiquark $k^P=(1/2)^+$ and $k^P=(3/2)^+$ (quark $k^P=(1/2)^-$ and $k^P=(3/2)^-$ ). 
The ground state heavy baryon $Qqq$ corresponds to light quarks $k^P=0^+$ and $k^P=1^+$, 
where we assume that the $k^P=0^+$ state is lower in energy.\footnote{
Our assumption is supported by the observation that $\Lambda$-baryons have lower mass than $\Sigma$-baryons.}
Each static heavy-light pair state or threshold is characterized by the spin and parity of the light quark states forming the two heavy-light states (heavy mesons or heavy baryons). 
We label the light quark states as $k^{P}_{\bar q}$ for heavy meson ($k^{P}_q$ for heavy antimeson) and $k^{P}_{qq}$ for heavy baryons. 
Combining the spin and parity of the LDF states for heavy mesons and baryons, we arrive at the total LDF quantum number of the light quarks characterizing the heavy-light pairs: 
Table~\ref{tab:qqbar-qq} for $Q\bar{q}$-$\bar{Q}q$ or $Q\bar{q}$-$Q\bar{q}$ pairs and Table~\ref{tab:qqq} for $Qqq$-$\bar{Q}q$ or $Qqq$-$Q\bar{q}$ pairs. 
The BO quantum numbers for heavy-light pairs corresponding to LDF quantum numbers $k^{PC}$ are shown in the last column of Tables~\ref{tab:qqbar-qq} and \ref{tab:qqq}.  Note that in Table~\ref{tab:qqbar-qq} for the $Q\bar{q}$-$\bar{Q}q$ pair, we considered symmetric and antisymmetric combinations under charge conjugation for the S-wave plus P-wave and the P-wave plus P-wave\footnote{ Our approach results in more entries compared to Table~2 in Ref.~\cite{TarrusCastella:2024zps} (see also Table~1 in Ref.~\cite{TarrusCastella:2022rxb}), where the states are not eigenstates under charge conjugation.}.
If we include the heavy-light pair state or threshold in the BOEFT, then each of the combinations in Tables~\ref{tab:qqbar-qq} and \ref{tab:qqq} will be represented as a field which will also depend on the LDF flavor quantum numbers (isospin) \cite{Soto:2020xpm}. 
The static energies for the heavy-light pair states  will also depend on the isospin or the light quark flavors. 

\begin{table}[th!]
\begin{center}
\small{\renewcommand{\arraystretch}{0.9}
\scriptsize
\begin{minipage}{.55\linewidth}
\begin{tabular}{|c|c||c|}  \hline\hline
\multirow{2}{*}{\hspace{2pt} $\begin{array}{c} k_{\bar{q}}^{P}\otimes k_q^{P}\end{array} $ \hspace{2pt}} & \multirow{2}{*}{\hspace{2pt} $\begin{array}{c}k^{PC}\end{array} $ \hspace{2pt}} &  \multirow{2}{*}{\hspace{2pt} $\begin{array}{c} \text{BO quantum \#}\\\Lambda^\sigma_\eta\end{array} $ \hspace{2pt}}\\& & \\
\hline\hline
$(1/2)^-\otimes(1/2)^+ $         & $0^{-+}$ & $\Sigma_u^-$ \\
                                & $1^{--}$ & $\Sigma_g^+,\,\Pi_g$ \\ \hline
$(1/2)^-\otimes(1/2)^-\,+\,(1/2)^+\otimes(1/2)^+$         & $0^{+-}$ & $\Sigma_u^+$\\                                 
                                & $1^{+-}$ & $\Sigma_u^-,\,\Pi_u$ \\
$(1/2)^-\otimes(1/2)^-\,-\,(1/2)^+\otimes(1/2)^+$         & $0^{++}$ & $\Sigma_g^+$\\                                 
                                & $1^{++}$ & $\Sigma_g^-,\,\Pi_g$ \\                                
$(1/2)^-\otimes(3/2)^-\,+\,(3/2)^+\otimes(1/2)^+ $      & $1^{+-}$ & $\Sigma_u^-,\,\Pi_u$  \\
                                & $2^{+-}$ & $\Sigma_u^+,\,\Pi_u,\,\Delta_u$\\
$(1/2)^-\otimes(3/2)^-\,-\,(3/2)^+\otimes(1/2)^+$         & $1^{++}$ & $\Sigma_g^-,\,\Pi_g$  \\
                                & $2^{++}$ & $\Sigma_g^+,\,\Pi_g,\,\Delta_g$\\ \hline
$(1/2)^+\otimes(1/2)^-$         & $0^{-+}$ & $\Sigma_u^-$  \\
                                & $1^{--}$ & $\Sigma_g^+,\,\Pi_g$ \\ 
$(1/2)^+\otimes(3/2)^-\,+\,(3/2)^+\otimes(1/2)^-$   & $1^{--}$ & $\Sigma_g^+,\,\Pi_g$ \\
                                & $2^{--}$ & $\Sigma_g^-,\,\Pi_g,\,\Delta_g$ \\
$(1/2)^+\otimes(3/2)^-\,-\,(3/2)^+\otimes(1/2)^-$   & $1^{-+}$ & $\Sigma_u^+,\,\Pi_u$ \\
                                & $2^{-+}$ & $\Sigma_u^-,\,\Pi_u,\,\Delta_u$ \\
$(3/2)^+\otimes(3/2)^-$         & $0^{-+}$ & $\Sigma_u^-$ \\
                                & $1^{--}$ & $\Sigma_g^+,\,\Pi_g$ \\
                                & $2^{-+}$ & $\Sigma_u^-,\,\Pi_u,\,\Delta_u$ \\
                                & $3^{--}$ & $\Sigma_g^+,\,\Pi_g,\,\Delta_g,\,\Phi_g$\\ \hline\hline
\end{tabular}
\end{minipage}%
\begin{minipage}{.45\linewidth}
\begin{tabular}{|c|c||c|}  \hline\hline
\multirow{2}{*}{\hspace{2pt} $\begin{array}{c}k_{\bar{q}}^{P}\otimes k_{\bar{q}}^{P}\end{array} $ \hspace{2pt}} & \multirow{2}{*}{\hspace{2pt} $\begin{array}{c}k^{P}\end{array} $ \hspace{2pt}} &  \multirow{2}{*}{\hspace{2pt} $\begin{array}{c} \text{BO quantum \#}\\\Lambda^\sigma_\eta\end{array} $ \hspace{2pt}}\\& & \\
\hline\hline
$(1/2)^-\otimes(1/2)^- $         & $0^{+}$ & $\Sigma_g^+$ \\
                                & $1^{+}$ & $\Sigma_g^-,\,\Pi_g$ \\ \hline
$(1/2)^-\otimes(1/2)^+$         & $0^{-}$ & $\Sigma_u^-$\\                                 
                                & $1^{-}$ & $\Sigma_u^+,\,\Pi_u$ \\
$(1/2)^-\otimes(3/2)^+$         & $1^{-}$ & $\Sigma_u^+,\,\Pi_u$  \\
                                & $2^{-}$ & $\Sigma_u^-,\,\Pi_u,\,\Delta_u$\\ \hline
$(1/2)^+\otimes(1/2)^+$         & $0^{+}$ & $\Sigma_g^+$  \\
                                & $1^{+}$ & $\Sigma_g^-,\,\Pi_g$ \\ 
$(1/2)^+\otimes(3/2)^+$         & $1^{+}$ & $\Sigma_g^-,\,\Pi_g$ \\
                                & $2^{+}$ & $\Sigma_g^+,\,\Pi_g,\,\Delta_g$ \\
$(3/2)^+\otimes(3/2)^+$         & $0^{+}$ & $\Sigma_g^+$ \\
                                & $1^{+}$ & $\Sigma_g^-,\,\Pi_g$ \\
                                & $2^{+}$ & $\Sigma_g^+,\,\Pi_g,\,\Delta_g$ \\
                                & $3^{+}$ & $\Sigma_g^-,\,\Pi_g,\,\Delta_g,\,\Phi_g$\\ \hline\hline
\end{tabular}
\end{minipage}%
\caption{The $k^{PC}$ quantum numbers of the light quark-antiquark $\left(\bar{q} q\right)$ pair and $k^{P}$ quantum numbers of the light antiquark-antiquark $\left(\bar{q} \bar{q}\right)$ pair for the three lightest light quark states contributing to static heavy mesons-antimeson and  static heavy meson-meson pair thresholds, respectively. 
The light quark states have quantum numbers $k_{q}^P$ and $k_{\bar{q}}^P$. 
The BO quantum number $\Lambda_{\eta}^{\sigma}$ ($D_{\infty h}$ representations) corresponding to $k^{P[C]}$ are shown in the third column.  
On the left table, for the $\left(\bar{q} q\right)$ pair, the linear combinations in some entries in the first column refer to symmetric and antisymmetric combination under charge conjugation. 
Each block of states, separated by a single horizontal line, corresponds to S-wave plus S-wave, S-wave plus P-wave, and P-wave plus P-wave  static heavy-light meson-antimeson and meson-meson pair thresholds.}
\label{tab:qqbar-qq}}
\end{center}
\end{table}

The  possibility to describe  exotic states at large distances as static heavy-light states  depends on the fact that they have the same BO quantum numbers $\Lambda^\sigma_\eta$. 
The quarkonium  adjoint tetraquark $Q\bar{Q}q\bar{q}$ and pentaquark $Q\bar{Q}qqq$ states will become at large distances the heavy meson-antimeson $\left(Q\bar{q}\right.$-$\left.\bar{Q}q\right)$ pair thresholds and heavy baryon-antimeson $\left(Qqq\right.$-$\left.\bar{Q}q\right)$ pair thresholds, respectively. 
The doubly heavy tetraquark $QQ\bar{q}\bar{q}$ and pentaquark $QQqq\bar{q}$ states will become the heavy meson-meson $\left(Q\bar{q}\right.$-$\left.Q\bar{q}\right)$ pair thresholds and heavy baryon-antimeson $\left(Qqq\right.$-$\left.Q\bar{q}\right)$ pair thresholds, respectively. 
This happens only if the BO quantum number $\Lambda^\sigma_\eta$ (see Tables~\ref{tab:qqbar-qq} and \ref{tab:qqq}) is the same as that of the exotic state.\footnote{
Also the flavor quantum numbers, such as isospin, have to be same for the LDF states in heavy-light pairs and exotic states.}

\begin{table}[h!]
\begin{center}
\small{\renewcommand{\arraystretch}{1.0}
\scriptsize
\begin{minipage}{.40\linewidth}
\begin{tabular}{|c|c||c|}  \hline\hline
\multirow{2}{*}{\hspace{2pt} $\begin{array}{c} k_{qq}^{P}\otimes k_q^{P}\end{array} $ \hspace{2pt}} & \multirow{2}{*}{\hspace{2pt} $\begin{array}{c}k^{P}\end{array} $ \hspace{2pt}}&  \multirow{2}{*}{\hspace{2pt} $\begin{array}{c} \text{BO quantum \#}\\D_{\infty h}\end{array} $ \hspace{2pt}}\\& & \\
\hline\hline
$0^+\otimes(1/2)^+ $         & $\left(1/2\right)^{+}$ & $\left(1/2\right)_g$ \\\hline
$1^+\otimes(1/2)^+$         & $\left(1/2\right)^{+}$ & $\left(1/2\right)'_{g}$\\                                 
                                & $\left(3/2\right)^{+}$ & $\left(3/2\right)_g$\\\hline\hline
\end{tabular}
\end{minipage}%
\begin{minipage}{.40\linewidth}
\begin{tabular}{|c|c||c|}  \hline\hline
\multirow{2}{*}{\hspace{2pt} $\begin{array}{c} k_{qq}^{P}\otimes k_{\bar{q}}^{P}\end{array} $ \hspace{2pt}} & \multirow{2}{*}{\hspace{2pt} $\begin{array}{c}k^{P}\end{array} $ \hspace{2pt}}&  \multirow{2}{*}{\hspace{2pt} $\begin{array}{c} \text{BO quantum \#}\\D_{\infty h}\end{array} $ \hspace{2pt}}\\& & \\
\hline\hline
$0^+\otimes(1/2)^- $         & $\left(1/2\right)^{-}$ & $\left(1/2\right)_u$ \\\hline
$1^+\otimes(1/2)^-$         & $\left(1/2\right)^{-}$ & $\left(1/2\right)'_{u}$\\                                 
                                & $\left(3/2\right)^{-}$ & $\left(3/2\right)_u$\\\hline\hline
\end{tabular}
\end{minipage}%
\caption{The $k^{P}$ quantum numbers of three light quarks: 
$qqq$ and $qq\bar{q}$ combinations of the lowest light quark states forming heavy baryon-antimeson and heavy baryon-meson pair thresholds, respectively. 
The light quark states have quantum numbers $k_{qq}^P$ and $k_{\bar{q}}^P$. 
For notation purposes,  the BO quantum number  corresponding to $k^{P}$ is shown in the third column, where $g, u$ is the parity eigenvalue of the light quark state. 
The prime in the second row indicates the excited static energy, assuming that the spin singlet $qq$ pair is lower in energy than the spin-triplet. 
Our assumption is supported by the observation that $\Lambda$-baryons have lower mass than $\Sigma$-baryons.}
\label{tab:qqq}}
\end{center}
\end{table}

\subsection{Complete set of coupled Schr\"odinger equations: short and long range mixing, decays to heavy-light, generation of molecular states} 
On the basis of what we have discussed, two facts appear to emerge.
First, the tetraquarks can be conveniently described as heavy-light static pairs at large distance. Second, there are cases in which two states with the same BO quantum numbers develop 
a non zero transition amplitude in some  distance interval and this originates 
the avoided level crossing phenomenon in the adiabatic description or an explicit mixing in the diabatic description. 
When we consider this second effect, we see that Eq.~\eqref{eq:LQQuar} is not sufficient
to describe quarkonium in all the distance regions. 
Close to distances typical of the string breaking, the mixing with the first tetraquark excitation 
with the same BO quantum number $\Sigma_{g}^{+\prime}$ should be considered. 
Then, since the $\Sigma_{g}^{+\prime}$ tetraquark mixes with the 
$\Pi_g$ at short distance [see Eqs.~\eqref{eq:M1} and \eqref{eq:diffeq3}],  the full Schr\"odinger equation to be considered is coupling  three states. 
We show this in the next subsection. 
We will show the same for a hybrid and tetraquark case in the following subsection.

\subsubsection{Quarkonium case}
Here, we present the coupled Schr\"odinger equations that describe quarkonium in the presence of light quarks. 
These equations involve the diabatic quarkonium BO potential with quantum number $\Sigma_g^+$ and the tetraquark BO potentials with quantum numbers $\Sigma_g^{+\prime}$ and $\Pi_g$, which correspond to the $1^{--}$ adjoint meson at short distances and approach the S-wave plus S-wave meson-antimeson pair (isospin singlet $\left(I=0\right)$) threshold at large distance (see Table~\ref{tab:qqbar-qq}). 
The S-wave plus S-wave meson-antimeson $M\bar{M}$ pair threshold is given by the sum of the meson masses: $m_{M}+m_{\bar{M}}$. 
The adiabatic static energies, i.e.\ the eigenvalues of this potential matrix with quantum numbers $1\Sigma_g^+$, $2\Sigma_g^+$, and $1\Pi_g$, which are shown in Fig.~\ref{tetra01}, are given by 
[see Eq.~\eqref{eq:A2}]
\begin{align}
   V_{1\Sigma_g^+ }(r) &= \frac{V_{\Sigma_g^+}\left(r\right)+ V_{\Sigma_g^{+\prime}}\left(r\right)} {2}-\sqrt{\left(\frac{V_{\Sigma_g^+}\left(r\right)- V_{\Sigma_g^{+\prime}}\left(r\right)}{2}\right)^2+\,V^2_{{\scaleto{\Sigma_g^+-\Sigma_g^{+\prime}\mathstrut}{6pt}}}(r)},\label{eq:V1S}\\
    V_{2\Sigma_g^+ }(r) &= \frac{V_{\Sigma_g^+}\left(r\right)+ V_{\Sigma_g^{+\prime}}\left(r\right)} {2}+\sqrt{\left(\frac{V_{\Sigma_g^+}\left(r\right)- V_{\Sigma_g^{+\prime}}\left(r\right)}{2}\right)^2+\,V^2_{{\scaleto{\Sigma_g^+-\Sigma_g^{+\prime}\mathstrut}{6pt}}}(r)},
    \label{eq:V2S}\\
   V_{1\Pi_g}(r) &= V_{\Pi_g}\left(r\right).
   \label{eq:V1P}
\end{align}
where $V_{\Sigma_{g}^{+}-\Sigma_{g}^{+\prime}}(r)$ is the mixing potential. 
The BO potentials $V_{\Sigma_g^+}\left(r\right)$, $V_{\Sigma_g^{+\prime}}\left(r\right)$, $V_{\Pi_g}\left(r\right)$, and the mixing potential $V_{\Sigma_{g}^{+}-\Sigma_{g}^{+\prime}}(r)$ can be expressed in terms of the correlators that we have introduced in Sec.~\ref{sec:characterization}. 
They are given by 
\begin{align}
    \begin{pmatrix} V_{\Sigma_{g}^{+}} & V_{\Sigma_{g}^{+}-\Sigma_{g}^{+\prime}} \\ V_{\Sigma_{g}^{+}-\Sigma_{g}^{+\prime}} & V_{\Sigma_{g}^{+\prime}} \end{pmatrix}(r)&=\lim_{T\to\infty} \frac{i}{T}\,\log\!\left[\bra{\text{vac}}\!\begin{pmatrix} \mathcal{O}_{0^{++},\,0}(T/2,\,\bm{r}) \\ \mathcal{O}^{(Q\bar{Q})_8}_{1^{--},\,0}(T/2,\,\bm{r}) \end{pmatrix}\!\!\begin{pmatrix} \mathcal{O}_{0^{++},\,0}(-T/2,\,\bm{r}) \\ \mathcal{O}^{(Q\bar{Q})_8}_{1^{--},\,0}(-T/2,\,\bm{r}) \end{pmatrix}^\dagger\!\ket{\text{vac}}\right],\notag\\
     V_{\Pi_{g}}(r) &=  \lim_{T\to\infty} \frac{i}{T}\,\log\left[ \langle \mathrm{vac}| \mathcal{O}^{(Q\bar{Q})_8}_{1^{--},\,\pm 1}(T/2,\,\bm{r})\,\mathcal{O}^{(Q\bar{Q})_8\dagger}_{1^{--},\,\pm1}(-T/2,\,\bm{r})|\mathrm{vac}\rangle\right]\!,
     \label{eq:corr-avoided-crossing-1}
    \end{align}
where 
\begin{equation}
    \mathcal{O}_{0^{++},\,0}({t},\,\bm{r}) = \chi^{\dagger}\left(t, \bm{r}/2\right)\phi\left(t; \bm{r}/2,-\bm{r}/2\right)\psi\left(t, -\bm{r}/2\right).
    \label{eq:corr-QQbar}
\end{equation}
 These expressions lead to the coupled Schr\"odinger equations for quarkonium and tetraquark (that includes the meson-antimeson threshold) in the diabatic basis:
\begin{align}
&\left[
-\frac{1}{m_Qr^2}\,\partial_rr^2\partial_r+\frac{1}{m_Qr^2} 
{\begin{pmatrix}  
l\left(l+1\right) & 0 & 0\\ 
0                 & l(l+1)+2        & -2\sqrt{l(l+1)} \\ 
0                 & -2\sqrt{l(l+1)} & l(l+1)  
\end{pmatrix}}\right.
\nonumber\\
&\hspace{4.0 cm}\left.
+\begin{pmatrix} V_{\Sigma_{g}^{+}}(r) &  V_{\Sigma_{g}^{+}-\Sigma_{g}^{+\prime}}(r) & 0 \\
    V_{\Sigma_{g}^{+}-\Sigma_{g}^{+\prime}}(r) & V_{\Sigma_{g}^{+\prime}}(r) & 0\\
      0 & 0 & V_{\Pi_g}(r)\end{pmatrix}
      \right]
      \hspace{-4pt}\begin{pmatrix} \psi_{\Sigma} \\ \psi_{\Sigma^{\prime}} \\ \psi_{\Pi }\end{pmatrix}={\mathcal{E}} \begin{pmatrix} \psi_{\Sigma} \\ \psi_{\Sigma^{\prime}} \\ \psi_{\Pi}\end{pmatrix}.
      \label{eq:VQQbar-threshold}
\end{align}
Equation \eqref{eq:VQQbar-threshold} contains both the mixing at short distance
between BO static energies with the same $\kappa$ and the mixing at long distance between
energies with the same BO quantum numbers. 
They are two different effects that dominate in different distance regions.

For quarkonium, the combined angular momentum is $l=l_Q$, where $l_Q\left(l_Q+1\right)$ is the eigenvalue of the orbital angular momentum of the heavy  quark-antiquark pair ${\bm L}_Q^2$. 
The  corresponding values of the combined angular momentum $l$ for the S-wave plus S-wave meson-antimeson threshold  are given in the fourth column of Table~\ref{tab:QQbarqqbar}.

The mixing potential $V_{\Sigma_{g}^{+}-\Sigma_{g}^{+\prime}}(r)$ in Eq.~\eqref{eq:VQQbar-threshold} has been computed on the lattice \cite{Bali:2003jq, Bulava:2019iut, Bulava:2024jpj} using the static heavy-light operators. 
The starting point of the lattice studies is the diabatic framework, which considers the matrix elements of the correlators of quarkonium and the one of the static heavy-light pair on the diagonal and the transition between the two on the off-diagonal entries. 
Such correlators are calculated on the lattice and then diagonalized in the adiabatic basis\footnote{
See Appendix~\ref{app_crossing} for how correlators with off-diagonal elements behave as a function of time in a simple two states example.}.
The result is a picture of avoided level crossing as shown in Fig.~\ref{fig:string-breaking}, where also the light strange quark has been considered, and so we have more static heavy-light pairs. 
Note that the static heavy-light operators 
usually used in lattice QCD computations overlap with quarkonium plus pion states and, therefore, 
are not suitable to 
investigations at 
short distance (in fact the shorter distance region is not considered for them in Fig.~\ref{fig:string-breaking}), 
while the (adjoint)
tetraquark operators that we introduced in Sec.~\ref{sec:characterization} 
have no overlap with quarkonium plus pion states and, therefore, 
give the possibility to describe the whole range of  distances when adopted in lattice QCD calculations.
The off-diagonal  mixing potential indicates that the quarkonium can decay to the S-wave plus S-wave meson-antimeson pair \cite{Bruschini:2020voj, Bruschini:2021cty, Bruschini:2021sjh, Bruschini:2023zkb, Braaten:2024stn, TarrusCastella:2022rxb}. 
Recently, model-independent selection rules based on Born--Oppenheimer for decays of heavy quarkonium into heavy-light pairs (S-wave plus S-wave or S-wave plus P-wave, etc) and analytic expressions of the relative partial decay rates have been obtained in Ref.~\cite{Braaten:2024stn}.

This BOEFT description is suggestive of the   $\chi_{c1}\left(3872\right)$  emerging as a solution of these coupled equations in the case when the bound state energy is close to zero.

\subsubsection{Hybrid case}
Here, we present the coupled Schr\"odinger equations that describe quarkonium hybrids corresponding to a $k^{PC}=1^{+-}$ gluelump in the presence of light quarks. 
These equations involve the  diabatic quarkonium hybrid BO potentials 
with quantum numbers $\Sigma_u^-$ and $\Pi_u$ and the tetraquark BO potential
with quantum number $\Sigma_u^{-\prime}$, which corresponds to the $0^{-+}$ adjoint meson at short distances and approaches the S-wave plus S-wave meson-antimeson pair (isospin singlet $\left(I=0\right)$) threshold at large distance (see Table~\ref{tab:qqbar-qq}). The S-wave plus S-wave meson-antimeson pair threshold is  given by the sum of the meson masses: $m_{M}+m_{\bar{M}}$. 

Since the hybrid BO potential $\Sigma_u^{-}$ (and even $\Pi_u$) is above the S-wave plus S-wave meson pair threshold, their mixing 
leads to significant effects only
if the energy gap between the two is less than the $\Lambda_{\mathrm{QCD}}$ scale, 
which is the scale of the mixing potential.\footnote{This also implies that there is no avoided level crossing between hybrid BO potential with quantum number $\Sigma_u^-$ and the tetraquark BO potential
with quantum number $\Sigma_u^{-\prime}$. Instead, the hybrid BO potentials will have avoided crossing with the S-wave plus P-wave meson-antimeson pair threshold based on Table~\ref{tab:qqbar-qq}.}  
The BO potentials $V_{\Sigma_{u}^{-}}(r)$, $V_{\Pi_{u}} (r)$, $V_{\Sigma_{u}^{-\prime}}(r)$ and the mixing potential $V_{\Sigma_{u}^{-}-\Sigma_{u}^{-\prime}}(r)$ can be expressed in terms of the correlators that we have introduced in Sec.~\ref{sec:characterization}. 
They are given by 
\begin{align}
    \begin{pmatrix} V_{\Sigma_{u}^{-}} & V_{\Sigma_{u}^{-}-\Sigma_{u}^{-\prime}} \\ V_{\Sigma_{u}^{-}-\Sigma_{u}^{-\prime}} & V_{\Sigma_{u}^{-\prime}} \end{pmatrix}(r)&=\lim_{T\to\infty} \frac{i}{T}\,\log\!\left[\bra{\text{vac}}\begin{pmatrix} \mathcal{O}^{(Q\bar{Q})_8}_{1^{+-},\,0}(T/2,\,\bm{r}) \\ \mathcal{O}^{(Q\bar{Q})_8}_{0^{-+},\,0}(T/2,\,\bm{r}) \end{pmatrix}\!\!\begin{pmatrix} \mathcal{O}^{(Q\bar{Q})_8}_{1^{+-},\,0}(-T/2,\,\bm{r}) \\ \mathcal{O}^{(Q\bar{Q})_8}_{0^{-+},\,0}(-T/2,\,\bm{r}) \end{pmatrix}^\dagger\!\ket{\text{vac}}\right]\!,\notag\\
    V_{\Pi_{g}}(r) &=  \lim_{T\to\infty} \frac{i}{T}\,\log\left[ \langle \mathrm{vac}| \mathcal{O}^{(Q\bar{Q})_8}_{1^{+-},\,\pm 1}(T/2,\,\bm{r})\,\mathcal{O}^{(Q\bar{Q})_8\dagger}_{1^{+-},\,\pm1}(-T/2,\,\bm{r})|\mathrm{vac}\rangle\right].
     \label{eq:corr-avoided-crossing-2}
    \end{align}
The coupled Schr\"odinger equations that follow from the mixing of the 
hybrid BO potential $\Sigma_u^-$  and the tetraquark BO potential $\Sigma_u^{-\prime}$ are
\begin{align}
&\left[-\frac{1}{m_Qr^2}\,\partial_rr^2\partial_r+\frac{1}{m_Qr^2}\begin{pmatrix} l(l+1)+2 & -2\sqrt{l(l+1)} & 0 \\ -2\sqrt{l(l+1)} & l(l+1) & 0 \\ 0 & 0 & l\left(l+1\right) \end{pmatrix}\right.\nonumber\\
&\hspace{4.0 cm}\left.+\begin{pmatrix} V_{\Sigma_{u}^{-}}(r) & 0 & V_{\Sigma_{u}^{-}-\Sigma_{u}^{-\prime}}(r) \\
      0 & V_{\Pi_{u}}(r) & 0\\
      V_{\Sigma_{u}^{-}-\Sigma_{u}^{-\prime}}(r) & 0 & V_{\Sigma_u^{-\prime}}(r)\end{pmatrix}\right]\hspace{-4pt}\begin{pmatrix} \psi_{\Sigma} \\ \psi_{\Pi} \\ \psi_{\Sigma^\prime}\end{pmatrix}={\cal E}\begin{pmatrix} \psi_{\Sigma} \\ \psi_{\Pi}\\ \psi_{\Sigma^\prime}\end{pmatrix}.
      \label{eq:Vhyb-threshold}
\end{align}
Equation~\eqref{eq:Vhyb-threshold} contains both the mixing at short distances
between BO potentials with the same $\kappa$ and the mixing at long distance between
energies with the same BO quantum numbers. 
They are two different effects that dominate in different distance regions.

The lattice computation of the  mixing potential $V_{\Sigma_{u}^{-}-\Sigma_{u}^{-\prime}}(r)$ is at present unknown.  The possible values of the combined  angular momentum $l$ for the lowest hybrid multiplets are 
$l=1$ for $H_1$ and $H_2$ multiplets, $l=0$ for $H_3$ multiplet, $l=2$ for $H_4$ and $H_5$ multiplets (see Table~II in Ref.~\cite{Berwein:2015vca}) and so on. 
The corresponding values of the combined angular momentum $l$ for the S-wave plus S-wave meson-antimeson threshold are given in the fourth column of Table~\ref{tab:QQbarqqbar}. 
The  selection rule on conservation of angular momentum was first derived in Ref.~\cite{TarrusCastella:2024zps}.  
  
The mixing potential $V_{\Sigma_{u}^{-}-\Sigma_{u}^{-\prime}}(r)$ in Eq.~\eqref{eq:Vhyb-threshold} implies that the hybrid can decay to the pair of S-wave plus S-wave meson-antimeson pair threshold contrary to the conventional wisdom \cite{Tanimoto:1982eh, Page:1996rj, Kou:2005gt, McNeile:2006bz, Woss:2020ayi, Farina:2020slb}.  
This was first observed in Ref.~\cite{Bruschini:2023tmm} and subsequently in Ref.~\cite{TarrusCastella:2024zps} based on BOEFT. 
Moreover, a recent lattice calculation of the decays of the lowest $1^{-+}$ charmonium hybrid
finds that the decay widths into the pairs of S-wave charm
mesons $D^*\bar{D}$ and $D^*\bar{D}^*$ are smaller than but comparable with the decay width into the P-wave plus S-wave charm meson pair $D_1\bar{D}$ \cite{Shi:2023sdy}.  
Moreover, model-independent selection rules for decays of heavy quarkonium hybrids into heavy-light pairs (S-wave plus S-wave or S-wave plus P-wave, etc) and analytic expressions of the relative partial decay rates have been obtained in Ref.~\cite{Braaten:2024stn}.

\subsection{Tetraquarks and pentaquarks}
Unlike quarkonium and quarkonium hybrids, the tetraquark and pentaquark states have light quarks as constituents. 
The presence of light quarks modifies the behavior of the Born--Oppenheimer potentials or static energies of tetraquark and pentaquark. 
In the short distance regime $r\rightarrow 0$, where $r$ is the separation between heavy quark-antiquark or heavy quark pairs, weakly-coupled pNRQCD dictates the behavior of the BO-potential. 
For $Q\bar{Q}$ systems, the BO-potential has repulsive octet behavior given by Eq.~\eqref{eq:QQbarpot-short} in the $r \rightarrow 0$ limit, which forms multiplets associated with adjoint mesons and baryons.  
For $QQ$ systems, the BO-potential is given by Eq.~\eqref{eq:QQpot-short}. 
It may have an attractive behavior in the $r\rightarrow 0$ limit, which forms multiplets associated with triplet mesons and  baryons or a repulsive behavior in the $r\rightarrow 0$ limit, which forms multiplets associated with sextet mesons and baryons. 
The BO quantum numbers $\Lambda_\sigma^\eta$ ($D_{\infty h}$ representations corresponding to LDF quantum numbers $k^{PC}$) for adjoint meson and  triplet or sextet mesons are given in Tables~\ref{tab:QQbarqqbar} and \ref{QQqqnumbers} respectively, and for the corresponding baryon configurations in Tables~\ref{tab:QQbarpenta} and \ref{QQpenta} respectively.

In the large $r$ region, the quark configurations in $Q\bar{Q}q\bar{q}$ tetraquarks rearrange to smoothly transition to static-light meson-antimeson $\left(Q\bar{q}\right.$-$\left.\bar{Q}q\right)$ states, 
while $QQ\bar{q}\bar{q}$ tetraquarks transition to static-light meson-meson $\left(Q\bar{q}\right.$-$\left.Q\bar{q}\right)$ states. 
Similarly, $Q\bar{Q}qqq$ pentaquarks rearrange to smoothly transition to static-light baryon-antimeson $\left(Qqq\right.$-$\left.\bar{Q}q\right)$ states, and $QQqq\bar{q}$ pentaquarks rearrange to smoothly transition to static-light meson-baryon $\left(Q\bar{q}\right.$-$\left.Qqq\right)$ states. 
The smooth transition implies no narrow-avoided crossing between the tetraquark and pentaquark BO potentials and the pair of static-light meson or baryon thresholds. 
The BO quantum numbers $\Lambda_\sigma^\eta$ for different pairs of static-light mesons or pairs of static-light mesons and baryons are shown in Tables~\ref{tab:qqbar-qq} and \ref{tab:qqq}. 
The BO quantum numbers $\Lambda^{\sigma}_{\eta}$ should be conserved at all $r$. 
This conservation of $\Lambda^{\sigma}_{\eta}$ between small $r$ and large $r$ implies that in $Q\bar{Q}q\bar{q}$ tetraquarks, the $\Sigma_u^-$ BO-potential (corresponding to LDF quantum numbers $0^{-+}$) and $\Sigma_g^+$ and $\Pi_g$ BO-potentials (corresponding to LDF quantum numbers $1^{--}$), which have repulsive color octet behavior at small $r$, mix with the S-wave plus S-wave static-light meson-antimeson threshold. 
The mixing leads to BO-potentials approaching the threshold at large $r$. Experimentally observed states like $\chi_{c1}(3872)$, $Z_c$, and $Z_b$ indicate that these BO-potentials can have a minimum at some intermediate distance, resembling a Lennard--Jones potential in molecular systems. 
The recent lattice computation  in Ref.~\cite{Prelovsek:2019ywc} of the isospin-1 $\Sigma_u^-$ tetraquark BO-potentials (see red dots in Fig.~2 of Ref.~\cite{Prelovsek:2019ywc}) agrees with our above description at large $r$.\footnote{In Ref.~\cite{Prelovsek:2019ywc}, the potential was calculated using lattice spacing at $8$ values of $r$ ranging from $0.12$~fm to $0.99$~fm.}  

In the case of doubly heavy tetraquark $QQ\bar{q}\bar{q}$, the $\Sigma_g^+$ BO-potential (corresponding to LDF quantum number $0^{+}$) and  $\Sigma_g^-$ and $\Pi_g$ BO-potentials (corresponding to LDF quantum number $1^{+}$), which have attractive color triplet or repulsive color sextet behavior at small $r$, can mix with the S-wave plus S-wave static meson pair threshold. The mixing leads to BO-potentials approaching the threshold at large $r$. 
This behavior is in qualitative agreement with the lattice computation of the BO-potentials in Ref.~\cite{Bicudo:2015kna, Mueller:2023wzd}. 
Similar conclusions can be drawn for mixing between pentaquark states and static-light meson-baryon thresholds based on the comparison of the static energies in Tables~\ref{tab:QQbarpenta}, \ref{QQpenta} and \ref{tab:qqq}.

A final comment goes to the crossover phenomenon. 
Looking at  Fig.~\ref{morning}, one sees that for the case best investigated, i.e., the static energies of the hybrids, some crossover exists. 
BO quantum numbers are conserved, and static energies with different BO quantum numbers can cross. 
The reason for which they cross has been investigated in \cite{Alasiri:2024nue}.
For what concerns the BOEFT and the Schr\"odinger equations that we present here, such crossing has no impact.

\section{Spin effects, decays and transitions, state mixing}
\label{sec:spin}
Up to now we have discussed how to use the BOEFT in order to obtain the leading order coupled Schr\"odinger equations containing mixing effects. 
From such equations, we have predicted the hybrid, tetraquark, pentaquark and doubly heavy baryon multiplets with their quantum numbers.
Solving the Schr\"odinger equations with the static potentials, obtained from a lattice calculation of our interpolating operators given in Sec.~\ref{sec:characterization}, gives  the masses of all these states. 
Some of these static potentials are already known, and some have still to be calculated on the lattice. 
For the hybrids case in which the static potentials are known, the 
multiplets originating from Eq.~\eqref{eq:VQQ} have been extensively studied in 
\cite{Berwein:2015vca,Brambilla:2022hhi}.

However, the BOEFT is a systematic procedure  and corrections to the leading order can be calculated, in particular, the relativistic corrections to the BO potentials.
Additionally, inclusive decays can be obtained from the imaginary part of the potential
and the decay to heavy-light pairs from the off-diagonal entries of the potential in the diabatic description of the long distance mixing. 
In the next subsections,
we briefly comment about already existing results and what could still be calculated.

\subsection{Relativistic corrections to the potentials}
In Eq.~\eqref{eq:VQQ}, we have given the general form of the BO static potential. Up to now, we have focused 
just on the static contribution, however, this potential gets relativistic corrections
 starting at order $1/m_Q$:
\begin{align}
&V_{\kappa, \lambda\lambda^{\prime}}(r) = V^{(0)}_{\kappa, \vert\lambda \vert}(r)\delta_{\lambda\lambda^{\prime}}+\frac{V^{(1)}_{\kappa, \lambda\lambda' }(r)}{m_Q}+\dots\, .
\end{align}
These can be calculated through the non-perturbative matching procedure,
entailing to  define appropriate generalized Wilson loops with the insertion of the NRQCD operators appearing at higher order in the expansion in the inverse of the quark mass.
This procedure has been carried out for quarkonium in strongly coupled pNRQCD \cite{Brambilla:2004jw,Brambilla:2000gk,Pineda:2000sz} and the corresponding Wilson loops 
have been calculated on the lattice giving all the spin dependent and momentum dependent corrections to the quarkonium potential. 
The most relevant difference between quarkonium and exotics is that  in this last case the 
spin-dependent corrections already appear at order $1/m_Q$, the reason being that the LDF 
supply a spin vector $\bf K$ that is not suppressed by an inverse power of the quark mass.
This fact has an important impact on the phenomenology: 
heavy quark spin-symmetry is less manifest than in the quarkonium case, 
which means that spin dependent effects are enhanced  both in the spectrum and in the transition widths.
The impact of the spin dependent corrections on the multiplet structure has been investigated in depth in the case of the hybrids \cite{Oncala:2017hop, Brambilla:2018pyn,Brambilla:2019jfi,Soto:2023lbh}.
In \cite{Brambilla:2018pyn,Brambilla:2019jfi}, the whole set of spin dependent contributions
in the potential up to order $1/m_Q^2$ was obtained and all the hybrid multiplets were calculated. 
Thanks to the BOEFT factorization, the low energy correlators are flavor independent and 
those extracted from charmonium hybrid lattice calculations could be used to predict bottomonium hybrid spin multiplets.
The momentum and spin-dependent relativistic corrections in terms of generalized Wilson loops have been obtained for all exotics in Ref.~\cite{Soto:2020xpm}.

\subsection{Decays and transitions}

\subsubsection{Inclusive or semi-inclusive decays and mixing}
The information on the inclusive decay rates is carried by the imaginary part of the potentials in  BOEFT. 
Inclusive quarkonium decays have been studied in this way, see e.g.~\cite{Brambilla:2002nu}.
On similar grounds, also inclusive decays  of exotics can be calculated. Using this 
method, semi-inclusive spin-conserving \cite{Oncala:2017hop,Brambilla:2022hhi}
and spin-flipping \cite{Brambilla:2022hhi}
transitions from hybrids  to quarkonium plus $X$, where $X$ are light hadrons, have been calculated. 
The same method can be used to calculate similar tetraquark, or other exotic, semi-inclusive decays. Input from the lattice on some correlators arising in the BOEFT calculation is needed.

Finally, for very excited states inside a BO potential it may happen that their energy grows such 
that it starts to overlap the energy range of the states living in the next BO potential.
This can happen, for example, for excited quarkonium states and the states living in the first hybrid BO potential characterized by the quantum numbers $\Sigma_u^-$ and $\Pi_u$.
Then, the phenomenon mentioned at the end of Sec.~\ref{sec:BO} appears:
the mixing between ultrasoft states that exist in the same energy range. 
For the low-lying hybrids corresponding to the  quantum numbers $\kappa=1^{+-}$, 
the mixing with quarkonium is through a heavy-quark spin dependent operator coupled 
to a chromomagnetic field, which is $1/m_Q$ suppressed. 
The mixing is contained in the appropriate generalized Wilson loop, which has not yet 
been calculated on the lattice. 
Meanwhile, a QCD effective  string  calculation  of such a
Wilson loop has been performed to start a phenomenological investigation of the 
hybrid-quarkonium mixing \cite{Oncala:2017hop}.
It has been found that the mixing effects may  be important and produce large spin symmetry violations. The reason for this is that 
the mixing term implies that the actual physical states are a superposition of spin zero (one) hybrids and spin one (zero) quarkonium. This facilitates the identification of certain  states as hybrids, because it gives a reason for the apparent spin symmetry violating decays.
A similar effect may be mediated at short distance 
by the chromoelectric dipole term with no suppression in the 
inverse of the mass of the quark, but this has not yet been investigated.
These effects need to be explored for all exotics.
It is crucial to obtain lattice calculations for the relevant low energy correlators that depend only on chromomagnetic and chromoelectric fields.

\subsubsection{Decays to heavy-light pairs}
In the diabatic description introduced for the long range mixing in Sec.~\ref{sec:LD},
the decay of  quarkonium and hybrids to heavy-light meson-antimeson pairs
(also called the string breaking) can be obtained directly from the off-diagonal elements
of the potential matrix given in Eqs.~\eqref{eq:corr-avoided-crossing-1} and \eqref{eq:corr-avoided-crossing-2}.

\section{Phenomenological identification of XYZ states} 
\label{sec:phen}
The system of coupled Schr\"odinger equations describes the spin-symmetry multiplets of quarkonium tetraquark  $\left(Q\bar{Q}q\bar{q}\right)$  and pentaquark $\left(Q\bar{Q}qqq\right)$ states in Tables~\ref{tab:QQbarqqbar} and \ref{tab:QQbarpenta} and doubly heavy tetraquark $\left(QQ\bar{q}\bar{q}\right)$ and  pentaquark $\left(QQqq\bar{q}\right)$ states in Tables~\ref{QQqqnumbers} and \ref{QQpenta} respectively. 

\begin{table}[th!]
	\resizebox{0.77\columnwidth}{!}{%
		\begin{tabular}{|c||c|}
			\hline
			$J^{PC}$ or $J^{P}$ (multiplets)
			& Exotic tetraquark and pentaquark candidates\\
			\hline\hline
			$0^{++} \left(T_1^0, T_1^1\right)$ & $\chi_{c0}(3915)$, $\chi_{c0}(4500)$, $\chi_{c0}(4700)$, $X(4350)$\footnote{ Based on quantum numbers assigned in Refs.~\cite{Olsen:2017bmm, Brambilla:2022hhi}.}\\
			$0^{-+} \left(T_2^0, T_3^1\right)$ & $X(3940)$\footnote{Based on quantum numbers assigned in Ref.~\cite{Olsen:2017bmm}.}\\
			$0^{--} \left(T_2^1\right)$ & $T_{c\bar{c}0}(4240)^+$\\
			$1^{++} \left(T_1^1\right)$ & $\chi_{c1}(3872)$, $\chi_{c1}\left(4140\right)$, $\chi_{c1}\left(4274\right)$, $\chi_{c1}\left(4685\right)$ \\
			$1^{--} \left(T_2^0, T_2^1, T_3^1, T_4^1\right)$ & $\psi(4230)$, $\psi(4360)$, $\psi(4390)$\footnote{This state is not listed in PDG \cite{Workman:2022ynf}. Its existence has been suggested by  the BESII experiment \cite{BESIII:2016adj}. For a critical review see Ref.~\cite{Brambilla:2019esw}.}, $\psi\left(4660\right)$, $Y(4500)$\footnote{State recently observed by the BESIII collaboration~\cite{BESIII:2022joj}. This state is not listed in PDG \cite{Workman:2022ynf}.}, $Y(4710)$\footnote{State recently observed by the BESIII collaboration~\cite{BESIII:2022kcv}. This state is not listed in PDG \cite{Workman:2022ynf}.}\\ & $\Upsilon\left(10753\right)$, $\Upsilon\left(10860\right)$, $\Upsilon\left(11020\right)$ \\
			$1^{-+} \left(T_2^0, T_2^1\right)$ & $X(4630)$\footnote{The $J^{PC}=1^{-+}$ assignment is favored over $J^{PC}=2^{-+}$
				with a $3\sigma$ significance and other assignments are disfavored by more than $5\sigma$ \cite{Workman:2022ynf}.} \\
			$1^{+-} \left(T_1^0, T_1^1, T_3^0\right)$ & $Z_c(3900)$, $Z_c(4020)$\footnote{Based on quantum numbers assigned in Ref.~\cite{Olsen:2017bmm, Brambilla:2019esw}.}, $Z_c(4200)$, $Z_c(4430)$\\ & $Z_b(10610)$, $Z_b(10650)$\\
			$2^{++} \left(T_3^0, T_1^1\right)$ & $X(4350)$\footnote{The quantum numbers of $X(4350)$ are $J^{PC}=(0 \, \mbox{or} \, 2)^{++}$~\cite{Belle:2009rkh}.}\\
			$2^{-+} \left(T_2^0, T_4^1\right)$ & $X(4160)$\footnote{The $J^{PC}=2^{-+}$ assignment is favored over other assignments with more than $5\sigma$ significance  \cite{Workman:2022ynf}.} \\
			\hline
			$1^+ \left(k^P=0^+, I=0\right)$ & $T_{cc}\left(3875\right)^+$\\
			\hline
			$(1/2)^- \left(k^P=\left(1/2\right)^+\right)$ & $P_{c\bar{c}}\left(4312\right)^+$\footnote {Based on quantum numbers assigned in Ref.~\cite{LHCb:2019kea}.}, $P_{c\bar{c}}\left(4440\right)^+$\footnote {Based on quantum numbers assigned in Ref.~\cite{PavonValderrama:2019nbk, Chen:2022asf}.}\\
			$(3/2)^- \left(k^P=\left(3/2\right)^+\right)$ & $P_{c\bar{c}}\left(4380\right)^+$\footnote {Based on quantum numbers assigned in Ref.~\cite{LHCb:2015yax,Olsen:2017bmm}.}, $P_{c\bar{c}}\left(4457\right)^+$\footnote {Based on quantum numbers assigned in Ref.~\cite{PavonValderrama:2019nbk, Chen:2022asf}.}\\
			
			\hline
	\end{tabular}}
	\caption{A list of potential exotic XYZ candidates for tetraquarks and pentaquarks corresponding to the quantum numbers $J^{PC}$ or $J^{P}$ in the spin-symmetry multiplets shown in  Tables~\ref{tab:QQbarqqbar}, \ref{QQqqnumbers}, \ref{tab:QQbarpenta}, and \ref{QQpenta}. 
		The left column displays the quantum numbers $J^{PC}$ or $J^{P}$ with the associated multiplet information enclosed in parentheses. 
		The second to last entry in the table corresponds to a doubly heavy tetraquark and the last entry corresponds to a pentaquark state.  In PDG,  the $Z_c$ and $Z_b$ states in the table are termed as $T_{c\bar{c}1}$ and $T_{b\bar{b}1}$, respectively, based on the new naming convention \cite{Workman:2022ynf}.}
	\label{tab:XYZ}
\end{table}

\begin{table}[th!]
	\begin{tabular}{|c||c|}
		\hline
		$J^{PC}$ or $J^{P}$ (multiplet [mass])
		& Exotic hybrid candidates\\
		\hline\hline
		$1^{--} \left(H_1\left[4507\right]\right)$ &  $\psi(4360)$, $\psi(4390)$\footnote{This state is not listed in PDG \cite{Workman:2022ynf}. 
			Its existence has been suggested by  the BESIII experiment \cite{BESIII:2016adj}. 
			For a critical review see Ref.~\cite{Brambilla:2019esw}.},  $Y(4710)$\footnote{State recently observed by the BESIII collaboration~\cite{BESIII:2022kcv}. 
			This state is not listed in PDG \cite{Workman:2022ynf}.}\\ 
		$1^{--} \left(H_1\left[10786\right]\right)$& $\Upsilon\left(10753\right)$ \\
		$1^{-+} \left(H_1\left[4507\right]\right)$ & $X(4630)$\footnote{The $J^{PC}=1^{-+}$ assignment is favored over $J^{PC}=2^{-+}$ with a $3\sigma$ significance, and other assignments are disfavored by more than $5\sigma$ \cite{Workman:2022ynf}.} \\
		$2^{-+} \left(H_1\left[4155\right]\right)$ & $X(4160)$\footnote{The $J^{PC}=2^{-+}$ assignment is favored over other assignments with more than $5\sigma$ significance  \cite{Workman:2022ynf}.} \\
		\hline
	\end{tabular}
	\caption{XYZ candidates for quarkonium hybrids based on mass, quantum numbers and decays to quarkonium taken from \cite{Brambilla:2022hhi}. The left column displays the quantum numbers $J^{PC}$ along with the multiplet and mass enclosed in parentheses. The results for the hybrid multiplets, spectrum, and decays to quarkonium can be found in Ref.~\cite{Brambilla:2022hhi}.}
	\label{tab:XYZ-hybrid}
\end{table}

For a specific LDF quantum number $k^{PC}$ or $k^P$, the $\Lambda$-doubling effect breaks the degeneracy between opposite parity spin-symmetry multiplets and lowers the mass of the multiplets that receive mixed contributions from various static energies. 
This has been shown specifically for quarkonium hybrids in Refs.~\cite{Berwein:2015vca, Oncala:2017hop, Brambilla:2022hhi}. 
Based on the quantum numbers $J^{PC}$ or $J^P$ in the  spin-symmetry multiplets in Tables~\ref{tab:QQbarqqbar}-\ref{QQpenta}, we show potential exotic XYZ candidates for tetraquark and pentaquark states that can belong to those multiplets in Table~\ref{tab:XYZ}. 
For completeness, we also show the XYZ candidates for quarkonium hybrids based on the mass, quantum numbers $J^{PC}$, and decays to quarkonium in Table~\ref{tab:XYZ-hybrid}  extracted from \cite{Brambilla:2022hhi}. 
For a given $k^{PC}$ or $k^P$ quantum number of the LDF, we know that the $1/m_Q$ terms in the BO-potential in Eq.~\eqref{eq:VQQ} include spin-dependent operators that break the degeneracy in the spin-symmetry multiplets \cite{Soto:2020xpm, Brambilla:2019jfi, Brambilla:2018pyn}. 
Moreover, there exist $1/m_Q$ terms that could potentially mix states associated with distinct $k^{PC}$ quantum numbers of the LDF. 
This  was particularly emphasized in the context of hybrid-quarkonium mixing \cite{Oncala:2017hop}. 
Ideally, the complete $1/m_Q$ potential terms should be again determined in lattice QCD, but such a computation is not available yet. 

For a
thorough analysis of the XYZs, 
it is left to future work the explicit solution of the coupled
Schr\"odinger 
equations 
for hybrids, tetraquarks, and pentaquarks, 
the inclusion of
spin-dependent corrections, 
and quarkonium-exotic
mixing 
should also be considered  whenever it becomes important. 
The calculation of transition and decay widths will allow to sharpen the states' characterization. 
In the BOEFT, such a comprehensive and systematic approach may become possible for the first time.

In this paper, we did not address states formed by four heavy quarks as it is e.g.\ the case of the $X(6900)$. 
Such states can be described using a BOEFT  once a potential has been defined, as it has been done in the case of triply heavy baryons $QQQ$ \cite{Brambilla:1993zw,Brambilla:2005yk,Brambilla:2009cd,Brambilla:2013vx}, and it will necessitate the calculation on the lattice of an appropriate generalized Wilson loop from where to extract the static energy.
Should such states be dominated by the short range regime, 
then weakly coupled pNRQCD could be used together with the methods introduced in \cite{Assi:2023cfo}.

\section{Conclusions} \label{sec:conclusion}
In this work, we have presented a QCD description of states with two heavy quarks (heavy quark-antiquark or heavy quark-quark pairs) including exotic hybrids, tetraquark and pentaquark states. 
The description is  based on the Born--Oppenheimer nonrelativistic effective field theory framework (BOEFT). 
The underlying principle is the systematic factorization of  the dynamics of heavy quarks and the light degrees of freedom (LDF) based on the  expansion $\Lambda_{\mathrm{QCD}}\gg E$.

Within the BOEFT, hybrid, tetraquark and pentaquark states are bound states of two heavy quarks in the BO-potentials that depend on the LDF quantum numbers. 
The leading order BOEFT potentials in the $1/m_Q$ expansion are the energies of the LDF in the presence of static heavy quark pairs. 
We have shown that the same description applies to quarkonium and doubly heavy baryons.

In this framework, the only parameters are the fundamental QCD parameters, such as the quark masses in a given scheme, the strong coupling constant $\als$, and $\L$. 
However, the theory needs the lattice calculation of some universal low energy correlators, whose precise form we have given in Sec.~\ref{sec:characterization} and Appendix~\ref{app:Overlap}. 
These correlators are universal (they are heavy flavor independent, because the flavor dependence is inside the NRQCD matching coefficients) and once calculated can be used for several applications both in the bottom and in the charm sector.

We describe 
exotic states as a superposition of different static states, as was done for quarkonium hybrids in Ref.~\cite{Berwein:2015vca}. 
At short distances, the orbital or angular part of the kinetic energy operator in the Schr\"odinger equation mixes the wavefunctions corresponding to different static states (an effect known as $\Lambda$-doubling in molecular physics), which leads to coupled Schr\"odinger equations. 
This is representative of the adiabatic Born--Oppenheimer framework. 
We derived 
a general expression of the mixing matrix for an arbitrary set of quantum numbers of the LDF. 
The specific form is given by Eqs.~\eqref{eq:Mlambdalambdap} and \eqref{eq:Mlambdalambdap-1} and the explicit expressions for the integer (hybrids, tetraquarks) and half-integer (doubly heavy baryons, pentaquarks) states are obtained in Eqs.~\eqref{eq:M1}-\eqref{eq:M3/2}.  
In Appendix~\ref{app:Schreodinger}, we give the explicit form of the coupled radial Schr\"odinger equations for the lowest hybrid, tetraquark and pentaquark states.

The BO-potentials (static energies) for exotics corresponding to  LDF states enter the coupled Schr\"odinger equations through a diagonal potential matrix in the adiabatic BO-framework [see Eqs.~\eqref{eq:VQQ} and \eqref{eq:Sch2}]. 
In Sec.~\ref{subsec:potential}, we describe the short distance behavior of these BO-potentials. 
As long as one does not need to consider higher order corrections in the multipole 
expansion, which are suppressed in the short range, 
the short range behavior of the BO-potentials depends only on two nonperturbative parameters: 
the mass dimension one constant $\Lambda_{\kappa}$ 
and the quadratic slope as shown in Eqs.~\eqref{eq:QQbarpot-short} and \eqref{eq:QQpot-short}.
The parameter $\Lambda_{\kappa}$ in the case of hybrids is called the gluelump mass and has been evaluated on the lattice by multiple collaborations \cite{Campbell:1985kp,Foster:1998wu,Marsh:2013xsa,Herr:2023xwg,Bali:2003jq},
while in the case of adjoint tetraquarks and pentaquarks
the adjoint meson and the baryon mass still need to be evaluated.
We have also listed a set of light quark operators at the NRQCD level that transform under gauge transformations in such a way to define the gauge-invariant interpolating operators shown in Eqs.~\eqref{eq:intop-1}-\eqref{eq:intop-3} for exotic hadrons with two heavy (anti)quarks (see also Appendix~\ref{app_Gauge}). 
The two-point correlators involving these interpolating operators may be expressed in terms of Wilson loops eventually to be computed on the lattice to extract the BO-potentials for tetraquarks and pentaquarks. 

In Sec.~\ref{sec:LD}, we describe the long distance behavior of the BO-potentials for exotic states focusing on the mixing with the heavy-light pair threshold. 
The mixing only happens if the LDF of the heavy-light pair threshold (see Tables~\ref{tab:qqbar-qq} and \ref{tab:qqq}) have the same BO quantum numbers as the LDF of the exotic state.
This implies, for example, that the $\Sigma_u^-$ BO-potential of quarkonium hybrids can mix with the $\Sigma_u^-$ static energy of the S-wave  meson-antimeson pair threshold 
(see first row in Table~\ref{tab:qqbar-qq}),  and the mixing is significant if the energy gap between the two is less than $\Lambda_{\mathrm{QCD}}$. 
We write the expression for the potential matrix that captures this mixing in Eq.~\eqref{eq:corr-avoided-crossing-2}. 
The mixing phenomenon implies that the hybrid can decay into an S-wave plus S-wave meson-antimeson pair, challenging conventional understanding. 
With respect to tetraquarks and pentaquarks, we conclude that the mixing with the heavy-light pair threshold proceeds without string-breaking, which is a new finding. 
In the diabatic framework, the BO potentials for tetraquark and pentaquark states that approach the static heavy-light meson or baryon pair threshold intersect with the quarkonium and hybrid BO potentials. This picture is particularly attractive for what concerns the 
avoided level crossing 
that happens
between quarkonium and the first tetraquark static energy with the same BO quantum numbers. 
We believe that this is the phenomenon underlying the existence of an exotic state like the $\chi_{c1}(3872)$ with prominent molecular characteristics.
In the case in which the states in different 
BO static energies 
overlap (for excited states or for small gaps between the static energies), 
the operators responsible for the mixing  are a spin-violating operator coupled to the chromomagnetic field, 
suppressed by a power of the mass of the heavy quark, 
and, at short distance, a
spin-conserving operator coupled to the chromoelectric field 
that is not suppressed in the heavy quark mass.

{\it The results of our findings are manifold}.
First, we have made available for the first time the set of coupled Schr\"odinger equations and related multiplets for all quarkonium exotics. 
Second, we have given appropriate gauge invariant operators to be used on the lattice to obtain the exotic static energies. 
An important feature of such operators is that in the
tetraquark case they have zero overlap with states of quarkonium and quarkonium plus pion(s), while they do have overlap with heavy-light states (see Appendix~\ref{app:Overlap}), a feature that could facilitate lattice calculations. 
The behavior of the static energies at short distances that we have given in Sec.~\ref{sec:characterization} may serve as an important test for lattice calculations.
Third, we have supplied a comprehensive framework in which to account for the different types of mixing as well as the role and the impact of relativistic and spin-dependent corrections, which can be calculated systematically in the BOEFT. 
Mixing can happen at short distances and is fully accounted for by the given coupled Schr\"odinger equations. 
It can also happen at large distances (avoided level crossing), 
in which case it needs the input of the potentials from lattice QCD calculations 
and it will be relevant close to the strong decay threshold and possibly for states with a more prominent molecular nature like the $\chi_{c1}(3872)$.

{\it More in general, 
	the method may be applied to a vast variety of observables.}
The BOEFT allows for the study of quarkonia, hybrids, tetraquarks and pentaquarks in an unified framework derived from QCD. 
Leveraging on scale separation, factorization and symmetries, it allows for greater simplifications with respect to direct lattice QCD evaluations of the XYZ properties.
Lattice QCD calculations are an important input, but the correlators requested by the BOEFT are simpler and universal, allowing to access a large number of observables. 
As we have shown, the BOEFT already has a large model independent predictive power based on its symmetries, 
for instance,  we can obtain the exotic multiplets and several decay selection rules only on the basis of them.
Then, once the lattice input has been incorporated at the level of static energies, gluelump and adjoint meson masses,
and generalized Wilson loops containing mixing and spin corrections, the predictive power of the theory is further enhanced. 
The BOEFT framework could allow for the study of exotic quarkonia production as well as propagation of exotic quarkonia in medium, which are processes not directly accessible to lattice computations. 
In fact, BOEFT could be applied to such studies following what has been done 
in NRQCD \cite{Bodwin:1994jh} and pNRQCD \cite{Brambilla:2022ayc,Brambilla:2022rjd} for quarkonium production 
and in \cite{Brambilla:2017zei,Brambilla:2021wkt,Brambilla:2023hkw} for quarkonium evolution in medium. 

The BOEFT can reconcile in a unique framework the different approaches used up to now in the literature 
to describe XYZ states: the tetraquark model and the molecular description.
These seemingly different and not reconciliable approaches appear to be both contained in the BOEFT.
In fact, our work indicates that the tetraquark static energies contain and evolve into the heavy-light static energies at large distances. 
Hence, the BOEFT contains, in a sense, both the tetraquark model, naturally supplying a repulsive barrier for short distances\footnote{Phenomenological compact tetraquark models in the
literature needs such a repulsive core, see Refs.~\cite{Maiani:2019cwl, Maiani:2019lpu}}  
and the molecular model, arising from the avoided level crossing close to the strong decay threshold region. 
It will be the dynamics of QCD, encoded in the static energies and in the form of the mixing, 
that decide which state has characteristics of a more prominent molecular or compact tetraquark nature.

In general, these results allow us to give a fresh new look to the strong force.
The confining properties of the potential between a heavy quark and a heavy antiquark have been studied since the paper of Wilson in 1974. 
However, now we finally understand the interplay of the full spectrum of 
static energies in (NR)QCD. 
We understand that the confining properties are typical of a quarkonium or a hybrid state up to the region in which an avoided level crossing with a tetraquark static energy with the same BO quantum numbers happens. 
We understand that the tetraquark static energies will show no confining properties.
The strongly interacting world interestingly resembles more the atomic and molecular realm.

\section*{Acknowledgements}
This work has been supported by the DFG Project-ID 196253076 TRR 110 and the NSFC through
funds provided to the Sino-German CRC 110 ``Symmetries and the Emergence of Structure in QCD". 
We acknowledge the DFG cluster of excellence ORIGINS funded by the Deutsche Forschungsgemeinschaft under Germany’s Excellence Strategy-EXC-2094-390783311 and the  STRONG-2020-Strong Interaction at the Frontier of Knowledge: Fundamental Research and Applications”. 
STRONG-2020 has received funding from the European Union’s Horizon 2020 research and innovation program under grant agreement No. 824093. 
N. B. acknowledges the European Union ERC-2023-ADG- Project EFT-XYZ.
We are grateful to Wai-Kin Lai for collaboration in the early stage of this work. We acknowledge useful discussions with Eric Braaten, Roberto Bruschini,  Francesco Knechtli, Lasse M\"uller, Sasa Prelovsek, Tommaso Scirpa and  Marc Wagner. 

\appendix

\section{Avoided level crossing} \label{app_crossing}
\begin{figure}[h!]
	\begin{minipage}[b]{0.48\textwidth}
		\includegraphics[width=0.95\textwidth]{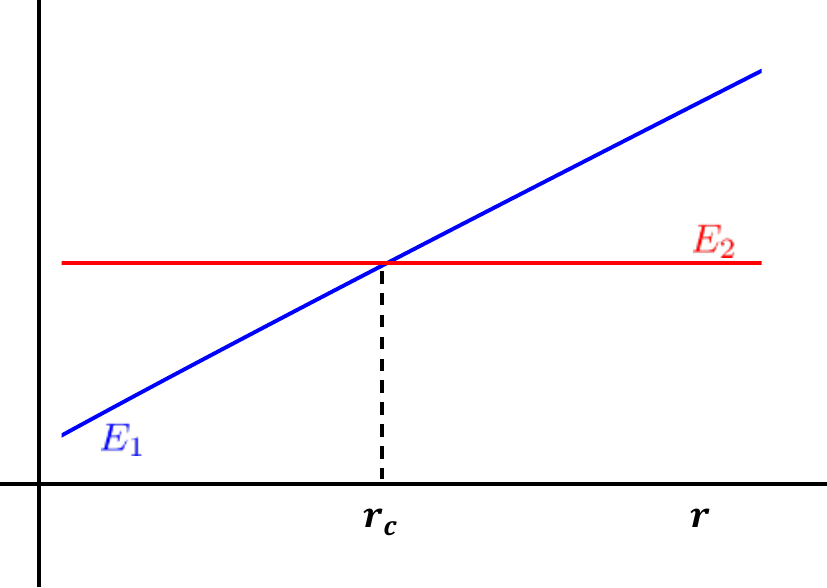}
	\end{minipage}
	\hfill
	\begin{minipage}[b]{0.48\textwidth}
		\hspace*{0mm}
		\vspace*{0.02in} 
		\includegraphics[width=0.95\textwidth]{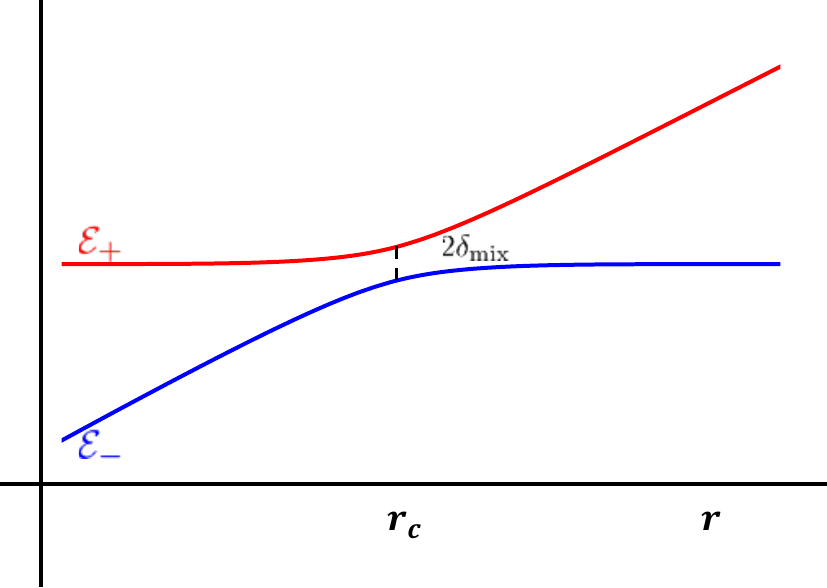}
	\end{minipage}
	\caption{Left: level crossing. Right: avoided level crossing. }
	\label{fig:crossing}    
\end{figure}

Avoided level crossing happens whenever the Hamiltonian of a system has diagonal elements that become equal at some point of the parameter space, while the off-diagonal elements are small but finite~\cite{PhysRevA.23.3107}.
For illustration consider the $2\times2$ Hamiltonian
\begin{equation}
	\bra{i}  H \ket{j}
	= \begin{pmatrix} E_1(r)  & \delta_{\textrm{mix}} \\ \delta_{\textrm{mix}} & E_2(r)  \end{pmatrix}_{ij}, \qquad \textrm{for} \quad i,j=1,2,
\end{equation}
constructed on a basis of orthonormal states $|1\rangle$ and $|2\rangle$ that clearly are not eigenstates of $H$ as long as $\delta_{\textrm{mix}} \neq 0$. 
The diagonal entries $E_1(r)$ and $E_2(r)$ depend on some parameter $r$ (e.g. the heavy quark-antiquark or quark-quark distance).
We assume that $E_1(r)$ and $E_2(r)$ behave like in Fig.~\ref{fig:crossing}, left panel, i.e. that they cross at some point $r=r_c$.
Furthermore, we assume that the off diagonal entry, $\delta_{\textrm{mix}}$, is constant, positive and much smaller than $|E_1(r)-E_2(r)|$ for $r\gg r_c$ and $r\ll r_c$.

Diagonalizing the matrix $H$ leads to the two eigenvalues
\begin{equation}
	\mathcal{E}_\pm(r) = \frac{E_1(r) + E_2(r)}{2} \pm \frac{1}{2}\sqrt{(E_1(r)-E_2(r))^2 + 4 \delta_{\textrm{mix}}^2 },
	\label{eq:A2}  
\end{equation}
where $\mathcal{E}_-(r)$ is the ground state and $\mathcal{E}_+(r)$ the excited state.
At $r=r_c$, $E_1(r_c)=E_2(r_c)$ and the two eigenvalues become
\begin{equation}
	\mathcal{E}_\pm(r_c) = E_1(r_c) \pm \delta_{\textrm{mix}}.
\end{equation}
This means that the two levels do not cross at $r=r_c$, but develop a gap, which is $2 \delta_{\textrm{mix}}$.
For $r \gg r_c$ or $r \ll r_c$, under the assumption $\delta_{\textrm{mix}} \ll |E_1(r)-E_2(r)|/2$, we obtain (at leading order in $\delta_{\textrm{mix}}$)
\begin{equation}
	\mathcal{E}_\pm(r) \approx \frac{E_1(r) + E_2(r)}{2} \pm \frac{|E_1(r) - E_2(r)|}{2} .
\end{equation}
This implies that $\mathcal{E}_+(r)$ behaves like $E_2(r)$ for $r\ll r_c$ and like $E_1(r)$ for $r \gg r_c$.
Viceversa, $\mathcal{E}_-(r)$ behaves like $E_1(r)$ for $r\ll r_c$ and like $E_2(r)$ for $r \gg r_c$.
The behaviour of $\mathcal{E}_\pm(r)$ is shown in Fig.~\ref{fig:crossing}, right panel.
It goes under the name of avoided level crossing.

The eigenstates of $H$, $H|\pm\rangle = \mathcal{E}_\pm(r)|\pm\rangle$, are a linear combination of the original states $|1\rangle$ and $|2\rangle$,
\begin{align}
	\begin{pmatrix}  \ket{+}\\ \ket{-} \end{pmatrix} = 
	U(\theta(r)) \begin{pmatrix}  \ket{1}\\ \ket{2} \end{pmatrix},
	\quad \textrm{with} \quad 
	U(\theta(r))=    \begin{pmatrix}  \cos\frac{\theta(r)}{2} &  \sin\frac{\theta(r)}{2}\\
		-\sin\frac{\theta(r)}{2} & \cos\frac{\theta(r)}{2} 
	\end{pmatrix}, 
	\label{A5}
\end{align}
where the mixing angle, $0\le \theta(r) < \pi$, has been defined as $\tan\theta(r) \equiv 2\delta_\mathrm{mix}/(E_1(r)-E_2(r))$.
For $r\ll r_c$, $\tan\theta(r)$ is small and negative, hence $\theta$ approaches $\pi^-$; 
for  $r\gg r_c$, $\tan\theta(r)$ is small and positive, hence $\theta$ approaches $0^+$;
at $r=r_c$, $\theta(r_c) = \pi/2$. 
Consistently with the behaviour of the energy eigenvalues $\mathcal{E}_\pm(r)$, 
the excited eigenstate $|+\rangle$ approaches $|2\rangle$ for $r\ll r_c$ and 
$|1\rangle$ for $r\gg r_c$.
Viceversa, the ground state $|-\rangle$ approaches $-|1\rangle$ for $r\ll r_c$ and 
$|2\rangle$ for $r\gg r_c$.

The evolution operator in the $\ket{1}$ and $\ket{2}$ basis reads
\begin{align}
	\bra{1} e^{-iHT} \ket{1} &= \cos^2\frac{\theta(r)}{2}\, e^{-i\mathcal{E}_+(r)T} + \sin^2\frac{\theta(r)}{2}\, e^{-i\mathcal{E}_-(r)T}, \\
	\bra{2} e^{-iHT} \ket{2} &= \sin^2\frac{\theta(r)}{2} \,  e^{-i\mathcal{E}_+(r)T} + \cos^2\frac{\theta(r)}{2} \, e^{-i\mathcal{E}_-(r)T}, \\
	\bra{1} e^{-iHT} \ket{2} &= \bra{2} e^{-iHT} \ket{1} = 
	\frac{\sin\theta(r)}{2}\left(e^{-i\mathcal{E}_+(r)T} - e^{-i\mathcal{E}_-(r)T} \right),
\end{align}
where $T$ is time.
For $r$ far away from $r_c$ (either $r\gg r_c$ or $r \ll r_c$) the first matrix element 
behaves like $e^{-iE_1T}$ and the second one like $e^{-iE_2T}$.
Determining the above three matrix elements means determining the three functions 
$\mathcal{E}_{\pm}(r)$ and $\theta(r)$. 
Once these are known the $2\times2$ matrix $H$ in the $\ket{1}$ and $\ket{2}$ basis 
may be reconstructed through the rotated basis \eqref{A5}.
Alternatively, we can write
\begin{align}
	&\frac{i}{T} \log 
	\begin{pmatrix}  \bra{1} e^{-iHT}\ket{1} & \bra{1} e^{-iHT}\ket{2} \\
		\bra{2} e^{-iHT}\ket{1} & \bra{2} e^{-iHT}\ket{2} 
	\end{pmatrix}
	= 
	\frac{i}{T} \log \left[ U(\theta(r))^{-1}
	\begin{pmatrix}  e^{-i\mathcal{E}_+(r)T}  &  0\\
		0 & e^{-i\mathcal{E}_-(r)T}  
	\end{pmatrix}
	U(\theta(r))\right]\nonumber\\
	& \qquad =  U(\theta(r))^{-1}
	\begin{pmatrix}  \mathcal{E}_+(r) & 0\\
		0 & \mathcal{E}_-(r)
	\end{pmatrix}
	U(\theta(r))
	= 
	\begin{pmatrix} E_1(r)  & \delta_{\textrm{mix}} \\ \delta_{\textrm{mix}} & E_2(r)  
	\end{pmatrix}.
\end{align}

\section{Wigner D matrices and mixing matrices} \label{app_Wigner_matrix}
The conventional spherical coordinate system  used for the relative coordinate is spanned by the vectors
\begin{equation}
	\hat{\bm{r}}=\begin{pmatrix} \sin\theta\cos\varphi \\ \sin\theta\sin\varphi \\ \cos\theta \end{pmatrix}\,,\qquad\hat{\bm{\theta}}=\begin{pmatrix} \cos\theta\cos\varphi \\ \cos\theta\sin\varphi \\ -\sin\theta \end{pmatrix}\,,\qquad\hat{\bm{\varphi}}=\begin{pmatrix} -\sin\varphi \\ \cos\varphi \\ 0 \end{pmatrix}\,.
\end{equation}
With these, the heavy quark angular momentum operator $\bm{L}_Q$ can be expressed as\footnote{To be more accurate, we should distinguish between $\bm{L}_Q$ as a differential operator and $\bm{L}_Q$ in terms of field operators. However, their relation is such that the field operator $\bm{L}_Q$ acting on $\ket{\theta,\varphi}$ has the effect of the differential operator $\bm{L}_Q$ acting on any function of the coordinates convoluted with $\ket{\theta,\varphi}$, so we simply use the same symbol for both out of convenience.} 
\begin{equation}
	\bm{L}_Q=-i\hat{\bm{\varphi}}\,\partial_\theta+\frac{i}{\sin\theta}\hat{\bm{\theta}}\,\partial_\varphi\,.
\end{equation}
However, only one out of these three coordinate vectors actually transforms as a vector under rotations:
\begin{align}
	\left[L_{Q}^i\,,\hat{r}^j\right]&=i\hat{\theta}^i\hat{\varphi}^j-i\hat{\varphi}^i\hat{\theta}^j=i\epsilon^{ijk}\hat{r}^k\,,\\
	\left[L_Q^i,\hat{\theta}^j\right]&=i\hat{\varphi}^i\hat{r}^j+i\cot\theta\,\hat{\theta}^i\hat{\varphi}^j\,,\\
	\left[L_Q^i,\hat{\varphi}^j\right]&=-i\hat{\theta}^i\hat{r}^j-i\cot\theta\,\hat{\theta}^i\hat{\theta}^j\,.
\end{align}
This can be remedied by introducing a new angle $\psi$ in the following way:\footnote{We adhere to the convention of active rotation, where the vector undergoes transformation as a result of rotation, while keeping the coordinate system fixed. An active rotation for the vector corresponds to a passive rotation for the coordinate system.} 
\begin{equation}
	\bm{L}_Q'=-i\left(\bm{\hat{r}}+\cot\theta\bm{\hat{\theta}}\right)\partial_\psi-i\hat{\bm{\varphi}}\,\partial_\theta+\frac{i}{\sin\theta}\hat{\bm{\theta}}\,\partial_\varphi\,,
\end{equation}
\begin{equation}
	\bm{\hat{x}}'=\cos\psi\bm{\hat{\theta}}+\sin\psi\bm{\hat{\varphi}}\,,\qquad\bm{\hat{y}}'=-\sin\psi\bm{\hat{\theta}}+\cos\psi\bm{\hat{\varphi}}\,,\qquad\bm{\hat{z}}'=\bm{\hat{r}}\,.
	\label{eq:x'-y'-z'}
\end{equation}
It can be easily checked that in this new coordinate system $\left(\bm{\hat{x}}',\bm{\hat{y}}',\bm{\hat{z}}'\right)$ each direction transforms as a vector under rotations. The new angle $\psi$ can be interpreted as the  Euler first angle. We follow the $z$-$y$-$z$ convention for the rigid-body rotation: the rigid body is first rotated around the space-fixed $z$-axis by an angle $\psi$, then around the space-fixed $y$-axis  by an angle $\theta$, and finally around the space-fixed $z$-axis by an angle $\varphi$. The new axes $\left(\bm{\hat{x}}',\bm{\hat{y}}',\bm{\hat{z}}'\right)$ are body-fixed axes, which naturally transform as vectors when the rigid body is rotated and the new angular momentum operator $\bm{L}_Q'$ is known as the generator for rotations of a rigid body.  In terms of space-fixed axes $\left(\bm{\hat{x}},\bm{\hat{y}},\bm{\hat{z}}\right)$, the rotation matrix is given by
\begin{equation}
	R\left(\psi, \varphi, \theta\right)=R_z\left(\psi\right)  R_y\left(\theta\right) R_z\left(\varphi\right) .
	\label{eq:rotationR}
\end{equation}
where $R_{n}\left(\theta\right)$ denotes the rotation matrix describing  a rotation about the $\hat{\bm{n}}$ axis by an angle $\theta$. 
We may introduce $\psi$ into the expression of the exotic states in the following way:
\begin{equation}
	\ket{l,m;k,\lambda}=\int \frac{d\Omega'}{\sqrt{2\pi}}\ket{\theta,\varphi}\ket{k,\lambda}'D_{lm}^\lambda(\psi,\theta,\varphi)\,,\label{Lstate-extended}
\end{equation}
where
\begin{equation}
	\int d\Omega'=\int d\Omega\int_{-\pi}^\pi d\psi\,,\qquad\ket{k,\lambda}'=e^{-i\lambda\psi}\ket{k,\lambda}\,,\quad\mathrm{and}\quad D_{lm}^\lambda(\psi,\theta,\varphi)=\frac{e^{i\lambda\psi}}{\sqrt{2\pi}}v_{lm}^\lambda(\theta,\varphi)\,.
\end{equation}
We see that the two phase factors in $\ket{k,\lambda}'$ and $D_{lm}^\lambda$ cancel each other, and the two normalization factors $1/\sqrt{2\pi}$ cancel against the integral over $\psi$, so this expression is evidently the same as Eq. \eqref{eq:angular}.
Since the whole expression does not depend on $\psi$, it is also no problem to replace $\bm{L}_Q$ by $\bm{L}_Q'$.

Since the static heavy quark-antiquark pair is a linear rigid rotator, there is no physical meaning for $\psi$, it does not relate to the position of a static particle, as opposed to $\theta$ and $\varphi$ (which is the reason why we could not promote $\ket{\theta,\varphi}$ to a $\ket{\psi,\theta,\varphi}$ state as well). However, there is a great advantage gained from its inclusion. The coordinates for the light degrees of freedom in $\ket{k,\lambda}$ are expressed relative to the $\left(\bm{\hat{x}}',\bm{\hat{y}}',\bm{\hat{z}}'\right)$ axes. An active rotation generated by $\bm{L}_Q'$ acting on the coordinates of the static quarks (including $\psi$) is, therefore, perceived as a passive rotation by the coordinates of the light degrees of freedom:
\begin{equation}
	\bm{L}_Q'\ket{k,\lambda}'=-\bm{K}\ket{k,\lambda}'\,.
\end{equation}
Consequently, when acting with the combined angular momentum operator $\bm{L}^2=\left(\bm{K}+\bm{L}_Q'\right)^2$ on the exotic states, 
we obtain:
\begin{align}
	\bm{L}^2\ket{l,m;k,\lambda}&=\int\frac{d\Omega'}{\sqrt{2\pi}} \left(\bm{K}^2+2\bm{K}\cdot\bm{L}_Q'+\bm{L}_Q'^2\right)\ket{\theta,\varphi}\ket{k,\lambda}'D_{lm}^\lambda(\psi,\theta,\varphi)\notag\\
	&=\int\frac{d\Omega'}{\sqrt{2\pi}}\ket{\theta,\varphi}\ket{k,\lambda}'\bm{L}_Q'^2\,D_{lm}^\lambda(\psi,\theta,\varphi)\,,
	\label{eq:Lsq}
\end{align}
where we have used
\begin{align}
	\bm{L}_Q^{\prime2}\ket{k,\lambda}'&={\bm K}^{2}\ket{k,\lambda}'-2\left(\bm{K}\ket{k,\lambda}'\right)\cdot\bm{L}_Q'+\ket{k,\lambda}'\bm{L}_Q^{\prime2},\nonumber\\
	\bm{K}\cdot\bm{L}_Q'\ket{k,\lambda}'&=\left(\bm{K}\ket{k,\lambda}'\right)\cdot\bm{L}_Q'-\bm{K}^2\ket{k,\lambda}'.
\end{align}
Hence, in Eq.~\eqref{eq:Lsq}, in order for $\ket{l,m;k,\lambda}$ to be an eigenstate of the combined angular momentum $\bm{L}^2$, it is sufficient for the orbital wave function $D_{lm}^\lambda$ to be an eigenfunction of $\bm{L}_Q'^2$.

The eigenfunctions of angular momentum in the case of a rotating rigid body are known in terms of Wigner's D-matrices (which is why we use the symbol $D$). It can be easily verified that the  components of $\bm{L}_Q'$ satisfy the same commutation relations as those of $\bm{L}_Q$. 
Moreover, the set of operators $\{\bm{\hat{x}}'\cdot\bm{L}_Q',\bm{\hat{y}}'\cdot\bm{L}_Q',\bm{\hat{z}}'\cdot\bm{L}_Q'\}$ commutes with $\bm{L}_Q'$ and exhibits almost similar commutation relations:
\begin{align}
	\left[\bm{\hat{x}}'\cdot\bm{L}_Q',\bm{\hat{y}}\cdot\bm{L}_Q'\right]&=\hat{x}^{\prime i}\left[L_Q^{\prime i},\hat{y}^{\prime j}\right]L_Q^{\prime j}+\hat{x}^{\prime i}\hat{y}^{\prime j}\left[L_Q^{\prime i}, L_Q^{\prime j}\right]+\hat{y}^{\prime j}\left[\hat{x}^{\prime i}, L_Q^{\prime j}\right]L_Q^{\prime i}\notag\\
	&=i\epsilon^{ijk}\hat{x}^{\prime i}\hat{y}^{\prime k}\,L_Q^{\prime j}+i\epsilon^{ijk}\hat{x}^{\prime i}\hat{y}^{\prime j}\,L_Q^{\prime k}+i\epsilon^{ijk}\hat{y}^{\prime j}\hat{x}^{\prime k}\,L_Q^{\prime i}=-i\bm{\hat{z}}'\cdot\bm{L}_Q',
\end{align}
and the ones obtained by cyclic permutations. 
We also have $\left(\bm{\hat{x}}'\cdot\bm{L}_Q'\right)^2+\left(\bm{\hat{y}}'\cdot\bm{L}_Q'\right)^2+\left(\bm{\hat{z}}'\cdot\bm{L}_Q'\right)^2=\bm{L}_Q^{\prime2}$. We therefore can construct the orbital wave functions as eigenfunctions of $\bm{L}_Q^{\prime2}$, $L_{Q, z}'= \hat{\bm {z}}\cdot{\bm L}_Q^{\prime}$, and $\tilde{\Lambda}=\bm{\hat{z}}'\cdot\bm{L}_Q'$, with eigenvalues $l(l+1)$, $m$, and $-\lambda$ respectively, and use 
\begin{align}
	L_{Q, \pm}'&=L_{Q, x}'\pm iL_{Q, y}',\label{eq:ladder-1}\\
	\tilde{\Lambda}_\pm &=\bm{\hat{x}}'\cdot\bm{L}_Q'\mp i\bm{\hat{y}}'\cdot\bm{L}_Q',\label{eq:ladder-2}
\end{align}
as ladder operators to raise(lower) the quantum numbers $m$ and $\lambda$ respectively. These ladder operators lead to the constraints on the quantum numbers $-l\leq m\leq l$ and $-l\leq \lambda\leq l$. An explicit expression for the eigenfunctions (out of several equivalent possibilities) is given by
\begin{align}
	D_{lm}^\lambda(\psi,\theta,\varphi)&=\frac{(-1)^{l+m}}{2^l}\sqrt{\frac{2l+1}{8\pi^2}\frac{(l-m)!}{(l+m)!(l-\lambda)!(l+\lambda)!}}e^{i\lambda\psi}P_{lm}^\lambda(\cos\theta)e^{im\varphi}\,,
	\label{CoeffExplicit}\\
	P_{lm}^\lambda(x)&=(1-x)^{(m-\lambda)/2}(1+x)^{(m+\lambda)/2}\partial_x^{l+m}(1-x)^{l+\lambda}(1+x)^{l-\lambda}\,.
\end{align}
The relation with the  Wigner's D-matrix is found in the expression of an angular momentum eigenstate polarized in the $\left(\bm{\hat{x}}',\bm{\hat{y}}',\bm{\hat{z}}'\right)$ system, such as the ${\bm K}^2$ and $K_{z^\prime}$ eigenstate $\ket{k,\lambda}'$, in terms of states polarized in an external reference frame $\left(\bm{\hat{x}},\bm{\hat{y}},\bm{\hat{z}}\right)$,  such as the ${\bm K}^2$, $K_z$ eigenstate $\ket{k, m_k}$:
\begin{equation}
	\ket{k,\lambda}'=\sqrt{\frac{8\pi^2}{2k+1}}\sum_{m_k=-k}^kD_{k\,m_k}^{\lambda*}(\psi,\theta,\varphi)\ket{k,m_k}\,.
	\label{eq:lambda-m-expansion}
\end{equation}
Accordingly, we have the following orthogonality relations:
\begin{align}
	\int d\Omega'\,D_{l'm'}^{\lambda'*}(\psi,\theta,\varphi)D_{lm}^\lambda(\psi,\theta,\varphi)&=\delta_{l'l}\delta_{m'm}\delta^{\lambda'\lambda}\,,\\
	\sum_{\lambda=-k}^kD_{km'}^{\lambda*}(\psi,\theta,\varphi)D_{km}^\lambda(\psi,\theta,\varphi)&=\frac{2k+1}{8\pi^2}\delta_{m'm}\,,\\
	\sum_{m=-k}^kD_{km}^{\lambda'*}(\psi,\theta,\varphi)D_{km}^\lambda(\psi,\theta,\varphi)&=\frac{2k+1}{8\pi^2}\delta^{\lambda'\lambda}\,,
\end{align}
where we have used the labels $k$ or $l$ for the first quantum number, depending on the context for which the relation is more relevant.

The expression of the  mixing matrix in the Schr\"odinger equation is given by the matrix elements of $\bm{L}_Q^2$ between the eigenstates defined in Eq.~\eqref{Lstate-extended}:
\begin{equation}
	M_{\lambda'\lambda}=\bra{l,m;k,\lambda'}\bm{L}_Q^2\ket{l,m;k,\lambda}=\int d\Omega'\,D_{lm}^{\lambda'*}(\psi,\theta,\varphi)\bra{k,\lambda'}'\bm{L}_Q^{\prime2}\ket{k,\lambda}'D_{lm}^\lambda(\psi,\theta,\varphi)\,.
	\label{eq:Sch-mixing}
\end{equation}
Since $\bm{L}_Q'$ acts on $\ket{k,\lambda}'$ like $-\bm{K}$, we can rewrite this as
\begin{align}
	\bm{L}_Q^{\prime2}\ket{k,\lambda}'&=\bm{L}_Q^{\prime2}\ket{k,\lambda}'-2\bm{K}\ket{k,\lambda}'\cdot\bm{L}_Q'+\ket{k,\lambda}'\bm{L}_Q^{\prime2}\notag\\
	&=\bm{K}^2\ket{k,\lambda}'-K_+'\ket{k,\lambda}'\tilde{\Lambda}_+-K_-'\ket{k,\lambda}'\tilde{\Lambda}_--2K_z'\ket{k,\lambda}'\tilde{\Lambda}+\ket{k,\lambda}'\bm{L}_Q^{\prime2}\,,
\end{align}
where $K_\pm'=\left(\bm{\hat{x}}'\pm i\bm{\hat{y}}'\right)\cdot\bm{K}$ and $K_z'=\bm{\hat{z}}'\cdot\bm{K}$. With this we have expressed $\bm{L}_Q^{\prime2}$ in terms of defining and ladder operators for both $\ket{k,\lambda}'$ and $D_{lm}^\lambda$, so we find
\begin{align}
	M_{\lambda'\lambda}=&\bigl(l(l+1)-2\lambda^2+k(k+1)\bigr)\delta^{\lambda'\lambda}-\sqrt{k(k+1)-\lambda(\lambda+1)}\sqrt{l(l+1)-\lambda(\lambda+1)}\delta^{\lambda'\lambda+1}\notag\\
	&-\sqrt{k(k+1)-\lambda(\lambda-1)}\sqrt{l(l+1)-\lambda(\lambda-1)}\delta^{\lambda'\lambda-1}\,.
\end{align}
We recall here that beyond the short-distance limit the action of $\bm{K}^2$ and $K_\pm'$ on $\ket{k,\lambda}'$ is only approximately given by the eigenvalue relations used above (while the other relations remain exact), and there will be corrections suppressed by higher powers in $r$ involving static states not included in the same short-range multiplet.

\section{Group factors}
\label{app:group}
The product of two triplet representations of $SU(3)$ can be decomposed into the sum of an antitriplet and a sextet representation: $3\otimes3=\bar{3}+6$. We use the $\bar{3}$ and $6$ tensor invariants from Ref.~\cite{Brambilla:2005yk}
\begin{align}
	&\underline{T}^{\ell}_{ij}  = \frac{1}{\sqrt{2}} \epsilon_{lij},\quad i,\,j,\,\ell=1,2,3,
	\label{Tabg}
	\\
	& \nn\\
	& \underline{\Sigma}^\sigma_{ij}\quad i,\,j=1,2,3\,\quad \sigma=1,..,6 ,\nn\\
	&\underline{\Sigma}^1_{11}  = \underline{\Sigma}^4_{22} = \underline{\Sigma}^6_{33} = 1,
	\nn\\
	&\underline{\Sigma}^2_{12}  = \underline{\Sigma}^2_{21} = 
	\underline{\Sigma}^3_{13}  = \underline{\Sigma}^3_{31} = 
	\underline{\Sigma}^5_{23}  = \underline{\Sigma}^5_{32} = \frac{1}{\sqrt{2}},
	\label{eq:csextet}
\end{align}
all other entries are zero. Both $\underline{T}^{\ell}_{ij}$ and $\underline{\Sigma}^\sigma_{ij}$ are real; $\underline{T}^{\ell}_{ij}$ is totally antisymmetric and $\underline{\Sigma}^\sigma_{ij}$  symmetric in the $i$ and $j$ indices.
They satisfy the orthogonality and normalization relations:
\begin{align}
	\sum_{i, j=1}^3 \underline{T}^{\ell_1}_{ij} \, \underline{T}^{\ell_2}_{ij} =
	\delta^{\ell_1\ell_2}\,, \qquad
	\sum_{i, j=1}^3 \underline{\Sigma}^{\sigma_1}_{ij} \, \underline{\Sigma}^{\sigma_2}_{ij} = 
	\delta^{\sigma_1\sigma_2}\,, \qquad
	\sum_{i, j=1}^3 \underline{T}^{\ell}_{ij} \, \underline{\Sigma}^\sigma_{ij} = 0\,.
\end{align}

\section{Transformation under reflection} \label{app_reflection}
The quantum number $\sigma$ in $\Lambda^\sigma_\eta$ is the eigenvalue under reflection (denoted by $M$) across a plane containing the quark axis. It is only relevant for $\Sigma$ states (i.e. $\Lambda=0$), which are invariant under rotations around the quark axis. 

Let us discuss the transformation behavior under reflections $M$. We may choose any reflection plane that includes $\bm{\hat{r}}$, so in the $(\bm{\hat{\theta}},\bm{\hat{\varphi}},\bm{\hat{r}})$ coordinate system we choose the $\bm{\hat{r}}$-$\bm{\hat{\theta}}$ plane, which also includes the $z$-axis from the external reference frame (i.e. the coordinate system in which the heavy quark positions are defined). 
This means that the coordinate along the $\bm{\hat{\varphi}}$-axis receives a minus under reflections, whereas the other two remain unchanged. Because the angular momentum is a pseudovector (as the cross product between two vectors), it transforms with an overall minus, which means
\begin{equation}
	\bm{\hat{r}}\cdot\bm{K}\xrightarrow{M}-\left(\bm{\hat{r}}\cdot\bm{K}\right)\,,\qquad\left(\bm{\hat{\theta}}\pm i\bm{\hat{\varphi}}\right)\cdot\bm{K}\xrightarrow{M}-\left(\bm{\hat{\theta}}\mp i\bm{\hat{\varphi}}\right)\cdot\bm{K}\,,
	\label{Kreflection}
\end{equation}
where the second relation describes the transformation of the ladder operators for the $\lambda$ quantum number given by Eq.~\eqref{eq:ladder-2} [with $\psi=0$ in Eq.~\eqref{eq:x'-y'-z'}]. It follows then that the static states transform as
\begin{equation}
	\ket{k,\lambda}\xrightarrow{M}\sigma(-1)^\lambda\ket{k,-\lambda}\,,
	\label{statereflection}
\end{equation}
where the factor $\sigma$ is independent of $\lambda$ and for integer $k$ corresponds to the $\Lambda_\eta^\sigma$ quantum number. 

The reflection operation can be related to parity through a rotation around the $\bm{\hat{\varphi}}$-axis by an angle $\pi$:
\begin{equation}
	M=\exp\left[ i\pi\bm{\hat{\varphi}}\cdot\bm{K}\right]P\,.
	\label{reflectionparity}
\end{equation}
The rotation matrix is straightforward to compute for a given $k$-representation, but even without a direct calculation for general $k$ we can say that its eigenvalues are given by $(-1)^{\lambda'}$, $-k\leq\lambda'\leq k$, which would be the diagonal entries in a basis where the quantization axis is $\bm{\hat{\varphi}}$. Consequently, its trace is given by $(-1)^k$. Combining this with the $\sigma_T(-1)^k$ factor from parity, and noting that the only diagonal entry in Eq.~\eqref{statereflection} is for $\lambda=0$, we obtain
\begin{equation}
	\text{Tr}\{M\}=(-1)^k\sigma_T(-1)^k=\sigma\,.
\end{equation}
The two $k$-dependent signs cancel and we see that we can identify the reflection factor $\sigma$ with the tensor sign $\sigma_T$. For half-integer states, the overall sign is arbitrary, so we may simply adopt a convention where the states transform according to Eq.~\eqref{statereflection} with $\sigma=\sigma_T$.

Since reflections remain a symmetry of the LDF system beyond the short distance limit, it was to be expected that the transformation is independent of $k$. An interesting result of this section is the dependence on $\sigma_T$, which implies that the multipole expansion of the LDF operator (i.e. giving the static state as a superposition of different $k$ states) will consist either of all tensor or all pseudotensor representations.

\section{Gauge invariance} \label{app_Gauge}
We show that the interpolating operators in Eqs.~\eqref{eq:intop-1}-\eqref{eq:intop-3} are gauge-invariant. Under gauge transformations, the quark field $q$ and the Wilson line $\phi\left(\bm{x}_1, \bm{x}_2\right)$, transforms as
\begin{equation}
	q \to {\cal U} q,
	\quad\quad \phi\left(\bm{x}_1, \bm{x}_2\right) \to {\cal U}\phi\left(\bm{x}_1, \bm{x}_2\right){\cal U}^\dagger\,,
	\label{SUN:Uq}
\end{equation}
where the matrix ${\cal U}$ is unitary and has determinant 1: ${\cal U}^{\dagger} {\cal U}={\cal U} {\cal U}^{\dagger}= \mathbbm{1}$, and $\det {\cal U} = 1$. The matrix ${\cal U}$ is given by ${\cal U}=\exp\left(i\alpha^a T^a\right)$, where $\alpha^a$ are parameters and $T^a$ are the generators of the fundamental representation of the gauge group $SU(3)$. 
For simplicity, we consider a general operator $A^a = \bar{q} T^a q'$ that transforms according to the adjoint representation 
of $SU(3)$, 
\begin{equation}
	A^a \to \bar{q}\,{\cal U}^\dagger\,T^a\,{\cal U}\,q' \equiv {\cal U}^{ab}_{\mathrm {adj}}\,\, A^b\,,
	\label{SUN:adj}
\end{equation}
where the infinitesimal transformation ${\cal U}^{ab}_{\mathrm {adj}}$ is given by
\begin{align}
	{\cal U}^{ab}_{\mathrm {adj}} & = \delta^{ab} + i \alpha^c (t^c)^{ab}\equiv \delta^{ab} - \alpha^c  f^{acb},
	\label{SUN:Uab}
\end{align}
and in the last line we have used the generators in the adjoint representation. 
Moreover, from the infinitesimal trasnformation ${\cal U}^{ab}_{\mathrm {adj}}$, it follows immediately that 
\begin{equation}
	{\cal U}^{ab}_{\mathrm {adj}}\,\, {\cal U}^{ac}_{\mathrm {adj}}=\delta^{bc}.
	\label{eq:Uid}
\end{equation}

Under the gauge transformations \eqref{SUN:Uq}, the interpolating operator $\mathcal{O}^{(Q\bar{Q})_8}_{\kappa, \lambda}\left(t, \bm{r}, \bm{R}\right)$ for $Q\bar{Q}q\bar{q}$ tetraquarks in Eq.~\eqref{eq:intop-1}, with light fields $H_{8,\,\kappa}^{\alpha,\,a}(t, \bm{R})$ given by  Table~\ref{tab:Reps-QQbar}, and explicitly writing the color indices, transforms as
\begin{align}
	\mathcal{O}^{(Q\bar{Q})_8}_{\kappa, \lambda}\left(t, \bm{r}, \bm{R}\right)&=\chi^{\dagger}(t,\,\bm{x}_2)\,\phi(t,\,\bm{x}_2,\bm{R})\, P^{\alpha \dag}_{\kappa \lambda}\,H_{8,\,\kappa}^{\alpha,\,a}(t,\,\bm{R})\,T^a\,\phi(t,\,\bm{R},\bm{x}_1)\,\psi(t,\,\bm{x}_1)\nonumber\\
	&= P^{\alpha \dag}_{\kappa \lambda}\,\Bigg[\chi^{\dagger}(t,\,\bm{x}_2)\,\phi(t,\,\bm{x}_2,\bm{R})\Bigg]_k\,\left[\bar{q}_i(t,\bm{x})\,\tilde{\Gamma}^{\alpha}\, T^a_{ij}\, q_j(t,\bm{x})\right]\,T^a_{kl}\,\Bigg[\phi(t,\,\bm{R},\bm{x}_1)\,\psi(t,\,\bm{x}_1)\Bigg]_l\nonumber\\
	&\longrightarrow P^{\alpha \dag}_{\kappa \lambda}\,\Bigg[\chi^{\dagger}(t,\,\bm{x}_2)\,\phi(t,\,\bm{x}_2,\bm{R})\Bigg]_s\, {\cal U}^{\dagger}_{sk}\,\left[\bar{q}_p(t,\bm{x})\,\tilde{\Gamma}^{\alpha}\,{\cal U}^{\dagger}_{pi}\, T^a_{ij}\, {\cal U}_{jm}\,  q_m(t,\bm{x})\right]\,T^a_{kl}\nonumber\\
	&\hspace{4.0cm}\times {\cal U}_{lr} \Bigg[\phi(t,\,\bm{R},\bm{x}_1)\,\psi(t,\,\bm{x}_1)\Bigg]_r\nonumber\\
	&={\cal U}^{ac}_{\rm adj}\,{\cal U}^{ab}_{\rm adj}\Bigg[\chi^{\dagger}(t,\,\bm{x}_2)\,\phi(t,\,\bm{x}_2,\bm{R})\Bigg]_s\,P^{\alpha \dag}_{\kappa \lambda}\,\left[\bar{q}_p(t,\bm{x})\,\tilde{\Gamma}^{\alpha}\, T^c_{pm} \,q_m(t,\bm{x})\right]\nonumber\\
	&\hspace{4.0 cm}\times\,T^b_{sr}\, \Bigg[\phi(t,\,\bm{R},\bm{x}_1)\,\psi(t,\,\bm{x}_1)\Bigg]_r\nonumber\\
	&=\mathcal{O}^{(Q\bar{Q})_8}_{\kappa, \lambda}\left(t, \bm{r}, \bm{R}\right),
	\label{eq:GaugeQQbar}
\end{align}
where $\tilde{\Gamma}^{\alpha}$ is one of the Dirac matrices from Table~\ref{tab:Reps-QQbar} 
and we have used Eqs.~\eqref{SUN:adj} and \eqref{eq:Uid} to show gauge invariance.

Under the gauge  transformations \eqref{SUN:Uq}, the interpolating operator $\mathcal{O}^{(QQ)_{\bar{3}}}_{\kappa, \lambda}\left(t, \bm{r}, \bm{R}\right)$ for $QQ\bar{q}\bar{q}$ tetraquark in Eq.~\eqref{eq:intop-2}, with $\left(QQ\right)_{\bar{3}}$ antitriplet in color and light fields $H_{3,\,\kappa}^{\alpha,\,\ell}\left(t, \bm{R}\right)$ given by Tables~\ref{tab:Reps-QQ-I_0} and \ref{tab:Reps-QQ-I_1}, and explicitly writing the color indices, transforms as
\begin{align}
	\mathcal{O}^{(QQ)_{\bar{3}}}_{\kappa, \lambda}\left(t, \bm{r}, \bm{R}\right)&=\psi^T(t,\,\bm{x}_2)\,\phi^T(t,\,\bm{x}_2,\bm{R})\,P^{\alpha \dag}_{\kappa \lambda}\,H_{3,\,\kappa}^{\alpha,\,\ell}\left(t, \bm{R}\right)\,\underline{T}^{\ell}\,\phi(t,\,\bm{R},\bm{x}_1)\,\psi(t,\,\bm{x}_1)\nonumber\\
	&=P^{\alpha \dag}_{\kappa \lambda}\,\Bigg[\psi^T(t,\,\bm{x}_2)\,\phi^T(t,\,\bm{x}_2,\bm{R})\Bigg]_k\,\left[\bar{q}_i(t,\bm{x})\,\underline{T}^{\ell}_{ij}\, \tilde{\Gamma}^{\alpha}\, q^*_j(t,\bm{x})\right]\,\underline{T}^{\ell}_{kn}\nonumber\\
	&\hspace{5.0 cm}\times\,\Bigg[\phi(t,\,\bm{R},\bm{x}_1)\,\psi(t,\,\bm{x}_1)\Bigg]_n\nonumber\\
	&\longrightarrow P^{\alpha \dag}_{\kappa \lambda}\,\Bigg[\psi^T(t,\,\bm{x}_2)\,\phi^T(t,\,\bm{x}_2,\bm{R})\Bigg]_s\,{\cal U}_{sk}^{T}\,\left[\bar{q}_p(t,\bm{x})\,{\cal U}^{\dagger}_{pi}\,\underline{T}^{\ell}_{ij}\, \tilde{\Gamma}^{\alpha}\,{\cal U}^*_{jm}\, q^*_m(t,\bm{x})\right]\,\underline{T}^{\ell}_{kn}\nonumber\\
	&\hspace{4.0cm}\times {\cal U}_{nr}\,\Bigg[\phi(t,\,\bm{R},\bm{x}_1)\,\psi(t,\,\bm{x}_1)\Bigg]_r,
	\label{eq:GaugeQQ-1}
\end{align}
where $\underline{T}^{\ell}$ is the antisymmetric tensor  defined in Eq.~\eqref{Tabg}. Using the  color identity relation for the antisymmetric tensor $\underline{T}^{\ell}$, 
\begin{equation}
	\sum_{\ell}\underline{T}^{\ell}_{ij}\,\underline{T}^{\ell}_{kn}=\frac{1}{2}\left(\delta_{ik}\delta_{jn}-\delta_{in}\delta_{jk}\right),
\end{equation} 
and the identity, 
\begin{align}
	&{\cal U}^T_{sk}\,{\cal U}^{\dagger}_{pi}\,\underline{T}^{\ell}_{ij}\,\underline{T}^{\ell}_{kn}\,{\cal U}^*_{jm}\,{\cal U}_{nr}\nonumber\\
	&={\cal U}_{ks}\,{\cal U}^\dagger_{pi}\,{\cal U}^\dagger_{mj}\,{\cal U}_{nr}\, \frac{1}{2} \left(\delta_{ik}\delta_{jn} - \delta_{in} \delta_{jk}\right)\nonumber\\
	&= \frac{1}{2} \left[ \left({\cal U}^\dagger \,{\cal U}\right)_{ps} \left({\cal U}^\dagger\,{\cal U}\right)_{mr} - \left({\cal U}^\dagger\, {\cal U}\right)_{pr}
	\left({\cal U}^\dagger\,{\cal U}\right)_{ms}\right]\nonumber\\
	&= \frac{1}{2} \left( \delta_{ps} \delta_{mr} - \delta_{pr} \delta_{ns} \right) =
	\underline{T}^{\ell}_{pm}\,\underline{T}^{\ell}_{sr},
	\label{eq:GaugeQQ-2}
\end{align}
we can show the gauge invariance of the interpolating operator $\mathcal{O}^{(QQ)_{\bar{3}}}_{\kappa, \lambda}\left(t, \bm{r}, \bm{R}\right)$ for $QQ\bar{q}\bar{q}$  tetraquark states.

Under the gauge transformations \eqref{SUN:Uq}, following the same steps as in Eq.~\eqref{eq:GaugeQQ-1} and using the identity
\begin{equation}
	\sum_{\sigma}\underline{\Sigma}^{\sigma}_{ij}\,\underline{\Sigma}^{\sigma}_{kn}=\frac{1}{2}\left(\delta_{ik}\delta_{jn}+\delta_{in}\delta_{jk}\right),
\end{equation} 
we can similarly show the gauge invariance of the  interpolating operator $\mathcal{O}^{(QQ)_{6}}_{\kappa, \lambda}\left(t, \bm{r}, \bm{R}\right)$ for $QQ\bar{q}\bar{q}$ tetraquark in Eq.~\eqref{eq:intop-3}, with color sextet $\left(QQ\right)_{6}$  and light fields $H_{\bar{6},\,\kappa}^{\alpha,\,\ell}\left(t, \bm{R}\right)$ given by Tables~\ref{tab:Reps-QQ-I_0} and \ref{tab:Reps-QQ-I_1}.

With regard to pentaquark states $Q\bar{Q}qqq$ or $QQqq\bar{q}$ and doubly heavy baryons $QQq$, we can similarly show the gauge invariance of the interpolating operator $\mathcal{O}^{(QQ)_{\bar{3}}}_{\kappa, \lambda}\left(t, \bm{r}, \bm{R}\right)$ by combining results in Eqs.~\eqref{eq:GaugeQQbar} and \eqref{eq:GaugeQQ-1} with the understanding that the two light quarks (antiquarks) in color antitriplet (triplet) transform in the same way as antiquarks (quarks) under gauge transformation. 
Indeed, under infinitesimal gauge transformation, two quarks in a color antitriplet representation transform as 
\begin{align}
	\left(qq\right)_{\bar{3}, \ell}&\equiv \epsilon_{\ell jk}\, q_j\, q_k\nonumber\\
	&\rightarrow \epsilon_{\ell jk}\, {\cal U}_{jm}\,{\cal U}_{kn}\,q_m\,q_n\nonumber\\
	&=\epsilon_{rjk}\,({\cal U} {\cal U}^\dagger)_{r\ell}\,{\cal U}_{jm}\,{\cal U}_{kn}\,q_m\,q_n\nonumber\\
	&=\epsilon_{rjk}\,{\cal U}_{rs}\,{\cal U}_{jm} {\cal U}_{kn}\,{\cal U}^\dagger_{s\ell}\,q_m\,q_n\nonumber\\
	& = \mathrm{det}\left({\cal U}\right)\,\epsilon_{smn}\,{\cal U}^\dagger_{s\ell}\,q_m\,q_n =  \epsilon_{smn}\,q_m\,q_n\,{\cal U}^\dagger_{s\ell}\,,
	\label{eq:qqGauge}
\end{align}
which is equivalent to the gauge transformation of an antiquark (color antitriplet representation), if we identity $\bar{q}_s \equiv \epsilon_{smn}\,q_m\,q_n$.

\section{Projection vectors}
\label{app:projectors}
The projection vectors ${\bm P}_{\kappa\lambda}\left(\theta, \varphi\right)$ project the fields or operators on 
representations of the cylindrical symmetry, $D_{\infty h}$. 
The projection vectors are defined as the eigenvectors (column vectors) with dimension $\left(2k+1\right)\times 1$ of $\hat{{\bm r}}\cdot{\bm K}$, where ${\bm K}$ is the total angular momentum of the LDF and $\hat{{\bm r}}$ is the heavy quark pair axis:
\begin{equation}
	\left(\hat{{\bm r}}\cdot{\bm K}\right)\,{\bm P}_{\kappa\lambda}= \lambda\,{\bm P}_{\kappa\lambda} \,.
	\label{eq:Pdef}
\end{equation}
The operator $\left(\hat{{\bm r}}\cdot{\bm K}\right)$ is a matrix of dimension $\left(2k+1\right)\times \left(2k+1\right)$ and eigenvalues $\lambda = - k, \dots, 0$, $\dots, k$. 
Solving the eigensystem in Eq.~\eqref{eq:Pdef}, a general expression for the components of the projection vectors can be given in terms of Wigner D-matrices:
\begin{align}
	P^{\alpha}_{\kappa\lambda}\left(\theta, \varphi\right)=\sqrt{\frac{8\pi^2}{2k+1}} D^{\lambda*}_{k\,\alpha}\left(0, \theta, \varphi\right)=\sqrt{\frac{4\pi}{2k+1}} v_{k\,\alpha}^{\lambda*}(\theta,\varphi),
	\label{eq:Pdef2}
\end{align}
where $\alpha = k, \dots, 0, \dots, -k$, as we are using a spherical basis.
Specific examples of projection operators are:
\begin{align}
	{\bm P}_{\frac{1}{2}\frac{1}{2}}\left(\theta, \varphi\right)&=2\pi\begin{pmatrix} D^{1/2*}_{1/2\,1/2}\left(0, \theta, \varphi\right) \\ D^{1/2*}_{1/2\,-1/2}\left(0, \theta, \varphi\right) \end{pmatrix}=\begin{pmatrix}
		\cos(\frac{\theta}{2})e^{-i\varphi/2} \\ \sin(\frac{\theta}{2})e^{i\varphi/2}
	\end{pmatrix} \label{eq:P1/2-1},\\
	{\bm P}_{\frac{1}{2}-\frac{1}{2}}\left(\theta, \varphi\right)&=2\pi\begin{pmatrix} D^{-1/2*}_{1/2\,1/2}\left(0, \theta, \varphi\right) \\ D^{-1/2*}_{1/2\,-1/2}\left(0, \theta, \varphi\right) \end{pmatrix}=\begin{pmatrix}
		-\sin(\frac{\theta}{2})e^{-i\varphi/2} \\ \cos(\frac{\theta}{2})e^{i\varphi/2}
	\end{pmatrix},\label{eq:P1/2-2}\\
	{\bm P}_{10}\left(\theta, \varphi\right)&=\sqrt{\frac{8\pi^2}{3}}\begin{pmatrix} D^{0*}_{1\,1}\left(0, \theta, \varphi\right) \\ D^{0*}_{1\,0}\left(0, \theta, \varphi\right) \\ D^{0*}_{1\,-1}\left(0, \theta, \varphi\right) \end{pmatrix}=\begin{pmatrix} -\frac{1}{\sqrt{2}}\sin(\theta)e^{-i\varphi} \\ \cos(\theta) \\ \frac{1}{\sqrt{2}}\sin(\theta)e^{i\varphi} \end{pmatrix},\\
	{\bm P}_{1\,\pm 1}\left(\theta, \varphi\right)&=\sqrt{\frac{8\pi^2}{3}}\begin{pmatrix} D^{\pm1*}_{1\,1}\left(0, \theta, \varphi\right) \\ D^{\pm1*}_{10}\left(0, \theta, \varphi\right) \\ D^{\pm1*}_{1\,-1}\left(0, \theta, \varphi\right) \end{pmatrix}=\begin{pmatrix} \frac{1}{2}(1\pm\cos(\theta))e^{-i\varphi} \\ \pm\frac{1}{\sqrt{2}}\sin(\theta) \\ \frac{1}{2}(1\mp\cos(\theta))e^{i\varphi}\end{pmatrix}.
\end{align}

In Refs.~\cite{Berwein:2015vca, Brambilla:2017uyf,  Brambilla:2018pyn, Brambilla:2019jfi}, the projection vectors for $k=1\,\left(\lambda =0, \pm 1\right)$  in the Cartesian basis were used for the construction of the BOEFT and the coupled Schr\"odinger equations for hybrids:
\begin{equation}
	{\bm P}_{10} = {\bm \hat{r}},\qquad\qquad
	{\bm P}_{1\pm 1}= {\bm{\hat{r}}}_{\pm}\equiv \mp\frac{1}{\sqrt{2}}\left({\bm{\hat{\theta}}}\pm i{\bm{\hat{\varphi}}}\right),
	\label{eq:P_1}
\end{equation}
where ${\bm{\hat{r}}}$, ${\bm{\hat{\theta}}}$, and ${\bm{\hat{\varphi}}}$ are the spherical unit vectors. 
However, they can be also expressed as
\begin{align}
	{\bm P}_{1\lambda} = \hat{{\bm r}}_\lambda=\sqrt{\frac{8\pi^2}{3}} \begin{pmatrix} \frac{D^{\lambda*}_{1\,-1}\left(0, \theta, \varphi\right)-D^{\lambda*}_{1\,1}\left(0, \theta, \varphi\right)}{\sqrt{2}} \\ \frac{D^{\lambda*}_{1\,-1}\left(0, \theta, \varphi\right)+D^{\lambda*}_{1\,1}\left(0, \theta, \varphi\right)}{i\sqrt{2}} \\ D^{\lambda*}_{0\,0}\left(0, \theta, \varphi\right) \end{pmatrix},\,\, \lambda = 0, \pm 1.
	\label{eq:P_1D}
\end{align}
In Ref.~\cite{Berwein:2015vca}, the mixing matrices in the coupled Schr\"odinger equations for hybrids  arise from the term $P^{\dagger \alpha}_{1\lambda}\left(\theta, \varphi\right)\,{\bm L}_Q^2\,P^{\alpha}_{1\lambda'}\left(\theta, \varphi\right)$,\footnote{This term originates from the kinetic energy operator: $\displaystyle P^{\dagger \alpha}_{1\lambda}\left(\theta, \varphi\right)\,{\bm \nabla}^2\,P^{\alpha}_{1\lambda'}\left(\theta, \varphi\right)=\frac{1}{r}\frac{\partial^2}{\partial r^2}r - \frac{1}{r^2} P^{\dagger \alpha}_{1\lambda}\left(\theta, \varphi\right)\,{\bm L}_Q^2\,P^{\alpha}_{1\lambda'}\left(\theta, \varphi\right)$.} which is  equivalent to  $\bra{k,\lambda}{\bm L}_Q^2\ket{k,\lambda'}$ in Eq.~\eqref{eq:Sch-mixing} of the current work. 
The resulting matrix of differential operators  (in the $\lambda \lambda'$ indices) is then contracted with a vector wave function: $\Psi\left({\bm r}\right) = \psi\left(\theta, \varphi\right)R(r)$, where $\psi\left(\theta, \varphi\right)$ is the angular wavefunction matrix acting on the radial wavefunction (column vector) $R(r)$. 
The angular parts are defined as eigenfunctions of  $P^{\dagger \alpha}_{1\lambda}\,{\bm L}_Q^2\, P^{\alpha}_{1\lambda'}$,
in the sense that the eigenvalues take matrix form. 
This angular eigenvalue matrix then gives the mixing matrix in the radial Schr\"odinger equation. 
In Ref.~\cite{Berwein:2015vca}, the angular wavefunctions were written as $\psi\left(\theta, \varphi\right)= \mathrm{diag}\left(v_{lm}^0\left(\theta, \varphi\right), v_{lm}^1\left(\theta, \varphi\right), v_{lm}^{-1}\left(\theta, \varphi\right)\right)$. 
They give the same mixing matrices as here, where we obtained them from the matrix elements between eigenstates of total angular momentum, $\bra{l,m;k,\lambda}{\bm L}_Q^2\ket{l,m;k,\lambda'}$. 
The linear combinations of states $\ket{l,m;k,\pm \lambda}$ for $\lambda\ne0$ have been introduced here to construct the parity eigenstates $\ket{l,m;k, |\lambda|;\epsilon}$, which are given in Eq.~\eqref{eq:CPstates}. 
With the parity eigenstates, the mixing matrix is block-diagonal, which leads to the decoupling of the opposite parity states in the Schr\"odinger equations. 
In Ref.~\cite{Berwein:2015vca},   an identical linear combination of states  corresponding to $\lambda = \pm 1$ (Eqs. (49) and (50) in \cite{Berwein:2015vca}) was introduced that resulted in the decoupling of opposite parity states by making the mixing matrix in the Schr\"odinger equations block-diagonal.

\section{Overlap of interpolating operator with color octet $\left(Q\bar{Q}\right)_8$}
\label{app:Overlap}

\subsection{Quarkonium}
When writing down the interpolating operator for exotic hadrons with a $Q\bar{Q}$ pair in Eq.~\eqref{eq:intop-1}, we considered a color octet $\left(Q\bar{Q}\right)_8$ configuration. In the absence of light quarks $\left(P^\alpha_{\kappa, \lambda}\,H_{8,\,\kappa}^{\alpha,\,a}(t, \bm{R})\,T^a\equiv \mathbbm{1}\right)$, the interpolating operator for the color singlet $\left(Q\bar{Q}\right)_1$ pair 
corresponds to a quarkonium state. 
With light quarks, the interpolating operator for the  color singlet  $\left(Q\bar{Q}\right)_1$ pair in the 
isospin $I=1$ case corresponds to a quarkonium plus a pion state \cite{Prelovsek:2019ywc} 
and in the isospin 
$I=0$ case corresponds to a quarkonium or a quarkonium plus two pions states \cite{Bulava:2019iut, Bali:2005fu}. 
In the short distance limit $r\rightarrow 0$, the  color singlet potential being attractive is lower in energy than the repulsive color octet potential. 
As a result, within lattice computations, any operator that overlaps with the quarkonium state will primarily manifest 
an
attractive color singlet
potential 
at large times and short distances. 

A generic quarkonium state $|{\cal Q}\rangle$ (with principal quantum number $n$, total angular momentum $J$, orbital angular momentum $L$, and spin $S$) can be described in NRQCD by a state (for simplification, we use center of mass frame ${\bm R}=0$)\footnote{For the purpose of the current work, we don't need to explicitly specify the one pion $(I=1)$ or two pion states $(I=0)$ along with the quarkonium. 
	We assume that pions are contained in the state $|\Omega\rangle$} 
\begin{align}
	|{\cal Q}\rangle 
	&={\cal N}\int d^3{\bm r}\,\Psi^{(n)}\left({\bm r}\right)\,\psi^{\dagger}_b\left(t,- \bm{r}/2\right)\phi_{bc}\left(t; -\bm{r}/2,\bm{r}/2\right)\chi_c\left(t, \bm{r}/2\right)|\Omega\rangle,
	\label{eq:Quarkonium}
\end{align}
where $\Psi^{(n)}\left({\bm r}\right)$ is the quarkonium wavefunction with quantum numbers $(n)=\{n, J, L, S\}$, ${\cal N}$ is a normalization constant, and  $|\Omega\rangle$ denotes vacuum with respect to the heavy quark fields $\psi\left({\bm x}\right)$, $\chi\left({\bm x}\right)$ but can contain ultrasoft modes such as gluons and pions. 
We have explicitly specified the color indices for the operators. 
The quark field with the Wilson line, $\phi\left(t; \bm{0},-\bm{r}/2\right)\psi\left(t, -\bm{r}/2\right)$, transforms under gauge transformation like a quark field in $\bm{0}$. 
The overlap of the interpolating operator for exotic hadrons with color octet $\left(Q\bar{Q}\right)_8$ pair (Eq.~\eqref{eq:intop-1}) on the state $|{\cal Q}\rangle$ is given by 
\begin{align}
	& \mathcal{O}^{(Q\bar{Q})_8}_{\kappa, \lambda}\left(t, \bm{r}\right) |\mathcal{Q}\rangle\nonumber\\
	&\hspace{1.0cm}= \chi^{\dagger}_m\left(t, \bm{r}/2\right)\phi_{mi}\left(t; \bm{r}/2,\bm{0}\right)\,P^{\alpha \dag}_{\kappa \lambda}\,H_{8,\,\kappa}^{\alpha,\,a}(t, \bm{0})\,T^a_{ij}\,\phi_{jn}\left(t; \bm{0},-\bm{r}/2\right)\psi_n\left(t, -\bm{r}/2\right)|\mathcal{Q}\rangle\nonumber\\
	&\hspace{1.0 cm}={\cal N}\delta^3\left({\bm 0}\right)\Psi^{(n)}\left({\bm r}\right)\,P^{\alpha \dag}_{\kappa \lambda}\,H_{8,\,\kappa}^{\alpha,\,a}(t, \bm{0})\,T^a_{ii}\,|\mathrm{vac}\rangle=0,
	\label{eq:overlap}
\end{align}
since the generators $T^a$ are traceless matrices and we have used the equal time anticommutation relation for the fields $\psi$ and $\chi$. 
We get, therefore, that the interpolating operator in Eq.~\eqref{eq:intop-1} with the color octet $\left(Q\bar{Q}\right)_8$ pair has zero overlap on the state $|{\cal Q}\rangle$. 
From a practical point of view, this could be useful for lattice calculations of the static energies of the adjoint $Q\bar{Q}$ tetraquark and pentaquark states since the interpolating operator in Eq.~\eqref{eq:intop-1} will explicitly probe the repulsive color octet behavior of the $Q\bar{Q}$ pair at short distance.

\subsection{Pair of heavy-light meson state}
A generic heavy-light meson pair state $|M\bar{M}\rangle$  (with total angular momentum quantum number $J=|J_1-J_2|,\dots, J_1+J_2$, where $J_1$ is the angular momentum of meson $M$ and $J_2$ is the angular momentum of meson $\bar{M}$) can be described in NRQCD by a state (for simplification, we use center of mass frame ${\bm R}=0$)
\begin{align}
	|M\bar{M}\rangle &= \Bigg[ {\cal N}\int d^3{\bm r}\,\Psi_{J}\left({\bm r}\right)\nonumber\\
	&\hspace{0.5 cm}\times\int d^3{\bm y}\,\varphi_{J_1}\left({\bm y}+{\bm r}/2\right)\,\psi_c^{\dagger}\left(t, -{\bm r}/2\right)\phi_{cd}\left(t;-{\bm r}/2, {\bm y}+{\bm r}/2\right)\left[P_+\,\Gamma_1\,q_d\left(t,{\bm y}+{\bm r}/2\right)\right]\nonumber\\
	&\hspace{0.5 cm}\times\int d^3{\bm z}\,\varphi_{J_2}\left({\bm z}-{\bm r}/2\right)\,\left[\bar{q}_b\left(t,{\bm z}-{\bm r}/2\right)\,\Gamma_2\,P_-\right]\phi_{be}\left(t;{\bm z}-{\bm r}/2, {\bm r}/2 \right)\chi_e\left(t, {\bm r}/2\right)\Bigg]|\mathrm{vac}\rangle,
	\label{eq:meson-meson}
\end{align}
where $\varphi_{J_1}\left({\bm y}+{\bm r}/2\right)$ is the light antiquark wavefunction in the meson $M$, 
$\varphi_{J_2}\left({\bm z}-{\bm r}/2\right)$ is the light quark wavefunction in the meson $\bar{M}$, 
$\Psi_{J}\left({\bm r}\right)$ is the wavefunction of the heavy-light meson pair state $M\bar{M}$, 
$\Gamma_1$ and $\Gamma_2$ are (combinations of) Dirac gamma matrices such as $\gamma^0, \gamma^5, \gamma^0\gamma^5, \gamma^i, \dots$, $P_{\pm}=\left(1\pm\gamma^0\right)/2$, $|\mathrm{vac}\rangle$ denotes the vacuum state, $ {\cal N}$ is a normalization factor 
and we have explicitly specified the color indices.

The overlap of the interpolating operator for exotic hadrons with color octet $\left(Q\bar{Q}\right)_8$ pair in Eq.~\eqref{eq:intop-1} on the state $ |M\bar{M}\rangle$ is given by 
\begin{align}
	\mathcal{O}^{(Q\bar{Q})_8}_{\kappa, \lambda}\left(t, \bm{r}\right) |M\bar{M}\rangle& = P^{\alpha \dag}_{\kappa \lambda}\,\Bigg[ \chi^{\dagger}_m\left(t, \bm{r}/2\right)\phi_{mi}\left(t; \bm{r}/2,\bm{0}\right)\,\left[\bar{q}_p\left(t,{\bm 0}\right)\,\tilde{\Gamma}^{\alpha}\,T^a_{pk}\,q_k\left(t,{\bm 0}\right)\right]\nonumber\\
	&\hspace{1.5 cm}\times\,T^a_{ij}\,\phi_{jn}\left(t; \bm{0},-\bm{r}/2\right)\psi_n\left(t, -\bm{r}/2\right)\Bigg]|M\bar{M}\rangle\nonumber\\ 
	& \propto {\cal N}\delta^3\left({\bm 0}\right)\Psi_{J}\left({\bm r}\right)\varphi_{J_1}\left({\bm 0}\right)\varphi_{J_2}\left({\bm 0}\right)\mathrm{Tr}\left[T^a\,T^a\right]\nonumber\\
	&=\frac{1}{2}\delta^{aa}{\cal N}\delta^3\left({\bm 0}\right)\Psi_{J}\left({\bm r}\right) \varphi_{J_1}\left({\bm 0}\right)\varphi_{J_2}\left({\bm 0}\right),
\end{align}
where $\tilde{\Gamma}^{\alpha}$ in the first line denotes gamma matrices, $\gamma^0, \gamma^5, \gamma^0\gamma^5, \gamma^i, \dots$, 
and we have used the equal time anticommutation relation for the fields $\psi$ and $\chi$.  The proportionality symbol in the third line denotes that we have not explicitly computed the Dirac gamma matrices. 
We get that the interpolating operator in Eq.~\eqref{eq:intop-1} with the color octet $\left(Q\bar{Q}\right)_8$ pair has non-zero overlap on the state $|M\bar{M}\rangle$. 
From a practical point of view, this implies that in lattice calculations of the static energies of $\left(Q\bar{Q}\right)_8$ tetraquark and pentaquark states, the static energies will approach the heavy-light meson pair threshold at long distance, $r\gg\Lambda_{\mathrm{QCD}}^{-1}$, 
which is consistent with the lattice results in Ref.~\cite{Prelovsek:2019ywc}.

\section{Spin-isospin-color combinations for pentaquarks}
\label{SICcombinations}

\subsection{\texorpdfstring{$Q\bar{Q}qqq$ pentaquarks}{QQbarqqq pentaquarks}}
In the (iso)spin sector, the quartet is fully symmetric, while the two doublets have mixed symmetry. 
Ignoring color and isospin indices, the explicit spin combinations are given by
\begin{align}
	\ket{3/2,+3/2}&=\ket{\uparrow\uparrow\uparrow}\,, & \ket{1/2,+1/2}_S&=\frac{-1}{\sqrt{6}}\left(\ket{\downarrow\uparrow\uparrow}+\ket{\uparrow\downarrow\uparrow}-2\ket{\uparrow\uparrow\downarrow}\right)\,,\notag\\
	\ket{3/2,+1/2}&=\frac{1}{\sqrt{3}}\left(\ket{\downarrow\uparrow\uparrow}+\ket{\uparrow\downarrow\uparrow}+\ket{\uparrow\uparrow\downarrow}\right)\,, & \ket{1/2,-1/2}_S&=\frac{1}{\sqrt{6}}\left(\ket{\uparrow\downarrow\downarrow}+\ket{\downarrow\uparrow\downarrow}-2\ket{\downarrow\downarrow\uparrow}\right)\,,\notag\\
	\ket{3/2,-1/2}&=\frac{1}{\sqrt{3}}\left(\ket{\uparrow\downarrow\downarrow}+\ket{\downarrow\uparrow\downarrow}+\ket{\downarrow\downarrow\uparrow}\right)\,, & \ket{1/2,+1/2}_A&=\frac{-1}{\sqrt{2}}\left(\ket{\downarrow\uparrow\uparrow}-\ket{\uparrow\downarrow\uparrow}\right)\,,\notag\\
	\ket{3/2,-3/2}&=\ket{\downarrow\downarrow\downarrow}\,, & \ket{1/2,-1/2}_A&=\frac{1}{\sqrt{2}}\left(\ket{\uparrow\downarrow\downarrow}-\ket{\downarrow\uparrow\downarrow}\right)\,.
	\label{3spins}
\end{align}
The labels $S$ and $A$ on the spin-$1/2$ states denote symmetric or antisymmetric in the first two entries. 
The choice about which two entries to make symmetric or antisymmetric is arbitrary, as all three particles are indistinguishable. 
Accordingly, the full color-spin-isospin combination lifts this ambiguity.

For the color representations, we write
\begin{align}
	\ket{1}&=\frac{1}{\sqrt{6}}\epsilon_{ijk}|q_i, q_j, q_k\rangle\,, & \ket{10}^a&=\Delta_{ijk}^a|q_i, q_j, q_k\rangle\,,\notag\\
	\ket{8}^a_A&=\epsilon_{ijl}T_{lk}^a|q_i, q_j, q_k\rangle\,, & \ket{8}^a_S&=\frac{1}{\sqrt{3}}\left(\epsilon_{ljk}T_{li}^a-\epsilon_{ilk}T_{lj}^a\right)|q_i, q_j, q_k\rangle\,.
\end{align}
For the decuplet projector there is no expression in terms of other known tensors, so we just use the symbol $\Delta^a$ with $a=1,\dots,10$, and keep in mind that it is fully symmetric in its tensor indices. 
Again, the mixed symmetry of the color octets has been put in the form of a symmetric and an antisymmetric combination under exchange of the first two particles. 
Since later we have to combine this mixed symmetry with other mixed symmetries, it is more convenient to introduce the projectors
\begin{align}
	&\left(T_1\right)^a_{ijk}=\epsilon_{ljk}T_{li}^a=\sqrt{2}\,\underline{T}^l_{jk}T_{li}^a,\nonumber\\
	&\left(T_2\right)^a_{ijk}=\epsilon_{ilk}T_{lj}^a=\sqrt{2}\,\underline{T}^l_{ki}T_{lj}^a,\nonumber\\
	&\left(T_3\right)^a_{ijk}=\epsilon_{ijl}T_{lk}^a=\sqrt{2}\,\underline{T}^l_{ij}T_{lk}^a\,,
	\label{eq:T1T2T3}
\end{align}
such that 
\begin{align}
	&\ket{8}^a_A=(T_3)_{ijk}^a|q_i, q_j, q_k\rangle \equiv |T_3^a\rangle,\nonumber\\
	&\ket{8}^a_S=\frac{1}{\sqrt{3}}\,(T_1-T_2)_{ijk}^a|q_i, q_j, q_k\rangle \equiv \frac{1}{\sqrt{3}}\,\left(|T_1^a\rangle-|T_2^a\rangle\right),
\end{align}
where $\underline{T}^l_{ij}$ is given by Eq.~\eqref{Tabg}. 
Concerning the normalization of $T_i\,\left(i=1, 2, 3\right)$, we have
\begin{equation}
	(T_m)_{ijk}^{a\,*}(T_n)_{ijk}^b=\frac{1}{2}\left(3\delta_{mn}-1\right)\delta^{ab}\,.
\end{equation}
Each of these three projectors generates a color octet, but they are not independent, as shown through the identity
\begin{equation}
	T_1^a+T_2^a+T_3^a=\epsilon_{ljk}T_{li}^a+\epsilon_{ilk}T_{lj}^a+\epsilon_{ijl}T_{lk}^a=\epsilon_{ijk}\mathrm{Tr}\left[T^a\right]=0\,.
	\label{Tidentity}
\end{equation}
This is why there can only be two independent octets, and $\ket{8}_{A/S}$ are two possible orthonormal choices.

In the color singlet sector, the allowed configurations are the same as for light hadrons. The color projection is already fully antisymmetric, so the spin-isospin combination needs to be fully symmetric. 
This can either be achieved by having both spin and isospin parts be symmetric on their own, or by combining their mixed symmetry configurations. 
Note that the mixed symmetry spin or isospin configurations by themselves vanish under symmetrization of all three particle entries, which is why there is no $I=3/2$ and $k=1/2$ combination or viceversa in the color singlet sector; 
it is only the combination of spin and isospin that allows a full symmetrization also in the mixed symmetry case.

In the case where both spin and isospin are symmetric (i.e. $I=k=3/2$), the relevant combinations can be obtained by taking the tensor product of the respective spin or isospin configurations from Eq.~\eqref{3spins}. 
In the case of isospin, $\ket{\uparrow}$ should be understood as $\ket{u}$ and $\ket{\downarrow}$ as $\ket{d}$. 
By tensor product, we mean the combination of spin, isospin and color indices in the way of the following example:
\begin{equation}
	\ket{I=3/2,m_I=3/2;k=3/2,m_k=3/2}_S=\ket{uuu}\otimes\ket{\uparrow\uparrow\uparrow}\otimes\ket{1}=\frac{1}{\sqrt{6}}\epsilon_{ijk}\ket{u_i^\uparrow u_j^\uparrow u_k^\uparrow}\,.
\end{equation}

In case of the mixed symmetries, it is convenient to write the spin-isospin combinations in bilinear form:
\begin{align}
	\ket{1/2,+1/2;1/2,+1/2}_S&=\frac{1}{3\sqrt{2}}\begin{pmatrix} \ket{duu} & \ket{udu} & \ket{uud} \end{pmatrix}\begin{pmatrix} -2 & 1 & 1 \\ 1 & -2 & 1 \\ 1 & 1 & -2 \end{pmatrix}\begin{pmatrix} \ket{\downarrow\uparrow\uparrow} \\ \ket{\uparrow\downarrow\uparrow} \\ \ket{\uparrow\uparrow\downarrow} \end{pmatrix}\otimes\ket{1}\,,\notag\\
	\ket{1/2,+1/2;1/2,-1/2}_S&=\frac{-1}{3\sqrt{2}}\begin{pmatrix} \ket{duu} & \ket{udu} & \ket{uud} \end{pmatrix}\begin{pmatrix} -2 & 1 & 1 \\ 1 & -2 & 1 \\ 1 & 1 & -2 \end{pmatrix}\begin{pmatrix} \ket{\uparrow\downarrow\downarrow} \\ \ket{\downarrow\uparrow\downarrow} \\ \ket{\downarrow\downarrow\uparrow} \end{pmatrix}\otimes\ket{1}\,,\notag\\
	\ket{1/2,-1/2;1/2,+1/2}_S&=\frac{-1}{3\sqrt{2}}\begin{pmatrix} \ket{udd} & \ket{dud} & \ket{ddu} \end{pmatrix}\begin{pmatrix} -2 & 1 & 1 \\ 1 & -2 & 1 \\ 1 & 1 & -2 \end{pmatrix}\begin{pmatrix} \ket{\downarrow\uparrow\uparrow} \\ \ket{\uparrow\downarrow\uparrow} \\ \ket{\uparrow\uparrow\downarrow} \end{pmatrix}\otimes\ket{1}\,,\notag\\
	\ket{1/2,-1/2;1/2,-1/2}_S&=\frac{1}{3\sqrt{2}}\begin{pmatrix} \ket{udd} & \ket{dud} & \ket{ddu} \end{pmatrix}\begin{pmatrix} -2 & 1 & 1 \\ 1 & -2 & 1 \\ 1 & 1 & -2 \end{pmatrix}\begin{pmatrix} \ket{\uparrow\downarrow\downarrow} \\ \ket{\downarrow\uparrow\downarrow} \\ \ket{\downarrow\downarrow\uparrow} \end{pmatrix}\otimes\ket{1}\,.
\end{align}
It is straightforward to see that their combinations are still (iso)spin doublet configurations, since each row and column of the bilinear matrix corresponds to an (iso)spin doublet. 
Particle exchange corresponds to exchanging two rows and two columns with the same indices, under which these bilinear matrices are clearly invariant.

In the color octet sector, either spin or isospin may be fully symmetric, meaning that the other two mixed symmetries of (iso)spin and color have to be in an antisymmetric combination, or all three may be mixed. 
The former case can be obtained by combining the partially symmetric (iso)spin $1/2$ with the partially antisymmetric color octet or vice versa, and then performing all particle exchanges, both of which lead to the same result:
\begin{align}
	\ket{3/2,+3/2;1/2,+1/2}_O&=\frac{-1}{3\sqrt{6}}\ket{uuu}\otimes\begin{pmatrix} \ket{\downarrow\uparrow\uparrow} & \ket{\uparrow\downarrow\uparrow} & \ket{\uparrow\uparrow\downarrow} \end{pmatrix}\begin{pmatrix} -2 & 1 & 1 \\ 1 & -2 & 1 \\ 1 & 1 & -2 \end{pmatrix}\begin{pmatrix} \ket{T_1} \\ \ket{T_2} \\ \ket{T_3} \end{pmatrix}\,,\notag\\
	\ket{3/2,+3/2;1/2,-1/2}_O&=\frac{1}{3\sqrt{6}}\ket{uuu}\otimes\begin{pmatrix} \ket{\uparrow\downarrow\downarrow} & \ket{\downarrow\uparrow\downarrow} & \ket{\downarrow\downarrow\uparrow} \end{pmatrix}\begin{pmatrix} -2 & 1 & 1 \\ 1 & -2 & 1 \\ 1 & 1 & -2 \end{pmatrix}\begin{pmatrix} \ket{T_1} \\ \ket{T_2} \\ \ket{T_3} \end{pmatrix}\,,
\end{align}
where for the other isospin projections one just has to replace $\ket{uuu}$ by the respective expression from Eq.~\eqref{3spins}, and the expressions for $I=1/2$ and $k=3/2$ are completely analogous. 
Exchanging two particles $m$ and $n$ involves swapping the rows $m$ 
and $n$ for spin and swapping the columns $m$ and $n$ for color.
From the  projectors, we get $\ket{T_m} \leftrightarrow -\ket{T_n}$  and
$\ket{T_k} \to -\ket{T_k}$ for $k \neq m$, $n$ and $m\neq n$.
Thus, we have an additional minus sign from the color projectors,
which explains why the bilinear matrices remain symmetric under the
exchange of rows and columns, 
while the states are antisymmetric under particle exchange.

The remaining color octet configurations come from both spin and isospin doublet combinations, which means that all symmetries are mixed. With the use of Eq.~\eqref{Tidentity}, we can put them in the convenient form
\begin{align}
	\ket{1/2,+1/2;1/2,+1/2}_O&=\frac{1}{3\sqrt{2}}\begin{pmatrix} \ket{duu} & \ket{udu} & \ket{uud} \end{pmatrix}\begin{pmatrix} \ket{T_1} & \ket{T_3} & \ket{T_2} \\ \ket{T_3} & \ket{T_2} & \ket{T_1} \\ \ket{T_2} & \ket{T_1} & \ket{T_3} \end{pmatrix}\begin{pmatrix} \ket{\downarrow\uparrow\uparrow} \\ \ket{\uparrow\downarrow\uparrow} \\ \ket{\uparrow\uparrow\downarrow} \end{pmatrix}\,,\notag\\
	\ket{1/2,+1/2;1/2,-1/2}_O&=\frac{-1}{3\sqrt{2}}\begin{pmatrix} \ket{duu} & \ket{udu} & \ket{uud} \end{pmatrix}\begin{pmatrix} \ket{T_1} & \ket{T_3} & \ket{T_2} \\ \ket{T_3} & \ket{T_2} & \ket{T_1} \\ \ket{T_2} & \ket{T_1} & \ket{T_3} \end{pmatrix}\begin{pmatrix} \ket{\uparrow\downarrow\downarrow} \\ \ket{\downarrow\uparrow\downarrow} \\ \ket{\downarrow\downarrow\uparrow} \end{pmatrix}\,,\notag\\
	\ket{1/2,-1/2;1/2,+1/2}_O&=\frac{-1}{3\sqrt{2}}\begin{pmatrix} \ket{udd} & \ket{dud} & \ket{ddu} \end{pmatrix}\begin{pmatrix} \ket{T_1} & \ket{T_3} & \ket{T_2} \\ \ket{T_3} & \ket{T_2} & \ket{T_1} \\ \ket{T_2} & \ket{T_1} & \ket{T_3} \end{pmatrix}\begin{pmatrix} \ket{\downarrow\uparrow\uparrow} \\ \ket{\uparrow\downarrow\uparrow} \\ \ket{\uparrow\uparrow\downarrow} \end{pmatrix}\,,\notag\\
	\ket{1/2,-1/2;1/2,-1/2}_O&=\frac{1}{3\sqrt{2}}\begin{pmatrix} \ket{udd} & \ket{dud} & \ket{ddu} \end{pmatrix}\begin{pmatrix} \ket{T_1} & \ket{T_3} & \ket{T_2} \\ \ket{T_3} & \ket{T_2} & \ket{T_1} \\ \ket{T_2} & \ket{T_1} & \ket{T_3} \end{pmatrix}\begin{pmatrix} \ket{\uparrow\downarrow\downarrow} \\ \ket{\downarrow\uparrow\downarrow} \\ \ket{\downarrow\downarrow\uparrow} \end{pmatrix}\,.
\end{align}
Also here we see that replacing two rows and columns and also the color projectors with the same indices leaves the whole expression invariant, and the overall minus sign from the particle exchange in the color projectors ensures that these are indeed fully antisymmetric spin-isospin-color configurations. 
It is slightly harder to see in this form that these are indeed (iso)spin doublet configurations, but replacing any of the color projectors through the other two (e.g., $T_3=-T_1-T_2$) makes it clear that the coefficient of each remaining projector corresponds to an antisymmetric (iso)spin doublet configuration in each row or column.

Finally, we also give here the color decuplet configurations for the sake of completeness.
These cannot be directly coupled to the $Q\bar{Q}$ and form a color neutral state, but with the help of additional gluon fields it would be possible, as the combination of octet and decuplet contains another octet, although it is questionable whether such a highly excited configuration may form a bound state at all.

Since the decuplet color configuration is fully symmetric, the combination of spin and isospin needs to be fully antisymmetric. 
This can only be achieved by combining two mixed symmetries, as only spin or isospin alone cannot be fully antisymmetric. 
A straightforward way to achieve this is by combining the partially symmetric spin configuration with the partially antisymmetric isospin configuration and consequently performing all particle exchanges, which leads to the following expressions:
\begin{align}
	\ket{1/2,+1/2;1/2,+1/2}_D&=\frac{1}{\sqrt{6}}\begin{pmatrix} \ket{duu} & \ket{udu} & \ket{uud} \end{pmatrix}\begin{pmatrix} 0 & 1 & -1 \\ -1 & 0 & 1 \\ 1 & -1 & 0 \end{pmatrix}\begin{pmatrix} \ket{\downarrow\uparrow\uparrow} \\ \ket{\uparrow\downarrow\uparrow} \\ \ket{\uparrow\uparrow\downarrow} \end{pmatrix}\otimes\ket{10}\,,\notag\\
	\ket{1/2,+1/2;1/2,-1/2}_D&=\frac{-1}{\sqrt{6}}\begin{pmatrix} \ket{duu} & \ket{udu} & \ket{uud} \end{pmatrix}\begin{pmatrix} 0 & 1 & -1 \\ -1 & 0 & 1 \\ 1 & -1 & 0 \end{pmatrix}\begin{pmatrix} \ket{\uparrow\downarrow\downarrow} \\ \ket{\downarrow\uparrow\downarrow} \\ \ket{\downarrow\downarrow\uparrow} \end{pmatrix}\otimes\ket{10}\,,\notag\\
	\ket{1/2,-1/2;1/2,+1/2}_D&=\frac{-1}{\sqrt{6}}\begin{pmatrix} \ket{udd} & \ket{dud} & \ket{ddu} \end{pmatrix}\begin{pmatrix} 0 & 1 & -1 \\ -1 & 0 & 1 \\ 1 & -1 & 0 \end{pmatrix}\begin{pmatrix} \ket{\downarrow\uparrow\uparrow} \\ \ket{\uparrow\downarrow\uparrow} \\ \ket{\uparrow\uparrow\downarrow} \end{pmatrix}\otimes\ket{10}\,,\notag\\
	\ket{1/2,-1/2;1/2,-1/2}_D&=\frac{1}{\sqrt{6}}\begin{pmatrix} \ket{udd} & \ket{dud} & \ket{ddu} \end{pmatrix}\begin{pmatrix} 0 & 1 & -1 \\ -1 & 0 & 1 \\ 1 & -1 & 0 \end{pmatrix}\begin{pmatrix} \ket{\uparrow\downarrow\downarrow} \\ \ket{\downarrow\uparrow\downarrow} \\ \ket{\downarrow\downarrow\uparrow} \end{pmatrix}\otimes\ket{10}\,.
\end{align}
It is immediately apparent that this combination is antisymmetric under exchange of two particles (i.e., exchanging two rows and columns of the same indices), and each row or column does indeed correspond to an (iso)spin doublet configuration.

As a consistency check, we compare the total number of spin-isospin-color configurations with the expected $\displaystyle\binom{12}{3}=220$. 
In the color singlet sector, we had the (iso)spin doublet-doublet and quartet-quartet combinations, which give $1\times\left(2^2+4^2\right)=20$ components. 
In the color octet sector, either spin or isospin could be a quartet with the other being a doublet, or both could be doublets, giving $8\times\left(2\times4+4\times2+2^2\right)=160$ components. And in the color decuplet sector, both spin and isospin are doublets, from which we obtain the last $10\times2^2=40$ components. 
The sum over these does indeed give $220$, meaning that we have exhausted all possible spin-isospin-color combinations allowed by the Pauli principle.

\subsection{\texorpdfstring{$QQqq\bar{q}$ pentaquarks}{QQqqqbar pentaquarks}}
The (iso)spin combinations given in Eq.~\eqref{3spins} are already constructed with a unique symmetry in the first two entries, so we associate them with the quark indices. 
The light quarks and antiquark may form a color triplet, an antisextet, or a $15$-plet, where one triplet and the antisextet are antisymmetric in the $qq$ indices, and the other triplet and the $15$-plet are symmetric. Explicitly, we have
\begin{equation}
	\ket{3}_{S/A}^{\ell}=\frac{1}{\sqrt{6\pm2}}\left(\delta_{{\ell}j}\delta_{kl}\pm\delta_{{\ell} k}\delta_{jl}\right)|q_j, q_k,\bar{q}_l\rangle\,,\qquad\mathrm{and}\qquad\ket{\bar{6}}^{\sigma}=\frac{1}{\sqrt{2}}\underline{\Sigma}_{il}^{\sigma}\,\epsilon_{ijk}|q_j, q_k, \bar{q}_l\rangle\,,
\end{equation}
where $\underline{\Sigma}^{\sigma}$ is the color sextet tensor in Eq.~\eqref{eq:csextet}, and the $15$-plet requires a new projector which we do not specify here further. The triplets and the antisextet can couple to the antitriplet and the sextet from the $QQ$ respectively to form a color neutral state, while the $15$-plet can only contribute if additional gluons are included.

This means that in the color triplet sector any spin-isospin combination is allowed, including both symmetric and antisymmetric doublets, since any combination will be either symmetric or antisymmetric in the quark indices and can thus be combined with the antisymmetric or symmetric color triplet respectively to obtain an overall antisymmetric configuration. 
For the color antisextet, the spin-isospin combination needs to be symmetric, which can be a quartet-quartet combination, a doublet-quartet or quartet-doublet, where the doublet is symmetric, or a doublet-doublet, where the doublets are either both symmetric or antisymmetric. 
For the $15$-plet, the spin-isospin combination needs to be antisymmetric, which can be a doublet-quartet or quartet-doublet combination, with the doublet being antisymmetric, or a doublet-doublet, where one is symmetric and the other is antisymmetric. Counting all components, we have $3\times2^3\times2^3=192$ in the color triplet sector, $6\times(2^2+2^2+2\times4+4\times2+4^2)=240$ in the antisextet sector, and $15\times(2^2+2^2+2\times4+4\times2)=360$ in the $15$-plet sector, equaling the expected $\displaystyle\binom{12}{2}\times12=792$ components in total.

\section{Sch\"odinger equations for several cases}
\label{app:Schreodinger}
In this appendix, we explicitly write down the radial Schr\"odinger equations for the lowest hybrid, tetraquark and pentaquark states based on the Sec.~\ref{sec_mixing}.

\subsection{Hybrids ($Q\bar{Q}g$)}
The lowest hybrids are bound states in  the BO-potentials (static potentials) $\Pi_u$ and $\Sigma_u^-$ corresponding to the gluelump quantum numbers $k^{PC} = 1^{+-}$. 
Using Eq.~\eqref{eq:Sch2} and the expression of the mixing matrix in Eq.~\eqref{eq:M1}, the coupled radial Schr\"odinger equation for the parity state $\epsilon=\sigma_P$ is given by
\begin{align}
	&\Bigg[-\frac{1}{m_Qr^2}\,\partial_rr^2\partial_r+\frac{1}{m_Qr^2}\begin{pmatrix} l(l+1)+2 & -2\sqrt{l(l+1)} \\ -2\sqrt{l(l+1)} & l(l+1) \end{pmatrix} \nonumber\\ &\hspace{4.5 cm}+\begin{pmatrix} V_{\Sigma_{u}^{-}} & 0 \\ 0 & V_{\Pi_{u}} \end{pmatrix}\Bigg]\hspace{-4pt}\begin{pmatrix} \psi_{\Sigma, \sigma_P}^{(N)} \\ \psi_{\Pi, \sigma_P}^{(N)}\end{pmatrix}=\mathcal{E}_N\begin{pmatrix} \psi_{\Sigma, \sigma_P}^{(N)} \\ \psi_{\Pi, \sigma_P}^{(N)}\end{pmatrix}\,,
	\label{eq:diffeq1}
\end{align}
and the radial Schr\"odinger equation for the opposite parity state $\epsilon = -\sigma_P$ is given by 
\begin{align}
	\left[-\frac{1}{m_Qr^2}\,\partial_r\,r^2\,\partial_r+\frac{l(l+1)}{m_Qr^2}+V_{\Pi_{u}}\right]\psi_{\Pi, -\sigma_P}^{(N)}=\mathcal{E}_N\,\psi_{\Pi, -\sigma_P}^{(N)}\,,
	\label{eq:diffeq2}
\end{align}
where $V_{\Pi_u}(r)$ and $V_{\Sigma_u^-}(r)$ are the static potentials, $\psi_{\Sigma, \sigma_P}^{(N)}(r)$, $\psi_{\Pi, \sigma_P}^{(N)}(r)$, and $\psi_{\Pi, -\sigma_P}^{(N)}(r)$ are the radial wavefunctions labeled by the set of quantum numbers $(N)$. 
These coupled Schr\"odinger equations were first derived in Ref.~\cite{Berwein:2015vca}.

\subsection{Tetraquarks ($Q\bar{Q}q\bar{q}$ or $QQ\bar{q}\bar{q}$)}
The lowest quarkonium tetraquarks $Q\bar{Q}q\bar{q}$ are bound states in the BO-potentials (static potentials) $\Sigma_g^{+\prime}$ and $\Pi_g$ corresponding to the adjoint meson quantum numbers $k^{PC} = 1^{--}$, or $\Sigma_u^-$ corresponding to adjoint meson quantum numbers $k^{PC} = 0^{-+}$ (see Table~\ref{tab:QQbarqqbar}). 
As discussed in Sec.~\ref{subsec:potential}, the ordering of the adjoint meson masses for $0^{-+}$ and $1^{--}$ is not well established by lattice QCD, so we write down the Schr\"odinger equations for both cases.

For $Q\bar{Q}q\bar{q}$ tetraquarks corresponding to the adjoint meson $1^{--}$, the coupled radial Schr\"odinger equation for the parity state $\epsilon=\sigma_P$, after using Eq.~\eqref{eq:Sch2} and the expression of the mixing matrix in Eq.~\eqref{eq:M1}, is given by
\begin{align}
	&\Bigg[-\frac{1}{m_Qr^2}\,\partial_rr^2\partial_r+\frac{1}{m_Qr^2}\begin{pmatrix} l(l+1)+2 & -2\sqrt{l(l+1)} \\ -2\sqrt{l(l+1)} & l(l+1) \end{pmatrix}\nonumber\\
	&\hspace{4.5 cm}+\begin{pmatrix} V_{\Sigma_{g}^{+\prime}}^{(I)} & 0 \\ 0 & V_{\Pi_{g}}^{(I)} \end{pmatrix}\Bigg]\hspace{-4pt}\begin{pmatrix} \psi_{\Sigma, \sigma_P}^{(N)} \\ \psi_{\Pi, \sigma_P}^{(N)}\end{pmatrix}=\mathcal{E}_N\begin{pmatrix} \psi_{\Sigma, \sigma_P}^{(N)} \\ \psi_{\Pi, \sigma_P}^{(N)}\end{pmatrix}\,,
	\label{eq:diffeq3}
\end{align}
and the radial Schr\"odinger equation for the opposite parity state $\epsilon = -\sigma_P$ is given by 
\begin{align}
	\left[-\frac{1}{m_Qr^2}\,\partial_r\,r^2\,\partial_r+\frac{l(l+1)}{m_Qr^2}+V_{\Pi_{g}}^{(I)}\right]\psi_{\Pi, -\sigma_P}^{(N)}=\mathcal{E}_N\,\psi_{\Pi, -\sigma_P}^{(N)}\,,
	\label{eq:diffeq4}
\end{align}
where $V_{\Pi_g}^{(I)}(r)$ and $V_{\Sigma_g^{+\prime}}^{(I)}(r)$ are the static potentials that could depend on the isospin $I$ and $\psi_{\Sigma, \sigma_P}^{(N)}(r)$, $\psi_{\Pi, \sigma_P}^{(N)}(r)$, and $\psi_{\Pi, -\sigma_P}^{(N)}(r)$ are the radial wavefunctions labeled by the set of quantum numbers $(N)$ that includes isospin.

For $Q\bar{Q}q\bar{q}$ tetraquarks corresponding to the adjoint meson $0^{-+}$, the radial Schr\"odinger equation is given by
\begin{align}
	\left[-\frac{1}{m_Qr^2}\,\partial_r\,r^2\,\partial_r+\frac{l(l+1)}{m_Qr^2}+V_{\Sigma_{u}^-}^{(I)}\right]\psi_{\Sigma_u^-}^{(N)}=\mathcal{E}_N\,\psi_{\Sigma_u^-}^{(N)}\,,
	\label{eq:diffeq5}
\end{align}

The lowest doubly heavy tetraquarks $QQ\bar{q}\bar{q}$ are bound states in the BO-potentials (static potentials) $\Sigma_g^+$ corresponding to the triplet or sextet meson quantum numbers $k^{PC} = 0^{+}$, 
or $\Sigma_g^{-}$ and $\Pi_g$ corresponding to the triplet or sextet meson quantum numbers  $k^{PC} = 1^{+}$ (see Table~\ref{QQqqnumbers}). 
The ordering of the triplet or sextet meson masses for $0^{+}$ and $1^{+}$ is not known from lattice QCD. 
For $QQ\bar{q}\bar{q}$ tetraquarks corresponding to the sextet or triplet meson $1^{+}$,  the coupled radial Schr\"odinger equation is given by Eqs.~\eqref{eq:diffeq3} and \eqref{eq:diffeq4} with the static potentials $V_{\Pi_g}^{(I)}(r)$ and $V_{\Sigma_g^{-}}^{(I)}(r)$. 
For $QQ\bar{q}\bar{q}$ tetraquarks corresponding to the triplet or sextet meson $0^{+}$,  the radial Schr\"odinger equation is given by Eq.~\eqref{eq:diffeq5} with the static potential $V_{\Sigma_{g}^+}^{(I)}(r)$. 
We expect the $QQ\bar{q}\bar{q}$ tetraquarks corresponding to triplet configurations to be lower in energy than the ones corresponding to sextet (see Table~\ref{QQqqnumbers}) 
because of the attractive short distance behavior of the potential.

\subsection{Doubly heavy baryons ($QQq$)}
The lowest doubly heavy baryons $QQq$ are bound states in the BO-potentials (static potentials) $\left(1/2\right)_g$ corresponding to the light quark quantum numbers $k^{P} = (1/2)^{+}$. 
The next lowest BO-potentials are $\left(1/2\right)_u$ and $\left(3/2\right)_u$ corresponding to $k^P=(3/2)^-$, 
and $\left(1/2\right)_u^{\prime}$ corresponding to $k^P = (1/2)^-$, where the prime indicates an excited configuration  within the same representation. 
The radial Schr\"odinger equation for parity state $\epsilon = \sigma_P$ is given by
\begin{align}
	\left[-\frac{1}{m_Qr^2}\,\partial_r\,r^2\,\partial_r+\frac{(l-1/2)(l+1/2)}{m_Qr^2}+V_{k_\eta}\right]\psi_{ \sigma_P}^{(N)}=\mathcal{E}_N\,\psi_{ \sigma_P}^{(N)}\,,
	\label{eq:diffeq6}
\end{align}
and for parity state $\epsilon = -\sigma_p$ is given by
\begin{align}
	\left[-\frac{1}{m_Qr^2}\,\partial_r\,r^2\,\partial_r+\frac{(l+1/2)(l+3/2)}{m_Qr^2}+V_{k_\eta}\right]\psi_{-\sigma_P}^{(N)}=\mathcal{E}_N\,\psi_{-\sigma_P}^{(N)}\,,
	\label{eq:diffeq7}
\end{align}
where $V_{k_\eta} = V_{\left(1/2\right)_g}$ for $k^P = (1/2)^+$ and $V_{k_\eta} = V_{\left(1/2\right)_u^{\prime}}$ for $k^P = (1/2)^-$. 

For doubly heavy baryons, which are bound states in the BO-potentials $\left(1/2\right)_u$ and $\left(3/2\right)_u$ corresponding to $k^P=3/2^-$, we have two sets of coupled Schr\"odinger equations corresponding to the two possible parities. 
They are given by:
\begin{align}
	&\hspace{-2.0 cm}\Bigg[-\frac{1}{m_Q r^2}\,\partial_rr^2\partial_r+\frac{1}{m_Qr^2}\begin{pmatrix} l(l-1)+\frac{9}{4} & -\sqrt{3l(l+1)-\frac{9}{4}} \\ -\sqrt{3l(l+1)-\frac{9}{4}} & l(l+1)-\frac{3}{4} \end{pmatrix} \nonumber\\
	&\hspace{3.0 cm}+\begin{pmatrix} V_{(1/2)_u} & 0 \\ 0 & V_{(3/2)_u} \end{pmatrix}\Bigg]\hspace{-2pt}\begin{pmatrix} \psi_{1/2,\sigma_P}^{(N)} \\ \psi_{3/2,\sigma_P}^{(N)}\end{pmatrix}
	=\mathcal{E}_n\begin{pmatrix} \psi_{1/2, \sigma_P}^{(N)} \\ \psi_{3/2, \sigma_P}^{(N)}\end{pmatrix}\,,
	\label{eq:diffeq8}
\end{align}

\begin{align}
	& \hspace{-2.0 cm}\Bigg[-\frac{1}{m_Q r^2}\,\partial_rr^2\partial_r+\frac{1}{m_Qr^2}\begin{pmatrix} l(l+3)+\frac{17}{4} & -\sqrt{3l(l+1)-\frac{9}{4}} \\ -\sqrt{3l(l+1)-\frac{9}{4}} & l(l+1)-\frac{3}{4} \end{pmatrix}\nonumber\\
	&\hspace{2.0 cm}+\begin{pmatrix} V_{(1/2)_u} & 0 \\ 0 & V_{(3/2)_u}(r) \end{pmatrix}\Bigg]\hspace{-2pt}\begin{pmatrix} \psi_{1/2, -\sigma_P}^{(N)} \\ \psi_{3/2, -\sigma_P}^{(N)}\end{pmatrix}
	=\mathcal{E}_n\begin{pmatrix} \psi_{1/2, -\sigma_P}^{(N)} \\ \psi_{3/2, -\sigma_P}^{(N)}\end{pmatrix}\,,
	\label{eq:diffeq9}
\end{align}
where $\psi_{1/2, \sigma_P}^{(N)}(r)$, $\psi_{3/2, \sigma_P}^{(N)}(r)$, $\psi_{1/2, -\sigma_P}^{(N)}(r)$ and $\psi_{3/2, -\sigma_P}^{(N)}(r)$ are radial wavefunctions labeled by the set of quantum numbers $(N)$. 
These coupled Schr\"odinger equations were first derived in Ref.~\cite{Soto:2020pfa}.

\subsection{Pentaquarks ($Q\bar{Q}qqq$ or $QQqq\bar{q}$)}
The lowest quarkonium pentaquarks $Q\bar{Q}qqq$ are bound states in the BO-potentials (static potentials) $(1/2)_g$ corresponding to the adjoint baryon quantum numbers $k^{P} = (1/2)^{+}$, 
or $(1/2)_g^{\prime}$ and $(3/2)_g$ corresponding to the adjoint baryon quantum numbers $k^{P} = (3/2)^{+}$ (see Table~\ref{tab:QQbarpenta}). 
The ordering of the adjoint baryon masses for $(1/2)^{+}$ and $(3/2)^{+}$ is not known from lattice QCD. 
For $Q\bar{Q}qqq$ pentaquarks corresponding to the adjoint baryon $(1/2)^{+}$, the radial Schr\"odinger equation is given by Eqs.~\eqref{eq:diffeq6} or \eqref{eq:diffeq7} with the static potentials $V_{(1/2)_g}(r)$. 
For $Q\bar{Q}qqq$ pentaquarks corresponding to the adjoint baryon $(3/2)^{+}$, the radial Schr\"odinger equation is given by Eqs.~\eqref{eq:diffeq8} and  \eqref{eq:diffeq9} with the static potentials $V_{(1/2)^\prime_g}(r)$ and $V_{(3/2)_g}(r)$.

Following the above argument, we can similarly obtain the radial Schr\"odinger equations for the lowest doubly heavy pentaquark states, 
which are bound states in the BO-potentials (static potentials) $(1/2)_u$ corresponding to triplet or sextet baryon quantum numbers $k^{P} = (1/2)^{-}$, 
or $(1/2)_u^{\prime}$ and $(3/2)_u$ corresponding to triplet or sextet baryon quantum numbers $k^{P} = (3/2)^{-}$ (see Table~\ref{QQpenta}). 
The ordering of the triplet or sextet baryon masses for $(1/2)^{-}$ and $(3/2)^{-}$ is not known from lattice QCD. 
We expect the $QQqq\bar{q}$ pentaquarks corresponding to triplet configurations to be lower in energy than the ones corresponding to sextet (see Table~\ref{QQpenta}) 
because of the attractive short distance behavior of the potential.

\bibliographystyle{apsrev4-2}
\renewcommand*{\bibfont}{\footnotesize}
\bibliography{tetra}

\end{document}